\pdfoutput=1
\documentclass[12pt,a4paper,english,square,numbers,sort&compress,nodisplayskipstretch,BCOR=21mm]{scrbook}

\usepackage[T1]{fontenc}
\usepackage[utf8]{inputenc}
\usepackage{color}
\usepackage{babel}
\usepackage{array}
\usepackage{booktabs}
\usepackage{units}
\usepackage{multirow}
\usepackage{amsmath}
\usepackage{amssymb}
\usepackage{mathdots}
\usepackage{graphicx}
\usepackage{setspace}
\usepackage{esint}
\usepackage[numbers]{natbib}
\usepackage[unicode=true,
 bookmarks=true,bookmarksnumbered=false,bookmarksopen=true,bookmarksopenlevel=2,
 breaklinks=true,pdfborder={0 0 0},pdfborderstyle={},backref=false,colorlinks=true]
 {hyperref}
\hypersetup{pdftitle={Theoretical and Computational Aspects of New Lattice Fermion Formulations},
 pdfauthor={Christian Zielinski},
 pdfsubject={Lattice gauge theory},
 pdfkeywords={Lattice gauge theory, lattice fermions, staggered Wilson fermions, staggered overlap fermions, staggered domain wall fermions},
 linkcolor=BlueViolet,anchorcolor=BlueViolet,citecolor=BlueViolet,filecolor=BlueViolet,urlcolor=BlueViolet}

\makeatletter

\pdfpageheight\paperheight
\pdfpagewidth\paperwidth

\providecommand{\tabularnewline}{\\}

\numberwithin{equation}{section}
\numberwithin{figure}{section}
\numberwithin{table}{section}

\@ifundefined{date}{}{\date{}}
\usepackage[scaled]{helvet}
\usepackage[proportional,space=1.1,scaled=0.98]{erewhon}
\usepackage[type1,scaled=.95]{cabin}
\usepackage[utopia,vvarbb,bigdelims]{newtxmath}

\usepackage[tracking=false,kerning=true,protrusion=true,expansion,spacing=true,babel=true,final]{microtype}

\usepackage[usenames,dvipsnames]{xcolor}
\usepackage{bbm}        
\usepackage{geometry}   
\usepackage{bibentry}   
\usepackage{soul}       

\usepackage[above,below,section]{placeins}

\g@addto@macro\bfseries{\boldmath}

\allowdisplaybreaks
\usepackage{tocloft}
\usepackage{scrhack}

\usepackage[hang]{footmisc}
\DeclareOldFontCommand{\rm}{\normalfont\rmfamily}{\mathrm}
\DeclareOldFontCommand{\sf}{\normalfont\sffamily}{\mathsf}
\DeclareOldFontCommand{\tt}{\normalfont\ttfamily}{\mathtt}
\DeclareOldFontCommand{\bf}{\normalfont\bfseries}{\mathbf}
\DeclareOldFontCommand{\it}{\normalfont\itshape}{\mathit}
\DeclareOldFontCommand{\sl}{\normalfont\slshape}{\@nomath\sl}
\DeclareOldFontCommand{\sc}{\normalfont\scshape}{\@nomath\sc}
\DeclareRobustCommand*\cal{\@fontswitch\relax\mathcal}
\DeclareRobustCommand*\mit{\@fontswitch\relax\mathnormal}

\colorlet{chapter}{black!75}
\setkomafont{chapter}{\LARGE\color{chapter}\normalfont\bfseries\scshape\selectfont}

\renewcommand*{\chapterformat}{%
\begingroup%
\setlength{\unitlength}{1mm}%
\begin{picture}(5,10)(0,5)%
\setlength{\fboxsep}{0pt}%
\put(5,15){\line(1,0){\dimexpr
\textwidth-5\unitlength\relax\@gobble}}%
\put(0,0){\makebox(5,20)[r]{%
\fontsize{28\unitlength}{28\unitlength}\selectfont\thechapter
\kern-.04em
}}%
\put(5,15){\makebox(\dimexpr
\textwidth-20\unitlength\relax\@gobble,\ht\strutbox\@gobble)[l]{%
\ \normalsize\color{black}\chapapp~\thechapter\autodot
}}%
\end{picture} %
\endgroup
}

\recalctypearea

\@ifundefined{showcaptionsetup}{}{%
 \PassOptionsToPackage{caption=false}{subfig}}
\usepackage{subfig}
\AtBeginDocument{

}

\makeatother

\begin{document}
\begin{center}
\thispagestyle{empty}
\pagenumbering{roman}
\par\end{center}

\begin{center}
\includegraphics[width=0.5\columnwidth]{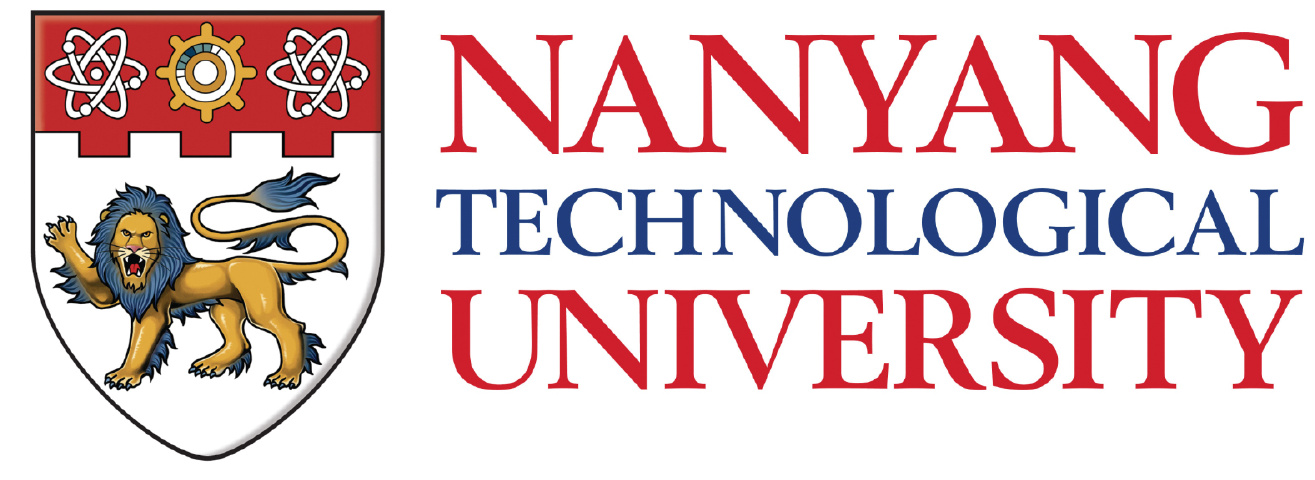}
\par\end{center}

\begin{singlespace}
\begin{center}
\smallskip{}
\par\end{center}
\end{singlespace}

\begin{onehalfspace}
\begin{center}
\begingroup
\fontfamily{phv}\selectfont

\begin{doublespace}
\begin{center}
\sodef\so{}{.05em}{.7em}{0em}
\so{\textbf{\Large{}Theoretical and Computational Aspects}}\\
\so{\textbf{\Large{}of New Lattice Fermion Formulations}}
\end{center}
\end{doublespace}

\begin{singlespace}
\begin{center}
\smallskip{}
\end{center}

\begin{center}
\textbf{\large{}Christian Zielinski}
\end{center}
\end{singlespace}

\begin{center}
\vfill{}
\end{center}

\begin{singlespace}
\begin{center}
Division of Mathematical Sciences\\
School of Physical and Mathematical Sciences\\
Nanyang Technological University
\end{center}

\begin{center}
\smallskip{}
\end{center}

\begin{center}
\textbf{\large{}2016}
\end{center}
\end{singlespace}

\endgroup
\par\end{center}
\end{onehalfspace}

\begin{center}
\textbf{\Large{}\clearpage{}}
\par\end{center}{\Large \par}

\begin{singlespace}
\begin{center}
\thispagestyle{empty}\textbf{\Large{}\clearpage{}}
\par\end{center}{\Large \par}

\begin{center}
\thispagestyle{empty}
\par\end{center}

\begin{center}
\bigskip{}
\par\end{center}
\end{singlespace}

\begin{doublespace}
\begin{center}
\textbf{\textsc{\Large{}Theoretical and Computational Aspects}}\\
\textbf{\textsc{\Large{}of New Lattice Fermion Formulations}}
\par\end{center}{\Large \par}
\end{doublespace}

\begin{singlespace}
\begin{center}
\smallskip{}
\par\end{center}

\begin{center}
\textsc{\Large{}Christian Zielinski}
\par\end{center}{\Large \par}
\end{singlespace}

\vfill{}

\begin{singlespace}
\begin{center}
\textbf{Division of Mathematical Sciences}\\
\textbf{School of Physical and Mathematical Sciences}\\
\textbf{Nanyang Technological University}
\par\end{center}

\begin{center}
\smallskip{}
\par\end{center}

\begin{center}
A thesis submitted to the Nanyang Technological University\\
in partial fulfillment of the requirement for the degree of\\
Doctor of Philosophy in the Mathematical Sciences
\par\end{center}

\begin{center}
\smallskip{}
\par\end{center}

\begin{center}
\textbf{\Large{}2016}
\par\end{center}{\Large \par}

\begin{center}
\textbf{\Large{}\clearpage{}}
\par\end{center}{\Large \par}
\end{singlespace}

\begin{onehalfspace}
\begin{center}
\thispagestyle{empty}\textbf{\Large{}\clearpage{}}
\par\end{center}{\Large \par}
\end{onehalfspace}

\phantomsection
\addcontentsline{toc}{chapter}{Abstract}
\begin{singlespace}
\noindent \begin{center}
\textsc{\large{}Abstract}
\par\end{center}{\large \par}
\end{singlespace}

\begin{singlespace}
In this work we investigate theoretical and computational aspects
of novel lattice fermion formulations for the simulation of lattice
gauge theories. The lattice approach to quantum gauge theories is
an important tool for studying quantum chromodynamics, where it is
the only known framework for calculating physical observables from
first principles. In our investigations we focus on staggered Wilson
fermions and the related staggered domain wall and staggered overlap
formulations. Originally proposed by Adams, these new fermion discretizations
bear the potential to reduce the computational costs of state-of-the-art
Monte Carlo simulations. Staggered Wilson fermions combine aspects
of both staggered and Wilson fermions while having a reduced number
of fermion doublers compared to usual staggered fermions. Moreover,
they can be used as a kernel operator for the domain wall fermion
construction with potentially significantly improved chiral properties
and for the overlap operator with its exact chiral symmetry. This
allows the implementation of chirality on the lattice in a controlled
manner at potentially significantly reduced costs. The practical potential
and limitations of these new lattice fermions are also critically
discussed.
\end{singlespace}

\cleardoublepage{}

\phantomsection
\addcontentsline{toc}{chapter}{Acknowledgements}
\begin{singlespace}
\noindent \begin{center}
\textsc{\large{}Acknowledgements}
\par\end{center}{\large \par}
\end{singlespace}

\begin{singlespace}
Over the course of the four years of my Ph.D.\ I had the pleasure
to work with many brilliant people in my field. My special thanks
go to Dr.\ David H.\ Adams for guiding me in my first three years,
for plenty of insightful discussions, for all the constructive feedback
and his confidence in my abilities. I also want to thank our collaborator
Asst.\ Prof.\ Daniel Nogradi from the Eötvös Loránd University for
patiently answering many questions by email. My thanks also go to
Assoc.\ Prof.\ Wang Li-Lian for his supervision in my final year.
Both in private discussions and in his excellent lectures I learned
a lot about the numerical foundations for my research.

At the Nanyang Technological University I want to thank my lecturers,
who all contributed in widening my horizons. Moreover, I had the pleasure
working together with many great colleagues and having plenty of interesting
discussions with friends. My special thanks go here to Dr.\ Reetabrata
Har, Jia Yiyang, Sai Swaroop Sunku, Andrii Petrashyk, Christian Engel,
Ido David Polak, Nguyen Thi Phuc Tan, Liu Zheng, Chen Penghua and
Lim Kim Song. During my time at the Nanyang Technological University
I got to know many great people, for whom I am very thankful.

I spent the final year of my Ph.D.\ in the Theoretical Particle Physics
Group at the University of Wuppertal (Bergische Universität Wuppertal).
Here my special thanks go to PD Dr.\ Christian Hoelbling and Prof.\ Zoltán
Fodor, who made my stay in Wuppertal possible. I greatly enjoyed working
together with Christian Hoelbling, as not only did I learn a lot in
our discussions, but he was also always encouraging and highly motivated.
My thanks also go to PD Dr.\ Stephan Dürr, who made a lot of great
suggestions and gave helpful feedback on our projects and manuscripts,
and to Prof.\ Ting-Wai Chiu at the National Taiwan University for
answering many questions by email. I was lucky to share my office
with two great colleagues, Dr.\ Attila Pásztor and Dr.\ Thomas Rae.
My stay in Wuppertal was very enjoyable and productive and my thanks
go also to the rest of the group.

I want to thank PD Dr.\ Christian Hoelbling, Assoc.\ Prof.\ Wang
Li-Lian, Jia Yiyang, Christian Engel and Pooi Khay Chong for providing
feedback on various parts of a draft of this thesis. I also thank
the three anonymous examiners of this thesis for their constructive
criticisms and valuable comments. Furthermore, I want to acknowledge
financial support from the Singapore International Graduate Award
(SINGA) and Nanyang Technological University.

Finally, I want to express my utmost gratitude to my family and my
beloved wife for their unconditional support over all these years.
\end{singlespace}

\cleardoublepage{}

\phantomsection
\addcontentsline{toc}{chapter}{List of publications}
\noindent \begin{center}
\textsc{\large{}List of publications \label{page:list-of-publications}}
\par\end{center}{\large \par}

\noindent As part of this Ph.D.\ the following work was published:
\nobibliography* \vspace{-2\topsep}
\begin{enumerate}
\begin{singlespace}
\item \sloppy \bibentry{Hoelbling:2016dug}
\item \sloppy \bibentry{Hoelbling:2016qfv}
\item \sloppy \bibentry{Adams:2013lpa}
\item \sloppy \bibentry{Adams:2013tya}
\end{singlespace}
\end{enumerate}
\begin{singlespace}
\noindent \textbf{Note:} In lattice field theory authors are traditionally
ordered by name.
\end{singlespace}

\cleardoublepage{}

\phantomsection
\addcontentsline{toc}{chapter}{List of presentations}
\begin{singlespace}
\noindent \begin{center}
\textsc{\large{}List of presentations }\label{page:conferences-presented}
\par\end{center}
\end{singlespace}

\begin{singlespace}
\noindent Parts of this work have been presented by the author at
the following venues: \vspace{-2\topsep}
\end{singlespace}
\begin{description}
\begin{singlespace}
\item [{{\small{}2016}}] Workshop of the Collaborative Research Center
Transregio SFB/TR-55,\\
University of Wuppertal, Wuppertal, Germany
\item [{{\small{}2016}}] Spring Meeting of the German Physical Society,\\
University of Hamburg, Hamburg, Germany
\item [{{\small{}2015}}] Invited Talk at the Theoretical Particle Physics
Seminar,\\
University of Wuppertal, Wuppertal, Germany
\item [{{\small{}2015}}] 33\textsuperscript{rd} International Symposium
on Lattice Field Theory,\\
Kobe International Conference Center, Kobe, Japan
\item [{{\small{}2015}}] Conference on 60 Years of Yang-Mills Gauge Field
Theories,\\
Institute of Advanced Studies, Nanyang Technological University, Singapore
\item [{{\small{}2015}}] 9\textsuperscript{th} International Conference
on Computational Physics,\\
National University of Singapore, Singapore
\item [{{\small{}2014}}] 5\textsuperscript{th} Singapore Mathematics Symposium,\\
National University of Singapore, Singapore
\item [{{\small{}2014}}] 32\textsuperscript{nd} International Symposium
on Lattice Field Theory,\\
Columbia University, New York City, United States
\end{singlespace}
\end{description}
\begin{singlespace}
\cleardoublepage{}

\global\long\def\One{\mathbbm{1}}

\end{singlespace}

\global\long\def\ii{\textrm{i}}

\global\long\def\diag{\operatorname{diag}}

\global\long\def\Dslash{\slashed{D}}

\global\long\def\spec{\operatorname{spec}}

\global\long\def\sn{\operatorname{sn}}

\global\long\def\sign{\operatorname{sign}}

\global\long\def\erf{\operatorname{erf}}

\global\long\def\tr{\operatorname{tr}}

\global\long\def\Tr{\operatorname{Tr}}

\global\long\def\index{\operatorname{index}}

\global\long\def\myRe{\operatorname{Re}}

\global\long\def\myIm{\operatorname{Im}}

\global\long\def\Mod{\operatorname{mod}}

\global\long\def\Uone{\mathrm{U}\!\left(1\right)}

\global\long\def\Ufour{\mathrm{U}\!\left(4\right)}

\global\long\def\SUtwo{\mathrm{SU}\!\left(2\right)}

\global\long\def\SUthree{\mathrm{SU}\!\left(3\right)}

\global\long\def\SUfour{\mathrm{SU}\!\left(4\right)}

\global\long\def\SOfour{\mathrm{SO}\!\left(4\right)}

\global\long\def\SUn{\mathrm{SU}\!\left(N_{\mathsf{c}}\right)}

\begin{singlespace}
\tableofcontents{}
\end{singlespace}

\cleardoublepage{}\pagenumbering{arabic}

\chapter{Introduction \label{chap:Introduction}}

Gauge theories are some of the most fundamental building blocks in
modern physics. In a gauge theory, physical observables are invariant
under transformations of some fundamental fields. This property is
referred to as gauge symmetry. Currently all known fundamental forces
in nature can be described as classical or quantum gauge theories.
General relativity is invariant under arbitrary coordinate transformations,
quantum electrodynamics (QED) is an Abelian gauge theory, while quantum
chromodynamics (QCD) and the electroweak interaction are of the Yang-Mills
type.

The generally accepted theory of the strong nuclear force is quantum
chromodynamics, which describes the interactions between quarks and
gluons. The strong nuclear force binds quarks and gluons to form hadrons,
among them nuclear particles such as the proton and neutron. In the
case of quantum chromodynamics, we deal with a non-Abelian gauge theory
with gauge group $\SUthree$. The associated force carrier is the
so-called gluon, the associated charge is called color. Quantum chromodynamics
is the primary setting of this thesis and has some peculiar properties,
in particular confinement and asymptotic freedom.

Confinement refers to the property that we can only observe color-singlet
states and, thus, quarks and gluons cannot propagate freely. When
quarks are pulled apart, the energy of the gluonic field in between
them increases until a new quark pair is created. While confinement
is phenomenologically well-established, no rigor mathematical proof
of confinement is known. In fact, confinement would follow from proving
the existence of a positive mass gap in quantum Yang-Mills theory,
which is one of the famous Millennium Problems of the Clay Mathematics
Institute.

Asymptotic freedom, on the other hand, refers to the fact that strong
nuclear interactions become weak at high energy scales. In experiments,
such as heavy ion collisions, one can then find a new state of matter,
the quark-gluon plasma. Due to the extremely high energy densities
in the quark-gluon plasma, quarks become deconfined.

In general, the non-Abelian nature of quantum chromodynamics makes
its treatment with analytical methods difficult. While in certain
regimes quantum chromodynamics can be treated by perturbative methods
much in the same way quantum electrodynamics is treated, in the low-energy
regime of bound hadrons the coupling constant of the strong nuclear
interaction is of order $\mathcal{O}\left(1\right)$ and perturbation
theory is rendered inapplicable.

An alternative and non-perturbative approach is the framework of lattice
quantum chromodynamics. Originally proposed by Wilson in Refs.\ \citep{Wilson:1974sk,Wilson:1975id},
one discretizes the theory by replacing continuous space-time by a
hypercubic space-time grid, the so-called lattice. The resulting numerical
formulation can, then, be simulated on computers. As the resulting
lattices usually have a very large number of sites, in particular
in the case of four or more space-time dimensions, extracting accurate
predictions from the theory is computationally extremely challenging.
Enormous amounts of computational resources are, thus, required for
lattice field theoretical simulations and the interest in finding
more efficient and accurate lattice formulations is high. As a consequence
the subject has been and remains a very active topic of research.
The vast numbers of proposed lattice formulations suggest that no
``optimal formulation'' is known, but that some are more suitable
than others depending on the application.

As part of these ongoing research efforts, in this thesis we study
theoretical and computational aspects of staggered Wilson, staggered
overlap and staggered domain wall fermions as proposed by Adams. These
novel lattice fermion discretizations have the potential of reducing
computational costs in numerical simulations and allow the generalization
of the overlap and domain wall fermion formulation to the case of
a staggered kernel. While the construction of these lattice fermions
are very interesting from a theoretical point of view, we also evaluate
them under practical aspects and try to understand in what setting
their use can be advantageous.

Although the lattice approach is traditionally concerned with quantum
chromodynamics, it is also used to study a large spectrum of other
quantum field theories. In this thesis, for example, we consider both
four-dimensional quantum chromodynamics and the Schwinger model, i.e.\ two-dimen\-sional
quantum electrodynamics.

\section{Outline of this thesis}

This thesis is structured as follows. In Chapter\ \ref{chap:GaugeTheories},
we give a short review of Dirac fermions and gauge theories both in
the continuum and on the lattice. In Chapter\ \ref{chap:Staggered-Wilson-fermions},
we recapitulate the construction and properties of flavored mass terms
and staggered Wilson fermions. In Chapter\ \ref{chap:Computational-efficiency},
we discuss the computational efficiency of staggered Wilson fermions
and compare them to the one of Wilson fermions. In Chapter\ \ref{chap:Pseudoscalar-mesons},
we discuss the feasibility of spectroscopy calculations with staggered
Wilson fermions and explicitly compute the pseudoscalar meson spectrum.
In Chapter\ \ref{chap:Overlap-fermions}, we review the formulation
of usual and staggered overlap fermions on the lattice and analytically
evaluate the continuum limit of the index and the axial anomaly. In
Chapter\ \ref{chap:Eigenvalue-spectra}, we investigate the eigenvalue
spectra of both the Wilson and staggered Wilson kernel and relate
them to the computational efficiency of their corresponding overlap
operators. In Chapter\ \ref{chap:Staggered-domain-wall}, we discuss
the formulation of usual and staggered domain wall fermions in detail
and compare spectral properties and chiral symmetry violations in
the case of the Schwinger model. Finally, in Chapter\ \ref{chap:Conclusions}
we summarize the central results of this thesis and provide an outlook.

\section{Original contributions}

Apart from an extensive review of the lattice fermion formulations
considered in this thesis, there are various novel contributions as
listed in the following:
\begin{itemize}
\item The generalized flavored mass terms for arbitrary mass splittings
in arbitrary even dimensions in Sec.\ \ref{sec:SW-Generalized-mass-terms}
as proposed in Ref.~\citep{Hoelbling:2016qfv} (done in collaboration
with C.\ Hoelbling).
\item The quantitative analysis of the computational efficiency and memory
bandwidth requirements of the staggered Wilson kernel in Chapter\ \ref{chap:Computational-efficiency}
as discussed in Refs.\ \citep{Adams:2013tya,ZielinskiLat14,ZielinskiICCP9}
(done in collaboration with D.\ H.\ Adams, D.\ Nogradi and A.\ Petrashyk).
\item The adaption of staggered spectroscopy methods to the two-flavor case
and the computation of the pseudoscalar meson spectrum with staggered
Wilson fermions in Chapter\ \ref{chap:Pseudoscalar-mesons} as presented
in Refs.\ \citep{AdamsLat14,ZielinskiLat15} (done in collaboration
with D.\ H.\ Adams).
\item The analysis of the continuum limit of the index and axial anomaly
for the staggered overlap Dirac operator\footnote{Note here also R.\ Har's thesis in Ref.~\citep{HarThesis}, elaborating
on this derivation in more detail.} in Sec.\ \ref{sec:OV-Continuum-limit} as given in Ref.~\citep{Adams:2013lpa}
(done in collaboration with D.\ H.\ Adams, R.\ Har and Y.\ Jia).
\item The discussion of the usual Wilson and staggered Wilson eigenvalue
spectrum and its connection to the computational efficiency of the
overlap construction in Chapter\ \ref{chap:Eigenvalue-spectra} as
presented in Refs.\ \citep{Adams:2013tya,ZielinskiLat14} (done in
collaboration with D.\ H.\ Adams).
\item The study of spectral properties and chiral symmetry violations of
various known and new variants of usual and staggered domain wall
fermions in Chapter\ \ref{chap:Staggered-domain-wall} as discussed
in Refs.\ \citep{Hoelbling:2016qfv,ZielinskiDPG16,Hoelbling:2016dug}
(done in collaboration with C.\ Hoelbling).
\end{itemize}
Some results presented in this thesis have already been published
in the literature as listed on page\ \pageref{page:list-of-publications},
while others were presented at international conferences as listed
on page \pageref{page:conferences-presented}. Throughout the thesis
the author uses the inclusive ``we'', to include both the reader
and collaborators in the discussion.

\chapter{Gauge theories in a nutshell \label{chap:GaugeTheories}}

Quantum gauge theories play a central role in modern physics. Their
defining property is that the action is invariant under a continuous
group of local transformations. The standard model can be understood
as a gauge theory with non-Abelian gauge group $\SUthree\times\SUtwo\times\Uone$.
It includes quantum chromodynamics (QCD) with gauge group $\SUthree$
and the electroweak sector with gauge group $\SUtwo\times\Uone$.
The electroweak interaction describes a unified theory of the weak
and electromagnetic interactions.

Throughout this thesis we are dealing primarily with two specific
gauge theories. The first being quantum chromodynamics, which is a
Yang-Mills theory with non-Abelian symmetry group $\SUthree$. The
other one is quantum electrodynamics, which is an Abelian gauge theory
with gauge group $\Uone$.

In this chapter, we review the formulation of gauge theories in the
continuum in Sec.\ \ref{sec:GT-Continuum} and on the lattice in
Sec.\ \ref{sec:GT-Lattice}. We also briefly discuss the use of Monte
Carlo methods for the simulation of lattice gauge theories in Sec.\ \ref{sec:GT-Numerical-simulations}.
For notational ease, we only discuss the case of four space-time dimensions.
For a detailed discussion of the continuum case, we refer the reader
to the many textbooks in the field such as Refs.\ \citep{Weinberg:1995mt,Weinberg:1996kr,Peskin:1995ev}.
For the lattice formulation, see e.g.\ Refs.\ \citep{Gattringer:2010zz,DeGrand:2006zz,Montvay:1994cy,Rothe:1992nt}.

\section{Gauge theories in the continuum \label{sec:GT-Continuum}}

We begin by reviewing some aspects of gauge theories in continuous
Minkowski space-time. Throughout this thesis we are using natural
units, where 
\begin{equation}
\hbar=c=1
\end{equation}
with the reduced Planck constant $\hbar\equiv h/\left(2\pi\right)$
and the speed of light $c$. In this unit system, energy is the only
remaining independent dimension. For converting back to conventional
units, we note that $\hbar c=\unit[197.3]{MeV\!\cdot fm}$.

In our convention the Minkowski metric tensor reads
\begin{equation}
\eta_{\mu\nu}=\eta^{\mu\nu}=\left(\begin{array}{cccc}
1\\
 & -1\\
 &  & -1\\
 &  &  & -1
\end{array}\right)
\end{equation}
and we follow Einstein's sum convention by summing over repeated indices.
Greek indices run over $0$, $1$, $2$, $3$, while roman indices
are restricted to the spacial components $1$, $2$, $3$. Indices
of four-vectors can be lowered and raised using the metric tensor.
If $\boldsymbol{x}$ denotes the spatial part of a four-vector, we
have
\begin{equation}
x^{\mu}=\left(x^{0},\boldsymbol{x}\right)^{\intercal},\qquad x_{\mu}=\eta_{\mu\nu}x^{\nu}=\left(x^{0},-\boldsymbol{x}\right)^{\intercal}.
\end{equation}
Furthermore, we introduce the derivative operator as
\begin{equation}
\partial_{\mu}=\frac{\partial}{\partial x^{\mu}}.
\end{equation}
In the following, we discuss the basic framework for the quantum field
theories of interest in continuous Minkowski space-time. We begin
by discussing fermions in Subsec.\ \ref{subsec:GT-Fermions} and
gauge theories in Subsec.\ \ref{subsec:GT-Gauge-theories}, followed
by chiral symmetry in Subsec.\ \ref{subsec:GT-Chiral-symmetry} and
the path integral formalism in Subsec.\ \ref{subsec:GT-Path-integrals}.
In our notation and discussion we follow Ref.~\citep{Peskin:1995ev}.

\subsection{Fermions \label{subsec:GT-Fermions}}

In the discussion of fermions we restrict ourselves to the case of
spin-$\frac{1}{2}$ Dirac fermions, such as quarks and electrons.
They can be described by the Dirac equation, a relativistic first-order
partial differential equation first introduced in Ref.~\citep{Dirac:1928hu}.
In the free-field case it is of the form
\begin{equation}
\left(\ii\gamma^{\mu}\partial_{\mu}-m_{\mathsf{f}}\right)\Psi\left(x\right)=0,\label{eq:DiracEq}
\end{equation}
where $m_{\mathsf{f}}$ denotes the fermion mass and $\Psi$ is a
four-component spinor. As $\Psi$ is of fermionic nature, it is described
by anti-commuting Grassmann numbers \citep{berezin1966method}. The
$\gamma^{\mu}$ matrices are complex $4\times4$ matrices and satisfy
the Dirac algebra, given by the anticommutation relations
\begin{equation}
\left\{ \gamma^{\mu},\gamma^{\nu}\right\} \equiv\gamma^{\mu}\gamma^{\nu}+\gamma^{\nu}\gamma^{\mu}=2\eta^{\mu\nu}\One.\label{eq:DiracAlgebraMinkowski}
\end{equation}
An explicit representation of the $\gamma^{\mu}$ matrices is given
by the Weyl or chiral basis
\begin{equation}
\gamma^{0}=\left(\begin{array}{cc}
 & \One\\
\One
\end{array}\right),\qquad\gamma^{k}=\left(\begin{array}{cc}
 & \sigma_{k}\\
-\sigma_{k}
\end{array}\right),\qquad\gamma_{5}=\left(\begin{array}{cc}
-\One\\
 & \One
\end{array}\right),
\end{equation}
where $\One$ refers to the $2\times2$ identity matrix, the $\sigma_{k}$
refer to the Pauli matrices
\begin{equation}
\sigma_{1}=\left(\begin{array}{cc}
 & 1\\
1
\end{array}\right),\qquad\sigma_{2}=\left(\begin{array}{cc}
 & -\ii\\
\ii
\end{array}\right),\qquad\sigma_{3}=\left(\begin{array}{cc}
1\\
 & -1
\end{array}\right)
\end{equation}
and we introduced the chirality matrix $\gamma_{5}\equiv\ii\gamma_{0}\gamma_{1}\gamma_{2}\gamma_{3}$.
We note that $\gamma_{5}$ anticommutes with all the other $\gamma^{\mu}$
matrices.

The Dirac equation implies the Klein-Gordon equation
\begin{equation}
\left(\partial_{\mu}\partial^{\mu}+m_{\mathsf{f}}^{2}\right)\Psi=0.\label{eq:GordonKleinEq}
\end{equation}
This is a manifestly covariant second-order partial differential equation
and expresses the relativistic energy-momentum relation. Its solutions
describe spinless particles. The Dirac equation can be seen as a ``square
root'' of the Klein-Gordon equation, in the sense if Eq.~\eqref{eq:DiracEq}
is multiplied by its complex conjugate, one recovers Eq.~\eqref{eq:GordonKleinEq}.

Finally, the corresponding action of free Dirac fermions is given
by
\begin{equation}
S_{\mathsf{f}}=\intop\mathrm{d}^{4}x\,\overline{\Psi}\left(x\right)\left(\ii\gamma^{\mu}\partial_{\mu}-m_{\mathsf{f}}\right)\Psi\left(x\right),\label{eq:DiracActionMinkowski}
\end{equation}
where $\overline{\Psi}\equiv\Psi^{\dagger}\gamma_{0}$ denotes the
Dirac adjoint of $\Psi$. In the presence of a gauge field, the partial
derivatives $\partial_{\mu}$ are replaced by the covariant derivative
$D_{\mu}$ as discussed in the following subsection.

\subsection{Gauge theories \label{subsec:GT-Gauge-theories}}

We continue by briefly reviewing the formulation of the gauge sector.
We discuss both quantum electrodynamics as well as Yang-Mills theory.

\minisec{The action}

Quantum electrodynamics is a gauge theory with Abelian symmetry group
$\Uone$. The action is given by
\begin{equation}
S_{\mathsf{g}}=-\frac{1}{4}\intop\mathrm{d}^{4}x\,\left(F_{\mu\nu}\right)^{2},
\end{equation}
where $F_{\mu\nu}=\partial_{\mu}A_{\nu}-\partial_{\nu}A_{\mu}$ is
the field-strength tensor with vector potential $A_{\mu}$.

The general case of a compact, semi-simple Lie group is described
by a Yang-Mills theory \citep{Yang:1954ek}. Here, we restrict ourselves
to the case of the special unitary groups $\SUn$. The action reads
\begin{equation}
S_{\mathsf{g}}=-\frac{1}{4}\intop\mathrm{d}^{4}x\,\left(F_{\mu\nu}^{a}\right)^{2},
\end{equation}
where $F_{\mu\nu}^{a}=\partial_{\mu}A_{\nu}^{a}-\partial_{\nu}A_{\mu}^{a}+gf^{abc}A_{\mu}^{b}A_{\nu}^{c}$
is the field-strength tensor, the $A_{\mu}^{a}$ denote the gauge
fields with group index $a=1,\dots,N_{c}^{2}-1$ and $g$ is the coupling
constant.

\minisec{Coupling to matter}

We can couple the gauge field to matter by replacing the partial derivatives
in Eq.~\eqref{eq:DiracActionMinkowski} by covariant derivatives.
In the case of quantum electrodynamics, this replacement takes the
form
\begin{equation}
\partial_{\mu}\to D_{\mu}=\partial_{\mu}+\ii eA_{\mu},
\end{equation}
where $-e$ denotes the electrical charge of the electron. Note that
the field-strength tensor can be represented as $\left[D_{\mu},D_{\nu}\right]=\ii eF_{\mu\nu}$.
The action for quantum electrodynamics then takes the form
\begin{equation}
\begin{aligned}S_{\mathsf{QED}} & =\intop\mathrm{d}^{4}x\,\left[\mathcal{L}_{\mathsf{f}}+\mathcal{L}_{\mathsf{g}}\right],\\
\mathcal{L}_{\mathsf{f}} & =\overline{\Psi}\left(\ii\gamma^{\mu}D_{\mu}-m_{\mathsf{f}}\right)\Psi,\\
\mathcal{L}_{\mathsf{g}} & =-\frac{1}{4}\left(F_{\mu\nu}\right)^{2}.
\end{aligned}
\end{equation}
By construction, the action $S_{\mathsf{QED}}$ is invariant under
local $\Uone$ transformations, for which the infinitesimal form reads
\begin{equation}
\begin{aligned}\Psi\left(x\right) & \to\left(1+\ii\alpha\right)\Psi\left(x\right),\\
\overline{\Psi}\left(x\right) & \to\overline{\Psi}\left(x\right)\left(1-\ii\alpha\right),\\
A_{\mu}^{a}\left(x\right) & \to A_{\mu}^{a}\left(x\right)-\frac{1}{e}\partial_{\mu}\alpha
\end{aligned}
\label{eq:ContGaugeTransformU1}
\end{equation}
with a scalar function $\alpha\equiv\alpha\left(x\right)$.

For a Yang-Mills theory this generalizes to
\begin{equation}
\partial_{\mu}\to D_{\mu}=\partial_{\mu}-\ii gA_{\mu}^{a}t^{a},
\end{equation}
where $\left[D_{\mu},D_{\nu}\right]=-\ii gF_{\mu\nu}^{a}t^{a}$. The
$t^{a}$ denote the generators of the gauge group $\SUn$, which obey
commutation relations of the form $\left[t^{a},t^{b}\right]=\ii f^{abc}t^{c}$
with structure constants $f^{abc}$. The resulting action reads
\begin{equation}
\begin{aligned}S_{\mathsf{YM}} & =\intop\mathrm{d}^{4}x\,\left[\mathcal{L}_{\mathsf{f}}+\mathcal{L}_{\mathsf{g}}\right],\\
\mathcal{L}_{\mathsf{f}} & =\overline{\Psi}\left(\ii\gamma^{\mu}D_{\mu}-m_{\mathsf{f}}\right)\Psi,\\
\mathcal{L}_{\mathsf{g}} & =-\frac{1}{4}\left(F_{\mu\nu}^{a}\right)^{2}.
\end{aligned}
\end{equation}
The infinitesimal transformation law takes the form
\begin{align}
\Psi\left(x\right) & \to\left(1+\ii\alpha^{a}t^{a}\right)\Psi\left(x\right),\nonumber \\
\overline{\Psi}\left(x\right) & \to\overline{\Psi}\left(x\right)\left(1-\ii\alpha^{a}t^{a}\right),\label{eq:ContGaugeTransformYM}\\
A_{\mu}^{a}\left(x\right) & \to A_{\mu}^{a}\left(x\right)+\frac{1}{g}\partial_{\mu}\alpha^{a}+f^{abc}A_{\mu}^{b}\alpha^{c}\nonumber 
\end{align}
with scalar functions $\alpha^{a}\equiv\alpha^{a}\left(x\right)$.

\minisec{The case of quantum chromodynamics}

An important case throughout this thesis is quantum chromodynamics
and therefore in the following we shall introduce some relevant terminology.
Quantum chromodynamics is the gauge theory of strong interactions
with gauge group $\SUthree$. The fermions are referred to as quarks
and come in three colors. We say that the quarks are in the fundamental
representation of $\SUthree$, while the eight gluon fields $A_{\mu}^{a}$
($a=1,\dots,8$) transform under the adjoint representation of the
$\SUthree$ color group. Hadrons come as color-singlet states and
both quarks and gluons have never been observed as free particles.

\subsection{Chiral symmetry \label{subsec:GT-Chiral-symmetry}}

One of the central symmetries discussed in this thesis is chiral symmetry.
For the discussion of this symmetry we follow Ref.~\citep{Gattringer:2010zz}
and write the Lagrangian density
\begin{equation}
\mathcal{L}_{0}=\overline{\Psi}\,\ii\gamma^{\mu}D_{\mu}\,\Psi
\end{equation}
of a massless Dirac fermion. We observe the invariance of the Lagrangian
density under global chiral rotations of the form
\begin{equation}
\Psi\left(x\right)\to e^{\ii\alpha\gamma_{5}}\Psi\left(x\right),\qquad\overline{\Psi}\left(x\right)\to\overline{\Psi}\left(x\right)e^{\ii\alpha\gamma_{5}}
\end{equation}
and note that the presence of a mass term breaks this symmetry explicitly.
We can reformulate chiral symmetry also elegantly in the form of the
anticommu\-tation relation
\begin{equation}
\left\{ D,\gamma_{5}\right\} =0,
\end{equation}
where $D=\ii\gamma^{\mu}D_{\mu}$ is the massless Dirac operator.

When considering $N_{\mathsf{f}}$ massless fermion flavors, one finds
a larger group of chiral symmetry transformations of the form
\begin{align}
\Psi\left(x\right)\to e^{\ii\alpha\gamma_{5}\tau^{a}}\Psi\left(x\right), & \qquad\overline{\Psi}\left(x\right)\to\overline{\Psi}\left(x\right)e^{\ii\alpha\gamma_{5}\tau^{a}},\\
\Psi\left(x\right)\to e^{\ii\alpha\gamma_{5}\One}\Psi\left(x\right), & \qquad\overline{\Psi}\left(x\right)\to\overline{\Psi}\left(x\right)e^{\ii\alpha\gamma_{5}\One},
\end{align}
where the spinors now carry an implicit flavor index and the $\tau^{a}$
are the generators of the group $\mathrm{SU}\!\left(N_{\mathsf{f}}\right)$
with $a=1,\dots,N_{\mathsf{f}}^{2}-1$.

\subsection{Path integrals \label{subsec:GT-Path-integrals}}

A systematic approach to the study of quantum field theories is the
path integral formalism \citep{Feynman:1948ur,Feynman:1949zx}. It
is also the starting point for the lattice approach, giving rise to
a discrete formulation suitable for numerical simulations.

To illustrate the formalism, let us consider an action $S\left[\phi\right]=\intop\mathrm{d}^{4}x\,\mathcal{L}\left[\phi\right]$
of some fields $\phi_{i}$, which we collectively refer to as $\phi\equiv\left\{ \phi_{i}\right\} $,
and an observable $\mathcal{O}\left[\phi\right]$. If we ignore some
subtleties regarding the rigorous definition of the path integral
for the moment (cf.\ Ref.~\citep{Albeverio:1976je}), the time-ordered
vacuum expectation value can be written as
\begin{equation}
\left\langle \mathcal{O}\right\rangle =\frac{1}{Z}\intop\mathcal{D}\phi\,\mathcal{O}\left[\phi\right]\,e^{\ii S\left[\phi\right]},\qquad Z=\intop\mathcal{D}\phi\,e^{\ii S\left[\phi\right]}.
\end{equation}
The notation $\intop\mathcal{D}\phi\equiv\intop\prod_{i}\mathcal{D}\phi_{i}$
represents an infinite-dimensional integration over all possible field
configurations of the $\phi_{i}$ fields. The partition function is
denoted as $Z$ and ensures the normalization of the expectation value.

When used for the study of gauge theories, one has to integrate over
all gauge fields $A$ and fermion fields $\Psi$, $\overline{\Psi}$
and evaluate integrations of the form $\intop\mathcal{D}A\,\mathcal{D}\overline{\Psi}\,\mathcal{D}\Psi$.
Due to gauge invariance, expectation values of observables are ill-defined
due to over-counting of physical degrees of freedom. This can be overcome
by imposing a gauge fixing condition using the Faddeev-Popov trick
\citep{Faddeev:1967fc}. For more details of the path integral method,
we refer the reader to the widely available textbooks such as Refs.~\citep{Weinberg:1995mt,Weinberg:1996kr,Peskin:1995ev}.

\section{Lattice gauge theories \label{sec:GT-Lattice}}

One of the standard tools for studying quantum field theories is perturbation
theory. It has proven itself to be an extremely successful method,
in particular when applied to quantum electrodynamics. As one of the
major achievement of quantum physics, the theoretical prediction of
the electron's anomalous magnetic dipole moment agrees with experiment
to more than ten significant figures \citep{Aoyama:2012wj}. In quantum
electrodynamics, perturbative expansions take the form of a power
series in the fine-structure constant $\alpha\approx1/137$. As $\alpha\ll1$,
higher terms in the expansion are typically strongly suppressed and
one can derive accurate predictions by including only the first few
orders.

For quantum chromodynamics, the coupling constant $\alpha_{\mathsf{s}}$
at high energies or short distances is also small. In this setting
quantum chromodynamics can be studied by perturbative methods as well.
However, in the important low-energy regime of quantum chromodynamics
where bound hadron states are present, the coupling constant $\alpha_{\mathsf{s}}$
of the strong nuclear interaction is of order $\mathcal{O}\left(1\right)$
and perturbation theory is rendered inapplicable. Here lattice quantum
chromodynamics is the only known framework for calculating physical
observables, such as hadron masses, from first principles. After its
proposal by Wilson in Refs.~\citep{Wilson:1974sk,Wilson:1975id},
the lattice approach has become one of the standard tools for non-perturbative
studies of the strong nuclear interaction. In order to arrive at a
lattice formulation of a quantum field theory, one replaces continuous
Euclidean space-time by a discrete lattice. The discretization of
the fields, the operators and the path integral then makes the use
of numerical methods possible.

In this section, we begin by reviewing the Wick rotation in Subsec.\ \ref{subsec:GT-Wick-rotation},
followed by the introduction of naïve lattice fermions in Subsec.\ \ref{subsec:GT-Naive-fermions}.
We then discuss the problem of fermion doubling in Subsec.\ \ref{subsec:GT-Fermion-doubling}
and try to get an understanding of the implications of the Nielsen-Ninomiya
No-Go theorem in Subsec.\ \ref{subsec:GT-Nielsen-Ninomiya}. In Subsec.\ \ref{subsec:GT-Coupling-gauge-fields},
we discuss the coupling to gauge theories before we introduce Wilson
fermions in Subsec.\ \ref{subsec:GT-Wilson-fermions}, staggered
fermions in Subsec.\ \ref{subsec:GT-Staggered-fermions} and staggered
Wilson fermions in Subsec.\ \ref{subsec:GT-Staggered-Wilson}. Finally
in Subsec.\ \ref{subsec:GT-Gauge-fields}, we discuss the discretization
of the gauge action on the lattice.

\subsection{Wick rotation \label{subsec:GT-Wick-rotation}}

In Minkowski space-time the complex factor $\exp\left(\ii S\right)$
appears in the path integral, forbidding a probabilistic interpretation
of expectation values. In order for Monte Carlo techniques to be applicable,
we move to Euclidean space-time. To that end we analytically continue
the time-component of four-vectors to purely imaginary values. Instead
of the four-vector $x^{\mu}=\left(x^{0},\boldsymbol{x}\right)^{\intercal}$,
we consider $x_{\mu}=\left(\boldsymbol{x},x_{4}\right)^{\intercal}$
with $x_{4}\equiv\ii x_{0}$. After this so-called Wick rotation \citep{Wick:1954eu},
we find a weight factor $\exp\left(-S\right)$ with Euclidean action
$S$ in the path integral.

In Euclidean space-time the Lorentz group reduces to the four-dimensional
rotation group. For convenience we replace the Dirac algebra as stated
in Eq.~\eqref{eq:DiracAlgebraMinkowski} by
\begin{equation}
\left\{ \gamma_{\mu},\gamma_{\nu}\right\} =2\delta_{\mu,\nu}\One.\label{eq:DiracAlgebraEuclidean}
\end{equation}
An explicit representation of the $\gamma_{\mu}$ matrices in the
chiral basis is given by
\begin{equation}
\gamma_{k}=\left(\begin{array}{cc}
 & \ii\sigma_{k}\\
-\ii\sigma_{k}
\end{array}\right),\qquad\gamma_{4}=\left(\begin{array}{cc}
 & \One\\
\One
\end{array}\right),\qquad\gamma_{5}=\left(\begin{array}{cc}
-\One\\
 & \One
\end{array}\right)
\end{equation}
with the Euclidean chirality matrix $\gamma_{5}\equiv\gamma_{0}\gamma_{1}\gamma_{2}\gamma_{3}$.
In Euclidean space-time the action of a gauge theory as discussed
in Subsec.\ \ref{subsec:GT-Gauge-theories} takes the form
\begin{equation}
S=\intop\mathrm{d}^{4}x\,\left(\mathcal{L}_{\mathsf{g}}+\mathcal{L}_{\mathsf{f}}\right),\qquad\mathcal{L}_{\mathsf{f}}=\overline{\Psi}\left(\gamma_{\mu}D_{\mu}+m_{\mathsf{f}}\right)\Psi,\label{eq:DiracActionEuclidean}
\end{equation}
where $\mathcal{L}_{\mathsf{g}}=\frac{1}{4}F^{2}$ with $F$ being
the respective field-strength tensor of the theory. Note that in Euclidean
space-time the lowering or raising of indices has no effect, so we
consequently write lowered indices.

\subsection{Naïve lattice fermions \label{subsec:GT-Naive-fermions}}

After having moved to Euclidean time, we can now begin with the discretization
of the path integral. We replace the continuous space-time domain
by a discrete hypercubic lattice
\begin{equation}
\Omega=(a_{x}\mathbb{Z})\times(a_{y}\mathbb{Z})\times(a_{z}\mathbb{Z})\times(a_{t}\mathbb{Z})
\end{equation}
with lattice spacings $a_{x}$, $a_{y}$, $a_{z}$ in the spatial
directions and $a_{t}$ in the temporal direction. In numerical simulations
one considers a hypercubic subdomain $\Lambda\subseteq\Omega$ of
the form
\begin{equation}
\Lambda=\left\{ \left.\left(\begin{array}{c}
a_{x}n_{x}\\
a_{y}n_{y}\\
a_{z}n_{z}\\
a_{t}n_{t}
\end{array}\right)\quad\right|\;\begin{array}{c}
0\leq n_{x}<N_{x}\\
0\leq n_{x}<N_{y}\\
0\leq n_{x}<N_{z}\\
0\leq n_{x}<N_{t}
\end{array}\right\} 
\end{equation}
with $N_{x}$, $N_{y}$, $N_{z}$ and $N_{t}$ being the number of
slices in each respective dimension. On this finite domain one usually
imposes (anti-)periodic boundary conditions for the fields. To simplify
our discussion, we restrict ourselves to the most common special case
of an isotropic lattice
\begin{equation}
a\equiv a_{x}=a_{y}=a_{z}=a_{t},
\end{equation}
where the lattice spacing is the same along all axes.

Having introduced the lattice domain, we discretize the derivative
in Eq.~\eqref{eq:DiracActionEuclidean}, which can be done in a straightforward
manner by using the antihermitian symmetric difference operator
\begin{equation}
\partial_{\mu}\Psi\left(x\right)\to\nabla_{\mu}\Psi\left(x\right)\equiv\frac{1}{2a}\left[\Psi\left(x+a\hat{\mu}\right)-\Psi\left(x-a\hat{\mu}\right)\right],
\end{equation}
where $\hat{\mu}$ refers to a unit vector in $\mu$-direction. This
discretization has the correct continuum limit
\begin{equation}
\lim_{a\to0}\nabla_{\mu}\phi\left(x\right)=\partial_{\mu}\phi\left(x\right),
\end{equation}
namely it reproduces the derivative operator for $a\to0$.

In the action, the space-time integration is replaced by a sum over
the lattice sites, i.e.\ $\intop\mathrm{d}^{4}x\to a^{4}\sum_{x\in\Lambda}$.
The resulting action reads
\begin{equation}
S_{\mathsf{n}}=\overline{\Psi}D_{\mathsf{n}}\Psi\equiv a^{4}\sum_{x,y\in\Lambda}\overline{\Psi}\left(x\right)D_{\mathsf{n}}\left(x,y\right)\Psi\left(y\right)
\end{equation}
with the lattice Dirac operator
\begin{equation}
D_{\mathsf{n}}=\gamma_{\mu}\nabla_{\mu}+m_{\mathsf{f}}\One.
\end{equation}
The operator density $D\left(x,y\right)$ is here defined as $D\Psi\left(x\right)=\sum_{y}D\left(x,y\right)\Psi\left(y\right)$
for an operator $D$, which yields in our case
\begin{equation}
D_{\mathsf{n}}\left(x,y\right)=\frac{1}{2a}\gamma_{\mu}\left(\delta_{y,x+a\hat{\mu}}-\delta_{y,x-a\hat{\mu}}\right)+m_{\mathsf{f}}\delta_{x,y}.\label{eq:NaiveLatticeD}
\end{equation}
We note that the Dirac operator respects chiral symmetry in the sense
that $\left\{ D_{\mathsf{n}}^{0},\gamma_{5}\right\} =0$, where $D_{\mathsf{n}}^{0}$
denotes the massless naïve lattice Dirac operator.

In our discrete setting the path integral measure now takes the form
\begin{equation}
\intop\mathcal{D}\overline{\Psi}\,\mathcal{D}\Psi\to\prod_{x,\mu}\mathrm{d}\overline{\Psi}_{\mu}\left(x\right)\prod_{y,\nu}\mathrm{d}\Psi_{\nu}\left(y\right)
\end{equation}
and expectation values translate to
\begin{align}
\left\langle \mathcal{O}\right\rangle  & =\frac{1}{Z}\intop\prod_{x,\mu}\mathrm{d}\overline{\Psi}_{\mu}\left(x\right)\prod_{y,\nu}\mathrm{d}\Psi_{\nu}\left(y\right)\,\mathcal{O}\left[\overline{\Psi},\Psi\right]\,e^{-S_{\mathsf{f}}\left[\overline{\Psi},\Psi\right]},\\
Z & =\intop\prod_{x,\mu}\mathrm{d}\overline{\Psi}_{\mu}\left(x\right)\prod_{y,\nu}\mathrm{d}\Psi_{\nu}\left(y\right)\,e^{-S_{\mathsf{f}}\left[\overline{\Psi},\Psi\right]}.
\end{align}
We emphasize that we discretized here in the simplest possible way.
As the name suggests, naïve lattice fermions are not suitable for
actual simulations as we explain in the following subsection.

\subsection{Fermion doubling \label{subsec:GT-Fermion-doubling}}

To understand a fundamental problem of the naïve discretization, let
us compute the free-field propagator by inverting the lattice Dirac
operator in Eq.~\eqref{eq:NaiveLatticeD}. For this well-known derivation
we follow Ref.~\citep{Gattringer:2010zz}. We begin by Fourier transforming
$D_{\mathsf{n}}$ and writing the result as $\hat{D}_{\mathsf{N}}\left(p,q\right)=\delta\left(p-q\right)\hat{D}\left(p\right)$
with
\begin{equation}
\hat{D}\left(p\right)=\frac{\ii}{a}\gamma_{\mu}\sin\left(p_{\mu}a\right)+m_{\mathsf{f}}\One.
\end{equation}
Here the hat indicates a Fourier transformed expression. We can invert
$\hat{D}\left(p\right)$ to find
\begin{equation}
\hat{D}^{-1}\left(p\right)=\frac{m_{\mathsf{f}}\One-\ii\gamma_{\mu}\sin\left(p_{\mu}a\right)/a}{m_{\mathsf{f}}^{2}+\sum_{\nu}\sin^{2}\left(p_{\nu}a\right)/a^{2}}.\label{eq:NaivePropagator}
\end{equation}
Specializing it to the massless case, Eq.~\eqref{eq:NaivePropagator}
reduces to
\begin{equation}
\left.\hat{D}^{-1}\left(p\right)\right|_{m_{\mathsf{f}}=0}=\frac{-\ii\gamma_{\mu}\sin\left(p_{\mu}a\right)/a}{\sum_{\nu}\sin^{2}\left(p_{\nu}a\right)/a^{2}}.\label{eq:NaivePropagatorMassless}
\end{equation}
We can verify that Eq.~\eqref{eq:NaivePropagatorMassless} has the
correct continuum limit for fixed values of $p$. The resulting propagator
$-\ii\gamma_{\mu}p_{\mu}/p^{2}$ of a massless fermion has a pole
at $p=\left(0,0,0,0\right)^{\intercal}$ only. On the other hand,
we observe that the lattice propagator in Eq.~\eqref{eq:NaivePropagator}
has a pole whenever all components take up values $p_{\mu}\in\left\{ 0,\pi/a\right\} $.
In addition to the physical pole, we find additional 15 unphysical
poles in the corners of the Brillouin zone, among them e.g.\ $p=\left(\pi/a,0,0,0\right)^{\intercal}$
and $p=\left(\pi/a,\pi/a,\pi/a,\pi/a\right)^{\intercal}$.

The appearance of these spurious states on the lattice is the notorious
fermion doubler problem. The presence of these doubler states is related
to the use of the symmetric lattice derivative in Eq.~\eqref{eq:NaiveLatticeD}.
However, the use of left or right derivatives would give rise to non-covariant
contributions, which would render the theory non-renormalizable \citep{Sadooghi:1996ip}.

\subsection{The Nielsen-Ninomiya No-Go theorem \label{subsec:GT-Nielsen-Ninomiya}}

A more complete understanding of the phenomenon of fermion doublers
and their connection to chiral symmetry is given by the Nielsen-Ninomiya
No-Go theorem \citep{Nielsen:1981hk,Nielsen:1980rz,Nielsen:1981xu}.
It states that there can be no net chirality under a set of relatively
general conditions, namely the lattice Hamiltonian of a fermion being
local, quadratic in the fields and invariant under both lattice translations
as well as under a change of the phase of the fields \citep{Friedan:1982nk}.
In this case the number of left-handed and right-handed states must
be equal.

This theorem explains the difficulties in constructing fermion formulations
with all the properties one would like to have. Up to this day, no
theoretically sound lattice fermion formulation is known which satisfies
all of the following nontrivial conditions:
\begin{enumerate}
\item \setlength{\itemsep}{0pt} Absence of fermion doublers;
\item \setlength{\itemsep}{0pt} Respecting chiral symmetry;
\item \setlength{\itemsep}{0pt} Being computationally efficient.
\end{enumerate}
In a realistic setting (e.g.\ dynamical four-dimensional quantum
chromodynamics) one can generally satisfy at most two of the three
conditions and we evaluate the novel fermion discretizations discussed
in this thesis with respect to these points.

\subsection{Coupling to gauge fields \label{subsec:GT-Coupling-gauge-fields}}

Up to this point we only considered naïve lattice fermions in the
free-field case. If we seek to couple fermions to a gauge field, we
have to include a suitable interaction term in the fermion action.

A peculiarity of the lattice formulation is that one introduces the
gauge fields as elements of the gauge group $G$, rather than as elements
of the Lie algebra $\mathfrak{g}$ like in the continuum. In the following,
we restrict ourselves to the case of the (special) unitary groups,
i.e.\ for all $U\in G$ the defining property $U^{\dagger}=U^{-1}$
holds.

Let us now introduce (oriented) gauge links $U_{\mu}\left(x\right)\in G$
for $\mu=1,\dots,4$ and $x\in\Lambda$. We include interactions with
the gauge field by incorporating the gauge links in the finite difference
operator via
\begin{equation}
\nabla_{\mu}\Psi\left(x\right)\equiv\frac{1}{2a}\left[U_{\mu}\left(x\right)\Psi\left(x+a\hat{\mu}\right)-U_{\mu}^{\dagger}\left(x-a\hat{\mu}\right)\Psi\left(x-a\hat{\mu}\right)\right],
\end{equation}
resulting in
\begin{equation}
D_{\mathsf{n}}\left(x,y\right)=\frac{1}{2a}\gamma_{\mu}\left(U_{\mu}\left(x\right)\delta_{y,x+a\hat{\mu}}-U_{\mu}^{\dagger}\left(x-a\hat{\mu}\right)\delta_{y,x-a\hat{\mu}}\right)+m_{\mathsf{f}}\delta_{x,y}.
\end{equation}
Note that the fermionic degrees of freedom live on the lattice sites,
while the gauge degrees of freedoms are located on the links connecting
neighboring sites. If we now choose $\Omega\left(x\right)\in G$ and
do a gauge transformation of the form
\begin{equation}
\begin{aligned}\Psi\left(x\right) & \to\Omega\left(x\right)\Psi\left(x\right),\\
\overline{\Psi}\left(x\right) & \to\overline{\Psi}\left(x\right)\Omega^{\dagger}\left(x\right),\\
U_{\mu}\left(x\right) & \to\Omega\left(x\right)U_{\mu}\left(x\right)\Omega^{\dagger}\left(x+a\hat{\mu}\right),
\end{aligned}
\end{equation}
we find that the action $S_{\mathsf{n}}=\overline{\Psi}D_{\mathsf{n}}\Psi$
is invariant with respect to the symmetry group $G$. This transformation
is the lattice equivalent of the continuum gauge transformations as
described in Eqs.~\eqref{eq:ContGaugeTransformU1} and \eqref{eq:ContGaugeTransformYM}.

\subsection{Wilson fermions \label{subsec:GT-Wilson-fermions}}

Let us now return to the problem of fermion doublers. Wilson's original
proposal \citep{Wilson:1974sk,Wilson:1975id} for a lattice Dirac
operator can be written in the form
\begin{equation}
D_{\mathsf{w}}\left(m_{\mathsf{f}}\right)=\gamma_{\mu}\nabla_{\mu}+m_{\mathsf{f}}\One+W_{\mathsf{w}},\qquad W_{\mathsf{w}}=-\frac{ar}{2}\Delta\label{eq:DefWilsonDiracOp}
\end{equation}
and is the result from adding the so-called Wilson term $W_{\mathsf{w}}$
to the naïve lattice Dirac operator. The parameter $r\in\left(0,1\right]$
denotes the Wilson parameter and $\Delta$ is the covariant lattice
Laplacian. The corresponding action reads $S_{\mathsf{w}}=\overline{\Psi}D_{\mathsf{w}}\Psi$.
In terms of the parallel transporters
\begin{align}
T_{\mu}\Psi\left(x\right) & =U_{\mu}\left(x\right)\Psi\left(x+a\hat{\mu}\right),\\
T_{\mu}^{\dagger}\Psi\left(x\right) & =U_{\mu}^{\dagger}\left(x-a\hat{\mu}\right)\Psi\left(x-a\hat{\mu}\right)=T_{\mu}^{-1}\Psi\left(x\right),
\end{align}
we can write all terms concisely as
\begin{equation}
\nabla_{\mu}=\frac{1}{2a}\left(T_{\mu}-T_{\mu}^{\dagger}\right),\qquad\Delta=\frac{2}{a^{2}}\sum_{\mu}\left(C_{\mu}-\One\right),\qquad C_{\mu}=\frac{1}{2}\left(T_{\mu}+T_{\mu}^{\dagger}\right).
\end{equation}
We note that $D_{\mathsf{w}}^{\dagger}=\gamma_{5}D_{\mathsf{w}}\gamma_{5}$
and therefore $\det D_{\mathsf{w}}\in\mathbb{R}$. For the important
case of two degenerate fermion flavors, we can construct an action
in terms of the operator $D_{\mathsf{w}}^{\dagger}D_{\mathsf{w}}$
with $\det\left(D_{\mathsf{w}}^{\dagger}D_{\mathsf{w}}\right)\geq0$.

Due to the Wilson term, the free-field propagator in momentum space
changes from Eq.~\eqref{eq:NaivePropagator} to 
\begin{equation}
\hat{D}_{\mathsf{w}}^{-1}\left(p\right)=\frac{M\left(p\right)\One-\ii\gamma_{\mu}\sin\left(p_{\mu}a\right)/a}{M^{2}\left(p\right)+\sum_{\nu}\sin^{2}\left(p_{\nu}a\right)/a^{2}}\label{eq:WilsonProp}
\end{equation}
with $M\left(p\right)=m_{\mathsf{f}}+2ra^{-1}\sum_{\mu}\sin^{2}\left(p_{\mu}a/2\right)$,
see e.g.\ Ref.~\citep{Rothe:1992nt}. In the corners of the Brillouin
zone the mass term is effectively modified to $m_{\mathsf{f}}+2rn/a$,
where $n$ is the number of components of the lattice momentum $p$
with value $\pi/a$. While the physical mode is unaffected, the fermion
doublers acquire a mass of $\mathcal{O}\left(a^{-1}\right)$. In total
we find four doubler branches, where the $k^{\textrm{th}}$ branch
has a mass of $m_{\mathsf{f}}+2rk/a$ and contains $\binom{4}{k}$
modes. In the continuum limit the number of flavors is then reduced
from sixteen to a single physical flavor.

A shortcoming of Wilson fermions is that even for $m_{\mathsf{f}}=0$
chiral symmetry is broken, i.e.\ $\left\{ D_{\mathsf{w}},\gamma_{5}\right\} \neq0$.
Nevertheless, Wilson fermions and their improved versions remain some
of the most popular choices for lattice theoretical calculations due
to their conceptual simplicity.

\subsection{Staggered fermions \label{subsec:GT-Staggered-fermions}}

An alternative approach to the fermion doubling problem are staggered
fermions, also known as Kogut-Susskind fermions \citep{Kogut:1974ag,Banks:1975gq,Banks:1976ia,Susskind:1976jm}.
Here one lifts an exact four-fold degeneracy of naïve lattice fermions.
As a result, the number of fermion species is reduced from sixteen
to four (in four space-time dimensions).

Let us briefly review the construction of staggered fermion, where
we follow Ref.~\citep{Rothe:1992nt}. Our starting point is the naïve
fermion action
\begin{equation}
S_{\mathsf{n}}=a^{4}\sum_{x}\overline{\Psi}\left(x\right)\left(\frac{1}{2a}\gamma_{\mu}\left[\Psi\left(x+a\hat{\mu}\right)-\Psi\left(x-a\hat{\mu}\right)\right]+m_{\mathsf{f}}\Psi\left(x\right)\right)
\end{equation}
in the free-field case. We can do a local change of variables of the
form
\begin{equation}
\Psi\left(x\right)\to\Gamma\left(x\right)\chi\left(x\right),\qquad\overline{\Psi}\left(x\right)\to\overline{\chi}\left(x\right)\Gamma^{\dagger}\left(x\right),
\end{equation}
where $\Gamma\left(x\right)$ is a unitary matrix. By choosing $\Gamma\left(x\right)$
so that
\begin{equation}
\Gamma^{\dagger}\left(x\right)\gamma_{\mu}\Gamma\left(x+a\hat{\mu}\right)=\eta_{\mu}\left(x\right)\One,
\end{equation}
where $\eta_{\mu}\left(x\right)$ is a scalar function, we can achieve
a ``spin diagonalization''. That this is possible can be verified
explicitly for the choice
\begin{equation}
\Gamma\left(x\right)=\prod_{\mu}\gamma_{\mu}^{x_{\mu}/a},\qquad\eta_{\mu}\left(x\right)=\left(-1\right)^{\sum_{\nu=0}^{\mu-1}x_{\nu}/a}
\end{equation}
with $\eta_{1}\left(x\right)=1$. The action now takes the form
\begin{equation}
S_{\mathsf{n}}=a^{4}\sum_{\nu=1}^{4}\sum_{x}\overline{\chi}_{\nu}\left(x\right)\left(\frac{1}{2a}\eta_{\mu}\left(x\right)\left[\chi_{\nu}\left(x+a\hat{\mu}\right)-\chi_{\nu}\left(x-a\hat{\mu}\right)\right]+m_{\mathsf{f}}\chi_{\nu}\left(x\right)\right),
\end{equation}
where we wrote the spinor index $\nu$ explicitly. After this change
of variables, we can see that there are four identical copies in the
sum over $\nu$. We can lift this degeneracy manually by omitting
three of them, giving rise to the staggered fermion action
\begin{equation}
S_{\mathsf{st}}=a^{4}\sum_{x}\overline{\chi}\left(x\right)\left(\frac{1}{2a}\eta_{\mu}\left(x\right)\left[\chi\left(x+a\hat{\mu}\right)-\chi\left(x-a\hat{\mu}\right)\right]+m_{\mathsf{f}}\chi\left(x\right)\right).\label{eq:FreeActionStagFermions}
\end{equation}
We emphasize that $\chi\left(x\right)$ is now a one-component spinor
and all that remains from the $\gamma_{\mu}$ matrices is the staggered
phase factor $\eta_{\mu}\left(x\right)$. By gauging the derivative
operators, we can couple staggered fermions to a gauge field as discussed
in Subsec.\ \ref{subsec:GT-Coupling-gauge-fields}. In summary, the
staggered action and the staggered Dirac operator read
\begin{equation}
S_{\mathsf{st}}=\overline{\chi}D_{\mathsf{st}}\chi,\qquad D_{\mathsf{st}}=\eta_{\mu}\nabla_{\mu}+m_{\mathsf{f}}\One,
\end{equation}
where we again make use of the matrix-vector-notation. We note that
$D_{\mathsf{st}}^{\dagger}=\epsilon D_{\mathsf{st}}\epsilon$, where
\begin{equation}
\epsilon\left(x\right)\equiv\left(-1\right)^{\sum_{\mu}x_{\mu}/a},
\end{equation}
and $\det D_{\mathsf{st}}\geq0$ as all eigenvalues of $D_{\mathsf{st}}$
come in complex conjugate pairs.

As the action in Eq.~\eqref{eq:FreeActionStagFermions} describes
four fermion species, a natural problem is the one of flavor identification.
The two common approaches are in coordinate space \citep{KlubergStern:1983dg,Gliozzi:1982ib}
or in momentum space \citep{Sharatchandra:1981si,vandenDoel:1983mf,Golterman:1984cy}.
We note that both approaches are identical in the continuum limit
\citep{Daniel:1986zm} and quickly review the schemes in the following.

\subsubsection{Flavor identification in coordinate space}

We begin with the identification of flavors in coordinate space, following
Refs.~\citep{Gattringer:2010zz,Rothe:1992nt}. To this end let us
relabel the fields as $\chi_{\rho}\left(x\right)\equiv\chi\left(2x+a\rho\right)$,
where the multi-index takes values $\rho\in\left\{ 0,1\right\} ^{4}$.
We remark that the field $\chi_{\rho}\left(x\right)$ effectively
lives on a lattice with spacing $a^{\prime}=2a$ and that $\eta_{\mu}\left(2x+a\rho\right)=\eta_{\mu}\left(a\rho\right)$.
We then define the matrix-valued fermion fields
\begin{equation}
\begin{aligned}\Psi\left(x\right) & =\frac{1}{8}\sum_{\rho}\Gamma\left(a\rho\right)\chi_{\rho}\left(x\right),\\
\overline{\Psi}\left(x\right) & =\frac{1}{8}\sum_{\rho}\overline{\chi}_{\rho}\left(x\right)\Gamma^{\star}\left(a\rho\right).
\end{aligned}
\end{equation}
This relation can be inverted to give
\begin{equation}
\begin{aligned}\chi\left(2x+a\rho\right) & =2\tr\left[\Gamma^{\dagger}\left(a\rho\right)\Psi\left(x\right)\right],\\
\overline{\chi}\left(2x+a\rho\right) & =2\tr\left[\overline{\Psi}\left(x\right)\Gamma\left(a\rho\right)\right].
\end{aligned}
\end{equation}
Reformulating the free staggered fermion action using these new fields,
we eventually find
\begin{equation}
S_{\mathsf{st}}=a^{\prime2}\sum_{f,x}\overline{\Psi}^{f}\left(x\right)\left(\gamma_{\mu}\nabla_{\mu}\Psi^{f}\left(x\right)-\frac{a^{\prime}}{2}\gamma_{5}\left(\xi_{5}\xi_{\mu}\right)_{fg}\Delta_{\mu}\Psi^{g}\left(x\right)+m_{\mathsf{f}}\Psi^{f}\left(x\right)\right),\label{eq:ActionReconstStag}
\end{equation}
where we introduced a flavor index $f=1,\dots,4$ via $\Psi_{\mu}^{f}\left(x\right)\equiv\Psi_{\mu f}\left(x\right)$.
Furthermore, all finite difference operators are with respect to the
lattice spacing $a^{\prime}$ of the blocked lattice and we define
$\xi_{\mu}\equiv\gamma_{\mu}^{\intercal}$. While the first and third
term in Eq.~\eqref{eq:ActionReconstStag} is the usual kinetic and
mass term of the four fermion species $\Psi^{f}$, the second term
has the form of a Wilson term with an additional nontrivial spin-flavor
structure.

In contrast to Wilson's construction, staggered fermions partially
preserve chiral symmetry in the massless case. For $m_{\mathsf{f}}=0$,
the staggered fermion action has a $\Uone\times\Uone$ symmetry, given
by the transformations
\begin{eqnarray}
\Psi\left(x\right)\to e^{\ii\alpha}\Psi\left(x\right), &  & \overline{\Psi}\left(x\right)\to\overline{\Psi}\left(x\right)e^{-\ii\alpha},\\
\Psi\left(x\right)\to e^{\ii\beta\Theta_{55}}\Psi\left(x\right), &  & \overline{\Psi}\left(x\right)\to\overline{\Psi}\left(x\right)e^{\ii\beta\Theta_{55}}.
\end{eqnarray}
Here $\Theta_{55}\equiv\gamma_{5}\otimes\xi_{5}$ is the generator
of the remnant chiral symmetry, which can be also expressed as $\left\{ D_{\mathsf{st}}^{0},\epsilon\right\} =0$
with $D_{\mathsf{st}}^{0}$ being the massless staggered Dirac operator.

\subsubsection{Flavor identification in momentum space}

\global\long\def\chio{\overset{\phantom{\cdots}}{\chi}_{\circ}}

Another method of flavor identification is carried out in momentum
space. We follow Ref.~\citep{Rothe:1992nt} in the discussion and
refer the reader to this reference for a more complete discussion.

First note that we can write the staggered phase as
\[
\eta_{\mu}\left(x\right)=e^{\ii\delta^{\mu}\cdot x/a}
\]
with $\delta_{\nu}^{\mu}=\pi$ for $\nu<\mu$ and $\delta_{\nu}^{\mu}=0$
else. In momentum space we can write
\begin{equation}
S_{\mathsf{st}}=\intop_{-\pi}^{\pi}\frac{\mathrm{d}^{4}p}{\left(2\pi\right)^{4}}\frac{\mathrm{d}^{4}q}{\left(2\pi\right)^{4}}\,\overline{\chi}_{\circ}\left(q\right)\hat{D}\left(q,p\right)\chio\left(p\right),
\end{equation}
where the lattice momenta $p$ and $q$ are in dimensionless lattice
units, $\chio$ is $2\pi$-periodic, the Fourier transformed Dirac
operator reads
\begin{equation}
\hat{D}\left(q,p\right)=\left(2\pi\right)^{4}\left[\ii\sin\left(p_{\mu}\right)\delta_{\mathsf{p}}\left(p-q+\delta^{\mu}\right)+m_{\mathsf{f}}\delta_{\mathsf{p}}\left(p-q\right)\right],
\end{equation}
and $\delta_{\mathsf{p}}$ refers to the periodic Kronecker $\delta$.
If we define multi-indices $A$ and $B$, which each sum over all
elements in $\left\{ 0,1\right\} ^{4}$, we can rewrite the staggered
action as
\begin{equation}
S_{\mathsf{st}}=\sum_{A,B}\intop_{-\pi/2}^{\pi/2}\frac{\mathrm{d}^{4}p}{\left(2\pi\right)^{4}}\frac{\mathrm{d}^{4}q}{\left(2\pi\right)^{4}}\,\overline{\chi}_{A}\left(q\right)\hat{D}_{A,B}\left(q,p\right)\chi_{B}\left(p\right)\label{eq:StagActionMomAB}
\end{equation}
with
\begin{multline}
\hat{D}_{A,B}\left(q,p\right)=\left(2\pi\right)^{4}\left[e^{\ii\pi B_{\mu}}\ii\sin\left(p_{\mu}\right)\delta_{\mathsf{p}}\left(p-q+\pi B-\pi A+\delta^{\mu}\right)\right.\\
\left.+m_{\mathsf{f}}\delta_{\mathsf{p}}\left(p-q+\pi B-\pi A\right)\right].
\end{multline}
The fields are defined as $\overline{\chi}_{A}\left(p\right)\equiv\overline{\chi}_{\circ}\left(p+\pi A\right)$
and $\chi_{B}\left(p\right)\equiv\chio\left(p+\pi B\right)$. Now
note the relations
\begin{align}
\delta_{\mathsf{p}}\left(p-q+\pi B-\pi A\right) & =\delta_{A,B}\delta\left(p-q\right),\\
\delta_{\mathsf{p}}\left(p-q+\pi B-\pi A+\delta^{\mu}\right) & =\varphi_{A,B}^{\mu}\delta\left(p-q\right),
\end{align}
where 
\begin{equation}
\varphi_{A,B}^{\mu}\equiv\prod_{\nu}\frac{1}{2}\left[e^{\ii\left(\pi B-\pi A+\delta_{\nu}^{\mu}\right)}+1\right].
\end{equation}
If we now define $\Gamma_{A,B}^{\mu}\equiv\exp\left(\ii\pi B_{\mu}\right)\varphi_{A,B}^{\mu}$,
we can show that they satisfy the Dirac algebra $\left\{ \Gamma^{\mu},\Gamma^{\nu}\right\} =2\delta_{\mu,\nu}\One$
and that the $\Gamma^{\mu}$ are unitary equivalent to $\gamma_{\mu}\otimes\One$
in spin\,$\otimes$\,flavor language. By using the above relations
and adopting the spin\,$\otimes$\,flavor notation, Eq.~\eqref{eq:StagActionMomAB}
takes the form
\begin{equation}
S_{\mathsf{st}}=\intop_{-\pi/2}^{\pi/2}\frac{\mathrm{d}^{4}p}{\left(2\pi\right)^{4}}\,\overline{q}\left(p\right)\left[\left(\gamma_{\mu}\otimes\One\right)\ii\sin\left(p_{\mu}\right)+m_{\mathsf{f}}\left(\One\otimes\One\right)\right]q\left(p\right),
\end{equation}
where $q$ now has an implicit spin and flavor index. The action in
this form is invariant under the full $\Ufour\otimes\Ufour$ chiral
symmetry group, but this comes at the price of a non-local action
in position space \citep{Rothe:1992nt}.

\subsection{Staggered Wilson fermions \label{subsec:GT-Staggered-Wilson}}

The construction of staggered Wilson fermions was originally proposed
by Adams \citep{Adams:2009eb,Adams:2010gx} and later extended by
Hoelbling \citep{Hoelbling:2010jw}. Starting from the staggered fermion
action a suitable flavored mass term is added, reducing the number
of doublers and making them suitable kernel operators for the overlap
and domain wall fermion construction. Due to the central role of staggered
Wilson fermions in this thesis, we dedicate Chapter\ \ref{chap:Staggered-Wilson-fermions}
to the discussion of their derivation, properties and symmetries.

\subsection{Gauge fields \label{subsec:GT-Gauge-fields}}

In addition to the fermionic fields, we need to formulate gauge fields
on the lattice. To this end we have to discretize the continuum gauge
action given in Subsec.\ \ref{subsec:GT-Gauge-theories}. We require
the lattice gauge action to be formulated in terms of the link variables
and to respect gauge invariance. The simplest gauge invariant objects
we can form with gauge links are closed loops. This leads us to the
introduction of the so-called plaquette
\begin{equation}
U_{\mu\nu}\left(x\right)\equiv U_{\mu}\left(x\right)U_{\nu}\left(x+a\hat{\mu}\right)U_{\mu}^{\dagger}\left(x+a\hat{\nu}\right)U_{\nu}^{\dagger}\left(x\right),\label{eq:DefPlaquette}
\end{equation}
which is a minimal closed loop in the $\mu$-$\nu$-plane. We can
then formulate a lattice gauge action in terms of plaquettes only,
ensuring gauge invariance by construction.

In the case of an Abelian $\Uone$ gauge theory, let us consider
\begin{equation}
S_{\mathsf{g}}=\beta\sum_{x}\sum_{\mu<\nu}\myRe\left(1-U_{\mu\nu}\left(x\right)\right),
\end{equation}
where $\beta=1/e^{2}$ is the inverse coupling. One can verify that
in the limit of zero lattice spacing and by rescaling $\frac{1}{e}A_{\mu}\to A_{\mu}$,
one recovers the well-known continuum expression for the action.

Although in the context of non-Abelian gauge theories we are primarily
interested in quantum chromodynamics, it makes sense to discuss the
general case of the symmetry group $\SUn$ for $N_{\mathsf{c}}\geq2$.
An important difference compared to the Abelian case is that the ordering
in Eq.~\eqref{eq:DefPlaquette} is now important as link variables
do not commute. The lattice action now generalizes to
\begin{equation}
S_{\mathsf{g}}=\frac{\beta}{N_{\mathsf{c}}}\sum_{x}\sum_{\mu<\nu}\myRe\left(\One-U_{\mu\nu}\left(x\right)\right),\label{eq:WilsonGaugeAction}
\end{equation}
where $\beta=2N_{\mathsf{c}}/g^{2}$ and $\One$ is the $N_{\mathsf{c}}\times N_{\mathsf{c}}$
identity matrix. One can again easily verify the correct continuum
limit of the lattice discretization.

While the Wilson gauge action \citep{Wilson:1974sk,Wilson:1975id}
in Eq.~\eqref{eq:WilsonGaugeAction} is the simplest case of a gauge
action on the lattice, there are also improved actions in use. For
examples we refer the reader here to the Iwasaki \citep{Iwasaki:1984cj,Iwasaki:1996sn}
and DBW2 \citep{deForcrand:1999bi,Borici:1997wg} actions, see also
Ref.~\citep{Golterman:2005fe}.

\section{Numerical simulations \label{sec:GT-Numerical-simulations}}

One of the major applications of the lattice approach to quantum field
theory is the simulation on a computer using Monte Carlo methods.
Instead of giving a detailed discussion of how a continuum limit is
taken and physical observables can be extracted, we give a short conceptual
overview of the Monte Carlo method in the context of lattice gauge
theory and refer the reader to the many excellent textbooks in the
field, such as Refs.~\citep{Gattringer:2010zz,DeGrand:2006zz,Montvay:1994cy,Rothe:1992nt}.
For a more general overview of the Monte Carlo method, see e.g.\ Ref.~\citep{rubinstein2011simulation}.

\subsection{Monte Carlo method}

Assume we want to determine the expectation value of an observable
$\mathcal{O}$, e.g.\ in order to extract a hadron mass, in the setting
of a lattice gauge theory. We then have to evaluate the expression
\begin{align}
\left\langle \mathcal{O}\right\rangle  & =\frac{1}{Z}\intop\mathcal{D}\overline{\Psi}\mathcal{D}\Psi\mathcal{D}U\,\mathcal{O}\left[\overline{\Psi},\Psi,U\right]\,e^{-S\left[\overline{\Psi},\Psi,U\right]},\\
Z & =\intop\mathcal{D}\overline{\Psi}\mathcal{D}\Psi\mathcal{D}U\,e^{-S\left[\overline{\Psi},\Psi,U\right]},
\end{align}
where the combined fermion and gauge action reads
\begin{equation}
S\left[\overline{\Psi},\Psi,U\right]=S_{\mathsf{f}}\left[\overline{\Psi},\Psi,U\right]+S_{\mathsf{g}}\left[U\right]
\end{equation}
and the generalization to several fermion species is straightforward.
Assuming the fermionic degrees of freedom appear quadratically, we
can integrate them out exactly. We arrive at an integrand, which only
depends on the gauge field and reads
\begin{align}
\left\langle \mathcal{O}\right\rangle  & =\frac{1}{Z}\intop\mathcal{D}U\,\mathcal{O}^{\prime}\left[U\right]\,\det D\left[U\right]e^{-S_{\mathsf{g}}\left[U\right]},\\
Z & =\intop\mathcal{D}U\,\det D\left[U\right]e^{-S_{\mathsf{g}}\left[U\right]}.
\end{align}
We note that the integral over the spinors $\Psi$, $\overline{\Psi}$
is of the Grassmann / Berezin kind \citep{berezin1966method} and
gives rise to the fermion determinant $\det D\left[U\right]$ with
the lattice Dirac operator $D$. The operator $\mathcal{O}^{\prime}\left[U\right]$
follows from $\mathcal{O}\left[\overline{\Psi},\Psi,U\right]$ after
integrating out the fermions in the path integral with the help of
Wick's theorem \citep{Wick:1950ee}. We note that in general $\mathcal{O}^{\prime}\left[U\right]$
contains propagators of the form $D^{-1}\left[U\right]$.

To give an explicit example following Ref.~\citep{Gattringer:2010zz},
consider two Dirac fermion species $\Psi_{\mathsf{u}}$, $\Psi_{\mathsf{d}}$
and let 
\begin{equation}
\mathcal{O}=\overline{\Psi}_{\mathsf{d}}\left(x\right)\Gamma\,\Psi_{\mathsf{u}}\left(x\right)\overline{\Psi}_{\mathsf{u}}\left(y\right)\Gamma\,\Psi_{\mathsf{d}}\left(y\right)
\end{equation}
be an iso-triplet operator, where $\Gamma$ is a monomial of $\gamma_{\mu}$
matrices. After integrating out both the $\Psi_{\mathsf{u}}$ and
$\Psi_{\mathsf{d}}$ fields, one finds
\begin{equation}
\mathcal{O}^{\prime}=-\tr\left[\Gamma\,D_{\mathsf{u}}^{-1}\left(x,y\right)\Gamma\,D_{\mathsf{d}}^{-1}\left(y,x\right)\right],
\end{equation}
where $D_{\mathsf{u}}$ and $D_{\mathsf{d}}$ are the lattice Dirac
operators of the respective fermion species.

As the resulting integral is of extremely high dimension, one makes
use of Monte Carlo methods for the numerical evaluation. To this end
one interprets the factor
\begin{equation}
\frac{1}{Z}\det D\left[U\right]e^{-S_{\mathsf{g}}\left[U\right]},\label{eq:MarkovProbDistr}
\end{equation}
the so-called Gibbs-measure, as a probability measure. One first generates
a Markov chain of gauge configurations $U_{i}$ with respect to this
probability distribution. Under certain conditions we can then replace
the path integral expectation value by
\begin{equation}
\left\langle \mathcal{O}\right\rangle =\lim_{N\to\infty}\frac{1}{N}\sum_{i=1}^{N}\mathcal{O}\left[U_{i}\right],
\end{equation}
where we take an ensemble average.

The computationally most expensive part is the generation of the gauge
ensembles according to the probability distribution in Eq.~\eqref{eq:MarkovProbDistr}
due to the presence of the fermion determinant $\det D\left[U\right]$.
Moreover, as one approaches the continuum limit and one gets closer
to the physical point, the computational costs rapidly increase, see
e.g.\ Refs.~\citep{Jansen:2008vs,Aoki:2012oma}. While in the early
days of lattice quantum chromodynamics numerical simulations where
done in the so-called quenched approximation, i.e.\ one does the
replacement $\det D\left[U\right]\to1$, dynamical simulations are
nowadays state of the art.

The high computational costs of these Monte Carlo simulations explain
why research in novel lattice fermion formulations remains a very
active field, see e.g.\ Refs.~\citep{Misumi:2012eh,MisumiThesis}
for an overview.

\chapter{Staggered Wilson fermions \label{chap:Staggered-Wilson-fermions}}

The central topic of this thesis are staggered Wilson fermions and
their related formulations, namely staggered overlap and staggered
domain wall fermions. For this reason, we dedicate this chapter to
the discussion of flavored mass terms and the construction of staggered
Wilson fermions.

The staggered Wilson fermion formulation first arose in investigations
regarding the index theorem on the lattice using staggered fermions
\citep{Adams:2009eb}. The lattice Dirac operator which emerged from
these investigations \citep{Adams:2010gx} shares properties of both
staggered and Wilson fermions. Later Hoelbling expanded upon this
idea and proposed a related fermion formulation \citep{Hoelbling:2010jw}.
In Ref.~\citep{Hoelbling:2016qfv}, we went one step further and
introduced generalizations of these constructions for arbitrary flavor
splittings in arbitrary even dimensions.

\section{Introduction}

To understand the idea behind staggered Wilson fermions, we recall
that there are two traditional approaches to deal with the problem
of fermion doublers. Wilson's original approach is to add the Wilson
term as discussed in Subsec.\ \ref{subsec:GT-Wilson-fermions}, which
is a symmetric covariant discretization of the Laplacian. The effect
of this term is that the spurious fermion states acquire a mass of
$\mathcal{O}\left(a^{-1}\right)$ and, thus, decouple in the continuum
limit. The other approach is the one of staggered fermions as introduced
in Subsec.\ \ref{subsec:GT-Staggered-fermions}. Here one lifts an
exact degeneracy of naïve fermions using a ``spin diagonalization''.
As a result, not all fermion doublers are removed, but the number
of species is reduced from sixteen to four. We note that in practical
simulations one can further reduce the number of flavors to $N_{\mathsf{f}}$
by taking an appropriate root of the fermion determinant. Explicitly,
one can make a replacement of the form
\begin{equation}
\det D_{\mathsf{st}}\to\left(\det D_{\mathsf{st}}\right)^{N_{\mathsf{f}}/4}.
\end{equation}
Due to the lack of a formal proof of correctness, the method remained
controversial for a long time and one can find arguments in favor
\citep{Adams:2004mf,Hasenfratz:2006nw,Bernard:2006vv,Bernard:2007eh,Bernard:2008gr,Adams:2008db}
and against \citep{Creutz:2007yg,Creutz:2007pr,Creutz:2008kb} this
approach in the literature, see also Refs.~\citep{Sharpe:2006re,Golterman:2008gt}
for an overview.

Staggered Wilson fermions combine both approaches by taking the staggered
fermion action and adding a suitable ``staggered Wilson term'' to
further reduce the number of doublers. This staggered Wilson term
turns out to be a combination of the flavored mass terms as discussed
in detail in the classical paper by Golterman and Smit \citep{Golterman:1984cy}.
Due to the presence of this term, staggered Wilson fermions have technical
properties similar to Wilson fermions. By using them as a kernel operator,
they allow the construction of staggered overlap and staggered domain
wall fermions. Usual staggered fermions, on the other hand, do not
define a suitable kernel operator. This is because these constructions
rely on the property $\gamma_{5}^{2}=\One$, which does not hold for
its staggered equivalent as we discuss in more detail in Sec.\ \ref{sec:OV-Staggered-overlap}.

In order to understand the construction of the staggered Wilson term
and its variants, we begin by discussing symmetries of the staggered
fermion action in Sec.\ \ref{sec:SW-Symmetries} and properties of
flavored mass terms in Sec.\ \ref{sec:SW-Flavored-mass-terms}. In
Sec.\ \ref{sec:SW-Adams-mass-term}, we then review Adams' proposal,
followed by Hoelbling's construction in Sec.\ \ref{sec:SW-Hoelblings-mass-term}.
We end this chapter with the discussion of our generalized mass terms
in Sec.\ \ref{sec:SW-Generalized-mass-terms}.

\section{Symmetries of staggered fermions \label{sec:SW-Symmetries}}

In the following we discuss the symmetries of the massless staggered
fermion action
\begin{align}
S & =\frac{1}{2}\sum_{x,\mu}\eta_{\mu}\left(x\right)\left[\bar{\chi}\left(x\right)U_{\mu}\left(x\right)\chi\left(x+a\hat{\mu}\right)-\bar{\chi}\left(x\right)U_{\mu}^{\dagger}\left(x-a\hat{\mu}\right)\chi\left(x-a\hat{\mu}\right)\right]\nonumber \\
 & =\frac{1}{2}\sum_{x,\mu}\eta_{\mu}\left(x\right)\left[\bar{\chi}\left(x\right)U_{\mu}\left(x\right)\chi\left(x+a\hat{\mu}\right)-\bar{\chi}\left(x+a\hat{\mu}\right)U_{\mu}^{\dagger}\left(x\right)\chi\left(x\right)\right].
\end{align}
The action is invariant under the transformations listed below, where
we follow the discussion and notation of Ref.~\citep{Golterman:1984cy}
(cf.\ Ref.~\citep{vandenDoel:1983mf}). We note that in order for
the path integral measure to be invariant under all these transformations,
we consider an even number of lattice sites and (anti-)periodic boundary
conditions.

\paragraph{Shift invariance.}

The transformation is given by
\begin{equation}
\begin{aligned}\chi\left(x\right) & \to\zeta_{\rho}\left(x\right)\chi\left(x+a\hat{\rho}\right),\\
\bar{\chi}\left(x\right) & \to\zeta_{\rho}\left(x\right)\bar{\chi}\left(x+a\hat{\rho}\right),\\
U_{\mu}\left(x\right) & \to U_{\mu}\left(x+a\hat{\rho}\right)
\end{aligned}
\end{equation}
with
\begin{equation}
\zeta_{\mu}\left(x\right)=\left(-1\right)^{\sum_{\nu=\mu+1}^{4}x_{\nu}/a}
\end{equation}
and $\zeta_{4}\left(x\right)=1$.

\paragraph{Rotational invariance.}

The transformation is given by
\begin{equation}
\begin{aligned}\chi\left(x\right) & \to S_{R}\left(R^{-1}x\right)\chi\left(R^{-1}x\right),\\
\bar{\chi}\left(x\right) & \to S_{R}\left(R^{-1}x\right)\bar{\chi}\left(R^{-1}x\right),\\
U\left(x,y\right) & \to U\left(R^{-1}x,R^{-1}y\right),
\end{aligned}
\end{equation}
where $R\equiv R^{\rho\sigma}$ is the rotation $x_{\rho}\to x_{\sigma}$,
$x_{\sigma}\to-x_{\rho}$, $x_{\mu}\to x_{\mu}$ with $\mu\neq\sigma,\rho$.
We also introduced
\begin{equation}
S_{R}\left(R^{-1}x\right)=\frac{1}{2}\left[1\pm\eta_{\rho}\left(x\right)\eta_{\sigma}\left(x\right)\mp\zeta_{\rho}\left(x\right)\zeta_{\sigma}\left(x\right)+\eta_{\rho}\left(x\right)\eta_{\sigma}\left(x\right)\zeta_{\rho}\left(x\right)\zeta_{\sigma}\left(x\right)\right]
\end{equation}
for $\rho\lessgtr\sigma$ and note that $S_{R}\left(R^{-1}x\right)\eta_{\mu}\left(x\right)S_{R}\left(R^{-1}\left[x+a\hat{\mu}\right]\right)=R_{\mu\nu}\eta_{\nu}\left(R^{-1}x\right)$.
Moreover, we make use of the compact notation
\begin{equation}
U\left(x,y\right)=\begin{cases}
U_{\mu}\left(x\right), & y=x+a\hat{\mu},\\
U_{\mu}^{\dagger}\left(y\right), & x=y+a\hat{\mu},
\end{cases}
\end{equation}
for the gauge links.

\paragraph{Axis reversal.}

The transformation is given by
\begin{equation}
\begin{aligned}\chi\left(x\right) & \to\left(-1\right)^{x_{\rho}}\chi\left(I_{\rho}x\right),\\
\bar{\chi}\left(x\right) & \to\left(-1\right)^{x_{\rho}}\bar{\chi}\left(I_{\rho}x\right),\\
U\left(x,y\right) & \to U\left(I_{\rho}x,I_{\rho}y\right),
\end{aligned}
\end{equation}
where $I_{\rho}x$ is the reversal of the $\rho$-axis, i.e.\ $x_{\rho}\to-x_{\rho}$,
$x_{\mu}\to x_{\mu}$ for $\mu\neq\rho$.

\paragraph{$U\left(1\right)$ symmetry.}

The transformation is given by
\begin{equation}
\begin{aligned}\chi\left(x\right) & \to e^{\ii\alpha}\chi\left(x\right),\\
\bar{\chi}\left(x\right) & \to\bar{\chi}\left(x\right)e^{-\ii\alpha}.
\end{aligned}
\end{equation}
This symmetry is associated with the conversation of charge.

\paragraph{$U\left(1\right)_{\epsilon}$ symmetry.}

The transformation is given by
\begin{equation}
\begin{aligned}\chi\left(x\right) & \to e^{\ii\beta\epsilon\left(x\right)}\chi\left(x\right),\\
\bar{\chi}\left(x\right) & \to\bar{\chi}\left(x\right)e^{\ii\beta\epsilon\left(x\right)}.
\end{aligned}
\end{equation}
with
\begin{equation}
\epsilon\left(x\right)=\left(-1\right)^{\sum_{\mu}x_{\mu}/a}.
\end{equation}
This is the remnant chiral symmetry of staggered fermions.

\paragraph{Exchange symmetry.}

The transformation is given by
\begin{equation}
\begin{aligned}\chi\left(x\right) & \to\bar{\chi}^{\intercal}\left(x\right),\\
\bar{\chi}\left(x\right) & \to\chi^{\intercal}\left(x\right),\\
U_{\mu}\left(x\right) & \to U_{\mu}^{\star}\left(x\right),
\end{aligned}
\end{equation}
which interchanges fermion and anti-fermion field.

\paragraph{Remarks.}

For the following discussion, let us also define the parity transformation
as $I_{1}I_{2}I_{3}$ followed by a shift in the $4$-direction. We
note, that in addition to the position-space formulations of these
symmetries, one can also express them elegantly in momentum space.
For brevity we do not quote the momentum representation here explicitly,
but refer the reader to Ref.~\citep{Golterman:1984cy} for details.

\section{Flavored mass terms \label{sec:SW-Flavored-mass-terms}}

Following Refs.~\citep{vandenDoel:1983mf,Golterman:1984cy} closely,
we now discuss flavored mass terms. In the classical continuum limit
we can add a mass term
\begin{equation}
S_{M}^{\mathsf{cont}}=\intop\mathrm{d}^{4}x\,\overline{\Psi}M\Psi
\end{equation}
to the action. We demand that $M$ is trivial in spinor space and
that the action is rotationally invariant. The latter requires the
mass term to commute with $\gamma_{\mu}\gamma_{\nu}$. Using an explicit
spin\,$\otimes$\,flavor-notation, a general flavored mass term
now assumes the structure
\begin{equation}
M=\One\otimes\left[m\One+m_{\mu}\xi_{\mu}+\frac{1}{2}m_{\mu\nu}\sigma_{\mu\nu}+m_{\mu}^{5}\,\ii\xi_{\mu}\xi_{5}+m^{5}\xi_{5}\right],\label{eq:ContMassTerm}
\end{equation}
where $m_{\mu\nu}$ is antisymmetric, $\sigma_{\mu\nu}\equiv\ii\xi_{\mu}\xi_{\nu}$
and the $\xi_{\mu}$ are a representation of the Dirac algebra in
flavor space, i.e.\ $\left\{ \xi_{\mu},\xi_{\nu}\right\} =2\delta_{\mu\nu}\One$.
Furthermore, we require that $M^{2}\geq0$ in the free-field case,
so $M$ is taken to be Hermitian and the mass parameters are real.
We note that in Eq.~\eqref{eq:ContMassTerm} we could in principle
also add similar terms multiplied by a $\gamma_{5}\otimes\xi_{5}$,
but we do not consider them in the following discussion.

Let us now introduce the symmetric shift operator
\begin{equation}
E_{\mu}\left(x,y\right)=\frac{1}{2}\sum_{z}\zeta_{\mu}\left[U_{\mu}\left(z\right)\delta\left(x-z\right)\delta\left(y-z-a\hat{\mu}\right)+U_{\mu}^{\dagger}\left(z\right)\delta\left(y-z\right)\delta\left(x-z-a\hat{\mu}\right)\right],
\end{equation}
which we expect to result in a Hermitian transfer matrix. We can implement
Eq.~\eqref{eq:ContMassTerm} on the lattice as
\begin{align}
S_{M}= & \,a^{4}\sum_{x}m\overline{\chi}\left(x\right)\chi\left(x\right)+\sum_{x,y}m_{\mu}\overline{\chi}\left(x\right)E_{\mu}\left(x,y\right)\chi\left(y\right)\nonumber \\
 & +\frac{\ii}{2}a^{4}\sum_{x,y,z}m_{\mu\nu}\overline{\chi}\left(x\right)E_{\mu}\left(x,y\right)E_{\nu}\left(y,z\right)\chi\left(z\right)\nonumber \\
 & -\frac{\ii}{6}a^{4}\sum_{w,x,y,z}m_{\mu}^{5}\varepsilon_{\mu\alpha\beta\gamma}\overline{\chi}\left(w\right)E_{\alpha}\left(w,x\right)E_{\beta}\left(x,y\right)E_{\gamma}\left(y,z\right)\chi\left(z\right)\nonumber \\
 & -\frac{1}{24}a^{4}\sum_{v,w,x,y,z}m^{5}\varepsilon_{\alpha\beta\gamma\delta}\overline{\chi}\left(w\right)E_{\alpha}\left(v,w\right)E_{\beta}\left(w,x\right)E_{\gamma}\left(x,y\right)E_{\delta}\left(y,z\right)\chi\left(z\right).\label{eq:LatticeMassTerm}
\end{align}
In order to get a better understanding of the mass term in Eq.~\eqref{eq:LatticeMassTerm},
let us discuss which of the symmetries in Sec.\ \ref{sec:SW-Symmetries}
it breaks. First we note that in this general form the mass term is
not invariant under lattice rotations. However, violations of rotational
invariance are expected to be of order $\mathcal{O}\left(a\right)$,
so that the symmetry is eventually restored in the continuum limit.
For parity transformations we find invariance of our mass term, while
for the exchange symmetry $m$, $m_{\mu}$ and $m^{5}$ change signs.

For numerical applications, we require a real determinant and, thus,
we need to retain $\epsilon$ Hermiticity. Hence the flavor structure
of the mass term needs to be restricted to a sum of products of an
even number of $\xi_{\mu}$. Therefore from now on we restrict ourselves
to the case 
\begin{equation}
m_{\mu}=m_{\mu}^{5}=0,
\end{equation}
leaving us with the $m$, $m_{\mu\nu}$ and $m^{5}$ term.

In Sec.\ \ref{sec:SW-Adams-mass-term}, we follow Adams' original
two-flavor proposal and construct a staggered Wilson term using $m$
and $m^{5}$, while in Sec.\ \ref{sec:SW-Hoelblings-mass-term} we
discuss Hoelbling's one flavor construction using $m$ and $m_{\mu\nu}$.
We finish this chapter with generalizations of these flavored mass
terms in Sec.\ \ref{sec:SW-Generalized-mass-terms}.

\section{Adams' mass term \label{sec:SW-Adams-mass-term}}

Adams proposed a particular form of the flavored mass term, giving
rise to the original two-flavor staggered Wilson fermion formulation
introduced in Refs.~\citep{Adams:2009eb,Adams:2010gx}. In the following,
we discuss its construction, interpretation and properties.

\subsection{Construction}

The staggered Wilson action reads
\begin{equation}
S_{\mathsf{sw}}=\bar{\chi}D_{\mathsf{sw}}\chi=\bar{\chi}\left(D_{\mathsf{st}}+m_{\mathsf{f}}\One+W_{\mathsf{st}}\right)\chi\label{eq:AdamsAction}
\end{equation}
with the massless staggered Dirac operator $D_{\mathsf{st}}=\eta_{\mu}\nabla_{\mu}$.
The staggered Wilson term $W_{\mathsf{st}}$ reads
\begin{equation}
W_{\mathsf{st}}=\frac{r}{a}\left(\One-\Gamma_{55}\Gamma_{5}\right)\label{eq:AdamsMassTerm}
\end{equation}
with Wilson-like parameter $r>0$. Following Adams' original notation,
we introduced the operator
\begin{equation}
\Gamma_{55}\chi\left(x\right)=\left(-1\right)^{\sum_{\mu}x_{\mu}/a}\chi\left(x\right),
\end{equation}
where depending on the context we also use $\epsilon\left(x\right)=\left(-1\right)^{\sum_{\mu}x_{\mu}/a}$
for the sign factor. In Eq.~\eqref{eq:AdamsMassTerm} we introduced
\begin{equation}
\Gamma_{5}=\eta_{5}C,\qquad\eta_{5}=\eta_{1}\eta_{2}\eta_{3}\eta_{4},\qquad\eta_{5}\chi\left(x\right)=\left(-1\right)^{\left(x_{1}+x_{3}\right)/a}\chi\left(x\right)
\end{equation}
and the symmetrized product of the $C_{\mu}=\left(T_{\mu+}+T_{\mu-}\right)/2$
operators
\begin{equation}
C=\left(C_{1}C_{2}C_{3}C_{4}\right)_{\mathsf{sym}}\equiv\frac{1}{4!}\sum_{\alpha\beta\gamma\delta}^{\mathsf{sym}}C_{\alpha}C_{\beta}C_{\gamma}C_{\delta}.
\end{equation}
As pointed out earlier, Adams' flavored mass term in Eq.~\eqref{eq:AdamsMassTerm}
preserves $\epsilon$ Hermiticity of the Dirac operator, namely $D_{\mathsf{sw}}^{\dagger}=\epsilon D_{\mathsf{sw}}\epsilon$,
and is itself Hermitian. This implies a non-negative fermion determinant
$\det D_{\mathsf{sw}}\geq0$ for suitable choices of the fermion mass
$m_{\mathsf{f}}$ and ensures the applicability of importance sampling
techniques to Monte Carlo simulations.

\subsection{Interpretation}

To interpret Adams' staggered Wilson term, we note that in the spin$\,\otimes\,$flavor
interpretation \citep{Golterman:1984cy} we find
\begin{equation}
\Gamma_{55}\cong\gamma_{5}\otimes\xi_{5},\qquad\Gamma_{5}\cong\gamma_{5}\otimes\One+\mathcal{O}\left(a^{2}\right).
\end{equation}
While $\Gamma_{5}$ corresponds to the usual $\gamma_{5}$ up to discretization
effects, the $\Gamma_{55}$ is a flavored $\gamma_{5}$ in the sense
that it acts nontrivially in flavor space. As a result, we find for
Adams' staggered Wilson term the spin$\,\otimes\,$flavor structure
\begin{equation}
W_{\mathsf{st}}\cong\frac{r}{a}\One\otimes\left(\One-\xi_{5}\right)+\mathcal{O}\left(a\right).\label{eq:AdamsMassTermInterpretation}
\end{equation}
We note that $W_{\mathsf{st}}$ is of the form of a projector $\mathcal{P}_{\pm}=\left(1\pm\xi_{5}\right)/2$
in flavor space and allows for a simple interpretation. The staggered
Wilson term gives a mass $\mathcal{O}\left(a^{-1}\right)$ to the
two negative flavor-chirality species, similar to the effect the usual
Wilson term has on the doubler modes. In the continuum limit $a\to0$,
the fermion doublers become heavy and decouple. On the other hand,
the two positive flavor-chirality species do not acquire additional
mass contributions. Note that the minus sign in Eq.~\eqref{eq:AdamsMassTerm}
and Eq.~\eqref{eq:AdamsMassTermInterpretation} can be replaced by
a plus sign, thus interchanging the role of positive and negative
flavor-chirality species.

\subsection{Properties}

An important observation is that the equivalent of $\gamma_{5}$,
namely $\Gamma_{5}$, does not square to the identity operator. While
$\gamma_{5}^{2}=\One$ holds exactly, we find $\Gamma_{5}^{2}=\One+\mathcal{O}\left(a^{2}\right)\neq\One$.
This is at the heart of the problem why usual staggered fermions are
not suitable kernel operators for the overlap and domain wall constructions
(see Sec.\ \ref{sec:OV-Staggered-overlap} for a more in-depth discussion).
However, there is another operator which squares exactly to unity,
namely $\Gamma_{55}^{2}=\One$. While $\Gamma_{55}$ has a spin$\,\otimes\,$flavor
interpretation of $\gamma_{5}\otimes\xi_{5}$, for Adams' construction
we project out the positive flavor-chirality species in the continuum
limit and find
\begin{equation}
\Gamma_{55}=\Gamma_{5}+\mathcal{O}\left(a^{2}\right)\cong\gamma_{5}\otimes\One+\mathcal{O}\left(a^{2}\right)
\end{equation}
on the physical species. Therefore on the subspace of these modes
one can use $\Gamma_{55}$ instead of $\Gamma_{5}$. This crucial
insight by Adams allows the construction of staggered overlap and
staggered domain wall fermions, which we discuss in Chapter\ \ref{chap:Overlap-fermions},
\ref{chap:Eigenvalue-spectra} and \ref{chap:Staggered-domain-wall}.

To summarize some interesting properties of two-flavor staggered Wilson
fermions, we recall that compared to usual staggered fermions we are
left with a reduced number of tastes, namely two instead of four (in
$d=4$ dimensions). We also find a smaller fermion matrix and expect
a better condition number compared to Wilson fermions, potentially
increasing computational efficiency. In addition, staggered Wilson
fermions allow the construction of staggered versions of overlap and
domain wall fermions.

On the downside, we point out that the staggered Wilson term $W_{\mathsf{st}}$
breaks a subset of the staggered fermion symmetries. In particular,
the exact flavored chiral symmetry is broken in the massless case,
i.e.\ $\left\{ D_{\mathsf{sw}}^{0},\epsilon\right\} \neq0$. This
gives rise to an additive mass renormalization and there is the need
to fine-tune the bare mass $m_{\mathsf{f}}$ like for Wilson fermions.
Moreover, new fermionic counterterms are allowed \citep{Adams:2010gx,AdamsLat14},
but the only effect is a wave function renormalization for the physical
species. On the physical species the staggered Wilson terms vanished
up to order $\mathcal{O}\left(a\right)$, hence the order $\mathcal{O}\left(a^{2}\right)$
discretization error of usual staggered fermions is lost. Finally,
we note that the $\SUtwo$ vector and chiral symmetries of the two
physical flavors are broken, similarly to the broken $\SUfour$ symmetry
of usual staggered fermions. Nevertheless, the remaining symmetries,
such as the flavored rotation symmetries, are still enough to ensure
e.g.\ a degenerate triplet of pions \citep{SharpeWorkshop}. We also
note that the staggered Wilson term is a four-hop operator, making
it potentially very susceptible to fluctuations of the gauge field
\citep{deForcrand:2011ak,deForcrand:2012bm} and difficult to parallelize.
We discuss these aspects in more detail in Chapter\ \ref{chap:Computational-efficiency}.

\subsection{Aoki phase}

We conclude our discussion of the theoretical properties of staggered
Wilson fermions with the investigations of the Aoki phase \citep{Aoki:1983qi,Aoki:1985mk,Aoki:1986kt,Aoki:1986xr,Aoki:1987us,Creutz:1996bg,Sharpe:1998xm}
carried out in Refs.~\citep{Misumi:2011su,Misumi:2012sp,Nakano:2012wa}.
In these studies, the authors investigated strong-coupling lattice
quantum chromodynamics with staggered Wilson fermions using hopping
parameter expansions and effective potential analyses. The existence
of a non-vanishing pion condensate in certain mass parameter ranges
could be established, while massless pions and PCAC\footnote{PCAC: Partially Conserved Axial Current}
relations were found around a second-order phase boundary. These results
are in support of the idea that, like with Wilson fermions, staggered
Wilson fermions can be applied to lattice quantum chromodynamics by
taking a chiral limit.

\section{Hoelbling's mass term \label{sec:SW-Hoelblings-mass-term}}

One can build upon Adams' idea and construct flavored mass terms to
completely lift the degeneracy of staggered fermions. Originally proposed
by Hoelbling in four dimensions in Ref.~\citep{Hoelbling:2010jw},
we are going to review the construction in the following.

\subsection{Construction}

A possible approach to construct single-flavor staggered fermions
is to start from Adams' operator and add an additional term, which
lifts the degeneracy of species of the same flavor-chirality. A natural
candidate for this is the $m_{\mu\nu}$ term in Eq.~\eqref{eq:LatticeMassTerm}
with a spin$\,\otimes\,$flavor interpretation of $\One\otimes\sigma_{\mu\nu}$.
A possible candidate for a one-flavor action \citep{Hoelbling:2010jw}
is then given by
\begin{equation}
S=\bar{\chi}D\chi\equiv\bar{\chi}\left[D_{\mathsf{st}}+m_{\mathsf{f}}\One+\frac{r}{a}(2\cdot\One+W_{\mathsf{st}}+M_{\mu\nu})\right]\chi,\label{eq:NaiveHoelblingOp}
\end{equation}
where we introduced the mass term $M_{\mu\nu}=\ii\eta_{\mu\nu}C_{\mu\nu}$
(no sum) with operators
\begin{equation}
\begin{aligned}\eta_{\mu\nu}\chi\left(x\right) & =\left(-1\right)^{\Sigma_{\rho=\mu+1}^{\nu}x_{\rho}/a}\chi\left(x\right)\quad\textrm{for}\quad\mu<\nu,\\
\eta_{\mu\nu} & =-\eta_{\nu\mu}\quad\textrm{for}\quad\mu\geq\nu
\end{aligned}
\end{equation}
and $C_{\mu\nu}=\left\{ C_{\mu},C_{\nu}\right\} /2$. Here $W_{\mathsf{st}}$
refers to Adams' staggered Wilson term as discussed in Sec.\ \ref{sec:SW-Adams-mass-term}
and the choice of $\mu\neq\nu$ is arbitrary.

While the Dirac operator in Eq.~\eqref{eq:NaiveHoelblingOp} lifts
the degeneracy of the staggered fermion flavors completely, one can
consider a more symmetric combination of the mass terms $M_{\mu\nu}$,
resulting in better symmetry properties. Hoelbling's proposal \citep{Hoelbling:2010jw}
now takes the form
\begin{equation}
S_{\mathsf{\mathsf{1f}}}=\bar{\chi}D_{\mathsf{1f}}\chi\equiv\bar{\chi}\left[D_{\mathsf{st}}+m_{\mathsf{f}}\One+\frac{r}{a}(2\cdot\One+M_{\mathsf{1f}})\right]\chi\label{eq:HoelblingSymMassAction}
\end{equation}
with the symmetrized flavored mass term
\begin{multline}
M_{\mathsf{1f}}=\left[s_{12}\left(s_{1}s_{2}M_{12}+s_{3}s_{4}M_{34}\right)+s_{13}\left(s_{1}s_{3}M_{13}+s_{4}s_{2}M_{42}\right)\right.\\
\left.+\,s_{14}\left(s_{1}s_{4}M_{14}+s_{2}s_{3}M_{23}\right)\right]/\sqrt{3}\label{eq:HoelblingSymMassTerm}
\end{multline}
and $s_{\mu}=\pm1$, $s_{\mu\nu}=\pm1$ being arbitrary sign factors
\citep{Hoelbling:2010jw}. We note that in numerical applications
a commonly employed form of Eq.~\eqref{eq:HoelblingSymMassTerm}
is given by
\begin{equation}
M_{\mathsf{1f}}=\left(M_{12}+M_{34}+M_{13}-M_{24}+M_{14}+M_{23}\right)/\sqrt{3},
\end{equation}
see Refs.~\citep{deForcrand:2012bm,Durr:2013gp}. As $M_{\mathsf{1f}}$
already lifts the degeneracy completely by itself, there is no need
to introduce a $W_{\mathsf{st}}$ term in Eq.~\eqref{eq:HoelblingSymMassAction}.

When compared to Adams' staggered Wilson operator $W_{\mathsf{st}}$,
the mass term $M_{\mathsf{1f}}$ has a smaller symmetry group. One
can verify the invariance \citep{Hoelbling:2010jw} under diagonal
shifts given by $x\to x\pm\hat{1}\pm\hat{2}\pm\hat{3}\pm\hat{4}$,
shifted axis reversals given by $x_{\mu}\to-x_{\mu}+a\hat{\mu}$ and
double rotations given by $x\to R^{\alpha\beta}R^{\gamma\delta}x$,
where $\alpha$, $\beta$, $\gamma$ and $\delta$ are mutually distinct.

\subsection{Interpretation}

To understand the effect of Hoelbling's flavored mass term, we note
\citep{Hoelbling:2010jw} that
\begin{equation}
\begin{aligned}M_{\mathsf{1f}} & \cong\frac{r}{a}\One\otimes\xi_{\mathsf{f}}+\mathcal{O}\left(a\right),\\
\xi_{\mathsf{f}} & =\diag\left(4,0,2,2\right)\textrm{ or }\diag\left(2,2,4,0\right)
\end{aligned}
\end{equation}
in the spin$\,\otimes\,$flavor interpretation. One can now see that
in the presence of the mass term $M_{\mathsf{1f}}$ three flavors
are lifted to two distinct doublers points, namely two flavors to
$2r/a$ and one to $4r/a$. The remaining physical flavor remains
massless under the action of the mass term, while the doublers decouple
in the continuum limit. Similar to the two flavor case, the staggered
overlap and staggered domain wall constructions can be carried out
for Hoelbling's operator.

\subsection{Properties}

The Dirac operator $D_{\mathsf{1f}}$ breaks several symmetries of
the staggered fermion action. Among them is rotational symmetry, where
only a residual subgroup survives. It turns out that there are no
additional fermionic counterterms in the action, but new gluonic counterterms
appear as pointed out by Sharpe \citep{SharpeWorkshop}. They arise
from the fermion loop contributions to the gluonic two-, three- and
four-point functions, cf.\ Ref.~\citep{AdamsLat14}. This means
that in dynamical simulations one has to include and fine-tune these
terms, severely limiting the practical applicability of single flavor
staggered fermions.

For this reason, starting from Chapter\ \ref{chap:Computational-efficiency},
we restrict ourselves to the case of Adams' original two-flavor staggered
Wilson kernel.

\section{Generalized mass terms \label{sec:SW-Generalized-mass-terms}}

In the following, we want to further generalize the flavored mass
terms discussed in Sec.\ \ref{sec:SW-Adams-mass-term} and Sec.\ \ref{sec:SW-Hoelblings-mass-term}
to allow for arbitrary mass splittings in arbitrary even dimensions.
In this section, we follow our discussion\footnote{Discussion based with permission on \sloppy \bibentry{Hoelbling:2016qfv}.
Copyright 2016 by the American Physical Society.} given in Ref.~\citep{Hoelbling:2016qfv} closely, where we previously
presented the following results.

\subsection{The four-dimensional case}

Up to discretization terms, the staggered Wilson term is a mass term
which is trivial in spin-space, but splits the different staggered
flavors. We require that the determinant of the lattice fermion Dirac
operator is real, to allow the use of importance sampling techniques
in Monte Carlo simulations. Adopting a more convenient notation for
the following discussion, the original proposal by Adams \citep{Adams:2009eb,Adams:2010gx}
can be written as
\begin{equation}
W_{\mathsf{st}}=\frac{r}{a}\left(\One+\Gamma_{1234}C_{1234}\right).\label{eq:AdamsMassTermDim4}
\end{equation}
Here $r>0$ is the Wilson-like parameter and we define the operators
\begin{align}
\Gamma_{1234}\chi\left(x\right) & =\left(-1\right)^{\sum_{\mu}x_{\mu}/a}\chi\left(x\right),\\
C_{1234} & =\eta_{1}\eta_{2}\eta_{3}\eta_{4}\left(C_{1}C_{2}C_{3}C_{4}\right)_{\mathsf{sym}}.
\end{align}
We recall that this term has a spin$\,\otimes\,$flavor interpretation
of the form of Eq.~\eqref{eq:AdamsMassTermInterpretation} and splits
the four flavors of staggered fermions into two pairs with opposite
flavor chirality, that is, the eigenbasis of $\xi_{5}$. For the following
discussion, the notation $A\sim B$ means that $A$ has the spin$\,\otimes\,$flavor
interpretation $B$ up to proportionality.

It is also possible to split the flavors with respect to the eigenbasis
of other elements of the Dirac algebra in flavor space \citep{Golterman:1984cy,Hoelbling:2010jw}.
We restrict the structure of the mass term to a sum of products of
an even number of $\xi_{\mu}$ matrices, so that $\epsilon$ Hermiticity
and the reality of the fermion determinant is retained. In four dimensions
we can, thus, reduce the number of staggered fermion flavors to one
by e.g.
\begin{align}
W_{\mathsf{st}} & =\frac{r}{a}\left(2\cdot\One+W_{\mathsf{st}}^{12}+W_{\mathsf{st}}^{34}\right),\label{eq:DefD4MassTerm}\\
W_{\mathsf{st}}^{\mu\nu} & =\ii\,\Gamma_{\mu\nu}C_{\mu\nu},\label{eq:SingleMassTerm}
\end{align}
where the operators $\Gamma_{\mu\nu}$ and $C_{\mu\nu}$ are defined
as
\begin{align}
\Gamma_{\mu\nu}\chi\left(x\right) & =\varepsilon_{\mu\nu}\left(-1\right)^{\left(x_{\mu}+x_{\nu}\right)/a}\chi\left(x\right),\\
C_{\mu\nu} & =\eta_{\mu}\eta_{\nu}\cdot\frac{1}{2}\left(C_{\mu}C_{\nu}+C_{\nu}C_{\mu}\right)\quad\left(\textrm{no sum}\right).
\end{align}
Here $\varepsilon_{\mu_{1}\cdots\mu_{N}}$ is the totally antisymmetric
Levi-Civita symbol. In order to interpret the mass term defined in
Eq.~\eqref{eq:SingleMassTerm}, we note that
\begin{equation}
\Gamma_{\mu\nu}\sim\gamma_{\mu}\gamma_{\nu}\otimes\xi_{\mu}\xi_{\nu},\qquad C_{\mu\nu}\sim\gamma_{\mu}\gamma_{\nu}\otimes\One,\qquad\epsilon\sim\gamma_{5}\otimes\xi_{5},
\end{equation}
up to discretization terms. Therefore, we find that
\begin{equation}
W_{\mathsf{st}}^{\mu\nu}\sim\One\otimes\sigma_{\mu\nu}+\mathcal{O}\left(a\right)
\end{equation}
with $\sigma_{\mu\nu}=\ii\xi_{\mu}\xi_{\nu}$. We now see that the
number of physical flavors is reduced by $W_{\mathsf{st}}$ to one,
i.e.\ all but a single flavor acquire a mass of $\mathcal{O}\left(a^{-1}\right)$.

\subsection{The $d$-dimensional case}

We can now generalize our constructions to an arbitrary even number
of dimensions $d$. Here we write a single flavor mass term as
\begin{equation}
W_{\mathsf{st}}=\frac{r}{a}\sum_{k=1}^{d/2}\left(\One+W_{\mathsf{st}}^{\left(2k-1\right)\left(2k\right)}\right),\label{eq:DefGeneralMassTerm}
\end{equation}
so that Eq.~\eqref{eq:DefD4MassTerm} follows after the specialization
to $d=4$. We can generalize this mass term even further by introducing
\begin{equation}
W_{\mathsf{st}}^{\mu_{1}\cdots\mu_{2n}}=\ii^{n}\,\Gamma_{\mu_{1}\cdots\mu_{2n}}C_{\mu_{1}\cdots\mu_{2n}},
\end{equation}
for an arbitrary $n\le d/2$, where
\begin{align}
\Gamma_{\mu_{1}\cdots\mu_{2n}}\chi\left(x\right) & =\varepsilon_{\mu_{1}\cdots\mu_{2n}}\left(-1\right)^{\sum_{i=1}^{2n}x_{\mu_{i}}/a}\chi\left(x\right),\\
C_{\mu_{1}\cdots\mu_{2n}} & =\eta_{\mu_{1}}\cdots\eta_{\mu_{2n}}\left(C_{\mu_{1}}\cdots C_{\mu_{2n}}\right)_{\mathsf{sym}}.
\end{align}
The spin$\,\otimes\,$flavor interpretations of these operators read
\begin{align}
\Gamma_{\mu_{1}\cdots\mu_{2n}} & \sim\left(\gamma_{\mu_{1}}\cdots\gamma_{\mu_{2n}}\right)\otimes\left(\xi_{\mu_{1}}\cdots\xi_{\mu_{2n}}\right),\\
C_{\mu_{1}\cdots\mu_{2n}} & \sim\left(\gamma_{\mu_{1}}\cdots\gamma_{\mu_{2n}}\right)\otimes\One,\\
W_{\mathsf{st}}^{\mu_{1}\cdots\mu_{2n}} & \sim\One\otimes\left(\ii^{n}\,\xi_{\mu_{1}}\cdots\xi_{\mu_{2n}}\right),
\end{align}
up to discretization terms. These new mass terms are $\epsilon$ Hermitian
as well, that is
\begin{equation}
M^{\dagger}=\epsilon\,M\epsilon,\qquad M\equiv W_{\mathsf{st}}^{\mu_{1}\cdots\mu_{2n}},
\end{equation}
where
\begin{equation}
\epsilon\left(x\right)=\left(-1\right)^{\sum_{\mu=1}^{d}x_{\mu}/a}.\label{eq:EpsilonDDim}
\end{equation}
This allows us to introduce a very general mass term of the form
\begin{equation}
W_{\mathsf{st}}=\sum_{n=1}^{d/2}\sum_{\boldsymbol{\mu}_{n}}\frac{r_{\boldsymbol{\mu}_{n}}}{a}\left(\One+W_{\mathsf{st}}^{\boldsymbol{\mu}_{n}}\right)\label{eq:DefMoreGeneralMassTerm}
\end{equation}
with generalized Wilson-parameters $r_{\boldsymbol{\mu}_{n}}\geq0$.
Here the sum is over all multi-indices $\boldsymbol{\mu}_{n}=\left(\mu_{1},\dots,\mu_{2n}\right)$
with $1\leq\mu_{i}\leq d$ for all $i$ with $1\leq i\leq2n$. We
remark, that in Eq.~\eqref{eq:DefMoreGeneralMassTerm} not all of
the possible combinations of mass terms are useful in practical applications.
To reproduce Adams' staggered Wilson term in $d=4$ dimensions as
given in Eq.~\eqref{eq:AdamsMassTermDim4}, we set $r_{1234}=r>0$
and $r_{\boldsymbol{\mu}_{n}}=0$ otherwise.

In Chapter\ \ref{chap:Staggered-domain-wall} we deal with the $d=2$
case. We note that here the definition is essentially unique and the
mass term takes the form
\begin{equation}
W_{\mathsf{st}}=\frac{r}{a}\left(\One+W_{\mathsf{st}}^{12}\right),\label{eq:StagWilsonDiracTwoDim}
\end{equation}
which reduces the number of staggered flavors from two to one.

We note that all possible $W_{\mathsf{st}}$ terms, like the Wilson
term $W_{\mathsf{w}}$, break chiral symmetry. Moreover, if too many
of the symmetries of staggered fermions are broken, some mass terms
may give rise to additional counterterms, cf.\ Ref.~\citep{SharpeWorkshop}.

\chapter{Computational efficiency \label{chap:Computational-efficiency}}

Staggered Wilson fermions and usual Wilson fermions share several
technical properties. Both are constructed by adding a momentum-dependent
term to a fermion action suffering from fermion doubling. This additional
term breaks chiral symmetry explicitly, resulting in an additive mass
renormalization and the need for fine-tuning the bare mass parameter.

Besides the technical aspects, comparing the computational properties
of different lattice fermion actions is important for practical applications.
As the accuracy of physical predictions in Monte Carlo simulations
is limited by the available computational resources, the question
of computational efficiency is then of high importance. In this chapter,
we deal with the problem of quantifying and comparing the computational
efficiency of two-flavor staggered Wilson fermions with the one of
usual Wilson fermions, where some of the results\footnote{Parts of the discussion based on and some figures reprinted from \sloppy
\bibentry{Adams:2013tya}. Copyright owned by the authors under the
terms of the Creative Commons Attribution-NonCommercial-NoDerivatives
4.0 International License (CC BY-NC-ND 4.0).} were previously presented in Refs.~\citep{Adams:2013tya,ZielinskiLat14,ZielinskiICCP9}.
We try to answer the question if staggered Wilson fermions have a
computational advantage, potentially bringing down the enormous costs
of state-of-the-art numerical simulations.

\section{Introduction}

Staggered Wilson fermions, like usual staggered fermions, are formulated
using one-component spinor fields. One might, thus, hope for a better
computational efficiency of staggered Wilson fermions compared to
Wilson fermions due to a smaller fermion matrix and a reduced condition
number. The prospect of a kernel operator with increased computational
efficiency is especially interesting in the context of the overlap
and domain wall fermion formulations. These formulations respect chiral
symmetry, but come at the price of high computational costs. This
limits their practical applicability despite many attractive theoretical
properties.

In this chapter, we evaluate the computational efficiency of Adams'
two-flavor staggered Wilson fermions. As the performance of overlap
and domain wall fermions is tightly connected to the computational
properties of their kernel operator, this is also interesting with
respect to our discussion of staggered overlap fermions in Chapter\ \ref{chap:Overlap-fermions}
and \ref{chap:Eigenvalue-spectra} and staggered domain wall fermions
in Chapter\ \ref{chap:Staggered-domain-wall}. We follow our report
given in Ref.~\citep{Adams:2013tya} together with the updates given
in Refs.~\citep{ZielinskiLat14,ZielinskiICCP9}, where we discuss
our results for the computational efficiency in a well-defined benchmark
setting.

We compare the computational costs of inverting the Dirac matrix on
a source for staggered Wilson fermions and usual Wilson fermions in
quenched lattice quantum chromodynamics. We do this while either keeping
the physical volume or the lattice spacing fixed. These inversions
are typically the most time-consuming part in realistic simulations
and are, thus, expected to be a good proxy for studying the computational
performance. The resulting large and sparse linear systems are solved
using iterative methods, such as the conjugate gradient method \citep{Lanczos:1950zz,hestenes1952methods}.
We characterize the computational efficiency as the ratio of the computation
time for both fermion formulations. For this comparison to be meaningful,
the benchmarking has to be done at fixed values of some physical quantity,
where in general the measured efficiency depends on this fixed quantity.
A natural candidate for this is the pion mass $m_{\pi}$, which is
easy to measure and tune with high accuracy.

This chapter is organized as follows. In Sec.\ \ref{sec:CE-Theoretical-estimates},
we discuss our strategy to quantify the computational efficiency of
staggered Wilson fermions and derive some theoretical estimates. In
Sec.\ \ref{sec:CE-Numerical-study}, we present our numerical study
and critically discuss our results. In Sec.\ \ref{sec:CE-Arithmetic-intensities},
we characterize the memory bandwidth requirements of staggered Wilson
fermions by analyzing the arithmetic intensities of several lattice
fermion formulations. Finally, in Sec.\ \ref{sec:CE-Conclusions}
we summarize and conclude our study.

\section{Theoretical estimates \label{sec:CE-Theoretical-estimates}}

As we already pointed out, the computational costs of realistic simulations
are dominated by inverting the Dirac operator $D\equiv D\left(m_{\mathsf{q}}\right)$
on a source $\chi$, namely solving the linear system $D\Psi=\chi$.
As these linear systems are usually very large, one uses iterative
solvers to find a numerical solution for $\Psi$. A traditional approach
to this is the use of the conjugate gradient method for the normal
equation, where one solves the system
\begin{equation}
D^{\dagger}D\,\Psi=\chi^{\prime}\label{eq:NormalEq}
\end{equation}
with $\chi^{\prime}\equiv D^{\dagger}\chi$ instead. We form these
normal equations with the Hermitian and positive (semi-)definite operator
$D^{\dagger}D$ to ensure the applicability of the conjugate gradient
method, see Ref.~\citep{shewchuk1994introduction}.

In the following, we want to derive a theoretical estimate for solving
Eq.~\eqref{eq:NormalEq} with the conjugate gradient method. First
we note that the cost can be characterized by
\begin{equation}
\mathrm{cost}=\left(\textrm{\#iterations}\right)\times(\textrm{cost per iteration}).
\end{equation}
For most iterative methods, such as the conjugate gradient method,
the cost per iteration is dominated by the cost of the matrix-vector
multiplications. This gives us
\begin{equation}
\mathrm{cost}\propto\left(\textrm{\#iterations}\right)\times\left(\textrm{cost mat.-vec.\,mult.}\right).
\end{equation}
If we now take the ratio for the computational costs of Wilson and
staggered Wilson fermions, we find 
\begin{equation}
\frac{\mathrm{cost}_{\mathsf{w}}}{\mathrm{cost}_{\mathsf{sw}}}=\frac{\left(\textrm{\#iterations}\right)_{\mathsf{w}}}{\left(\textrm{\#iterations}\right)_{\mathsf{sw}}}\times\frac{\left(\textrm{cost mat.-vec.\,mult.}\right)_{\mathsf{w}}}{\left(\textrm{cost mat.-vec.\,mult.}\right)_{\mathsf{sw}}}.\label{eq:CostRatio}
\end{equation}
When considering the fermion propagator, an additional factor of four
should be included on the right hand side because in the staggered
Wilson case one has four times fewer sources due to a reduced number
of spinor components. To get an estimate for the overall cost ratio
of both fermion formulations, we have to estimate both ratios on the
right hand side of Eq.~\eqref{eq:CostRatio}. The first ratio in
general depends on the iterative solver and the fixed physical quantity,
in our case the pion mass $m_{\pi}$. The second ratio depends on
size, structure and sparsity of the corresponding fermion matrices.

\subsection{Number of iterations}

We begin by further analyzing the first ratio in Eq.~\eqref{eq:CostRatio}.
When using the conjugate gradient method for a specified small residual
$\varepsilon$, we make use of the relation
\begin{equation}
\textrm{\#iterations}\propto\sqrt{\kappa}\log\left(\frac{2}{\varepsilon}\right),
\end{equation}
where $\kappa=\left|\lambda_{\mathsf{max}}/\lambda_{\mathsf{min}}\right|$
is the condition number of $D^{\dagger}D$, see e.g.\ Ref.~\citep{shewchuk1994introduction}.
Hence one can approximate the ratio for the number of iterations by
\begin{equation}
\frac{\left(\textrm{\#iterations}\right)_{\mathsf{w}}}{\left(\textrm{\#iterations}\right)_{\mathsf{sw}}}\approx\sqrt{\frac{\kappa_{\mathsf{w}}}{\kappa_{\mathsf{sw}}}},\label{eq:CondNumberRatio}
\end{equation}
when keeping $\varepsilon$ fixed. To verify this, we also numerically
determine this ratio. In the free-field case the ratio in Eq.~\eqref{eq:CondNumberRatio}
can be computed analytically for the same bare mass $m_{\mathsf{q}}$
for both formulations. We find 
\begin{equation}
\lim_{m_{\mathsf{q}}\to0}\sqrt{\frac{\kappa_{\mathsf{w}}^{0}}{\kappa_{\mathsf{sw}}^{0}}}=\frac{8}{\sqrt{5}}\approx3.6,\label{eq:CondNumberRatioEstimate}
\end{equation}
where $\kappa^{0}\equiv\kappa^{0}\left(m_{\mathsf{q}}\right)$ refers
to the free-field condition number at bare mass $m_{\mathsf{q}}$.

\subsection{Cost of matrix-vector multiplications}

Let us now analyze the second ratio on the right hand side of Eq.~\eqref{eq:CostRatio}.
We estimate the costs for the matrix-vector multiplications from FLOP
(floating-point operations) counts. We find
\begin{align}
\frac{\left(\textrm{cost mat.-vec.\,mult.}\right)_{\mathsf{w}}}{\left(\textrm{cost mat.-vec.\,mult.}\right)_{\mathsf{sw}}} & \approx\frac{\left(\textrm{FLOPs}\right)_{\mathsf{w}}}{\left(\textrm{FLOPs}\right)_{\mathsf{sw}}},\nonumber \\
 & \approx\frac{4\times\left(\textrm{FLOPs/site}\right)_{\mathsf{w}}}{\left(\textrm{FLOPs/site}\right)_{\mathsf{sw}}}\nonumber \\
 & =\frac{4\times\unit[1392]{FLOPs/site}}{\unit[1743]{FLOPs/site}}\nonumber \\
 & \approx3.2.\label{eq:FLOPRatio}
\end{align}
Here the factor of four appears due to the four-component spinors
in Wilson's formulation, $1392$ is the number of FLOPs per site needed
for the application of Wilson's Dirac operator and $1743$ is the
number of FLOPs per site for the staggered Wilson Dirac operator.
Note that Eq.~\eqref{eq:FLOPRatio} is an estimate and besides the
number of floating point operations also other factors come into play
such as the computational intensities, which we discuss in Sec.\ \ref{sec:CE-Arithmetic-intensities}.

Alternatively one can estimate the ratio in Eq.~\eqref{eq:FLOPRatio}
with the approach of Ref.~\citep{deForcrand:2011ak}, which we now
briefly review. In practice Wilson fermions are used with the spin-projection
trick \citep{Alford:1996nx}, effectively reducing the number of spinor
components from four to two. The Wilson action couples each lattice
site to $2\times8$ other sites, where the factor of two represents
the effective number of spinor components and $8=4\times2$ comes
from a forward and a backward hop in each space-time direction. Staggered
Wilson fermions are formulated with one-component spinors, but the
staggered Wilson term couples each site to all $2^{4}=16$ corners
of a four-dimensional hypercube. This means that the staggered Wilson
action couples each lattice site to $8+16$ nearby sites. We can now
alternatively estimate the ratio by
\begin{equation}
\frac{4\times(8\times2)}{8+16}\approx2.7,
\end{equation}
where we included the above-mentioned factor of four. This estimate
is smaller than the one we gave in Eq.~\eqref{eq:FLOPRatio} as we
neglected here the cost of spin decomposition and reconstruction for
the spin projection trick.

In summary we conservatively estimate 
\begin{equation}
\frac{\left(\textrm{cost mat.-vec.\,mult.}\right)_{\mathsf{w}}}{\left(\textrm{cost mat.-vec.\,mult.}\right)_{\mathsf{sw}}}\approx2-3\label{eq:CostMatVev}
\end{equation}
for the cost-ratio for the matrix-vector multiplications, in agreement
with Ref.~\citep{deForcrand:2012bm}.

\section{Numerical study \label{sec:CE-Numerical-study}}

\begin{figure}[t]
\begin{centering}
\includegraphics[width=0.8\columnwidth]{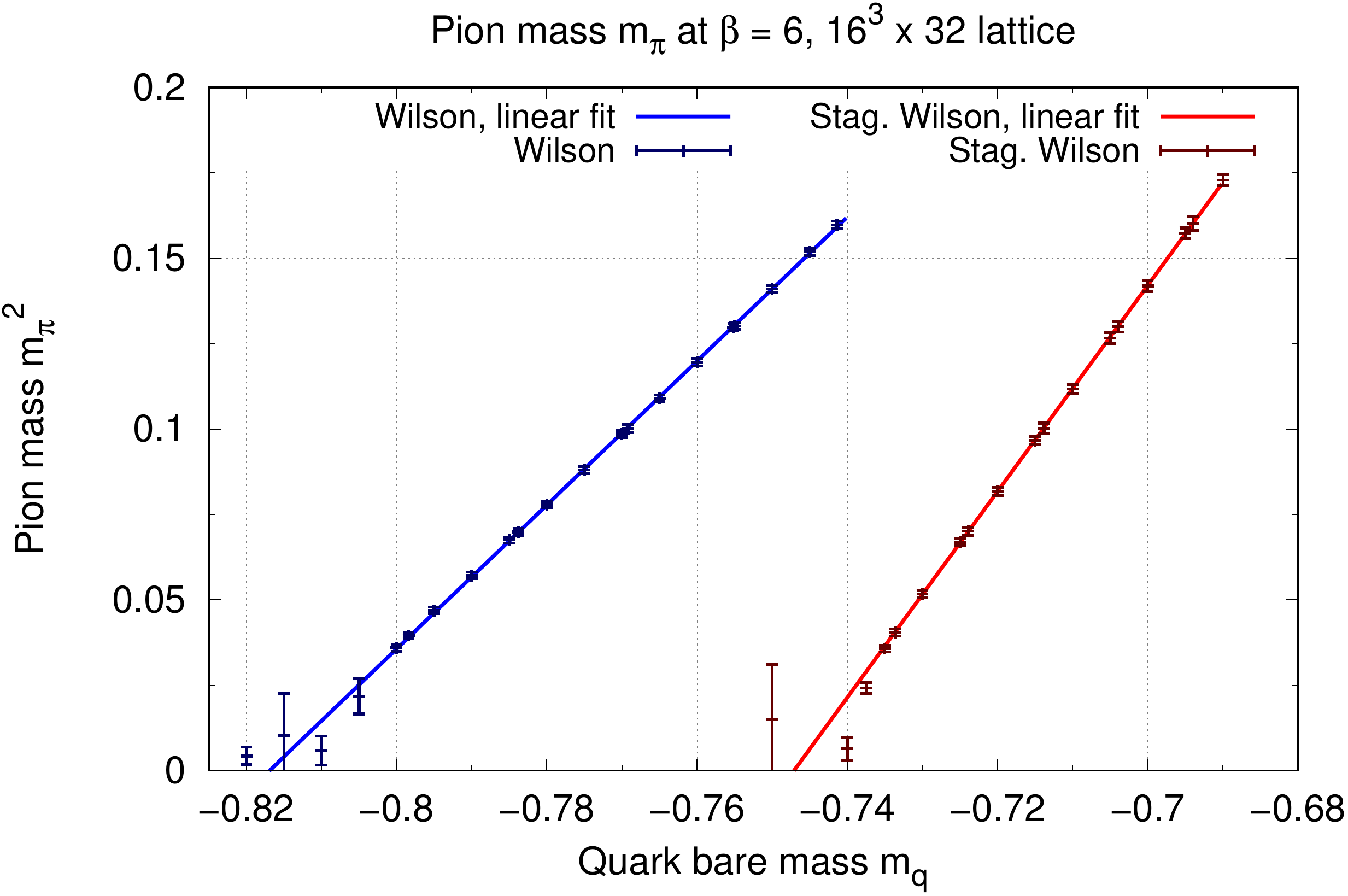}
\par\end{centering}
\caption{Squared pion mass $m_{\pi}^{2}$ as a function of the bare quark mass
$m_{\mathsf{q}}$ with a linear fit through the data points with $0.05\leq m_{\pi}^{2}$.
\label{fig:Pion-mass-1632-b6}}
\end{figure}
\begin{figure}[t]
\begin{centering}
\includegraphics[width=0.8\columnwidth]{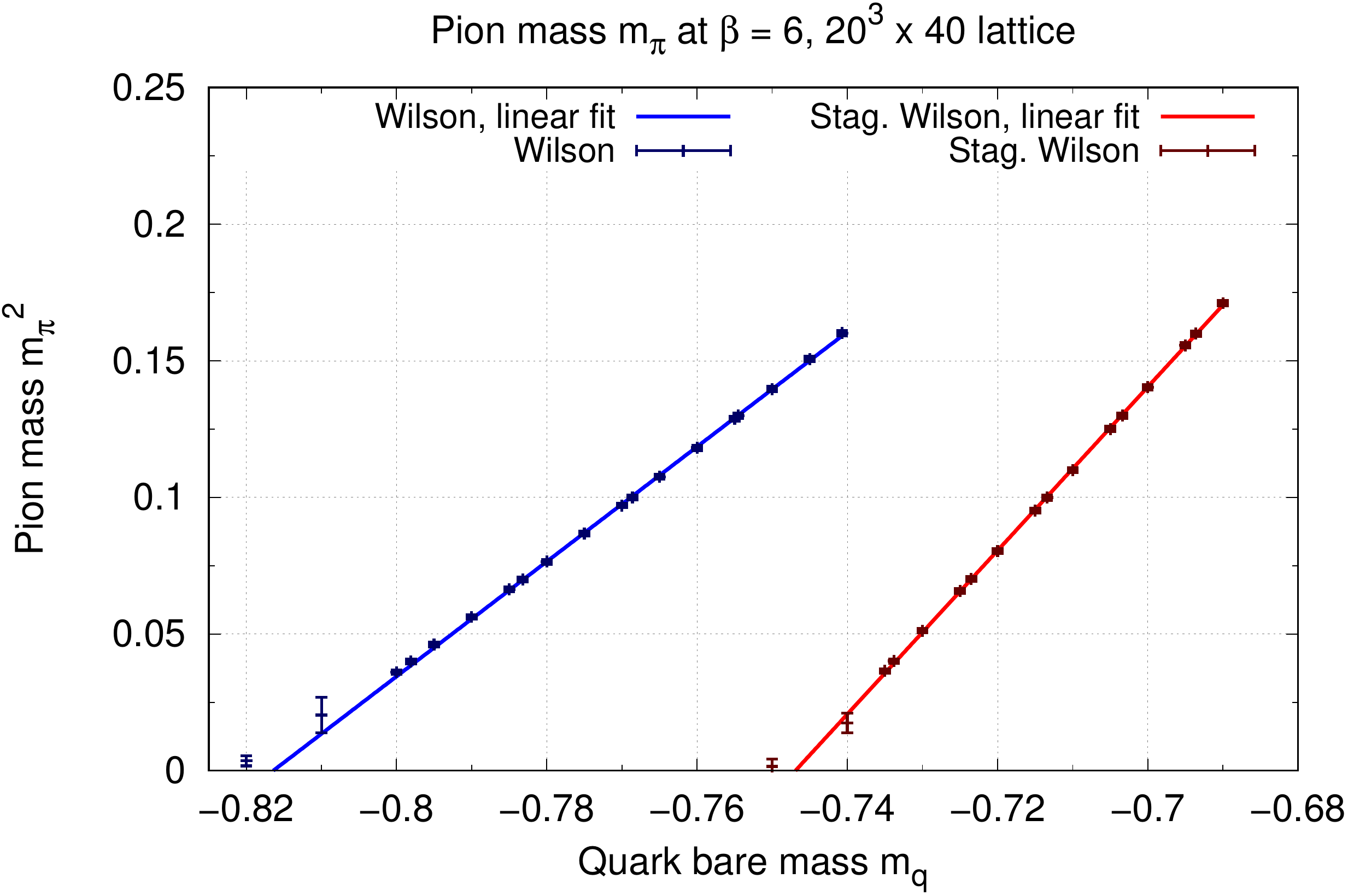}
\par\end{centering}
\caption{Squared pion mass $m_{\pi}^{2}$ as a function of the bare quark mass
$m_{\mathsf{q}}$ with a linear fit through the data points with $0.05\leq m_{\pi}^{2}$.
\label{fig:Pion-mass-2040-b6}}
\end{figure}
\begin{figure}[t]
\begin{centering}
\includegraphics[width=0.8\columnwidth]{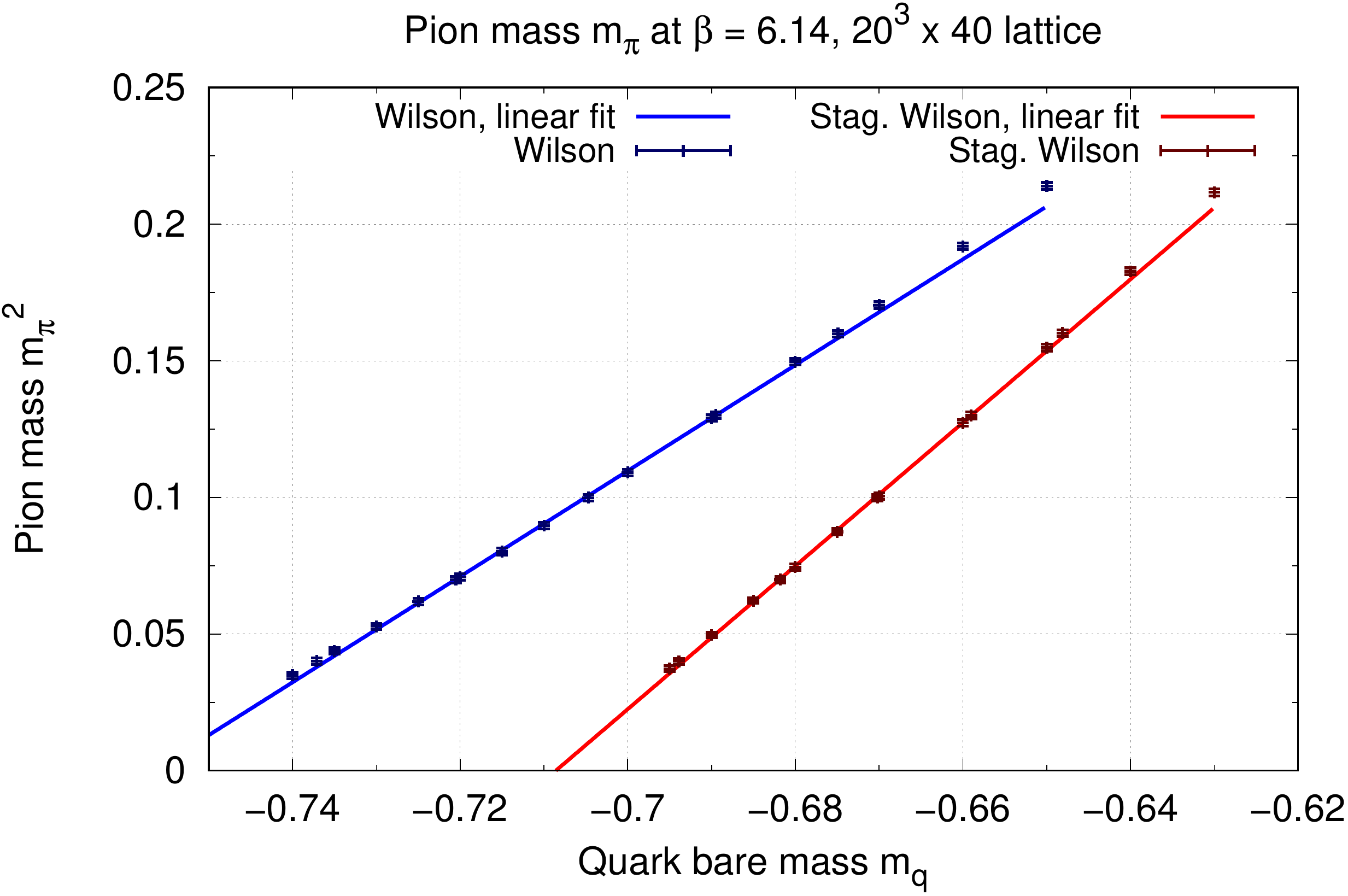}
\par\end{centering}
\caption{Squared pion mass $m_{\pi}^{2}$ as a function of the bare quark mass
$m_{\mathsf{q}}$ with a linear fit through the data points with $0.05\leq m_{\pi}^{2}\leq0.15$.
\label{fig:Pion-mass-2040-b614}}
\end{figure}
\begin{figure}[t]
\begin{centering}
\includegraphics[width=0.7\columnwidth]{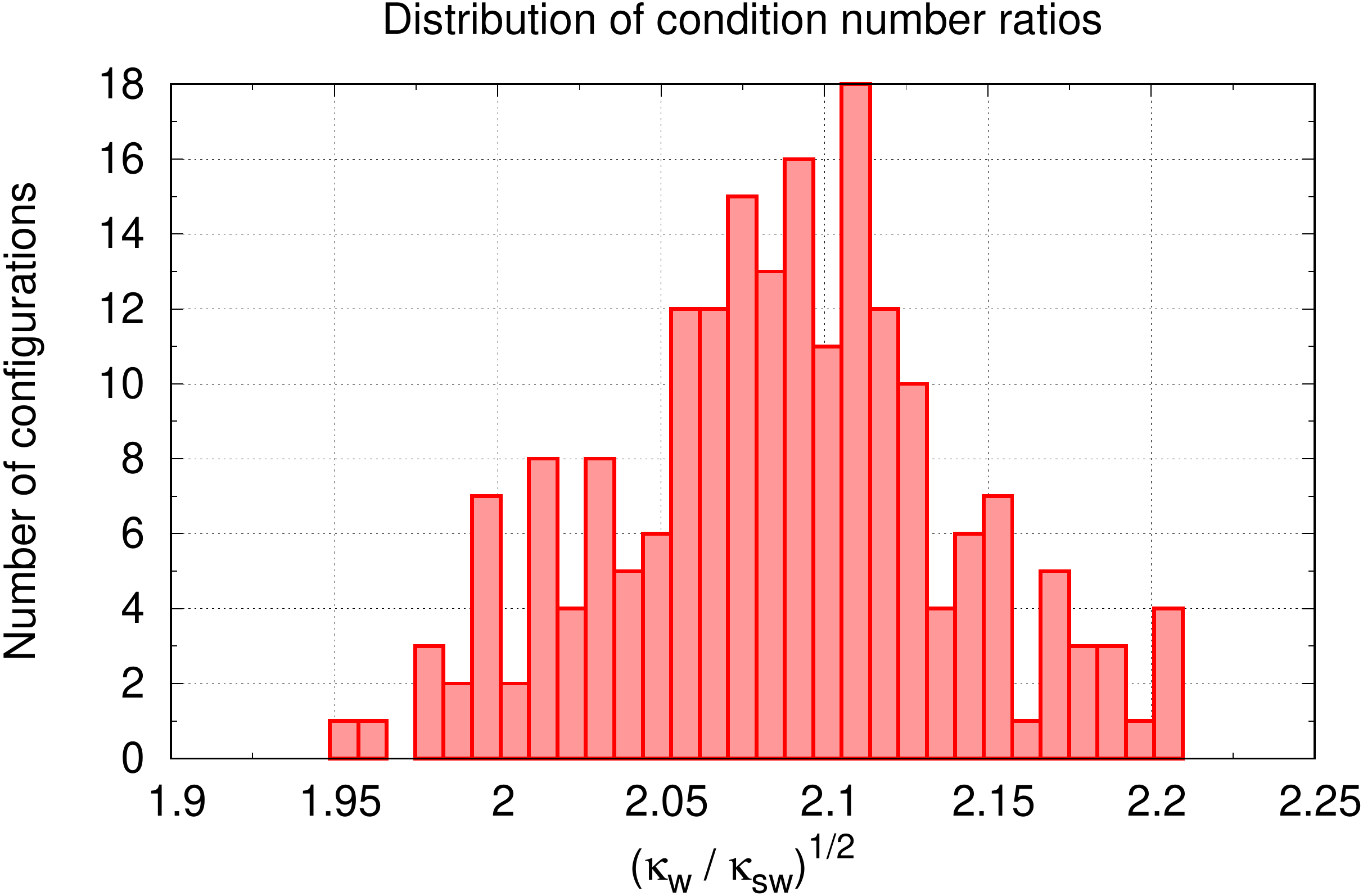}
\par\end{centering}
\caption{Distribution of $\sqrt{\kappa_{\mathsf{w}}/\kappa_{\mathsf{sw}}}$
for $m_{\pi}^{2}=0.10$ on the $20^{3}\times40$ lattice at $\beta=6$.
\label{fig:CondNumberRatio}}
\end{figure}
\begin{figure}[t]
\begin{centering}
\subfloat[Point sources]{\begin{centering}
\includegraphics[width=0.47\columnwidth]{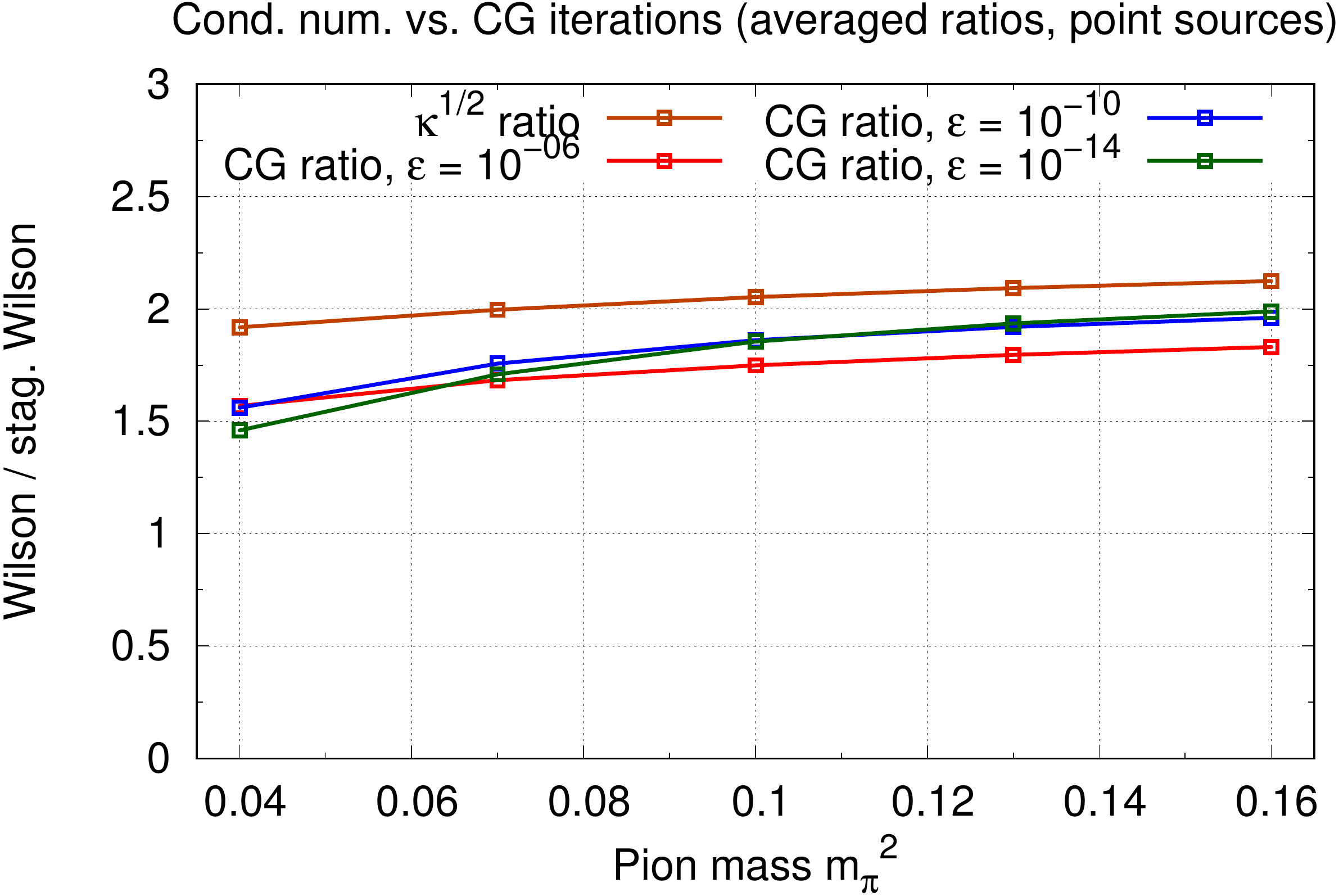}
\par\end{centering}
}\hfill{}\subfloat[Wall sources]{\begin{centering}
\includegraphics[width=0.47\columnwidth]{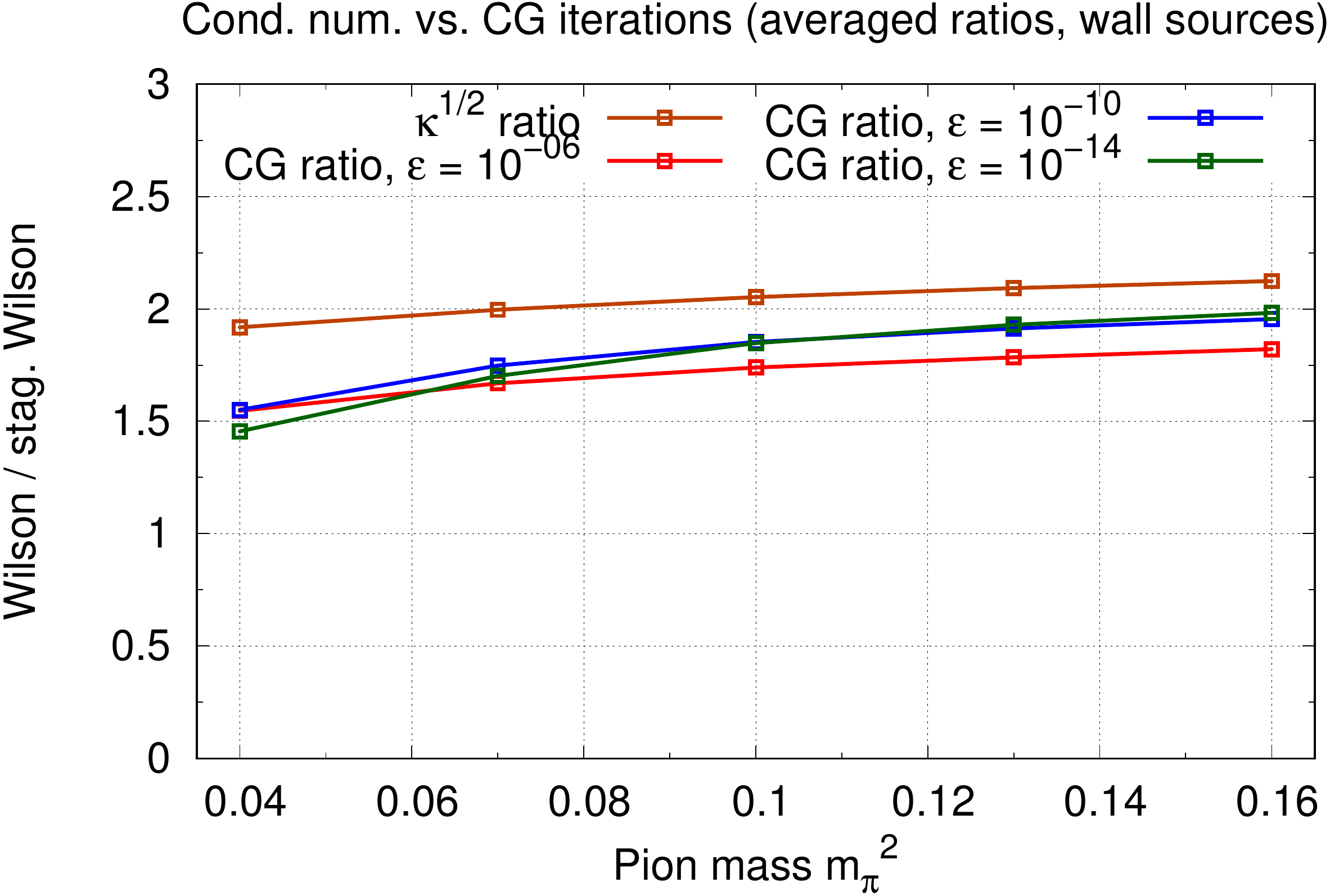}
\par\end{centering}
}
\par\end{centering}
\caption{Averages for the ratios of the number of iterations and $\sqrt{\kappa_{\mathsf{w}}/\kappa_{\mathsf{sw}}}$
as a function of the squared pion mass $m_{\pi}^{2}$ on a $16^{3}\times32$
lattice at $\beta=6$. \label{fig:IterationRatio1632b6}}
\end{figure}
\begin{figure}[t]
\begin{centering}
\subfloat[Point sources]{\begin{centering}
\includegraphics[width=0.47\columnwidth]{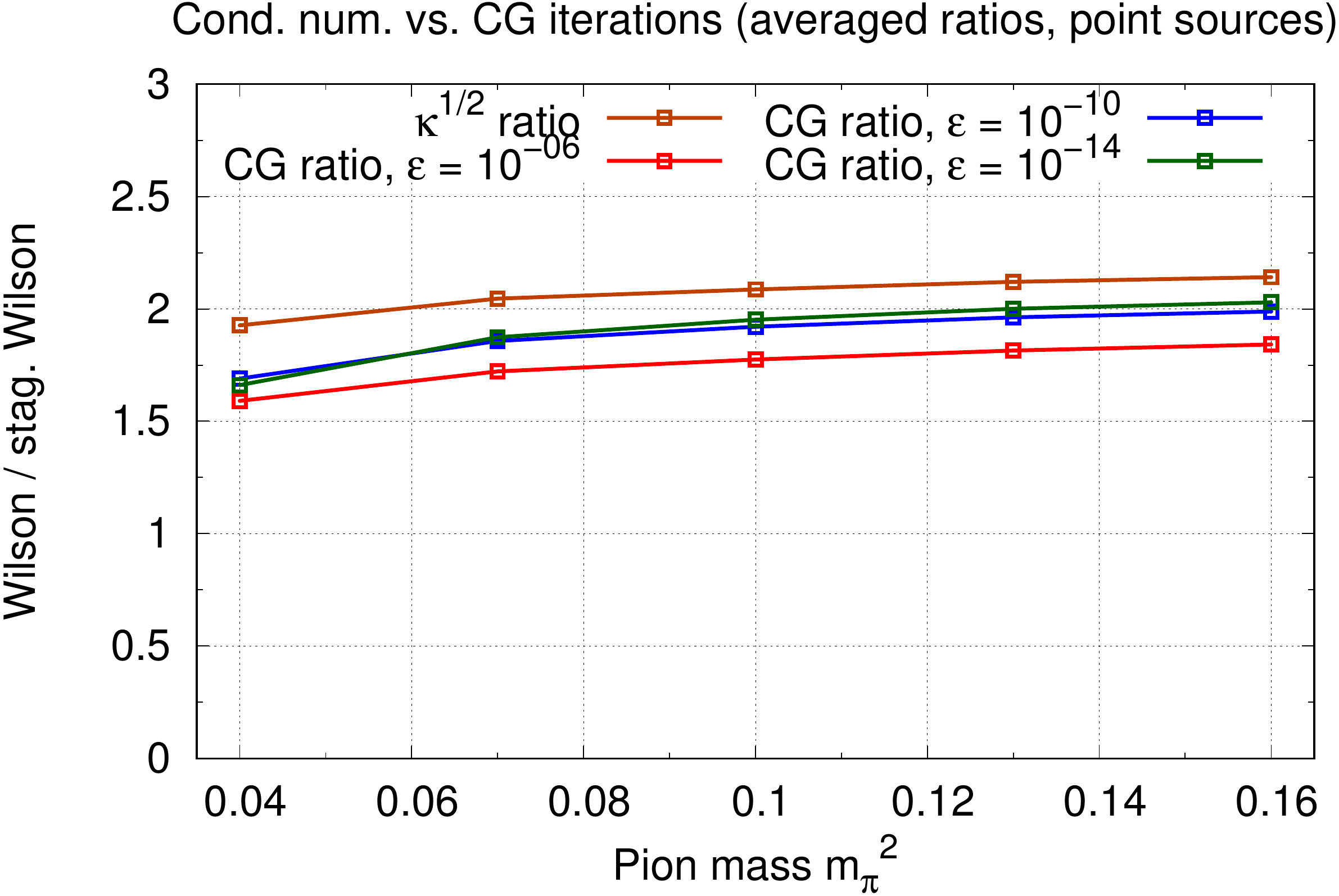}
\par\end{centering}
}\hfill{}\subfloat[Wall sources]{\begin{centering}
\includegraphics[width=0.47\columnwidth]{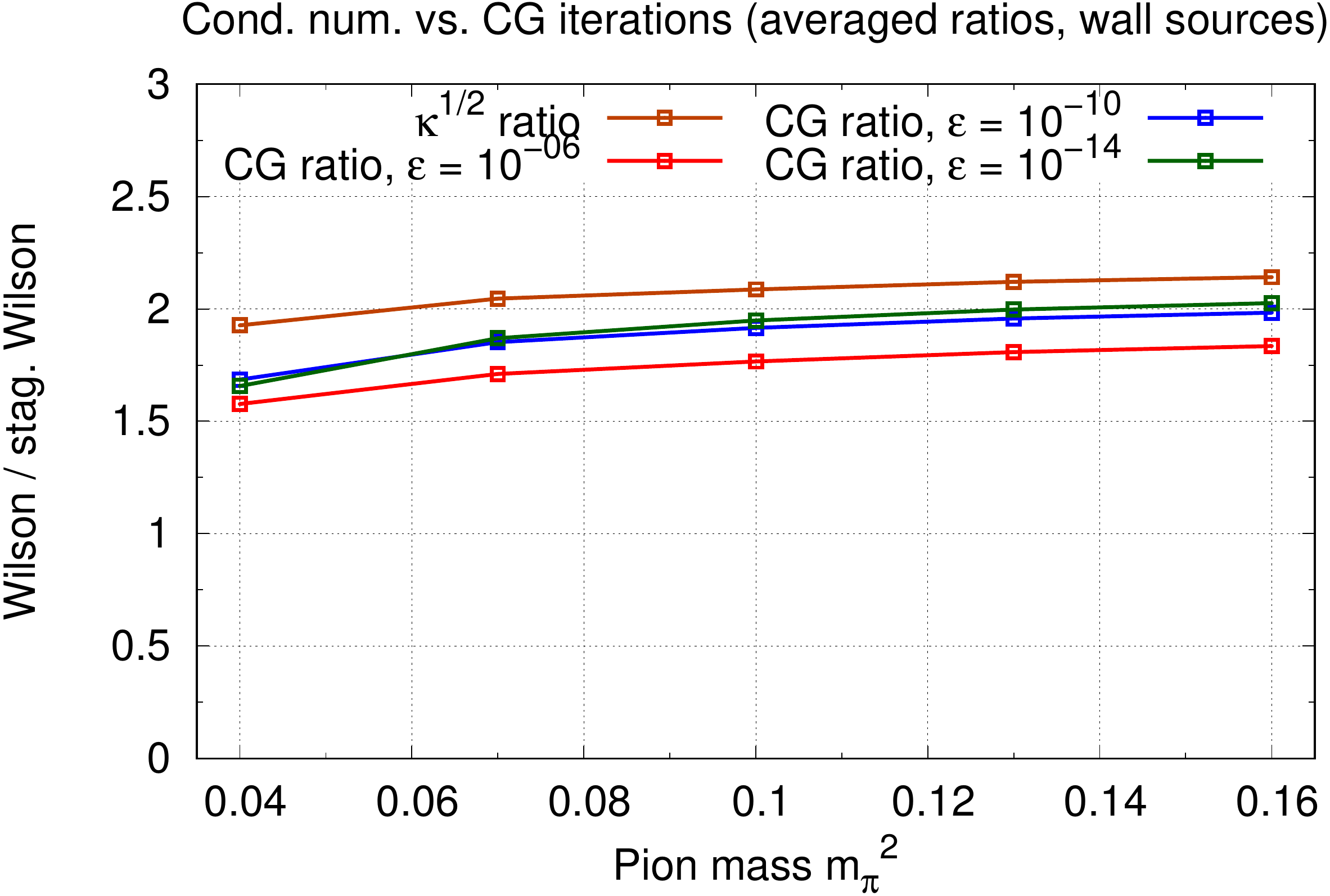}
\par\end{centering}
}
\par\end{centering}
\caption{Averages for the ratios of the number of iterations and $\sqrt{\kappa_{\mathsf{w}}/\kappa_{\mathsf{sw}}}$
as a function of the squared pion mass $m_{\pi}^{2}$ on a $20^{3}\times40$
lattice at $\beta=6$. \label{fig:IterationRatio2040b6}}
\end{figure}
\begin{figure}[t]
\begin{centering}
\subfloat[Point sources]{\begin{centering}
\includegraphics[width=0.47\columnwidth]{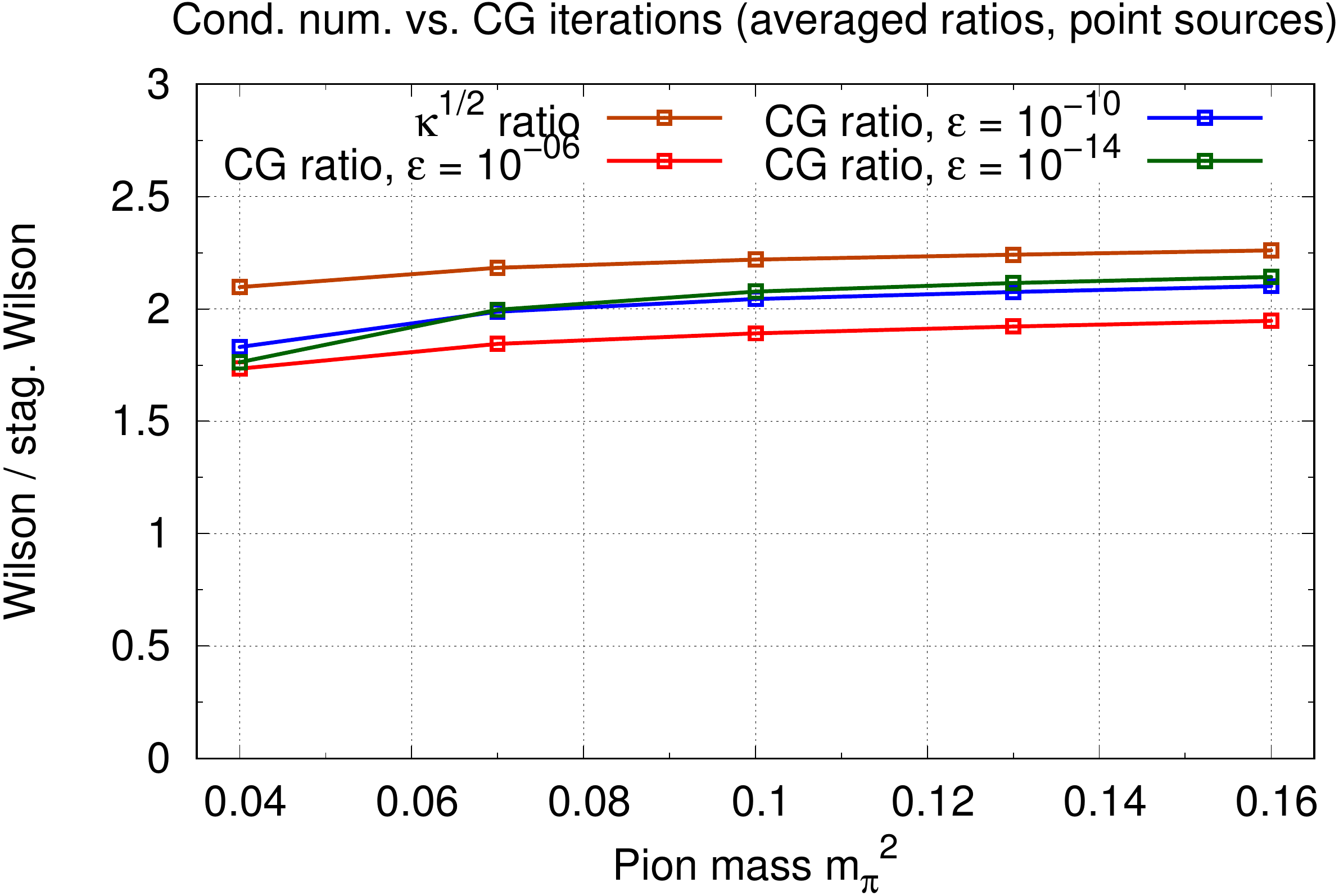}
\par\end{centering}
}\hfill{}\subfloat[Wall sources]{\begin{centering}
\includegraphics[width=0.47\columnwidth]{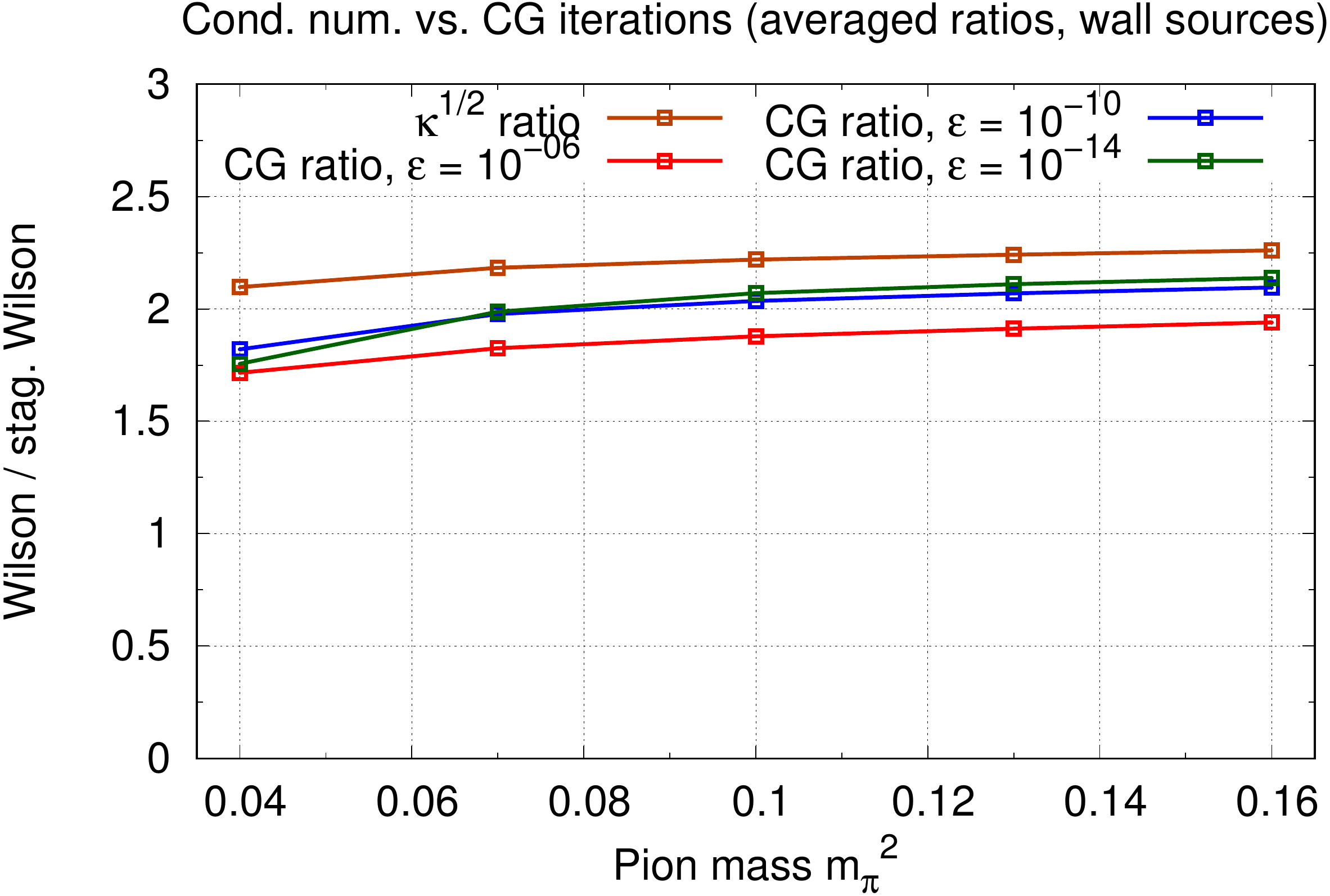}
\par\end{centering}
}
\par\end{centering}
\caption{Averages for the ratios of the number of iterations and $\sqrt{\kappa_{\mathsf{w}}/\kappa_{\mathsf{sw}}}$
as a function of the squared pion mass $m_{\pi}^{2}$ on a $20^{3}\times40$
lattice at $\beta=6.136716$. \label{fig:IterationRatio2040b614}}
\end{figure}

We did a numerical study with quenched QCD configurations on a $16^{3}\times32$
lattice at $\beta=6$, a $20^{3}\times40$ lattice at $\beta=6$ and
a $20^{3}\times40$ lattice at $\beta=6.136716$ with $200$ configurations
each. The $20^{3}\times40$ lattices are chosen in such a way to either
keep the lattice spacing ($\beta=6$) or the physical volume ($\beta=6.136716$)
fixed compared to the $16^{3}\times32$ lattice. For the simulations
we used the Chroma/QDP software package \citep{Edwards:2004sx}, which
we extended by implementing staggered Wilson fermions.

While the use of preconditioning can speed up the use of Wilson fermions
by roughly a factor of two and preconditioning is in principle also
possible for staggered Wilson fermions, our study here deals with
the unpreconditioned base case.

\subsection{Results \label{subsec:CE-Comp-eff-results}}

In Figs.~\ref{fig:Pion-mass-1632-b6}, \ref{fig:Pion-mass-2040-b6}
and \ref{fig:Pion-mass-2040-b614}, we find the squared pion mass
$m_{\pi}^{2}$ as a function of the bare quark mass $m_{\mathsf{q}}$
for the different ensembles. The relation between the pion mass and
bare quark mass for staggered Wilson fermions was previously also
presented in Refs.~\citep{AdamsWorkshop,Adams:2013tya}.

For the determination of the pion mass we follow the well-known standard
procedures, see Ref.~\citep{DeGrand:2006zz}. By inspecting the effective
mass plot of the propagator, we fix a suitable fit range where the
contributions of excited states is negligible and the signal-to-noise
ratio is acceptable. We then do an uncorrelated fit of the pion propagator
over the previously determined range using the function
\begin{equation}
p\left(n_{t}\right)=A\cosh\left(m_{\pi}\left[n_{t}-\frac{1}{2}N_{t}\right]\right)
\end{equation}
by minimization of $\chi^{2}$, where $t=n_{t}a$ and the fit quality
varies in the range $0.01\lesssim\chi/\textrm{d.o.f.}\lesssim0.75$.
Here the fit parameter $A$ is the amplitude, $m_{\pi}$ the pion
mass and $N_{t}\in\left\{ 32,40\right\} $ is the extent of the corresponding
lattice in temporal direction. The errors are combined statistical
and systematic errors and are determined with the jackknife method
\citep{quenouille1949,quenouille1956,tukey1958bias}, where we estimated
the systematic error by varying the range of the fit.

In agreement with chiral perturbation theory, we find an approximately
linear relationship between $m_{\pi}^{2}$ and $m_{\mathsf{q}}$.
Moreover, the lowest achievable pion mass for Wilson and staggered
Wilson fermions is comparable. We also note that the additive mass
renormalization is smaller for staggered Wilson fermions compared
to Wilson fermions. This could be a (somewhat weak) indicator for
a computational advantage in the construction of overlap and domain
wall fermions. Knowing the relation between the pion mass and the
bare quark mass, for a given value of $m_{\pi}$ we can use the respective
bare mass $m_{\mathsf{q}}$ for Wilson and staggered Wilson fermions.
In the following, we fix five values of the pion mass, namely $m_{\pi}^{2}\in\left\{ 0.04,\,0.07,\,0.10,\,0.13,\,0.16\right\} $,
as in general the efficiency is a function of the fixed physical parameter.

Using Chroma, we calculated the lowest and highest eigenvalue on each
configurations to determine the conditions number $\kappa$ of $D^{\dagger}D$.
With respect to Eq.~\eqref{eq:CondNumberRatio}, the ratio of the
condition numbers is an important factor in judging the computational
efficiency of staggered Wilson fermions. In Fig.~\ref{fig:CondNumberRatio},
we give an example for the distribution of $\sqrt{\kappa_{\mathsf{w}}/\kappa_{\mathsf{sw}}}$
on one of our ensembles. As one can see, the square root of the ratio
is of order $\mathcal{O}\left(2\right)$ and we conclude that the
condition number of Wilson fermions in typically a factor $\mathcal{O}\left(4\right)$
higher compared to staggered Wilson fermions.

In Figs.~\ref{fig:IterationRatio1632b6}, \ref{fig:IterationRatio2040b6}
and \ref{fig:IterationRatio2040b614}, we find the results for the
ratio of the number of conjugate gradient iterations as given in Eq.~\eqref{eq:CondNumberRatio}.
In the figures, we show both our theoretical estimate given by $\sqrt{\kappa_{\mathsf{w}}/\kappa_{\mathsf{sw}}}$
and the average measured ratios for the actual number of iterations
needed as a function of the pion mass. We did this measurements for
different values of the target residual $\varepsilon$ to check for
a possible dependence. We also used two different kinds of sources,
namely a point source and a wall source.

First we note that the choice of the source has no noticeable effect
on the relative computational performance as the figures are nearly
indistinguishable. Remarkably the dependence of the ratios on $m_{\pi}$
is also weak, although the number of conjugate gradient iterations
grows strongly with decreasing pion masses. We can see that our rough
theoretical estimate $\sqrt{\kappa_{\mathsf{w}}/\kappa_{\mathsf{sw}}}$
is a decent approximation for the actual measured ratios. Moreover,
for smaller values of $\varepsilon$ the measured ratios become closer
to our estimate as expected. In all cases we see that Wilson fermions
need roughly $\mathcal{O}\left(2\right)$ more iterations in the application
of the conjugate gradient method, which is in agreement with Ref.~\citep{deForcrand:2012bm}.

If we increase the physical volume in our simulation, i.e.\ when
we compare the $16^{3}\times32$ lattice at $\beta=6$ with the $20^{3}\times40$
lattice at $\beta=6$, we find an improvement at small $m_{\pi}$,
but otherwise almost unchanged results. If, on the other hand, we
decrease the lattice spacing, i.e.\ when we compare the $16^{3}\times32$
lattice at $\beta=6$ with the $20^{3}\times40$ lattice at $\beta=6.136716$,
we see a significant improvement in efficiency over the whole range
of $m_{\pi}$. As the staggered Wilson Dirac operator is more sensitive
to gauge fluctuations through its four-hop terms, we would expect
that on smoother gauge configurations its convergence properties improve
more compared to the Wilson Dirac operator.

\subsection{Overall speedup \label{subsec:CE-Overall-speedup}}

Combining the measured ratios from the previous discussion together
with Eq.~\eqref{eq:CostMatVev}, the product in Eq.~\eqref{eq:CostRatio}
takes the form
\begin{equation}
\frac{\mathrm{cost}_{\mathsf{W}}}{\mathrm{cost}_{\mathsf{SW}}}=\frac{\left(\textrm{\#iterations}\right)_{\mathsf{w}}}{\left(\textrm{\#iterations}\right)_{\mathsf{sw}}}\times\frac{\left(\textrm{cost mat-vec mult.}\right)_{\mathsf{w}}}{\left(\textrm{cost mat-vec mult.}\right)_{\mathsf{sw}}}\approx2\times(2\text{--}3)=4\text{--}6.\label{eq:CompEffFinal}
\end{equation}
This means that in our setting staggered Wilson fermions are a factor
of $4\text{--}6$ more efficient than usual Wilson fermions for inverting
the Dirac operator on a source. When computing the quark propagator,
an additional factor of four should be included in Eq.~\eqref{eq:CompEffFinal}
as for Wilson fermions one needs four times as many sources.

Before we move on with the discussion of arithmetic intensities, we
add some critical remarks for the interpretation of the estimate given
in Eq.~\eqref{eq:CompEffFinal}. Our cost ratio is for the case of
inverting the Dirac operator on a source and does not include the
setup phase. For staggered Wilson fermions, one would typically prepare
the averaged link products in the staggered Wilson term before applying
the conjugate gradient method. The factor in Eq.~\eqref{eq:CompEffFinal}
would therefore be a bit lower when taking the cost for this setup
phase into account. Also, if one intends to do hadron spectroscopy
with staggered Wilson fermions as described in Chapter\ \ref{chap:Pseudoscalar-mesons},
one has to take the additional number of sources needed into account,
effectively reducing the computational advantage.

Moreover, when doing lattice field theoretical computations on high-per\-for\-mance
computing clusters, communication over the network can become a bottleneck
for the performance. As large-scale simulations of lattice quantum
chromodynamics usually make use of a domain-decomposition approach
and the staggered Wilson term connects each site with all the corners
of the surrounding hypercube via four-hop terms, it is expected that
the resulting network traffic will be significant. How severely this
limits the performance of staggered Wilson fermions in large-scale
simulations has yet to be seen. Until then, Eq.~\eqref{eq:CompEffFinal}
should be taken as an estimate for the achievable performance gain
on shared memory machines.

\section{Arithmetic intensities \label{sec:CE-Arithmetic-intensities}}

\global\long\def\FLOPs{\mathtt{F}}
\global\long\def\floats{\mathtt{fl}}

On modern computer architectures the performance of lattice field
theoretical simulations is often not limited by the floating point
throughput, but by limited memory bandwidth. A measure for the bandwidth
requirements relative to the number of floating point operations (FLOPs)
is the so called arithmetic intensity
\begin{equation}
I=\frac{\textrm{FLOPs}}{\textrm{Memory transactions in byte}}.
\end{equation}
If for a given application this ratio is lower than what the respective
hardware can provide, characterized by
\begin{equation}
I_{\mathsf{hw}}=\frac{\textrm{max.\,FLOPs\thinspace/\thinspace s}}{\textrm{max.\,memory bandwidth in byte\thinspace/\thinspace s}},
\end{equation}
its performance is limited by the memory bandwidth. As a result, the
computing cores spend some time with idling while waiting for the
completion of memory access operations. On the other hand the performance
of problems with high arithmetic intensity is limited by the floating
point throughput of the hardware.

Simulations in the context of lattice quantum chromodynamics tend
to have a rather low computational intensity, i.e.\ a relatively
high number of memory transactions per floating point operation. If
optimized properly, lattice codes can archive a high sustained performance
on modern architectures. As the expected performance depends on the
arithmetic intensity of the lattice fermion formulation, it is worth
having a closer look.

In the following, we determine the arithmetic intensities for the
application of several lattice Dirac operators in the setting of quantum
chromodynamics, where we focus on the hopping-terms and exclude the
trivial diagonal mass term. Within derivations we abbreviate FLOPs
with $\FLOPs$. Before we begin, we note that for a complex addition
we need $\unit[2]{\FLOPs}$ and for a complex multiplication we need
$\unit[6]{\FLOPs}$. To multiply a complex $\SUthree$ matrix with
a complex three-component vector, we need $\unit[66]{\FLOPs}$. Regarding
memory transactions, we assume single-precision floating point numbers,
so-called floats, which are $\unit[32]{bit}$ or $\unit[4]{byte}$.
Where convenient, we abbreviate floats by $\floats$. Depending on
the application and hardware, in practice one might also use double-precision
floating point numbers with $\unit[64]{bit}$ or $\unit[8]{byte}$,
lowering the arithmetic intensity\footnote{While in this case the arithmetic intensity decreases, the number
of CPU cycles can potentially increase.} by a factor of two (see Ref.~\citep{jeffers2015high}). In some
cases even mixed-precision is used, consisting of a combination of
double-precision, single-precision and half-precision floating point
numbers.

\subsection{Wilson fermions}

We begin with Wilson fermions as defined in Subsec.\ \ref{subsec:GT-Wilson-fermions}.
We note that some fixed terms in the action can be precomputed, namely
terms like $\One\pm\gamma_{\mu}$, to save arithmetic operations.
We also have to take into account that a competitive implementation
of Wilson fermions employs the spin-projection trick \citep{Alford:1996nx},
which reduces the number of spin components by a factor of two.

\paragraph{Floating point operations.}

The spin projection takes $\unit[12\times2\times2]{\FLOPs}=\unit[96]{\FLOPs}$
(twelve operations $\times$ four directions $\times$ forward/backward)
and for the link multiplications we need $\unit[66\times2\times4\times2]{\FLOPs}=\unit[1056]{\FLOPs}$
(matrix-vector multiplication $\times$ two spin components $\times$
four directions $\times$ forward/backward). Accumulating the resulting
vector takes $\unit[12\times2\times7]{\FLOPs}=\unit[168]{\FLOPs}$
(twelve components $\times$ cost per complex addition $\times$ adding
eight vectors). In total we find that Wilson fermions take $\unit[1320]{\FLOPs/site}$.
This result is in agreement with Refs.~\citep{Boyle:2009vp,jeffers2015high,ChileSlides}.

\paragraph{Memory access operations.}

We need to read one spinor per direction, i.e.\ $\unit[12\times2\times4\times2]{\floats}=\unit[192]{\floats}$
(twelve complex numbers per spinor $\times$ floats per complex number
$\times$ four directions $\times$ forward/backward) and one $3\times3$
gauge link per direction, i.e.\ $\unit[9\times2\times4\times2]{\floats}=\unit[144]{\floats}$
(nine elements $\times$ floats per complex number $\times$ four
directions $\times$ forward/backward). Writing the resulting spinor
takes $\unit[12\times2]{\floats}=\unit[24]{\floats}$. In total we
find $\unit[360]{\floats/site}=\unit[1440]{byte/site}$ of memory
access operations per lattice site. This result is also in agreement
with Refs.~\citep{jeffers2015high,ChileSlides}.

\paragraph{Arithmetic intensity.}

We find
\begin{equation}
I_{\mathsf{w}}=\frac{\unit[1320]{FLOPs/site}}{\unit[1440]{byte/site}}\approx\unit[0.92]{FLOPs/byte}
\end{equation}
for the arithmetic intensity of Wilson fermions.

\subsection{Staggered fermions}

We are repeating the previous exercise now for usual staggered fermions,
which we discussed in Subsec.\ \ref{subsec:GT-Staggered-fermions}.

\paragraph{Floating point operations.}

For the link multiplications we have $\unit[66\times4\times2]{\FLOPs}=\unit[528]{\FLOPs}$
(matrix-vector multiplication $\times$ four directions $\times$
forward/backward). Accumulating the resulting vector takes $\unit[3\times2\times7]{\FLOPs}=\unit[42]{\FLOPs}$
(three color components $\times$ cost per complex addition $\times$
adding eight vectors). In total we find that usual staggered fermions
need $\unit[570]{\FLOPs/site}$, which is in agreement with Ref.~\citep{GPUStagTalk}.

\paragraph{Memory access operations.}

We read one spinor per direction, i.e.\ $\unit[3\times2\times4\times2]{\floats}=\unit[48]{\floats}$
(three complex numbers per spinor $\times$ floats per complex number
$\times$ four directions $\times$ forward/backward) as well as one
gauge link per direction, i.e.\ $\unit[9\times2\times4\times2]{\floats}=\unit[144]{\floats}$
(nine elements $\times$ floats per complex number $\times$ four
directions $\times$ forward/backward). Writing the resulting spinor
takes $\unit[3\times2]{\floats}=\unit[6]{\floats}$. In total we have
$\unit[198]{\floats/site}=\unit[792]{byte/site}$ of memory access
operations per lattice site, which is also in agreement with Ref.~\citep{GPUStagTalk}.

\paragraph{Arithmetic intensity.}

We find
\begin{equation}
I_{\mathsf{st}}=\frac{\unit[570]{FLOPs/site}}{\unit[792]{byte/site}}\approx\unit[0.72]{FLOPs/byte}
\end{equation}
for the arithmetic intensity of staggered fermions.

\subsection{Staggered Wilson fermions}

Compared to usual staggered fermions, staggered Wilson fermions have
an additional term in the action. In the following, we only analyze
the additional contributions from the staggered Wilson term and then
add the respective counts to the numbers of usual staggered fermions.

\paragraph{Floating point operations.}

Applying the $C$ operator takes $\unit[66\times16]{\FLOPs}+\unit[3\times2\times15]{\FLOPs}=\unit[1146]{\FLOPs}$
(matrix-vector multiplication $\times$ number of hybercube corners
$+$ three color components $\times$ cost per complex addition $\times$
adding sixteen vectors). Applying $\Gamma_{55}$ and $\eta_{5}$ takes
$\unit[\left(3+6\right)]{\FLOPs}$. The $r\One$ term can be combined
with the diagonal mass term $m\One$ and, thus, will be ignored as
in the preceding calculations. Accumulating the resulting vector takes
$\unit[3\times2\times1]{\FLOPs}=\unit[6]{\FLOPs}$ (three color components
$\times$ cost per complex addition $\times$ adding two vectors).
In total we find that staggered Wilson fermions take an additional
$\unit[1161]{\FLOPs/site}$ compared to staggered fermions, i.e.\ $\unit[1731]{\FLOPs/site}$
in total. Note that when the diagonal $\left(m+r\right)\One$ term
is included we need additional $\unit[12]{\FLOPs}$, i.e.\ $\unit[1743]{\FLOPs/site}$
in agreement with Ref.~\citep{Adams:2013tya}.

\paragraph{Memory access operations.}

We need to read one additional spinor per hypercube corner, i.e.\ $\unit[3\times2\times16]{\floats}=\unit[96]{\floats}$
(three complex numbers per spinor $\times$ floats per complex number
$\times$ number of corners) and one gauge link per hypercube corner,
i.e.\ $\unit[9\times2\times16]{\floats}=\unit[288]{\floats}$ (9
elements $\times$ floats per complex number $\times$ number of corners)
for the $C$ operator. Write the resulting spinor takes $\unit[3\times2]{\floats}=\unit[6]{\floats}$.
This means that we have an additional $\unit[390]{\floats/site}=\unit[1560]{byte/site}$,
i.e.\ in total we find $\unit[588]{\floats/site}=\unit[2352]{byte/site}$
for staggered Wilson fermions.

\paragraph{Arithmetic intensity.}

We find
\begin{equation}
I_{\mathsf{stw}}=\frac{\unit[1731]{FLOPs/site}}{\unit[2352]{byte/site}}\approx\unit[0.74]{FLOPs/byte}
\end{equation}
for the arithmetic intensity of staggered Wilson fermions.

\subsection{Summary}

\begin{table}[t]
\begin{centering}
\begin{tabular*}{1\columnwidth}{@{\extracolsep{\fill}}ccccc}
\toprule 
Formulation & Operations & I/O & Intensity & Source\tabularnewline
\midrule
\midrule 
Wilson & $1320$ & $1440$ & $0.92$ & Here and Refs.~\citep{Boyle:2009vp,jeffers2015high,ChileSlides}\tabularnewline
\midrule 
Wilson\,+\,Clover & $1824$ & $1728$ & $1.06$ & Ref.~\citep{ChileSlides}\tabularnewline
\midrule 
Staggered & $570$ & $792$ & $0.72$ & Here and Ref.~\citep{GPUStagTalk}\tabularnewline
\midrule 
HISQ / asqtad & $1146$ & $1560$ & $0.73$ & Ref.~\citep{ChileSlides}\tabularnewline
\midrule 
Staggered Wilson & $1731$ & $2352$ & $0.74$ & Here and Ref.~\citep{Adams:2013tya}\tabularnewline
\bottomrule
\end{tabular*}
\par\end{centering}
\caption{Arithmetic intensities of several lattice Dirac operators. Operations
are in units of FLOPs\,/\,site, I/O in bytes\,/\,site and Intensity
in FLOPs\,/\,byte. \label{tab:Arithmetic-intensities}}
\end{table}

We summarize our findings in Table\ \ref{tab:Arithmetic-intensities},
where we added Wilson fermions with a clover term (see e.g.\ Ref.~\citep{Gattringer:2010zz})
as well as HISQ \citep{Follana:2006rc} (highly improved staggered
quarks) and asqtad \citep{Orginos:1999cr} ($a$-squared tadpole improved)
staggered fermions from the literature. We note that all staggered
formulations have a very similar arithmetic intensity, while staggered
Wilson fermions have slightly lower memory bandwidth requirements
compared to the others. This means, that despite the large number
of memory access operations per site needed for the application of
the staggered Wilson Dirac operator, we do not expect a more severe
memory bandwidth bottleneck.

One possibility to increase arithmetic intensities and to lower memory
bandwidth requirements for all fermion formulations is to only load
the first two rows of each $\SUthree$ matrix and then reconstruct
the last one by using unitarity. This lowers memory bandwidth requirements
at the cost of additional FLOPs, thus increasing arithmetic intensity.

\section{Conclusions \label{sec:CE-Conclusions}}

We estimated the speedup factor of staggered Wilson fermions to be
$4\text{--}6$ compared to usual Wilson fermions for inverting the
Dirac matrix on a source. However, it is important to keep the critical
remarks from Subsec.\ \ref{subsec:CE-Overall-speedup} in mind when
interpreting this result.

To allow for an easier comparison, we discussed here the case of kernel
operators which are both unpreconditioned and unimproved. While for
Wilson fermions many preconditioned and improvement variants are known
and in use, there are many possibilities to further improve the computational
efficiency of staggered Wilson fermions as well. One could replace
the usual staggered part of staggered Wilson fermions by the HISQ
action to reduce $\mathcal{O}\left(a^{2}\right)$ effects and use
link smearing. While a certain form of a clover term for $\mathcal{O}\left(a\right)$
improvement was studied in Ref.~\citep{Durr:2013gp}, there is also
the possibility for the introduction of other clover terms.

Finally, comparing the arithmetic intensities of the different fermion
formulations in Sec.\ \ref{sec:CE-Arithmetic-intensities}, we note
that the memory bandwidth requirements of staggered Wilson fermions
are slightly lower compared to other staggered formulations. As memory
bandwidth is usually a limiting factor for the performance of lattice
codes, this is important in the context of writing highly optimized
implementations of the Dirac operator.

\chapter{Pseudoscalar meson spectrum \label{chap:Pseudoscalar-mesons}}

One of the traditional applications of lattice gauge theory is hadron
spectroscopy. In this setting not only computational efficiency, but
also the usability of the formalism is important. To show the feasibility
of doing spectroscopy with staggered Wilson fermions, in this chapter
we discuss the case of pions, i.e.\ pseudoscalar mesons.

For usual staggered fermions the four tastes give rise to sixteen
pions. In the following we study the pion spectrum of staggered fermions
when turning Adams' two-flavor staggered Wilson term on. The effect
of this term is that two of the four tastes acquire a mass of $\mathcal{O}\left(a^{-1}\right)$
and become heavy. As a result, we find that eight of the pions become
heavy as well, while the other eight remain light. In the physical
light part of the spectrum we find the degenerate two-flavor pion-triplet
and the flavor-singlet $\eta$ meson with each coming in two copies.
The $\eta$ meson remains light as we omit the disconnected piece
of the propagator in our study. Some of the results discussed here
were previously presented in Refs.~\citep{AdamsLat14,ZielinskiLat15}.

\section{Introduction}

From a practitioner's point of view, the usability of a lattice formulation
is important. To show that staggered Wilson fermions are not at a
disadvantage compared to other fermion formulations, we illustrate
spectroscopy calculations as one of the traditional applications of
lattice quantum chromodynamics (QCD). Like for other staggered formulations,
spin and flavor degrees of freedom are distributed over different
lattice sites. We will see, that one can easily adapt the spectroscopy
methods and operators of usual staggered fermions to the two flavor
case, as we are going to illustrate for the case of pions.

In this work we deal with Adams' original two flavor proposal \citep{Adams:2009eb,Adams:2010gx},
where the action reads
\begin{equation}
S_{\mathsf{sw}}=\bar{\chi}\left(D_{\mathsf{st}}+m_{\mathsf{q}}+W_{\mathsf{st}}\right)\chi,\qquad W_{\mathsf{st}}=\frac{r}{a}\left(\One-\Gamma_{55}\Gamma_{5}\right)\label{eq:StagWilsonAction}
\end{equation}
and $D_{\mathsf{st}}=\eta_{\mu}\nabla_{\mu}$ refers to the staggered
Dirac operator, $m_{\mathsf{q}}$ to the bare quark mass, $W_{\mathsf{st}}$
to the staggered Wilson term as defined in Sec.\ \ref{sec:SW-Adams-mass-term},
$a$ to the lattice spacing and $r>0$ is the equivalent of the Wilson
parameter. Here and in the following, the $\xi_{\mu}$ matrices refer
to a representation of the Dirac algebra in flavor space and in figures
we use the shorthand notation $\xi_{\mu\nu}\equiv\xi_{\mu}\xi_{\nu}$.

This chapter is organized as follows. In Sec.\ \ref{sec:PS-Pseudoscalar-Mesons},
we discuss pseudoscalar mesons in the framework of usual staggered
fermions and generalize the known methods to the case of staggered
Wilson fermions. In Sec.\ \ref{sec:PS-Numerical-results}, we discuss
our numerical results and compare them to our theoretical predictions.
Finally, in Sec.\ \ref{sec:PS-Conclusions} we summarize our results
and end with some concluding remarks.

\section{Pseudoscalar mesons \label{sec:PS-Pseudoscalar-Mesons}}

For usual staggered fermions, the meson and baryon rest-frame operators
are known \citep{Golterman:1985dz} and we can adapt these operators
to the two-flavor staggered Wilson case. However, as two of the four
tastes become heavy after the introduction of the staggered Wilson
term, the physical interpretation of these operators change. In the
following, we will discuss the case of pseudoscalar mesons both from
a theoretical and numerical point of view. We first begin with reviewing
the case of usual staggered fermions, before we move on to staggered
Wilson fermions.

Staggered fermions have four (quark) tastes, giving rise to sixteen
pseudoscalar mesons. They have a spin$\,\otimes\,$flavor structure
$\gamma_{5}\otimes\xi_{\mathsf{F}}$ with 
\begin{equation}
\xi_{\mathsf{F}}\in\left\{ \One,\,\xi_{5},\,\xi_{\mu},\,\xi_{\mu}\xi_{5},\,\xi_{\mu}\xi_{\nu}\right\} .
\end{equation}
On the lattice the states fall into eight irreducible representations
of the lattice timeslice group \citep{Golterman:1985dz}, namely 
\begin{equation}
\xi_{4}\xi_{5},\,\xi_{5},\,\xi_{i}\xi_{5},\,\xi_{i}\xi_{j},\,\xi_{i},\,\xi_{i}\xi_{4},\,\One,\,\xi_{4}.
\end{equation}
Here the $\gamma_{5}\otimes\xi_{5}$ state corresponds to the Goldstone
pion for the axial symmetry at vanishing $m_{\mathsf{q}}$. While
for staggered fermions the continuum $\SUfour$ flavor symmetry is
broken down to a discrete subgroup, by doing a joint expansion in
$m_{\mathsf{q}}$ and $a^{2}$ one can show that the pion masses are
$\SOfour$ symmetric up to leading discretization errors, see Ref.~\citep{Lee:1999zxa}.

Within these representations we find two distinct types of states:
while some states propagate with a $\left(-1\right)^{t}$-factor in
time, others do not. The timeslice operators, in general, excite both
these kinds of states, i.e.\ if an operator couples to $\gamma_{\mathsf{S}}\otimes\xi_{\mathsf{F}}$,
then it also couples to $\gamma_{4}\gamma_{5}\gamma_{\mathsf{S}}\otimes\xi_{4}\xi_{5}\xi_{\mathsf{F}}$.
In the case of pseudoscalar mesons we have $\gamma_{\mathsf{S}}\in\left\{ \gamma_{5},\,\gamma_{4}\gamma_{5}\right\} $.
The fact that we excite two states simultaneously does not pose a
problem as both states can be easily separated by their distinct time-dependence.
We can parametrize the time-time correlation function by 
\begin{equation}
f_{R_{\pm},m_{\pm}}\left(n_{t}\right)=R_{+}\cosh\left[m_{+}\left(n_{t}-\frac{1}{2}N_{t}\right)\right]+\left(-1\right)^{n_{t}}R_{-}\cosh\left[m_{-}\left(n_{t}-\frac{1}{2}N_{t}\right)\right],\label{eq:FitFunc}
\end{equation}
where $t=n_{t}a$ and we assume that one-particle states are dominant
\citep{Golterman:1985dz}. Here $R_{\pm}$ and $m_{\pm}$ are the
amplitudes and masses of both respective states and $N_{t}$ refers
to the number of timeslices of the lattice.

After the introduction of the staggered Wilson term, two tastes become
heavy and $\xi_{\mathsf{F}}$ can be thought of having the structure
\begin{equation}
\xi_{\mathsf{F}}\cong\left[\begin{array}{cc}
\textrm{light\thinspace+\thinspace light} & \textrm{light\thinspace+\thinspace heavy}\\
\textrm{heavy\thinspace+\thinspace light} & \textrm{heavy\thinspace+\thinspace heavy}
\end{array}\right].
\end{equation}
In the continuum limit, the heavy quark content decouples and the
``light\,+\,light'' part of $\xi_{\mathsf{F}}$ determines the
physical interpretation. To give an example, consider 
\begin{equation}
\xi_{\mathsf{F}}=\xi_{1}\xi_{2}=\left(\begin{array}{cc}
\ii\sigma_{3}\\
 & \ii\sigma_{3}
\end{array}\right).
\end{equation}
The physically relevant part is $\ii\sigma_{3}$, hence $-\ii\bar{\chi}\left(\gamma_{5}\otimes\xi_{3}\xi_{4}\right)\chi$
corresponds to a $\pi^{0}$ operator, see also Ref.~\citep{AdamsLat14}.

\begin{table}[t]
\begin{centering}
\begin{tabular*}{0.7\columnwidth}{@{\extracolsep{\fill}}>{\centering}p{0.25\columnwidth}cc}
\toprule 
Flavor matrix $\xi_{\mathsf{F}}$ & Composition & Particles\tabularnewline
\midrule
\midrule 
$\xi_{k}\xi_{4}$ & \multirow{2}{*}{$\overline{q}\sigma_{k}q$} & \multirow{2}{*}{$\pi$ triplet}\tabularnewline
\cmidrule{1-1} 
$\xi_{i}\xi_{j}$ &  & \tabularnewline
\midrule 
$\xi_{5}$ & \multirow{2}{*}{$\overline{q}q$} & \multirow{2}{*}{$\eta$ singlet}\tabularnewline
\cmidrule{1-1} 
$\One$ &  & \tabularnewline
\bottomrule
\end{tabular*}
\par\end{centering}
\caption{Pseudoscalar meson spectrum for staggered Wilson fermions ($i$, $j$
and $k$ mutually distinct). \label{tab:StagWilsonSpec}}
\end{table}

After analyzing all states, one finds that for $\xi_{\mathsf{F}}\in\left\{ \xi_{4}\xi_{5},\,\xi_{i}\xi_{5},\,\xi_{i},\,\xi_{4}\right\} $
there are only contributions from heavy tastes. For the states with
$\xi_{\mathsf{F}}\in\left\{ \xi_{5},\,\xi_{i}\xi_{j},\,\xi_{i}\xi_{4},\,\One\right\} $,
we find a mixture of purely light and purely heavy quark content,
where the heavy contributions decouple in the continuum limit. In
total we find that from the sixteen pseudoscalar mesons for usual
staggered fermions only eight remain light in the presence of the
staggered Wilson term. Among these eight light pions we find two copies
of the two-flavor pion triplet, which only differ by their heavy quark
content. We note that the pion triplet for staggered Wilson fermions
is degenerated as discussed in Ref.~\citep{SharpeWorkshop}. In addition
we find two copies of the flavor-singlet $\eta$ meson. These light
states and their physical interpretation are summarized in Table\ \ref{tab:StagWilsonSpec},
where $q=\left(q_{1},q_{2}\right)^{\intercal}$ denotes the two physical
tastes.

\section{Numerical results \label{sec:PS-Numerical-results}}

\begin{figure}[t]
\begin{centering}
\subfloat[Part I]{\begin{centering}
\includegraphics[width=0.47\textwidth]{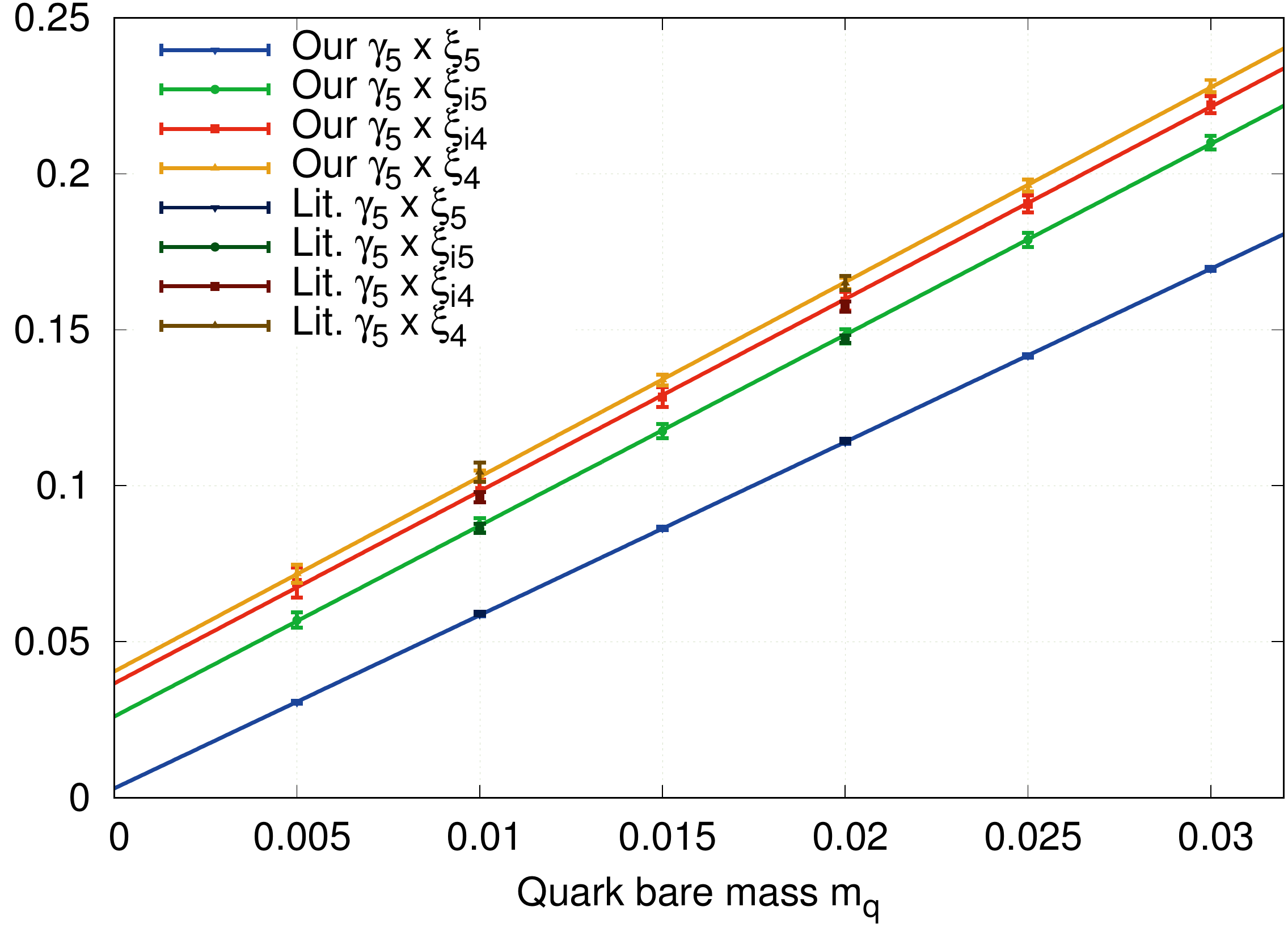}
\par\end{centering}
}\hfill{}\subfloat[Part II]{\begin{centering}
\includegraphics[width=0.47\textwidth]{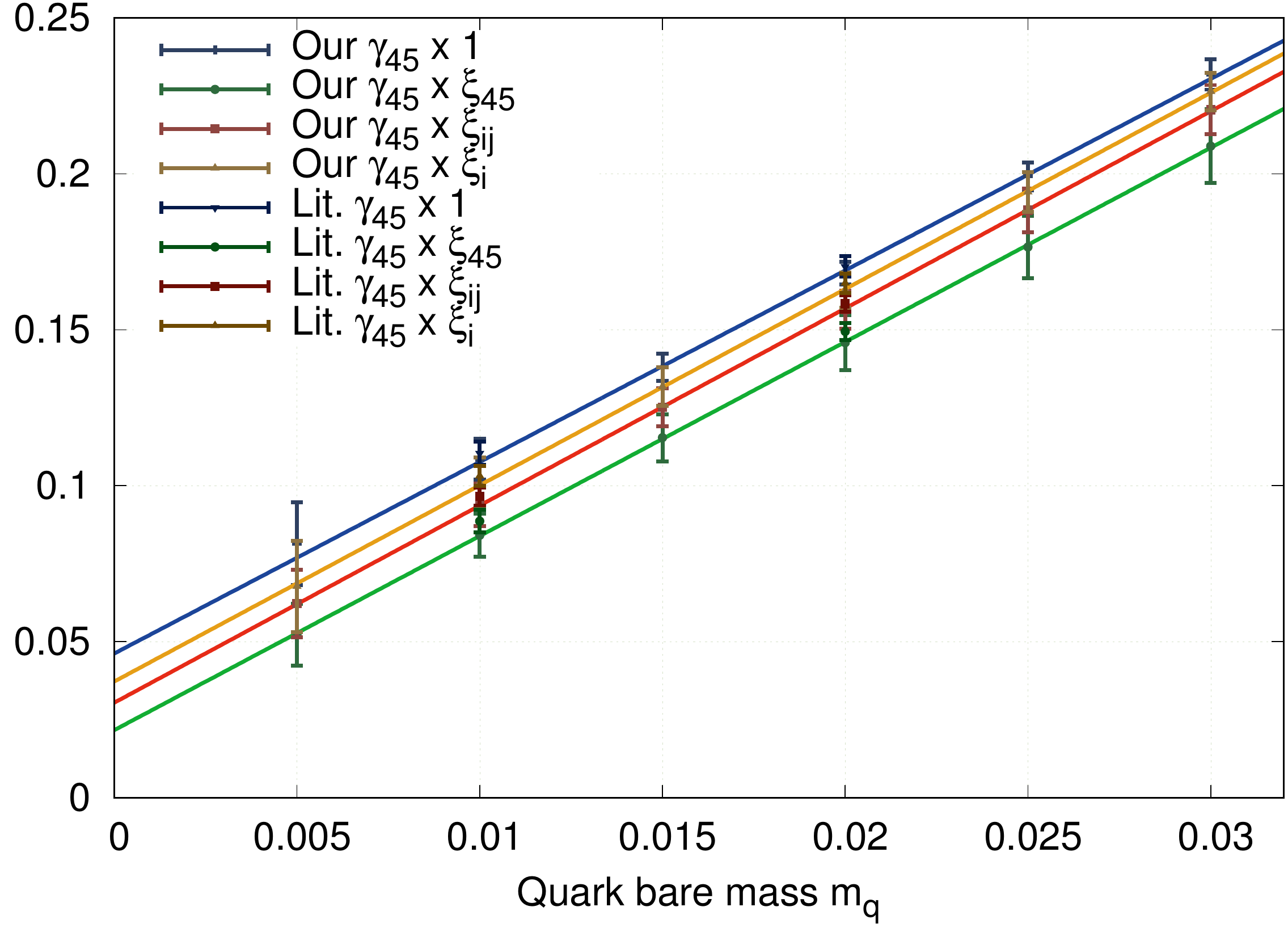}
\par\end{centering}
}
\par\end{centering}
\caption{Comparison between the pseudoscalar meson masses $m^{2}\left(m_{\mathsf{q}}\right)$
at $r=0$ in our study with the literature values from Ref.~\citep{Bae:2008qe}.
\label{fig:SpectrumUsualStag}}
\end{figure}
\begin{figure}[t]
\begin{centering}
\subfloat[Physical pseudoscalar mesons]{\begin{centering}
\includegraphics[width=0.47\columnwidth]{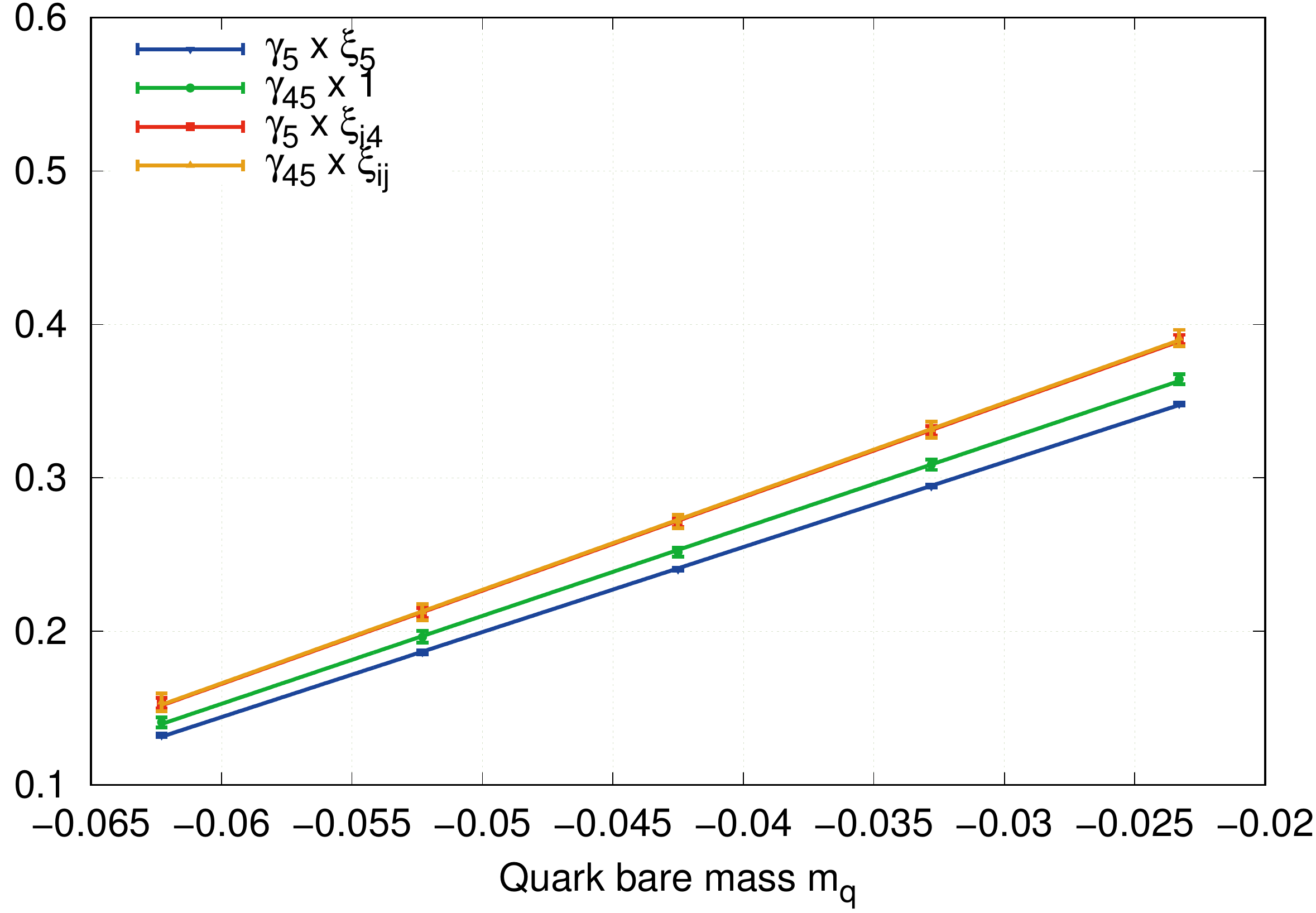}
\par\end{centering}
}\hfill{}\subfloat[Heavy pseudoscalar mesons]{\begin{centering}
\includegraphics[width=0.47\columnwidth]{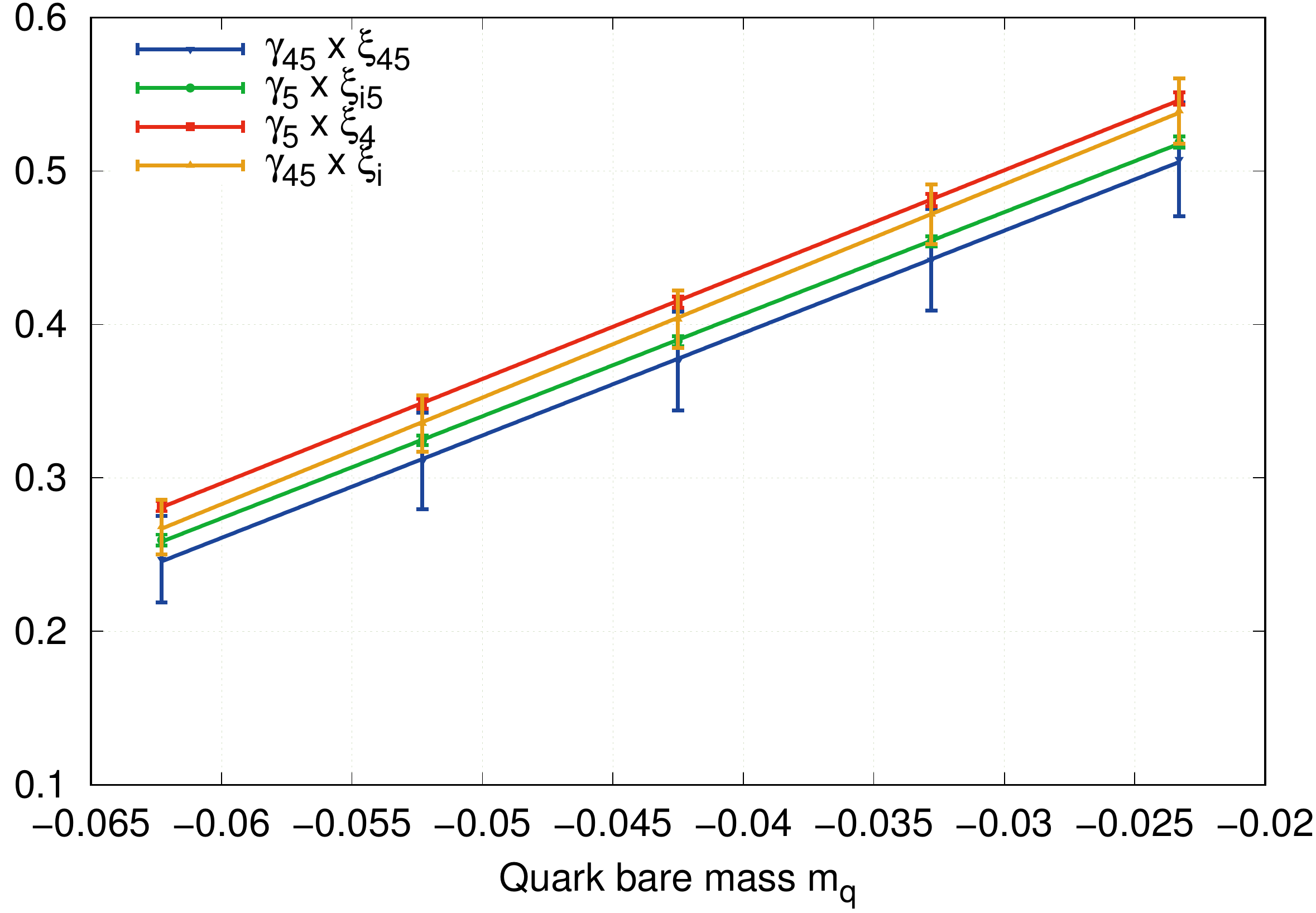}
\par\end{centering}
}
\par\end{centering}
\caption{Pseudoscalar meson masses $m^{2}\left(m_{\mathsf{q}}\right)$ at $r=0.1$.
\label{fig:StagWilsonSpectrumR01}}
\end{figure}
\begin{figure}[t]
\begin{centering}
\includegraphics[width=0.75\columnwidth]{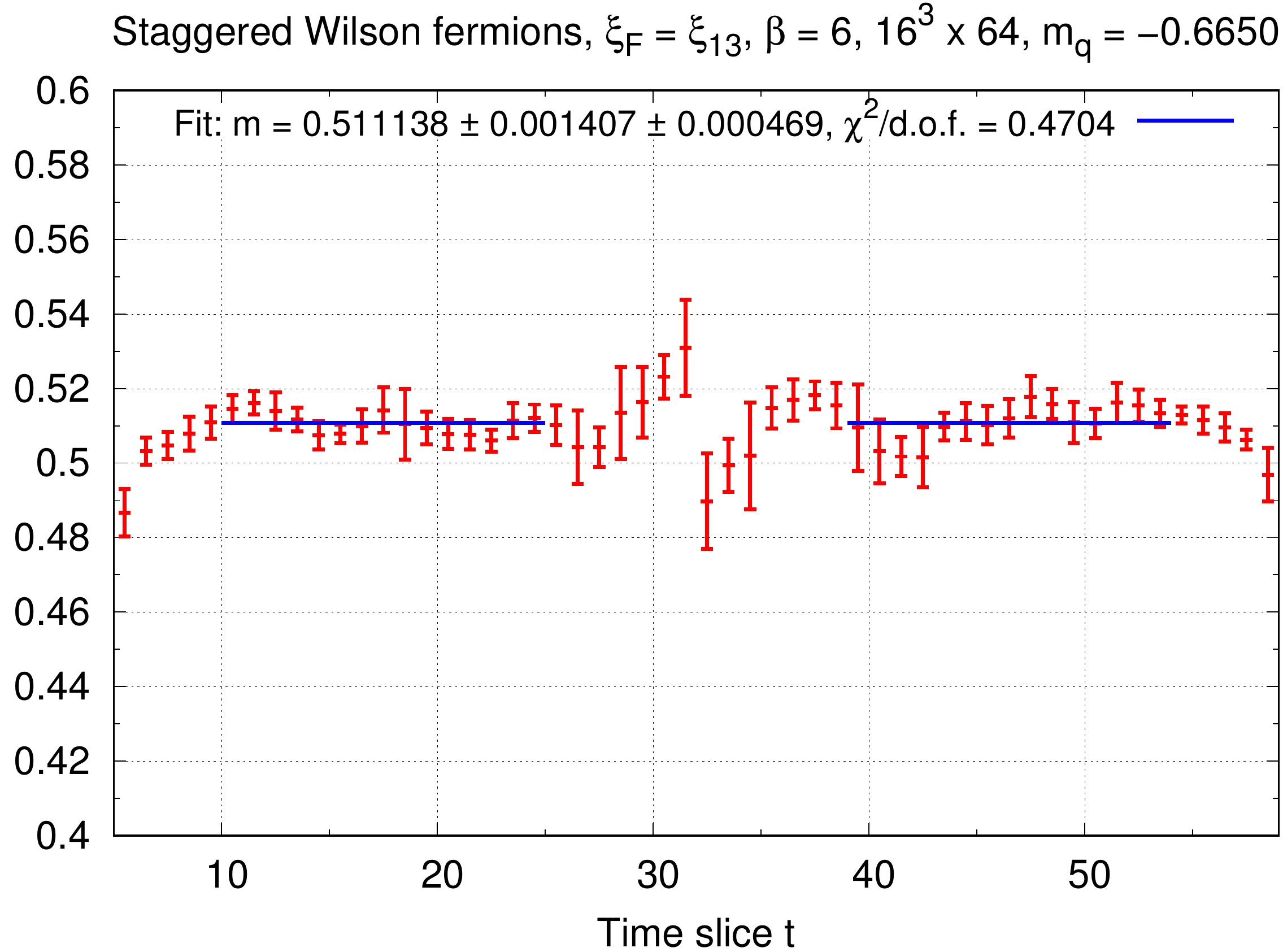}
\par\end{centering}
\caption{Effective mass plot $m_{\mathsf{eff}}\left(t\right)$ together with
the fitted mass $m$ and its statistical and systematic error at $r=1$
for the light pseudoscalar meson with $\xi_{\mathsf{F}}=\xi_{1}\xi_{3}$.
\label{fig:PSEffMassPlot}}
\end{figure}
\begin{figure}[t]
\begin{centering}
\includegraphics[width=0.8\columnwidth]{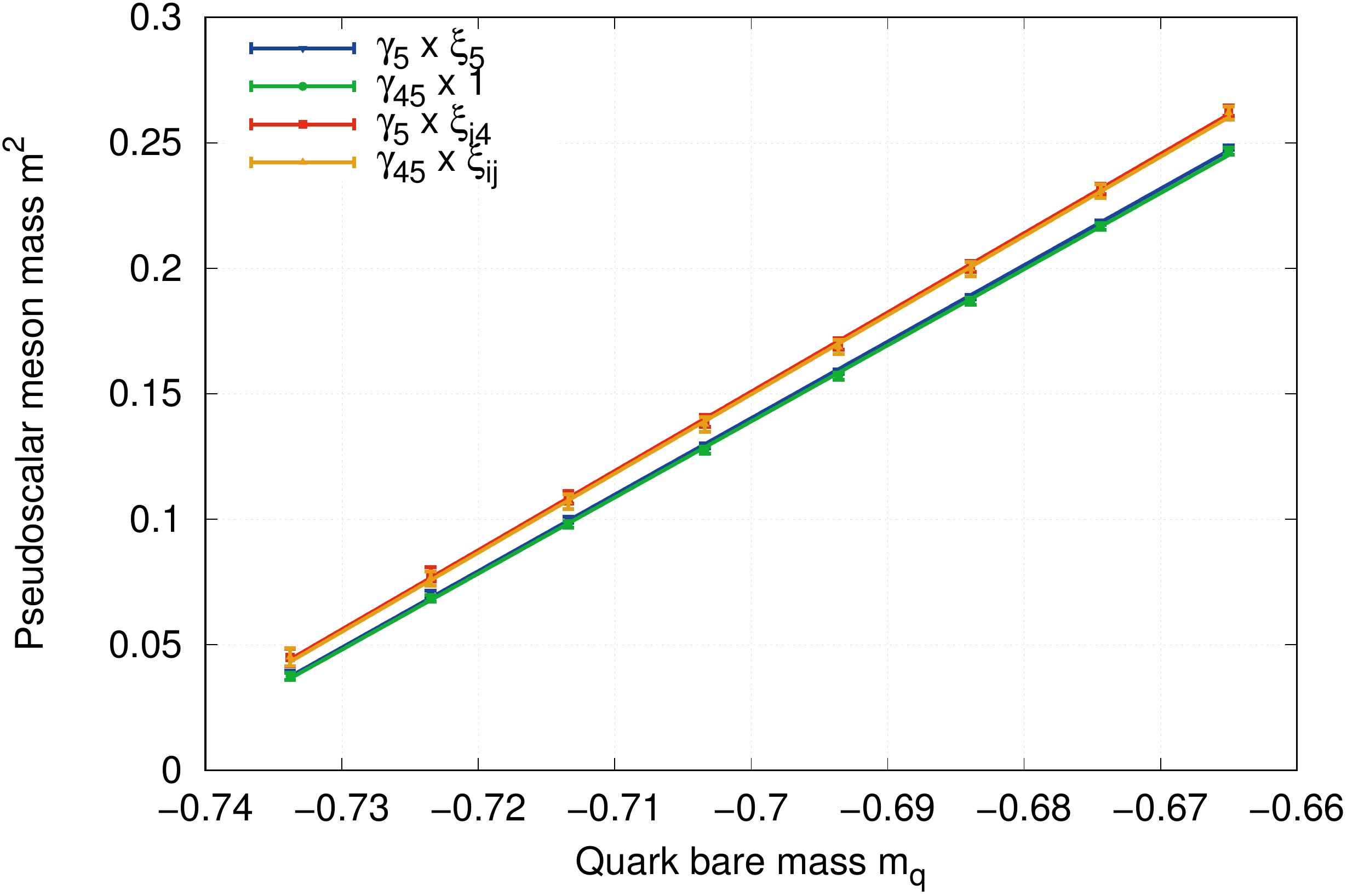}
\par\end{centering}
\caption{Pseudoscalar meson spectrum for two-flavor staggered Wilson fermions
($r=1$). \label{fig:StagWilsonSpectrumR10}}
\end{figure}

To illustrate the practicality of spectrum calculations with staggered
Wilson fermions and to verify the results from the previous section,
we implemented staggered Wilson fermions in the Chroma/QDP software
package \citep{Edwards:2004sx}. The software suite comes with an
application for spectrum calculations with staggered fermions\footnote{The program \texttt{spectrum\_s}.}.
For the pseudoscalar mesons operators we used the operators that are
already implemented and which match the ones used in Ref.~\citep{Bae:2008qe}.

Our quenched QCD study was carried out using $200$ gauge configurations
at a coupling of $\beta=6$ on a $16^{3}\times64$ lattice, where
we fixed the configurations to Coulomb gauge. On each configuration
we evaluated eight staggered wall sources\footnote{Note that this additional number of sources effectively reduces the
computational speedup factor as presented in Chapter\ \ref{chap:Computational-efficiency}. }, i.e.\ wall sources where within the wall there is a one in a single
corner of each of the spatial cubes and zero everywhere else. By combining
them appropriately, we can project onto the irreducible representations
of the timeslice group \citep{Bae:2008qe}. To extract the pseudoscalar
meson masses, we essentially follow the method described in Subsec.\ \ref{subsec:CE-Comp-eff-results}.
To compute effective masses, we can do a minimization of the form
$\min_{R_{\pm},m_{\pm}}\left\Vert F_{R_{\pm},m_{\pm}}\right\Vert $
with 
\begin{equation}
F_{R_{\pm},m_{\pm}}=\left(\begin{array}{c}
f_{R_{\pm},m_{\pm}}\left(n_{t}+0\right)-p\left(n_{t}+0\right)\\
f_{R_{\pm},m_{\pm}}\left(n_{t}+1\right)-p\left(n_{t}+1\right)\\
f_{R_{\pm},m_{\pm}}\left(n_{t}+2\right)-p\left(n_{t}+2\right)\\
f_{R_{\pm},m_{\pm}}\left(n_{t}+3\right)-p\left(n_{t}+3\right)
\end{array}\right),
\end{equation}
where $f_{R_{\pm},m_{\pm}}\left(n_{t}\right)$ refers to the fit function
defined in Eq.~\eqref{eq:FitFunc} and $p\left(n_{t}\right)$ to
the corresponding time-time correlation function. In the final four
parameter fit of Eq.~\eqref{eq:FitFunc} we fixed the fit range from
$n_{1}=10$ to $n_{2}=25$ together with the mirrored range $N_{t}-n_{2}$
to $N_{t}-n_{1}$ for all pions, where $N_{t}=64$ is the extent of
the lattice in temporal direction. Correlations are neglected in the
$\chi^{2}$-fit. The errors are estimated in the same way as described
in Subsec.\ \ref{subsec:CE-Comp-eff-results}.

As a check of the correctness of our method and fitting procedure,
we reproduced the known results from Ref.~\citep{Bae:2008qe} (cf.\ Ref.~\citep{Ishizuka:1993mt})
for the case of usual staggered fermions. By setting the staggered
Wilson parameter to $r=0$ in Eq.~\eqref{eq:StagWilsonAction} for
this test, we can calculate the pseudoscalar meson spectrum and find
an excellent match with the literature values, see Fig.~\ref{fig:SpectrumUsualStag}.
In this figure and the following mass plots, for every pseudoscalar
meson mass $m$ we add a linear fit of the form $m^{2}=c_{1}m_{\mathsf{q}}+c_{2}$.

As the negative flavor-chirality species acquire a mass of $\mathcal{O}\left(r/a\right)$
in the presence of the staggered Wilson term, we can use $r$ to control
the masses of the heavy pseudoscalar mesons. We can then consider
staggered Wilson fermions with a small $r$ to observe how a mass
splitting between the light and heavy pions forms. In Fig.~\ref{fig:StagWilsonSpectrumR01},
we can find the spectrum for $r=0.1$, where in the left part of the
figure we can find all light pions and in the right part the heavy
pions. Using the same scale for the vertical axes, one can see how
the heavy pseudoscalar mesons are systematically shifted upwards due
to contributions from heavy tastes.

Moving on to staggered Wilson fermions with $r=1$, the initially
small mass gap between the light and heavy pion states becomes large.
In Fig.~\ref{fig:PSEffMassPlot}, we can see a representative effective
mass plot for the state with $\xi_{\mathsf{F}}=\xi_{1}\xi_{3}$, which
remains light after the introduction of the staggered Wilson term
in agreement with Table\ \ref{tab:StagWilsonSpec} and the discussion
in Sec.\ \ref{sec:PS-Pseudoscalar-Mesons}. On the other hand, the
heavy states now have a mass of $m=\mathcal{O}\left(a^{-1}\right)$.
As the condition $am\ll1$ is violated, one would not expect to be
able to reliably describe these states on our lattices.

Finally, focusing on the physical (light) part of the pseudoscalar
mesons, we find the resulting spectrum of pseudoscalar mesons for
two-flavor staggered Wilson fermions in Fig.~\ref{fig:StagWilsonSpectrumR10}.
As predicted by chiral perturbation theory, we find that the squared
pseudoscalar meson masses $m^{2}$ are approximately linear in the
bare quark mass $m_{\mathsf{q}}$. Referring to Table\ \ref{tab:StagWilsonSpec},
in the continuum limit the pairs $\left\{ \xi_{5},\,\One\right\} $
and $\left\{ \xi_{k}\xi_{4},\,\xi_{i}\xi_{j}\right\} $ are expected
to be degenerated as the heavy quark content decouples. In the present
setting with a finite lattice spacing these degeneracies only hold
approximately, but already describe the observed spectrum well. Moreover,
we observe that both pairs separate well. We note that in our calculations
the flavor-singlet $\eta$ remains light due to the omission of the
disconnected piece of the propagator.

\section{Conclusions \label{sec:PS-Conclusions}}

The aim of this study was to demonstrate the feasibility of spectrum
calculations with Adams' two-flavor staggered Wilson fermions. To
this end we determined how the pseudoscalar spectrum can be computed
and changes with the presence of the staggered Wilson term.

In the physical part of the spectrum we find two copies for both the
degenerate two-flavor pion-triplet as well as the flavor-singlet $\eta$
meson. We conclude that two-flavor staggered Wilson fermions work
as expected for studying pseudoscalar mesons. In future work one can
investigate spectroscopy of other hadrons, in particular also baryons,
where our approach is expected to generalize.

\chapter{Staggered overlap fermions \label{chap:Overlap-fermions}}

Chiral symmetry plays an important role in understanding the low-energy
regime of quantum chromodynamics, in particular hadron phenomenology.
As discussed in Subsec.\ \ref{subsec:GT-Nielsen-Ninomiya}, the Nielsen-Ninomiya
theorem \citep{Nielsen:1980rz,Nielsen:1981hk,Nielsen:1981xu,Friedan:1982nk}
complicates the implementation of chiral symmetry on the lattice.
With the overlap construction \citep{Narayanan:1992wx,Narayanan:1993ss,Narayanan:1993sk,Narayanan:1994gw,Neuberger:1997fp,Neuberger:1998my,Neuberger:1998wv}
this problem could eventually be overcome and one can formulate a
lattice Dirac operator with exact chiral symmetry \citep{Ginsparg:1981bj,Hasenfratz:1998ri,Luscher:1998pqa}.
While the overlap formalism has many attractive theoretical properties,
its practical applications are limited due to the fact that overlap
fermions typically require a factor of $\mathcal{O}\left(10\text{--}100\right)$
more computational resources compared to Wilson fermions.

The prospect of a staggered overlap operator becomes interesting as
it can potentially bring down the enormous costs of lattice theoretical
simulations in which chiral symmetry is of importance. Usual staggered
fermions do not define a suitable kernel operator for the overlap
formalism due to the fact that the staggered equivalent of $\gamma_{5}$,
namely the $\Gamma_{5}\cong\gamma_{5}\otimes\One+\mathcal{O}\left(a^{2}\right)$
operator, does not square to the identity, i.e.\ $\Gamma_{5}^{2}\neq\One$.
As Adams pointed out in Refs.~\citep{Adams:2009eb,Adams:2010gx,Adams:2011xf},
this problem can be solved with the help of the staggered Wilson construction,
where on the physical species $\Gamma_{55}$ can be used for $\gamma_{5}$.
As $\Gamma_{55}^{2}=\One$ holds exactly, this allows the construction
of staggered variants of overlap and domain wall fermions.

In this chapter, we discuss Neuberger's overlap fermions in Sec.\ \ref{sec:OV-Neubergers-overlap-fermions}
and then introduce Adams' staggered overlap construction in Sec.\ \ref{sec:OV-Staggered-overlap}.
This is followed by a review of the index theorem in Sec.\ \ref{sec:OV-Index-theorem}.
In Sec.\ \ref{sec:OV-Continuum-limit}, we verify that the continuum
limit of the index and axial anomaly for the staggered overlap Dirac
operator is correctly reproduced. This is meant as a first step to
place staggered overlap fermions on solid theoretical foundations.
Our analytical work was first presented in Ref.~\citep{Adams:2013lpa},
where later a more comprehensive discussion was given by Har in Ref.~\citep{HarThesis}.

\section{Neuberger's overlap fermions \label{sec:OV-Neubergers-overlap-fermions}}

We begin by discussing the Ginsparg-Wilson relation in Subsec.\ \ref{subsec:OV-Ginsparg-Wilson-relation}
and Neuberger's overlap construction in Subsec.\ \ref{subsec:OV-overlap-operator},
following the outline of Ref.~\citep{Gattringer:2010zz}.

\subsection{The Ginsparg-Wilson relation \label{subsec:OV-Ginsparg-Wilson-relation}}

The Nielsen-Ninomiya theorem prevents a straight-forward implementation
of chiral symmetry on the lattice of the form $\left\{ D,\gamma_{5}\right\} =0$.
To circumvent this problem, Ginsparg and Wilson proposed in Ref.~\citep{Ginsparg:1981bj}
to replace the continuum expression by
\begin{equation}
\left\{ D,\gamma_{5}\right\} =aD\gamma_{5}D.\label{eq;GinspargWilsonRelation}
\end{equation}
For this so-called Ginsparg-Wilson relation in Eq.~\eqref{eq;GinspargWilsonRelation}
we see that the anticommutator is not vanishing on the lattice, but
is instead an order $\mathcal{O}\left(a\right)$ term, eventually
shrinking to zero in the continuum limit. A remarkable observation
is now that, while the continuum form of chiral symmetry is recovered
in the $a\to0$ limit, even at a finite lattice spacing a Dirac operator
$D$ satisfying Eq.~\eqref{eq;GinspargWilsonRelation} has an exact
chiral symmetry. However, for a long time no solution to the Ginsparg-Wilson
relation was known and the possibility of finding a closed solution
in the interacting case was doubted. Only much later solutions were
explicitly constructed, namely overlap fermions \citep{Neuberger:1998wv}
and fixed points fermions \citep{Hasenfratz:1997ft}.

Assuming that a Dirac operator $D$ satisfies the Ginsparg-Wilson
relation, one can define an infinitesimal chiral transformation of
the form 
\begin{equation}
\delta\overline{\Psi}=\overline{\Psi}\gamma_{5},\qquad\delta\Psi=\gamma_{5}\left(\One-aD\right)\Psi,\label{eq:OverlapChiralTransf}
\end{equation}
where the corresponding finite transformation takes the form
\begin{equation}
\overline{\Psi}\to\overline{\Psi}e^{\ii\alpha\gamma_{5}},\qquad\Psi\to e^{\ii\alpha\gamma_{5}\left(\One-aD\right)}\Psi
\end{equation}
with $\alpha\in\mathbb{R}$ being a continuous parameter. We note
that these transformations can also be written in a symmetric form,
where both $\overline{\Psi}$ and $\Psi$ transform with a factor
of $\gamma_{5}\left(\One-aD/2\right)$. One can easily verify the
invariance of the action $S=\overline{\Psi}D\Psi$ under these chiral
transformations, where in the $a\to0$ limit the continuum expressions
are reproduced.

Now let us impose that $D$ is also $\gamma_{5}$ Hermitian, i.e.\ $D^{\dagger}=\gamma_{5}D\gamma_{5}$.
This ensures that complex eigenvalues come in conjugated pairs and
$\det D\in\mathbb{R}$. Another important consequence is that for
an eigenvector $v_{\lambda}$ of $D$, corresponding to the eigenvalue
$\lambda$, one can verify that
\begin{equation}
\left\langle v_{\lambda},\gamma_{5}v_{\lambda}\right\rangle =0,
\end{equation}
unless $\lambda\in\mathbb{R}$. Here $\left\langle v,w\right\rangle =v^{\dagger}w$
denotes the usual inner product of $\mathbb{C}^{n}$. This means that
only eigenmodes with real eigenvalues can have definite chirality.

Going back to the Ginsparg-Wilson relation in Eq.~\eqref{eq;GinspargWilsonRelation},
we can see that by multiplying with $\gamma_{5}$ from either the
left or the right side and using $\gamma_{5}$ Hermiticity, one finds
that
\begin{equation}
aD^{\dagger}D=D^{\dagger}+D=D+D^{\dagger}=aDD^{\dagger},
\end{equation}
so $D$ is a normal operator satisfying $\left[D^{\dagger},D\right]=0$.
As a consequence one finds
\begin{equation}
\lambda+\lambda^{\star}=a\lambda\lambda^{\star}
\end{equation}
for the eigenvalues. Writing the eigenvalues explicitly with real
and imaginary part, this equation takes the form
\begin{equation}
\left(\myRe\lambda-\frac{1}{a}\right)^{2}+\left(\myIm\lambda\right)^{2}=\frac{1}{a^{2}}
\end{equation}
and one can see that they lie on a circle in the complex plane, commonly
referred to as the Ginsparg-Wilson circle.

As the Ginsparg-Wilson circle is going through the origin, $D$ can
have exact zero modes. If $v_{0}$ denotes such a zero mode, one can
verify that
\begin{equation}
Dv_{0}=\gamma_{5}Dv_{0}=D\gamma_{5}v_{0}=0.
\end{equation}
Thus on the eigenspace for the eigenvalue zero the Dirac operator
$D$ and the chirality operator $\gamma_{5}$ commute. One can now
choose the eigenvectors so that they are simultaneously eigenmodes
of both $D$ and $\gamma_{5}$. We can then write
\begin{equation}
\gamma_{5}v_{0}=\pm v_{0},\label{eq:Eigenmodeg5}
\end{equation}
where the eigenvalues are $\pm1$ due to $\gamma_{5}^{2}=\One$. This
means that the zero modes of $D$ have definite chirality. If we find
a plus sign in Eq.~\eqref{eq:Eigenmodeg5}, we say that $v_{0}$
has positive chirality or call it right-handed, otherwise we say that
$v_{0}$ has negative chirality or call it left-handed.

\subsection{The overlap operator \label{subsec:OV-overlap-operator}}

While for a long time the search for an explicit solution to the Ginsparg-Wilson
relation was not successful, eventually a solution in the form of
the overlap operator was found. The overlap formalism was developed
over the course of several papers \citep{Narayanan:1992wx,Narayanan:1993ss,Narayanan:1993sk,Narayanan:1994gw,Neuberger:1997fp,Neuberger:1998my,Neuberger:1998wv}
and finally brought into its present day form by Neuberger \citep{Neuberger:1997fp,Neuberger:1998my,Neuberger:1998wv}.
The Dirac operator reads
\begin{equation}
D_{\mathsf{ov}}=\frac{1}{a}\left(\One+A\left[A^{\dagger}A\right]^{-\frac{1}{2}}\right)=\frac{1}{a}\left(\One+\gamma_{5}H/\sqrt{H^{2}}\right)=\frac{1}{a}\left(\One+\gamma_{5}\sign H\right)\label{eq:NeubergerOverlapD}
\end{equation}
with kernel operator $A=D_{\mathsf{w}}\left(-M_{0}\right)$, Hermitian
operator $H=\gamma_{5}D_{\mathsf{w}}\left(-M_{0}\right)$ and negative
mass parameter $M_{0}$. In the free-field case valid choices are
in the range $M_{0}\in\left(0,2r\right)$, where $r$ denotes the
Wilson parameter, to give rise to a one flavor theory. In some contexts
there is an additional rescaling of $D_{\mathsf{ov}}$ by some factor
$\rho$, see our discussion in Chapter\ \ref{chap:Staggered-domain-wall}.
In principle one can replace the Wilson Dirac operator $D_{\mathsf{w}}$
also by some other suitable $\gamma_{5}$ Hermitian kernel operator.
Using the fact that $\sign^{2}H=\One$, one can easily verify that
the overlap Dirac operator satisfies the Ginsparg-Wilson relation
in Eq.~\eqref{eq;GinspargWilsonRelation}.

The overlap Dirac operator in Eq.~\eqref{eq:NeubergerOverlapD} is
not ultralocal due to the presence of the $1/\sqrt{H^{2}}$ term.
Here we call a lattice Dirac operator $D$ ultralocal if $\exists l\geq0$
so that $\left|D\left(x,\alpha,a;y,\beta,b\right)\right|=0$ for $\left\Vert x-y\right\Vert _{1}>l$,
where $D\left(x,\alpha,a;y,\beta,b\right)$ denotes the matrix element
of $D$ connecting the space-time coordinates $x$ and $y$ with respective
spin components $\alpha$, $\beta$ and color components $a$, $b$.
Under some general assumptions this lack of ultralocality holds in
fact for all solutions of the Ginsparg-Wilson relation \citep{Horvath:1998cm}.
However, locality of $D_{\mathsf{ov}}$ could be established in Ref.~\citep{Hernandez:1998et}
in the form of
\begin{equation}
\left|D_{\mathsf{ov}}\left(x,\alpha,a;y,\beta,b\right)\right|\leq\lambda\,e^{-\gamma/a\cdot\left\Vert x-y\right\Vert _{1}}.
\end{equation}
Here $\lambda\geq0$ and $\gamma>0$ are constants independent of
the gauge field. As a result, the decay constant $a/\gamma$ vanishes
in the continuum limit $a\to0$ and a local quantum field theory is
recovered.

The presence of the $\sign$-function in Eq.~\eqref{eq:NeubergerOverlapD}
makes the application of the overlap Dirac operator computational
expensive. As mentioned earlier, one typically needs an factor of
$\mathcal{O}\left(10\text{--}100\right)$ more computational resources
compared to Wilson fermions. As a consequence the use of overlap fermions
is limited in practical applications and a lot of past work deals
with the efficient numerical treatment of the $\sign$-function, see
e.g.\ Refs.~\citep{vandenEshof:2001hp,vandenEshof:2002ms,Arnold:2003sx,Cundy:2004pza,Cundy:2005pi}.

\section{Staggered overlap fermions \label{sec:OV-Staggered-overlap}}

When applying the overlap construction to the staggered Dirac operator,
one encounters some severe problems. For the resulting operator the
remnant chiral symmetry is lost, the Ginsparg-Wilson relation does
not hold and exact zero modes are absent \citep{Adams:2010gx}.

These problems can be traced back to the fact that the overlap construction
relies on the property $\gamma_{5}^{2}=\One$, while this does not
hold for the staggered equivalent $\Gamma_{5}\cong\gamma_{5}\otimes\One+\mathcal{O}\left(a^{2}\right)$.
The eigenvalues of $\gamma_{5}$ are $\pm1$, while the eigenvalues
of $\Gamma_{5}$ are distributed throughout the interval $\left[-1,1\right]$
and include zero as can be seen from the momentum representation $\prod_{\mu}\cos^{2}p_{\mu}$
of $\Gamma_{5}^{2}$ in the free-field case \citep{Adams:2011xf}.
Finally, we note that the operator $\Gamma_{5}D_{\mathsf{st}}$, unlike
its Wilson equivalent $\gamma_{5}D_{\mathsf{w}}$, is even lacking
Hermiticity \citep{Adams:2011xf}. For these reasons the overlap construction
with a usual staggered kernel does not give rise to a proper lattice
fermion formulation.

All these problems could be eventually overcome by Adams' construction
introduced in Refs.~\citep{Adams:2009eb,Adams:2010gx,Adams:2011xf}.
The idea is that, after the introduction of the staggered Wilson term,
one can use the $\Gamma_{55}$ operator instead of $\Gamma_{5}$.
Although $\Gamma_{55}$ has a spin$\,\otimes\,$flavor interpretation
\citep{Golterman:1984cy} of $\Gamma_{55}\cong\gamma_{5}\otimes\xi_{5}$,
the staggered Wilson term projects out the negative flavor-chirality
species and we find $\Gamma_{55}\cong\Gamma_{5}+\mathcal{O}\left(a^{2}\right)$
on the physical species. Furthermore, $\Gamma_{55}^{2}=\One$ holds
exactly and $H_{\mathsf{sw}}=\Gamma_{55}D_{\mathsf{sw}}$ defines
a Hermitian operator.

As a consequence, one can apply the overlap construction to the staggered
Wilson kernel with the help of a simple replacement rule given by
\begin{equation}
D_{\mathsf{w}}\to D_{\mathsf{sw}},\qquad\gamma_{5}\to\Gamma_{55}\label{eq:StagOvReplRule}
\end{equation}
as derived in Refs.~\citep{Adams:2009eb,Adams:2010gx,Adams:2011xf}.
Explicitly, the staggered overlap operator $D_{\mathsf{so}}$ reads
\begin{equation}
D_{\mathsf{so}}=\frac{1}{a}\left(\One+A\left[A^{\dagger}A\right]^{-\frac{1}{2}}\right)=\frac{1}{a}\left(\One+\Gamma_{55}\frac{H_{\mathsf{sw}}}{\sqrt{H_{\mathsf{sw}}^{2}}}\right)=\frac{1}{a}\left(\One+\Gamma_{55}\sign H_{\mathsf{sw}}\right)\label{eq:StagOvDop}
\end{equation}
with kernel operator $A=D_{\mathsf{sw}}\left(-M_{0}\right)$, Hermitian
operator $H=\Gamma_{55}D_{\mathsf{sw}}\left(-M_{0}\right)$ and negative
mass parameter $M_{0}$ within the range $M_{0}\in\left(0,2r\right)$
in the free-field case. We find that the action $S_{\mathsf{so}}=\overline{\chi}D_{\mathsf{so}}\chi$
is invariant under the infinitesimal transformation 
\begin{equation}
\delta\overline{\chi}=\overline{\chi}\,\Gamma_{55},\qquad\delta\chi=\Gamma_{55}\left(\One-aD_{\mathsf{so}}\right)\chi,
\end{equation}
cf.\ Eq.~\eqref{eq:OverlapChiralTransf} for the usual overlap operator.
Furthermore, the staggered overlap operator satisfies the Ginsparg-Wilson
relation 
\begin{equation}
\left\{ \Gamma_{55},D_{\mathsf{so}}\right\} =\frac{a}{r}D_{\mathsf{so}}\Gamma_{55}D_{\mathsf{so}}.
\end{equation}
As in four space-time dimensions $D_{\mathsf{sw}}$ describes two
tastes, the staggered overlap operator $D_{\mathsf{so}}$ represents
two degenerate chiral fermion species on the lattice, which can be
used to describe e.g.\ the up and down quark.

\section{The index theorem and the axial anomaly \label{sec:OV-Index-theorem}}

With respect to the topological aspects we are discussing throughout
this thesis, let us introduce some important concepts from topology.
If $n_{\mp}$ denote the number of left-handed and right-handed zero
modes, then we define the integer-valued $\index$ of the Dirac operator
$D$ as
\begin{equation}
\index D=n_{-}-n_{+}.
\end{equation}
In the $d$-dimensional ($d$ even) continuum setting of quantum chromodynamics,
the famous Atiyah-Singer index theorem \citep{AtiyahI,AtiyahIII,AtiyahIV,AtiyahV}
states that
\begin{equation}
n_{-}-n_{+}=\left(-1\right)^{d/2}Q_{\mathsf{cont}},\label{eq:ContIndexTheorem}
\end{equation}
where $Q_{\mathsf{cont}}$ is the so-called topological charge given
by
\begin{equation}
Q_{\mathsf{cont}}=\intop\mathrm{d}^{4}x\,q_{\mathsf{cont}}\left(x\right),\qquad q_{\mathsf{cont}}\left(x\right)=\frac{1}{32\pi^{2}}\varepsilon_{\mu\nu\rho\sigma}\tr\left[F_{\mu\nu}\left(x\right)F_{\rho\sigma}\left(x\right)\right]
\end{equation}
and $F_{\mu\nu}\left(x\right)$ is the corresponding field strength
tensor. The index theorem in Eq.~\eqref{eq:ContIndexTheorem} is
remarkable given the fact that the right hand side appears to be a
continuous function of the gauge field, while the left hand side is
an integer number. On the lattice one can formulate a similar index
theorem \citep{Hasenfratz:1998ri}, which reads
\begin{equation}
n_{-}-n_{+}=\left(-1\right)^{d/2}Q
\end{equation}
with
\begin{equation}
Q=a^{4}\sum_{x\in\Lambda}q\left(x\right),\qquad q\left(x\right)=\frac{1}{2a^{3}}\tr\left[\gamma_{5}D\left(x,x\right)\right].
\end{equation}
We note that in practice one never finds both $n_{-}$ and $n_{+}$
to be non-vanishing at the same time on thermalized configurations,
i.e.\ one finds either $n_{-}=\left(-1\right)^{d/2}Q$ or $n_{+}=\left(-1\right)^{1+d/2}Q$.
Moreover, in two dimensions one can show the Vanishing Theorem \citep{Kiskis:1977vh,Nielsen:1977aw,Ansourian:1977qe}.
That is, one can prove that if $Q\neq0$, then either $n_{-}$ or
$n_{+}$ necessarily vanishes.

While a chiral flavor-singlet rotation is a symmetry of the action
for a Dirac operator satisfying the Ginsparg-Wilson relation, the
path integral itself is not invariant due to the nontrivial transformation
of the integration measure \citep{Hasenfratz:1998ri,Luscher:1998pqa},
cf.\ Refs.~\citep{Adler:1969gk,Bell:1969ts,Fujikawa:1979ay,Fujikawa:1980eg,Fujikawa:1980vr}
for the continuum case. This so-called axial anomaly results in an
anomalous continuity equation for the singlet axial vector current.
The presence of the axial anomaly implies that there is no flavor-singlet
Goldstone particle and explains why the $\eta$-meson is significantly
heavier than the pions assuming an approximate chiral symmetry for
two flavors, see e.g.\ Ref.~\citep{Gattringer:2010zz}. In the three
flavor case the Witten-Veneziano formula \citep{Witten:1979vv,Veneziano:1979ec,Veneziano:1980xs,Giusti:2001xh,Seiler:2001je,Giusti:2004qd,Luscher:2004fu,Aguado:2005kp}
connects the mass of the $\eta^{\prime}$-meson to the topological
susceptibility $\chi_{\mathsf{top}}=\left\langle Q^{2}\right\rangle /V$,
where $V$ is the space-time volume.

\section{Continuum limit of the index and axial anomaly \label{sec:OV-Continuum-limit}}

Massless overlap fermions have exact zero modes with definite chirality
and one can define an index for the overlap Dirac operator as discussed
in Subsec.\ \ref{sec:OV-Index-theorem}. Overlap fermions implement
an exact chiral symmetry on the lattice, which is anomalously broken
in the quantum case and gives rise to the axial anomaly. This axial
anomaly and the index density of the overlap Dirac operator are proportional,
while the continuum axial anomaly is proportional to the topological
charge density and by the Atiyah-Singer index theorem \citep{AtiyahI,AtiyahIII,AtiyahIV,AtiyahV}
we find equality between the topological charge and the index.

In the setting of lattice quantum chromodynamics a natural question
is then if the continuum axial anomaly and index are recovered in
the continuum limit in sufficiently smooth gauge background fields.
In the past several investigations \citep{Ginsparg:1981bj,Hasenfratz:1998ri,Kikukawa:1998pd,Fujikawa:1998if,Suzuki:1998yz,Adams:1998eg,Adams:2000rn}
dealt with this question. Here, as a first step to secure the theoretical
foundations for Adams' construction, we want to check that the continuum
axial anomaly and index is correctly reproduced for the staggered
overlap Dirac operator.

Our strategy for evaluating the continuum limit is similar to the
Wilson case \citep{Adams:1998eg,Adams:2000rn}, although one encounters
some technical complications due to the nontrivial spin-flavor structure
of staggered fermions. In Ref.~\citep{Adams:2013lpa}, we found that
the index correctly reproduces the index in the continuum (cf.\ the
numerical studies in Refs.~\citep{PetrashykWorkshop,deForcrand:2011ak,deForcrand:2012bm}),
while for the anomaly first an averaging over the sites of a lattice
hypercube is needed. This is not unexpected as in the staggered formalism
spin and flavor degrees of freedom are distributed over lattice hypercubes,
while the distance between neighboring lattice sites is of order $\mathcal{O}\left(a\right)$,
shrinking to zero in the continuum limit.

In this section, we briefly review our discussion\footnote{Discussion based on \sloppy \bibentry{Adams:2013lpa}. Copyright
owned by the authors under the terms of the Creative Commons Attribution-NonCommercial-NoDerivatives
4.0 International License (CC BY-NC-ND 4.0).} given in Ref.~\citep{Adams:2013lpa}, where we previously presented
the following results. For a more elaborate discussion, we refer the
reader to the thesis in Ref.~\citep{HarThesis}.

\subsection{The index and axial anomaly}

We now give an overview of the derivation of the continuum limit of
the index and anomaly of the staggered overlap operator in the setting
of lattice quantum chromodynamics. We begin by recalling the definition
of the staggered overlap operator as defined in Eq.~\eqref{eq:StagOvDop}
for the case of $r=1$, which reads
\begin{equation}
D_{\mathsf{so}}=\frac{1}{a}\left(\One+\Gamma_{55}\frac{H}{\sqrt{H^{2}}}\right),\qquad H=\Gamma_{55}D_{\mathsf{sw}}\left(-m\right).
\end{equation}
Using the index formula in Ref.~\citep{Hasenfratz:1998ri}, we can
write
\begin{equation}
\index D_{\mathsf{so}}=-\frac{1}{2}\mbox{Tr}\left(\frac{H}{\sqrt{H^{2}}}\right).\label{eq:IndexTr}
\end{equation}
In the following, our setting is a four-dimensional box of length
$L$ with lattice spacing $a$ and $N$ sites in each direction. The
lattice transcripts of a smooth continuum gauge field $A_{\mu}\left(x\right)$
are taken as the link variables, taking values in the Lie algebra
$\mathfrak{su}\!\left(3\right)$ of $\SUthree$. Furthermore, we impose
boundary conditions as described in Ref.~\citep{Adams:2000rn}.

Let us now express Eq.~\eqref{eq:IndexTr} as a sum over the index
density
\begin{equation}
\index D_{\mathsf{so}}=a^{4}\sum_{x}q\left(x\right),\qquad q\left(x\right)=-\frac{1}{2}\mbox{tr}_{\mathsf{c}}\left(\frac{H}{\sqrt{H^{2}}}\right)\left(x,x\right).
\end{equation}
Here $\mbox{tr}_{\mathsf{c}}$ denotes a color trace and $\Theta\left(x,y\right)$
is defined for an operator $\Theta$ by $\Theta\chi\left(x\right)=\sum_{y}\Theta\left(x,y\right)\chi\left(y\right)$.
Like in the case of usual overlap fermions, we find that the index
density $q\left(x\right)$ is related to the axial anomaly via ${\cal A}\left(x\right)=2\ii q\left(x\right)$,
where ${\cal A}\left(x\right)$ is the divergence of the axial current.

In the case of the overlap construction with a Wilson kernel, the
central problem in verifying the correct continuum limits of the index
and axial anomaly is to show that
\begin{equation}
\lim_{a\to0}q\left(x\right)=-\frac{1}{32\pi^{2}}\varepsilon_{\mu\nu\sigma\rho}\mbox{tr}_{\mathsf{c}}\left[F_{\mu\nu}\left(x\right)F_{\sigma\rho}\left(x\right)\right],\label{eq:ContLimitIndexDensity}
\end{equation}
where $\mbox{tr}_{\mathsf{c}}$ denotes a trace in color space, as
it was done in Refs.~\citep{Adams:1998eg,Adams:2000rn}. The $a\to0$
limit is taken by consecutive symmetric refinements of the underlying
lattice. As mentioned earlier, in the case of the staggered overlap
operator Eq.~\eqref{eq:ContLimitIndexDensity} only holds if we average
over the sites of a lattice hypercube which contains $x$. For the
remainder of this section we are going to outline the proof of Eq.~\eqref{eq:ContLimitIndexDensity},
where we have to include the averaging procedure and a factor of two
on the right hand side to account for the two physical tastes of the
staggered Wilson kernel.

\subsection{Continuum limit of the index density}

Following the derivation for the usual overlap case given in Ref.~\citep{Adams:2000rn},
we use the identity
\begin{equation}
\lim_{a\to0}\frac{H}{\sqrt{H^{2}}}\left(x,x\right)=\lim_{a\to0}\intop_{-\pi/2a}^{3\pi/2a}\mathrm{d}^{4}p\,e^{-\ii px}\frac{H}{\sqrt{H^{2}}}e^{\ii px},\label{eq:HHsqTermLimitA}
\end{equation}
where expressions like $px$ are understood as scalar products. Let
us begin by introducing some useful notation. We write the lattice
momentum $p_{\mu}\in\left[-\pi/2,3\pi/2\right]$ as $p_{\mu}=\frac{\pi}{a}B_{\mu}+q_{\mu}$,
where $B_{\mu}\in\left\{ 0,1\right\} $ and $q_{\mu}\in\left[-\pi/2,\pi/2\right]$.
Furthermore, we define the sum of two vectors $A,B\in\left\{ 0,1\right\} ^{4}$
as the componentwise sum $\mathrm{mod}\,2$ and let $e_{B}\left(x\right)\equiv\exp\left(\ii\pi Bx/a\right)$
for $B\in\left\{ 0,1\right\} ^{4}$. On the vector space $V$ spanned
by the $e_{B}$, we can define two commuting representations of the
Dirac algebra \citep{Golterman:1984cy} given by
\begin{align}
\left(\hat{\Gamma}_{\mu}\right)_{AB} & =\left(-1\right)^{A_{\mu}}\delta_{A,B+\eta_{\mu}},\\
\left(\hat{\Xi}_{\mu}\right)_{AB} & =\left(-1\right)^{A_{\mu}}\delta_{A,B+\zeta_{\mu}},
\end{align}
where $\left(\eta_{\mu}\right)_{\sigma}=1$ for $\sigma<\mu$, $\left(\zeta_{\mu}\right)_{\sigma}=1$
for $\sigma>\mu$ and all remaining components being zero. Here we
use the notation with hat as the versions without the hat denote the
extensions from the space $V$ to the one-component spinor field space.

In the following, we use the notation
\begin{equation}
\Xi_{5}\equiv\Gamma_{55}\Gamma_{5},\qquad C_{5}\equiv{\textstyle \sum}_{\alpha\beta\gamma\delta}^{\mathsf{sym}}C_{\alpha}C_{\beta}C_{\gamma}C_{\delta}/4!,
\end{equation}
where $C_{5}$ is the symmetric product of the $C_{\mu}=\left(T_{\mu}+T_{\mu}^{\dagger}\right)/2$
operators. In our notation, the staggered Wilson term reads $W_{\mathsf{st}}=\frac{r}{a}\left(\One-\Xi_{5}\right)$,
cf.\ Sec.\ \ref{sec:SW-Adams-mass-term}. Writing a plane wave as
\begin{equation}
\exp\left(\ii px\right)=\exp\left(\ii\left[\pi B/a+q\right]x\right)=e_{B}\left(x\right)\exp\left(\ii qx\right),
\end{equation}
we can understand the action of the staggered Wilson kernel $D_{\mathsf{sw}}$
in $AB$-language to be
\begin{align}
D_{\mathsf{sw}}e^{\ii px} & =D_{\mathsf{sw}}\left(e_{B}\left(x\right)e^{\ii qx}\right)\nonumber \\
 & =e_{A}\left(x\right)\left[\left(\hat{\Gamma}_{\mu}\right)_{AB}\nabla_{\mu}+\frac{r}{a}\left(\delta_{AB}-\left(\hat{\Xi}_{5}\right)_{AB}C\right)\right]e^{\ii qx}\nonumber \\
 & =e_{A}\left(x\right)\left[\hat{\Gamma}_{\mu}\nabla_{\mu}+\frac{r}{a}\left(\One-\hat{\Xi}_{5}C\right)\right]_{AB}e^{\ii qx}\nonumber \\
 & \equiv e_{A}\left(x\right)\left(\tilde{D}_{\mathsf{sw}}\right)_{AB}e^{\ii qx},
\end{align}
where we sum over all sixteen possible values of the repeated $A$-index.
Similarly, we find for the $\Gamma_{55}$ operator
\begin{equation}
\Gamma_{55}e^{\ii px}=\Gamma_{55}\left(e_{B}\left(x\right)e^{\ii qx}\right)=e_{A}\left(x\right)\left(\hat{\Gamma}_{5}\hat{\Xi}_{5}\right)_{AB}e^{\ii qx}.
\end{equation}
We then conclude that
\begin{equation}
\frac{H}{\sqrt{H^{2}}}e^{\ii px}=\frac{H}{\sqrt{H^{2}}}\left(e_{B}\left(x\right)e^{\ii qx}\right)=e_{A}\left(x\right)\left(\frac{\tilde{H}}{\sqrt{\tilde{H}^{2}}}\right)_{AB}e^{\ii qx},
\end{equation}
where $\tilde{H}$ follows from $H$ by replacing $D_{\mathsf{sw}}\to\tilde{D}_{\mathsf{sw}}$
and $\Gamma_{55}\to\hat{\Gamma}_{5}\hat{\Xi}_{5}$. This allows us
to rewrite Eq.~\eqref{eq:HHsqTermLimitA} as
\begin{eqnarray}
\lim_{a\to0}\frac{H}{\sqrt{H^{2}}}\left(x,x\right) & = & \lim_{a\to0}\sum_{B}\intop_{-\pi/2a}^{\pi/2a}\mathrm{d}^{4}q\,e^{-\ii qx}e_{B}\left(x\right)\frac{H}{\sqrt{H^{2}}}e_{B}\left(x\right)e^{\ii qx}\nonumber \\
 & = & \lim_{a\to0}\sum_{A,B}e^{\ii\pi\left(A-B\right)x/a}\intop_{-\pi/2a}^{\pi/2a}\mathrm{d}^{4}q\,e^{-\ii qx}\left(\frac{\tilde{H}}{\sqrt{\tilde{H}^{2}}}\right)_{AB}e^{\ii qx}.\label{eq:IndexTwoContr}
\end{eqnarray}
In the last sum in Eq.~\eqref{eq:IndexTwoContr}, we now separately
analyze the contributions from the case $A=B$ and $A\neq B$. While
the former reproduces the continuum topological charge density $q\left(x\right)$,
the latter contributions vanish after averaging over a lattice hypercube
containing $x$.

\minisec{The $\boldsymbol{A=B}$ case}

Let us begin with the $A=B$ case in Eq.~\eqref{eq:IndexTwoContr}.
In this case we find
\begin{equation}
\lim_{a\to0}\intop_{-\pi/2a}^{\pi/2a}\mathrm{d}^{4}q\,e^{-\ii qx}\Tr\left(\frac{\tilde{H}}{\sqrt{\tilde{H}^{2}}}\right)e^{\ii qx},\label{eq:CL-AeqBContr}
\end{equation}
where the trace is taken in the vector space $V$. Following Ref.~\citep{Golterman:1984cy},
there exist an isomorphism such that $\hat{\Gamma}_{\mu}\cong\gamma_{\mu}\otimes\One$
and $\hat{\Xi}_{\nu}\cong\One\otimes\xi_{\nu}$. In a basis where
$\xi_{5}$ is diagonal, we find 
\begin{equation}
\tilde{D}_{\mathsf{sw}}=\left(\gamma_{\mu}\otimes\One\right)\nabla_{\mu}+\left(\One\otimes\One\right)\frac{r}{a}\left(1\mp C_{5}\right)\label{eq;CL-DswInterpretation}
\end{equation}
with signs $\mp$ for the flavors on which $\xi_{5}=\pm\One$. On
each of these two flavor subspaces $\tilde{D}_{\mathsf{sw}}$ is a
hypercubic lattice Dirac operator as discussed in Ref.~\citep{Adams:2003hy}
and $\Gamma_{55}\cong\gamma_{5}\otimes\xi_{5}=\pm\left(\gamma_{5}\otimes{\bf 1}\right)$.
Noting that the momentum representation of $C_{5}$ in the free-field
case reads
\begin{equation}
C_{5}\left(aq\right)={\textstyle \prod}_{\mu}\cos\left(aq_{\mu}\right)=1+\mathcal{O}\left(a^{2}\right),
\end{equation}
we see that $\tilde{D}_{\mathsf{sw}}$ describes a single physical
flavor for each of the two species of the subspace where $\xi_{5}=\One$.
Finally, note that in this case Eq.~\eqref{eq:CL-AeqBContr} equals
to $\tr\left(H/\sqrt{H^{2}}\right)\left(x,x\right)$ with $H=\left(\gamma_{5}\otimes\One\right)\left(\tilde{D}_{\mathsf{sw}}-m\right)$.
Using the general result of Ref.~\citep{Adams:2003hy} and the fact
that the trace over the flavor subspace produces a factor of two,
we find that $q\left(x\right)$ yields the continuum topological charge
density with an additional factor of two due to the two physical species
of the staggered Wilson kernel.

At the same time, the contribution from the other flavor subspace,
where $\xi_{5}=-\One$, vanishes as there are no physical fermion
species, which again follows from Ref.~\citep{Adams:2003hy}.

\minisec{The $\boldsymbol{A\protect\neq B}$ case}

We are left with analyzing the $A\neq B$ contributions in Eq.~\eqref{eq:IndexTwoContr}
when using the mentioned averaging procedure. Rewriting the prefactor
$e^{\ii\pi\left(A-B\right)x/a}=\left(-1\right)^{\left(A-B\right)n}$
with $n=x/a$, we see that the sum over lattice sites gives zero if
not all components of $A-B$ are zero. Hence the problem is reduced
to showing that
\begin{equation}
\intop_{-\pi/2a}^{\pi/2a}\mathrm{d}^{4}q\,e^{-\ii qx}\left(\frac{\tilde{H}}{\sqrt{\tilde{H}^{2}}}\right)_{AB}e^{\ii qx}\label{eq;CL-VaryingPos}
\end{equation}
changes by order $\mathcal{O}\left(a\right)$ when $x$ is moved within
a lattice hypercube. This can be shown by expanding the integrand
in powers of the continuum gauge field as done in Refs.~\citep{Adams:1998eg,Adams:2000rn}.
While the integral formally diverges as $\mathcal{O}\left(a^{-4}\right)$,
one can show that the lowest non-vanishing contribution in the expansion
is of order $\mathcal{O}\left(a^{4}\right)$. As the smooth continuum
gauge field varies by $\mathcal{O}\left(a\right)$ when moving $x$
inside a lattice hypercube, we conclude that Eq.~\eqref{eq;CL-VaryingPos}
changes by an order $\mathcal{O}\left(a\right)$ term as well, completing
our proof.

\minisec{Further discussion}

For more details on this derivation we refer the reader to Refs.~\citep{Adams:2013lpa,HarThesis}.
While Ref.~\citep{Adams:2013lpa} deals with the above derivation,
the thesis in Ref.~\citep{HarThesis} also discusses the index of
lattice Dirac operators in a wider setting.

\chapter{Spectral properties \label{chap:Eigenvalue-spectra}}

An important motivation for studying staggered overlap fermions \citep{Adams:2009eb,Adams:2010gx,Adams:2011xf}
is the prospect of bringing down the enormous computational costs
of simulations with lattice fermions with exact chiral symmetry. A
first study on this was done in Refs.~\citep{deForcrand:2011ak,deForcrand:2012bm},
where in the free-field limit a large speedup factor of $\mathcal{O}\left(10\right)$
compared to usual overlap fermions was found. However, after the introduction
of a thermalized QCD background field this factor dropped to $\mathcal{O}\left(2\right)$.
At the same time, we observed a large speedup factor of $4\text{--}6$
for the staggered Wilson kernel compared to the Wilson kernel as discussed
in Chapter\ \ref{chap:Computational-efficiency}.

In this chapter, we try to explain the somewhat surprising observation
that the overlap construction shows a much smaller speedup factor
compared to the underlying kernel operator. We try to explain this
discrepancy with the spectral and chiral properties of the staggered
Wilson Dirac operator for different lattices sizes. To this end, we
note that the studies of staggered overlap fermions in Refs.~\citep{deForcrand:2011ak,deForcrand:2012bm}
were carried out on relatively small lattices, where the eigenvalue
spectrum has a small gap and diffuse branches. Our investigations
of the computational efficiency of the staggered Wilson kernel operator
in Chapter\ \ref{chap:Computational-efficiency}, on the other hand,
were done on significantly larger lattices.

In the following, we present indicators in favor of the idea that
the spectral and chiral properties of the staggered Wilson kernel
improve on larger lattices and, thus, the performance of staggered
overlap fermions on small lattices is not representative of the expected
performance in a realistic setting. As only an actual benchmark on
larger lattices will eventually clarify this point, the following
discussion should be taken as a proposed explanation. Some of the
results discussed in this chapter were previously presented in Refs.~\citep{Adams:2013tya,ZielinskiLat14}.

\section{Introduction}

While usual overlap fermions \citep{Narayanan:1992wx,Narayanan:1993ss,Narayanan:1993sk,Narayanan:1994gw,Neuberger:1997fp,Neuberger:1998my,Neuberger:1998wv}
are very attractive due to the presence of chiral zero modes and having
a well-defined index, their high computational costs make their practical
use for high-precision studies limited. By using staggered overlap
fermions, which use the staggered Wilson Dirac operator as a kernel
operator, one can potentially reduce these costs significantly. Initial
numerical tests confirmed this expectation only partially \citep{deForcrand:2011ak,deForcrand:2012bm},
see also Ref.~\citep{Durr:2013gp}. While in the free-field case
an impressive speedup factor of $\mathcal{O}\left(10\right)$ compared
to usual overlap fermions was observed, in a quenched QCD background
field this factor reduced to $\mathcal{O}\left(2\right)$.

Naturally the computational performance of the overlap operator depends
on the spectral and chiral properties of the underlying kernel operator.
In particular, a more chiral kernel is expected to result in lowered
computational costs \citep{Bietenholz:1998ut}. The reduction of the
speedup factor for staggered overlap fermions, when moving from the
free-field case to thermalized configurations, was explained in Refs.~\citep{deForcrand:2011ak,deForcrand:2012bm}
by the spectral properties of the staggered Wilson kernel. While in
the free-field limit the eigenvalue spectrum has a shape close to
the Ginsparg-Wilson circle, in the case of a $\beta=6$ background
field the spectrum contracts, the gap of the spectrum becomes narrow
and the branches become diffuse. At $\beta=5.8$, these effects become
even more severe. Due to these changes in the eigenvalue spectrum
significantly more computational time is needed, partly offsetting
the advantage of a smaller fermion matrix.

At the same time we could show in Refs.~\citep{Adams:2013tya,ZielinskiLat14},
that at $\beta=6$ the staggered Wilson kernel has a significantly
reduced condition number and is $4\text{--}6$ times more efficient
for inverting the Dirac operator on a source compared to the Wilson
kernel. The difference in the relative performance of staggered overlap
vs.\ usual overlap fermions compared to the one of their kernel operators
as well as the connection to spectral properties then has to be clarified.

Looking at Refs.~\citep{deForcrand:2011ak,deForcrand:2012bm}, we
note the use of small lattices of up to a size of $12^{4}$ for benchmarking
the staggered overlap operator. Generally, the use of large lattices
for overlap fermions is difficult due to the high computational costs
and might be difficult to justify for exploratory studies. On the
other hand, the studies of the staggered Wilson Dirac operator in
Refs.~\citep{Adams:2013tya,ZielinskiLat14} were carried out on lattice
sizes of up to $20^{3}\times40$.

This gives rise to the question if the observed large differences
of the computational efficiency of overlap vs.\ staggered overlap
fermions on small lattices compared to the respective kernel operators
on larger lattices are connected to changes in the eigenvalue spectrum
and chiral properties when varying the lattice size. To answer this
question, we investigate the eigenvalue spectrum of the staggered
Wilson Dirac operator. We investigate how the spectrum changes when
moving from small to larger lattices and discuss how this can impact
the performance of staggered overlap fermions.

This chapter is organized as follows. In Sec.\ \ref{sec:EV-Free-field},
we begin by discussing the free-field eigenvalue spectrum of the relevant
fermion discretizations. In Sec.\ \ref{sec:EV-Quenched-QCD}, we
investigate eigenvalue spectra in the case of quenched QCD background
fields and discuss how changes in the spectra affect computational
performance. Finally, in Sec.\ \ref{sec:EV-Conclusions} we make
some concluding remarks.

\section{Free-field case \label{sec:EV-Free-field}}

We begin by examining the free-field eigenvalue spectra of usual Wilson
and staggered Wilson fermions. Throughout this chapter, we set the
lattice spacing to $a=1$ and, in order to avoid ambiguities, write
summations explicitly.

\subsection{Wilson fermions}

Wilson fermions differ from the naïve discretization by the introduction
of the Wilson term as discussed in Subsec.\ \ref{subsec:GT-Wilson-fermions}.
The Wilson term is an additional covariant Laplacian term and we can
write the resulting Dirac operator in the free-field case as
\begin{equation}
D_{\mathsf{w}}\left(x,y\right)=\left(m+4r\right)\delta_{x,y}\One-\frac{1}{2}\sum_{\mu}\left[\left(r\One-\gamma_{\mu}\right)\delta_{x+\hat{\mu},y}+\left(r\One+\gamma_{\mu}\right)\delta_{x-\hat{\mu},y}\right].
\end{equation}
Here $m$ refers to the bare fermion mass, $r\in\left(0,1\right]$
to the Wilson parameter, $\One$ to a $4\times4$ unit matrix in spinor
space, the $\gamma_{\mu}$ matrices to a representation of the Dirac
algebra $\left\{ \gamma_{\mu},\gamma_{\nu}\right\} =\delta_{\mu,\nu}\One$
and $\delta_{a,b}$ to the Kronecker delta. Going to momentum space
we find
\begin{equation}
D_{\mathsf{w}}\left(p\right)=\left(m+4r\right)\One+\ii\sum_{\mu}\gamma_{\mu}\sin p_{\mu}-r\sum_{\nu}\cos p_{\nu}\One.
\end{equation}
The eigenvalues of this matrix can be easily computed as
\begin{equation}
\lambda_{\mathsf{w}}\left(p\right)=m+2r\sum_{\mu}\sin^{2}\left(p_{\mu}/2\right)\pm\ii\sqrt{\sum_{\nu}\sin^{2}p_{\nu}}.
\end{equation}
On a finite lattice the values of $p_{\mu}$ are discretized. The
allowed values are
\begin{equation}
p_{\mu}=2\pi\left(n_{\mu}+\varepsilon_{\mu}\right)/N_{\mu}
\end{equation}
with $n_{\mu}=0,1,\dots,N_{\mu}-1$. In the case of periodic boundary
conditions we have $\varepsilon_{\mu}=0$, for antiperiodic boundary
conditions $\varepsilon_{\mu}=1/2$. In both cases $N_{\mu}$ denotes
the number of slices in $\mu$-direction.

\subsection{Staggered Wilson fermions}

Two-flavor staggered Wilson fermions were discussed in Sec.\ \ref{sec:SW-Adams-mass-term}.
In the free-field limit, the Dirac operator has a spin$\,\otimes\,$flavor
interpretation \citep{Golterman:1984cy,Adams:2011xf} of
\begin{equation}
D_{\mathsf{sw}}\left(p\right)=m\left(\One\otimes\One\right)+\ii\sum_{\mu}\sin p_{\mu}\left(\gamma_{\mu}\otimes\One\right)+r\One\otimes\left(\One-\gamma_{5}\prod_{\nu}\cos p_{\nu}\right).
\end{equation}
Explicitly multiplying out the Kronecker products we find a $16\times16$
matrix, whose eigenvalues can be calculated as
\begin{equation}
\lambda_{\mathsf{sw}}\left(p\right)=m\pm\ii\sqrt{\sum_{\mu}\sin^{2}p_{\mu}}+r\left(1\pm\prod_{\nu}\cos p_{\nu}\right).
\end{equation}
Here the two ``$\pm$'' are to be chosen independently and the allowed
range for $p_{\mu}$ is as for Wilson fermions, but with $n_{\mu}=0,1,\dots,N_{\mu}/2-1$.
This expression was also derived in Ref.~\citep{deForcrand:2012bm}.

\begin{figure}[t]
\begin{centering}
\includegraphics[width=1\textwidth]{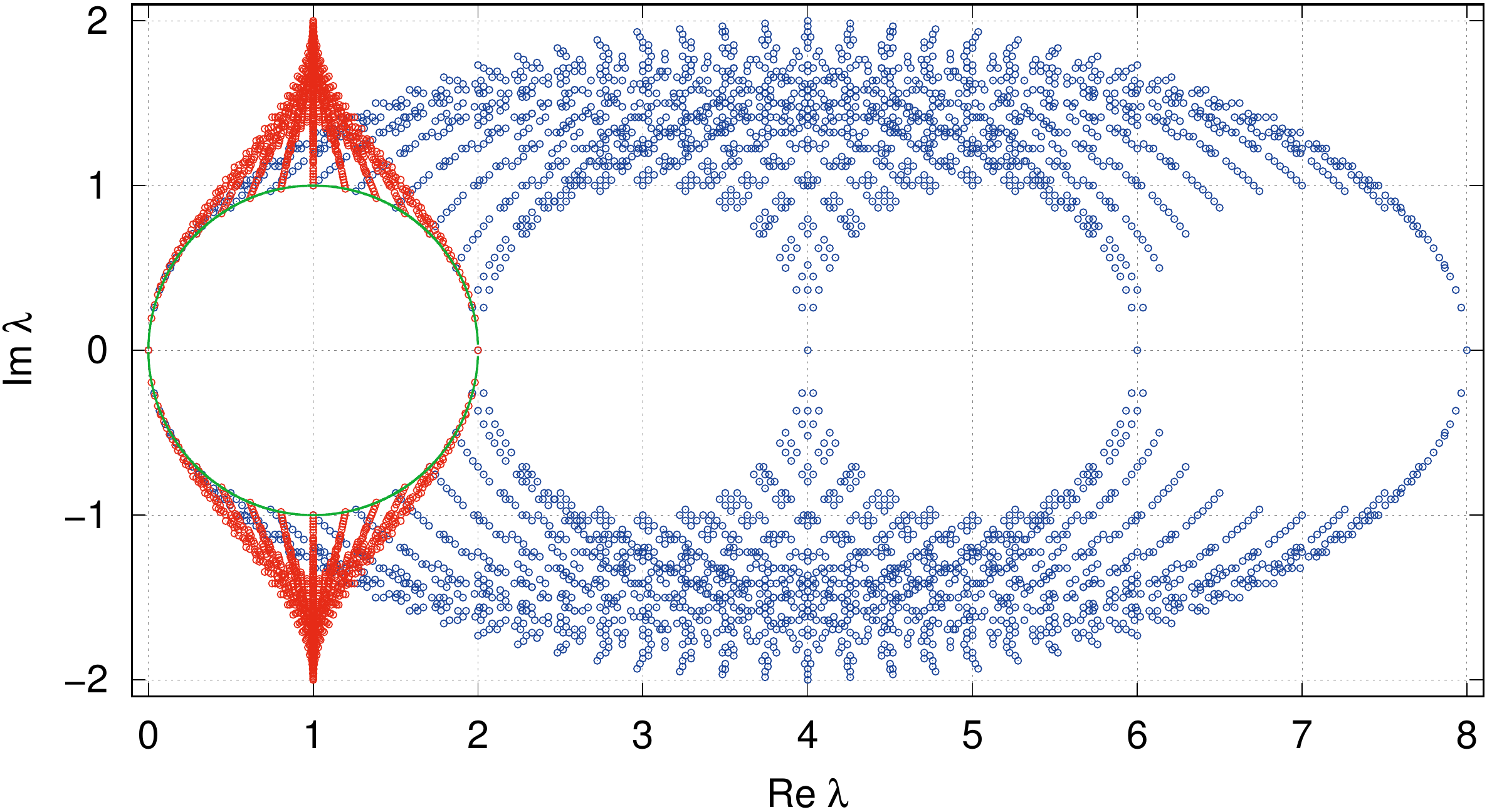}
\par\end{centering}
\caption{Eigenvalue spectrum of the staggered Wilson (red) and the Wilson Dirac
operator (blue) in the free-field case together with the Ginsparg-Wilson
circle (green). \label{fig:FreeFieldSpec}}
\end{figure}

For Wilson and staggered Wilson fermions, the free-field eigenvalue
spectrum can be found in Fig.~\ref{fig:FreeFieldSpec}. As one can
see, the staggered Wilson spectrum is close to the Ginsparg-Wilson
circle and we have a clear separation between the physical and the
doubler branch. Together with the fact that the staggered Wilson fermion
matrix is smaller than the Wilson fermion matrix, one expects a significant
computational advantage for staggered overlap fermions. This was confirmed
by the free-field speedup factor of $\mathcal{O}\left(10\right)$
in the studies of Refs.~\citep{deForcrand:2011ak,deForcrand:2012bm}.

\section{Quenched quantum chromodynamics \label{sec:EV-Quenched-QCD}}

\begin{figure}[t]
\begin{centering}
\subfloat[$32^{4}$ lattice, free field]{\begin{centering}
\includegraphics[width=0.3\textwidth]{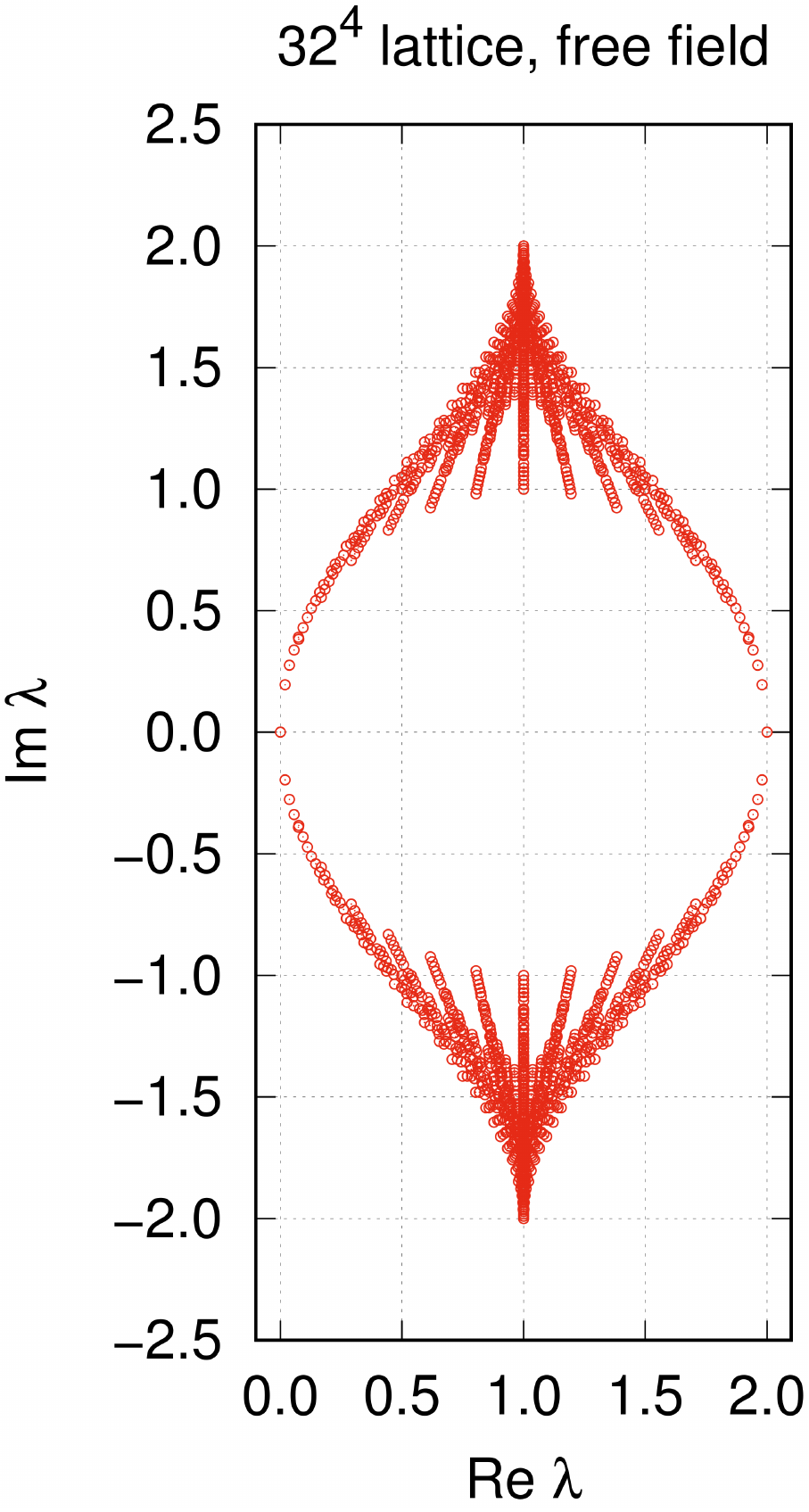}
\par\end{centering}
}\hfill{}\subfloat[$8^{4}$ lattice at $\beta=6$]{\begin{centering}
\includegraphics[width=0.3\textwidth]{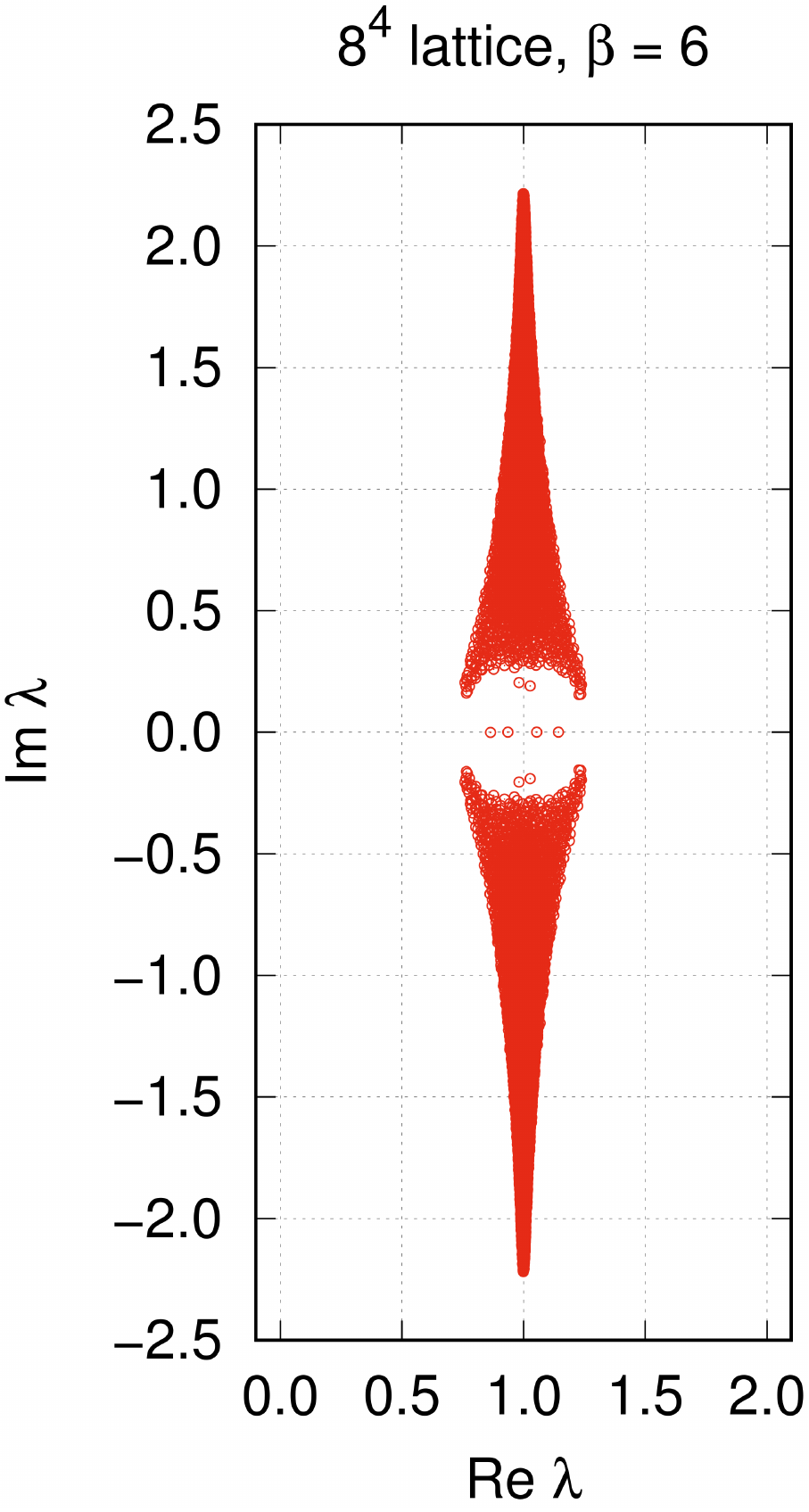}
\par\end{centering}
}\hfill{}\subfloat[$10^{4}$ lattice at $\beta=6$]{\begin{centering}
\includegraphics[width=0.3\textwidth]{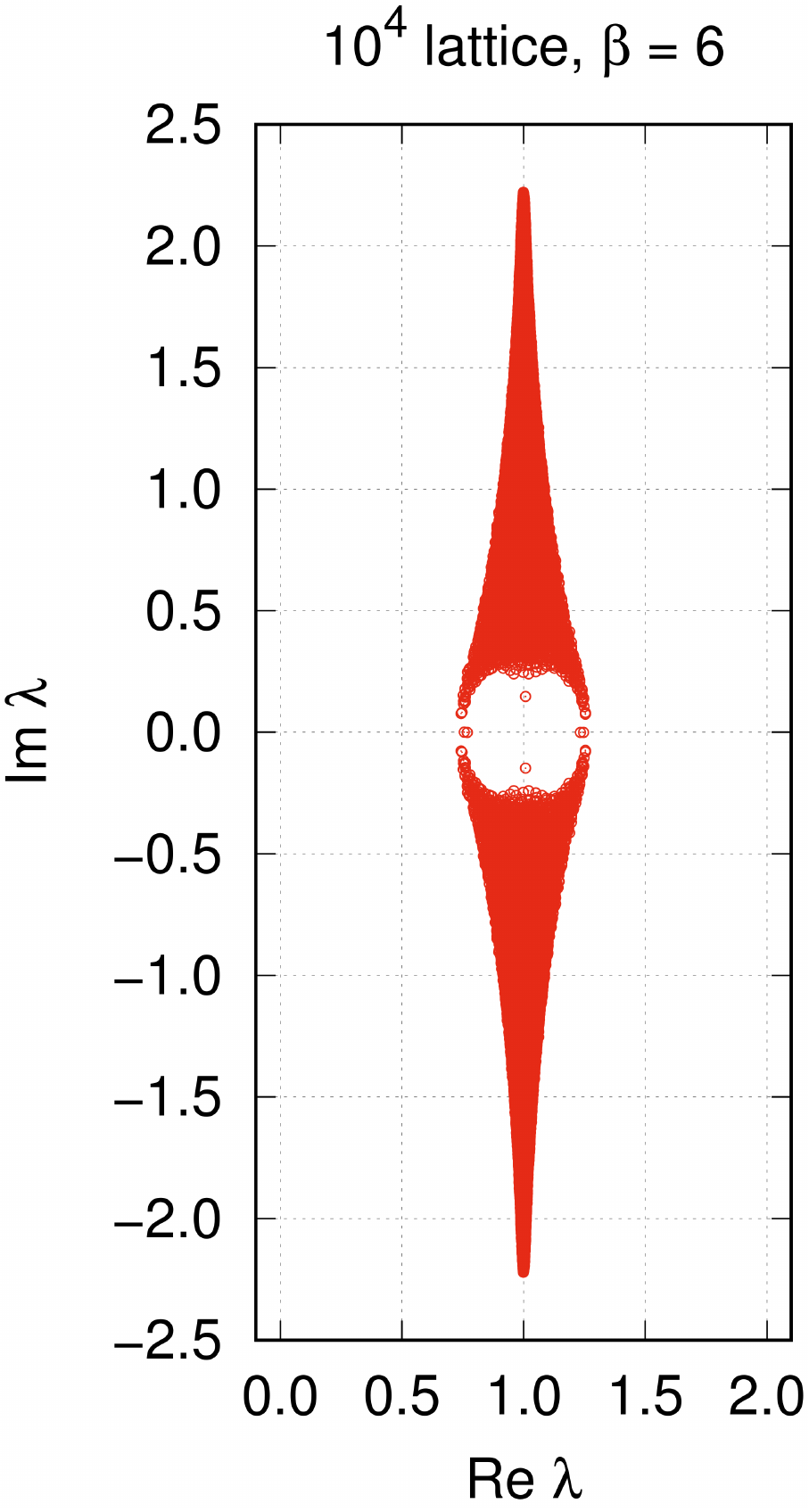}
\par\end{centering}
}
\par\end{centering}
\caption{The staggered Wilson eigenvalue spectrum at $\beta=6$. \label{fig:FullSpecBeta6}}
\end{figure}
\begin{figure}[t]
\begin{centering}
\subfloat[$32^{4}$ lattice, free field]{\begin{centering}
\includegraphics[width=0.3\textwidth]{figures/stag_wilson_spectrum/stwd_free_32x32}
\par\end{centering}
}\hfill{}\subfloat[$8^{4}$ lattice at $\beta=5.8$]{\begin{centering}
\includegraphics[width=0.3\textwidth]{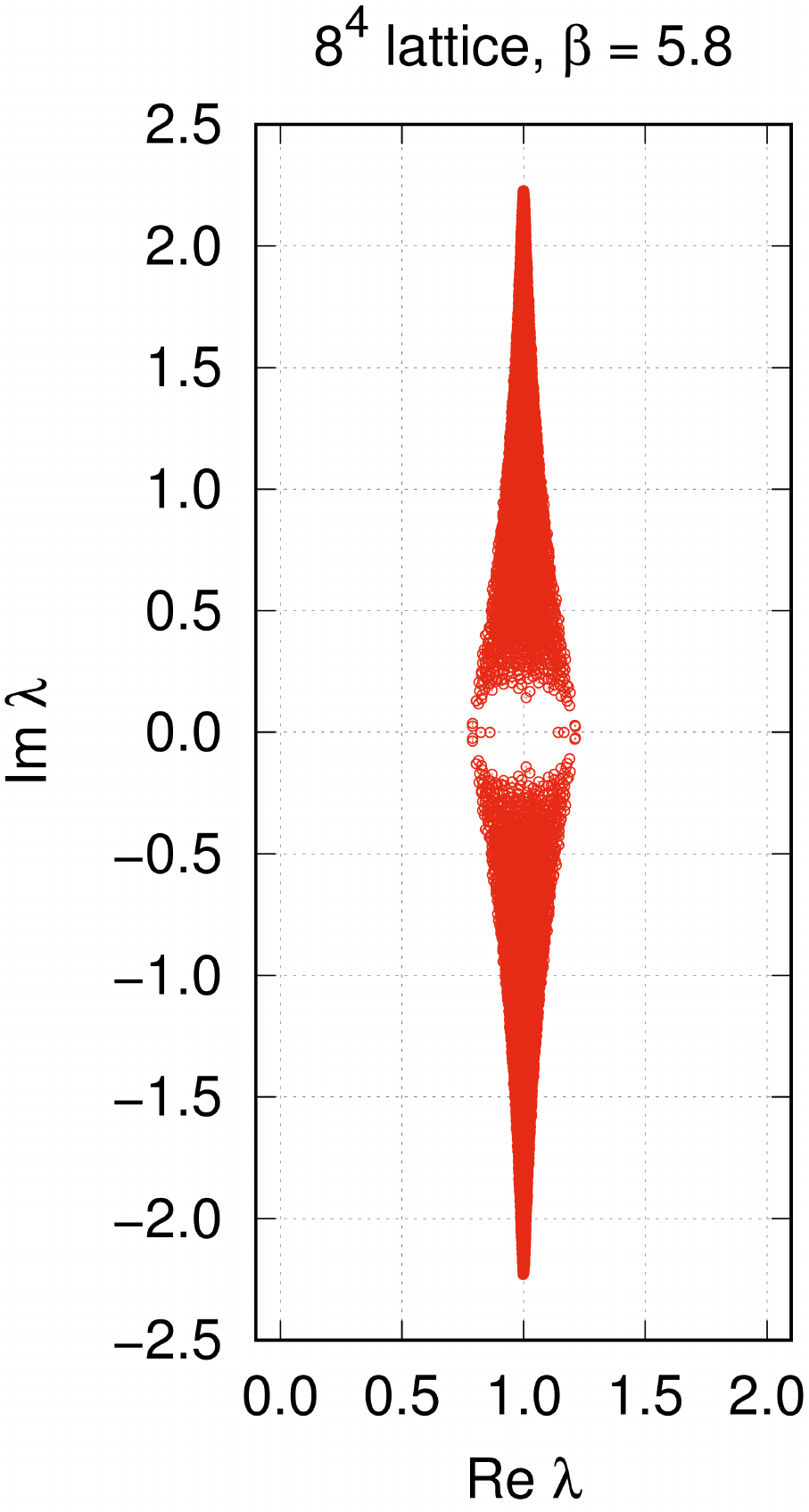}
\par\end{centering}
}\hfill{}\subfloat[$10^{4}$ lattice at $\beta=5.8$]{\begin{centering}
\includegraphics[width=0.3\textwidth]{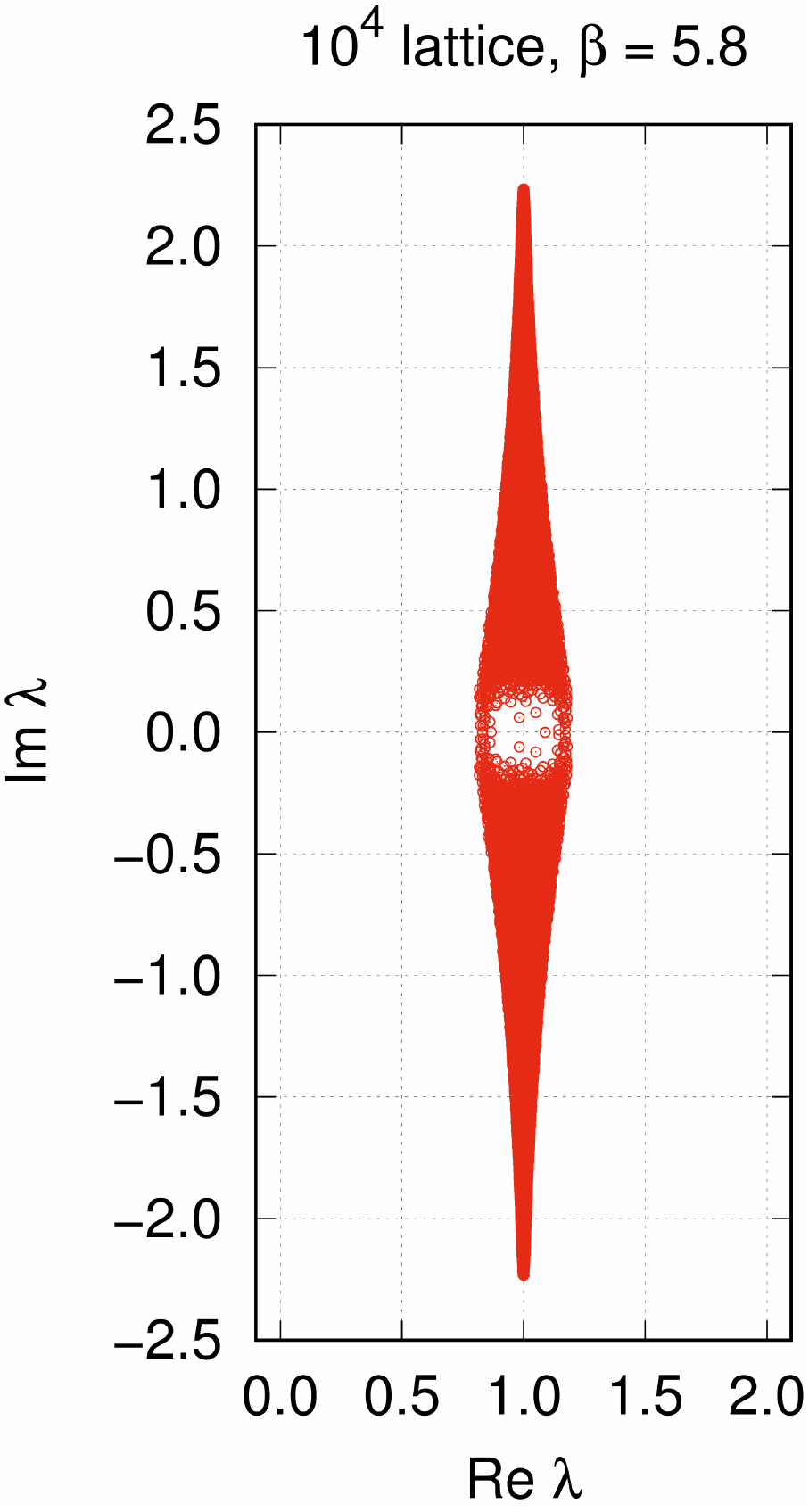}
\par\end{centering}
}
\par\end{centering}
\caption{The staggered Wilson eigenvalue spectrum at $\beta=5.8$. \label{fig:FullSpecBeta58}}
\end{figure}
\begin{figure}[t]
\begin{centering}
\subfloat[$8^{4}$ lattice at $\beta=6$]{\begin{centering}
\includegraphics[height=0.65\textwidth]{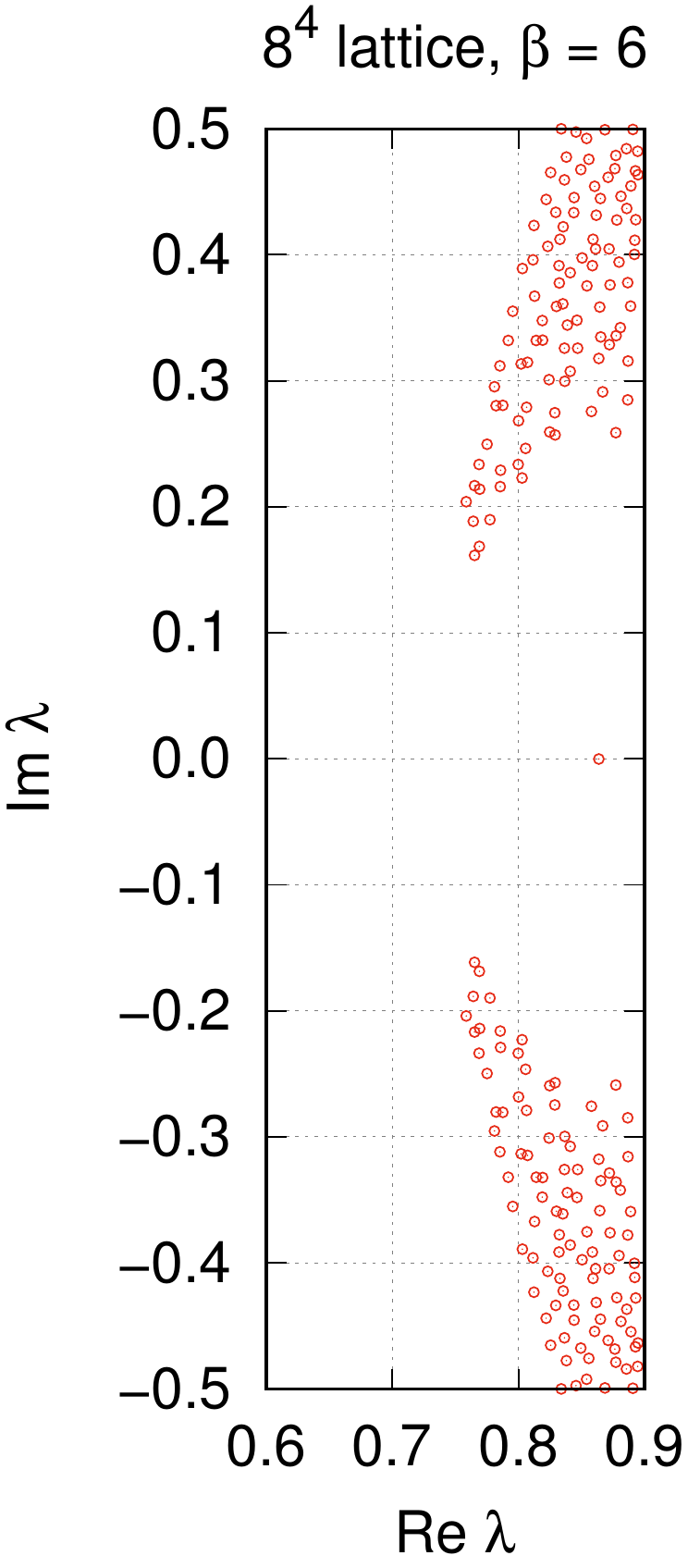}
\par\end{centering}
}\hfill{}\subfloat[$10^{4}$ lattice at $\beta=6$]{\begin{centering}
\includegraphics[height=0.65\textwidth]{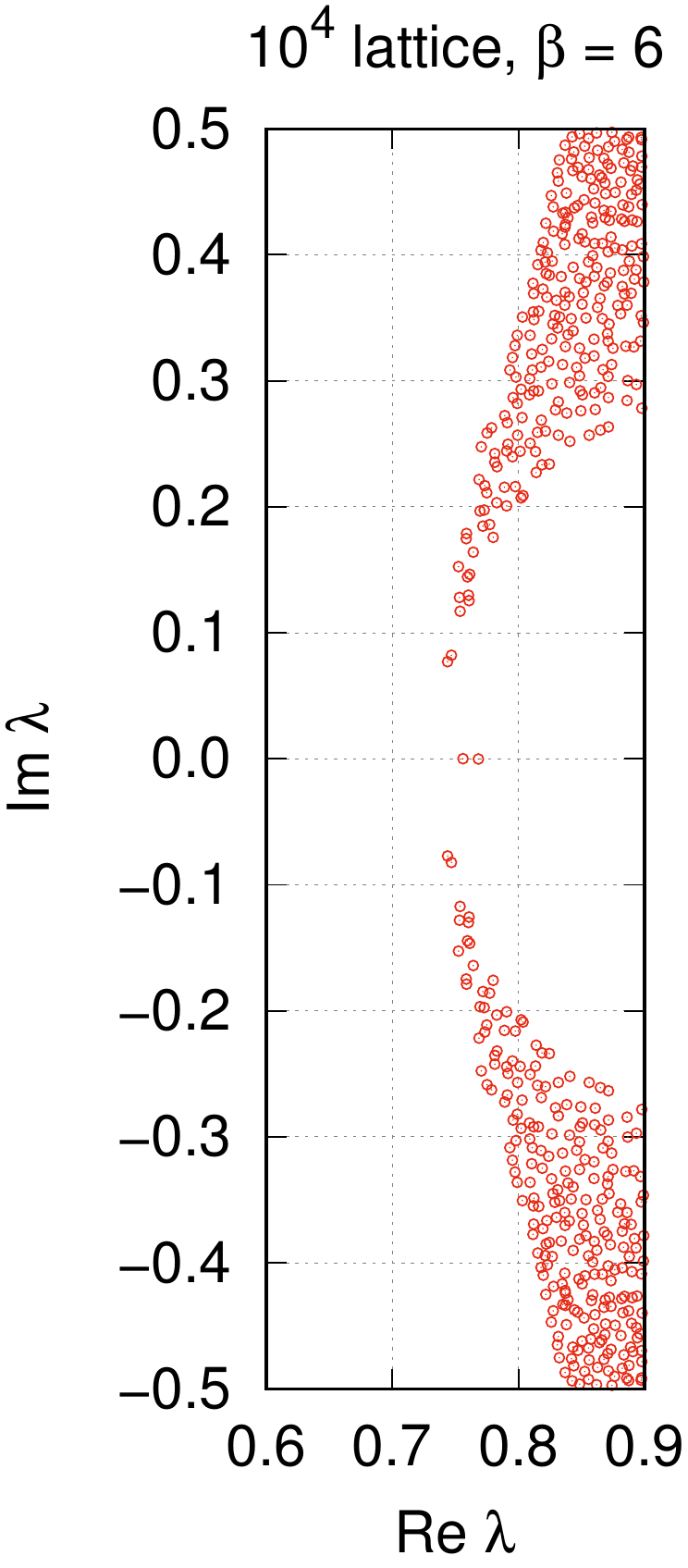}
\par\end{centering}
}\hfill{}\subfloat[$12^{3}\times16$ lattice at $\beta=6$]{\begin{centering}
\includegraphics[height=0.65\textwidth]{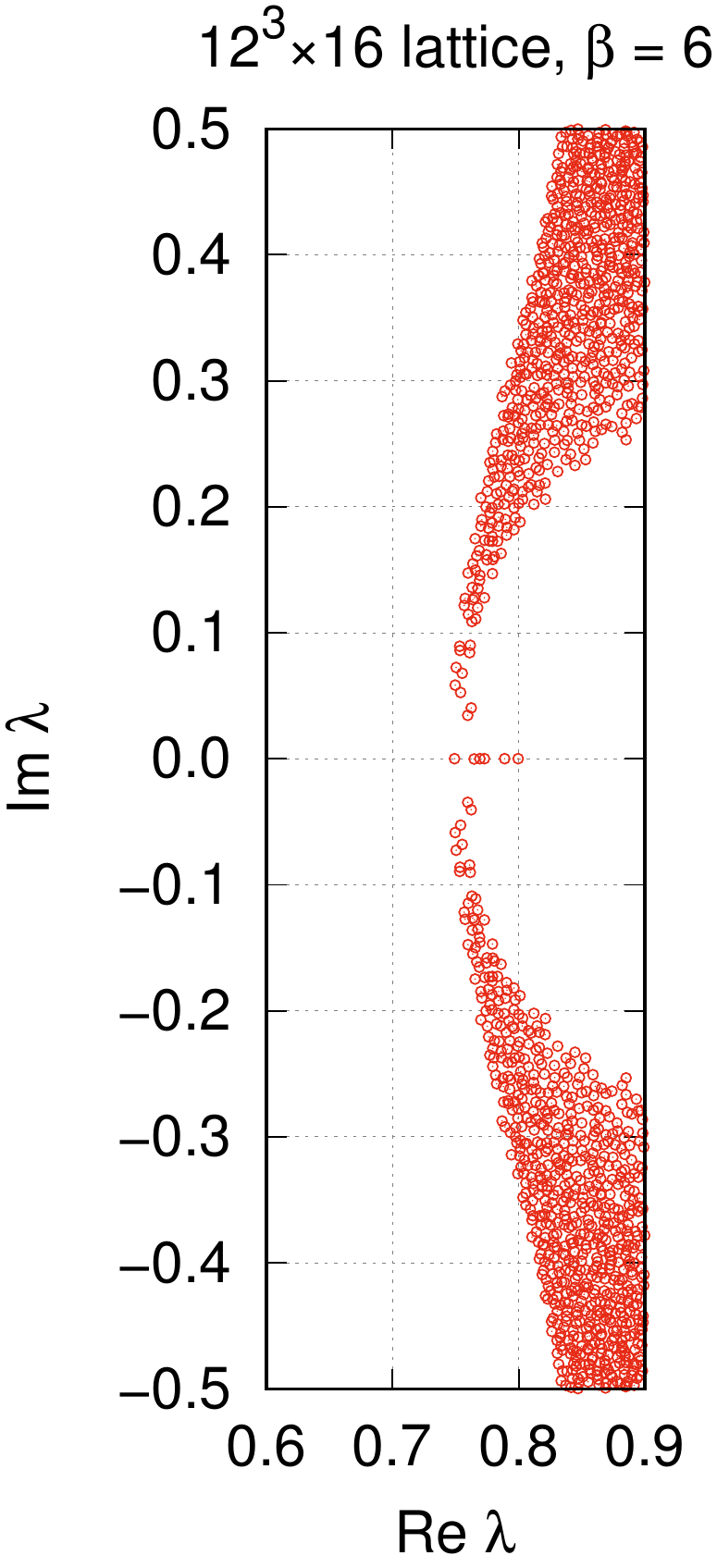}
\par\end{centering}
}
\par\end{centering}
\caption{Close-up of the physical branch of the staggered Wilson Dirac operator
for increasing lattice sizes. \label{fig:CloseUpPhysBranch}}
\end{figure}
\begin{figure}[t]
\begin{centering}
\subfloat[$12^{3}\times16$ lattice at $\beta=6$]{\begin{centering}
\includegraphics[width=0.3\textwidth]{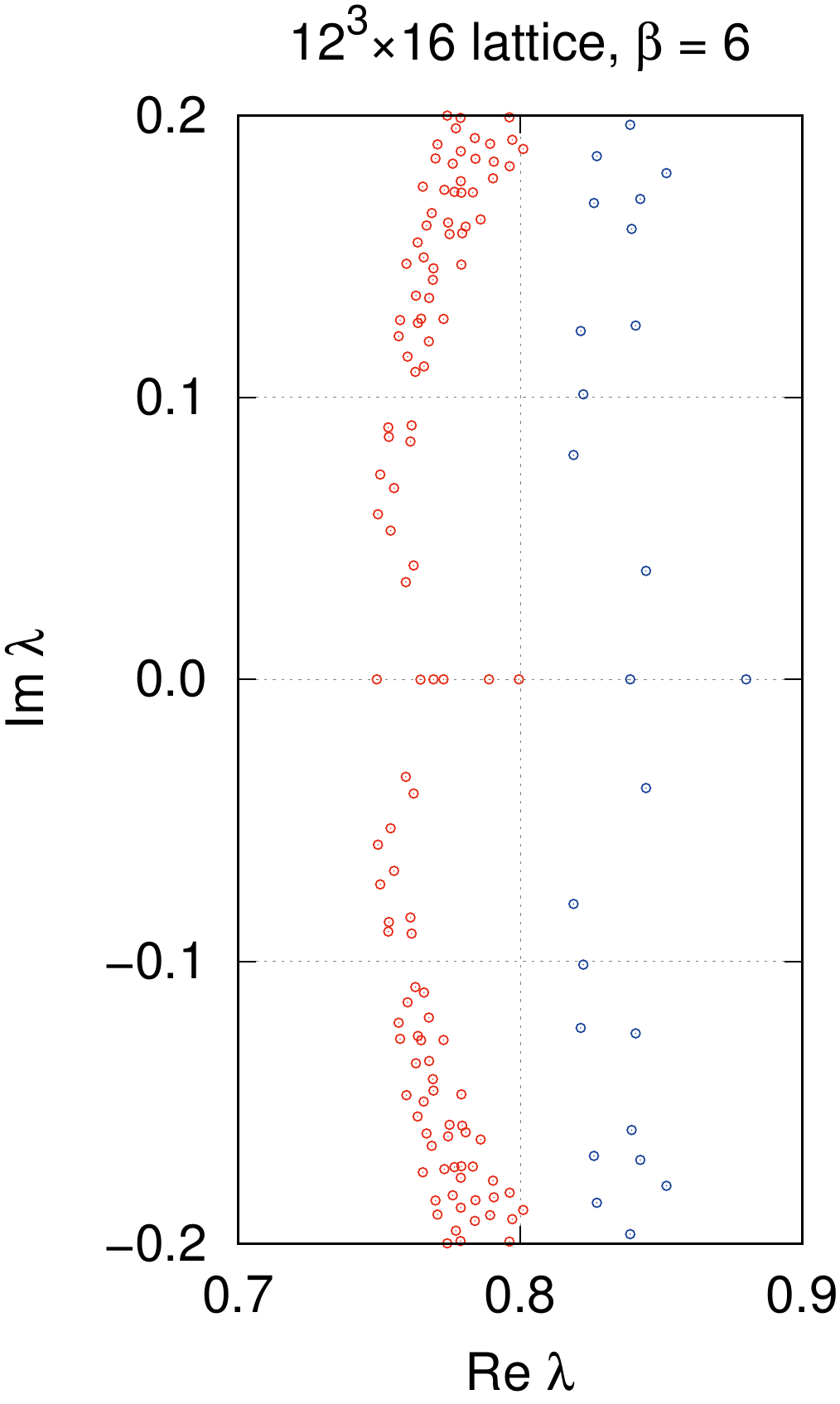}
\par\end{centering}
}\hfill{}\subfloat[$12^{3}\times32$ lattice at $\beta=6$]{\begin{centering}
\includegraphics[width=0.3\textwidth]{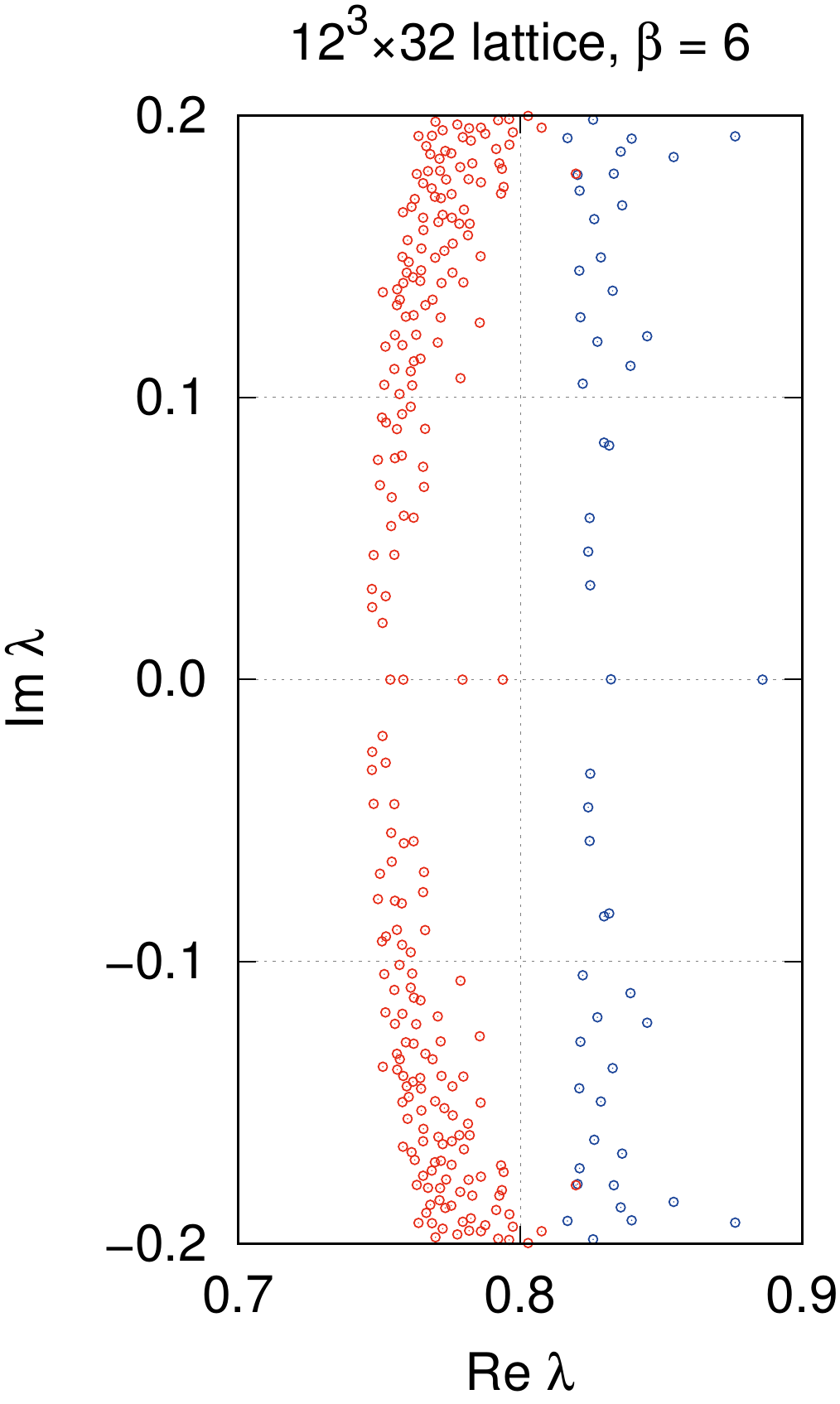}
\par\end{centering}
}\hfill{}\subfloat[$16^{3}\times32$ lattice at $\beta=6$]{\begin{centering}
\includegraphics[width=0.3\textwidth]{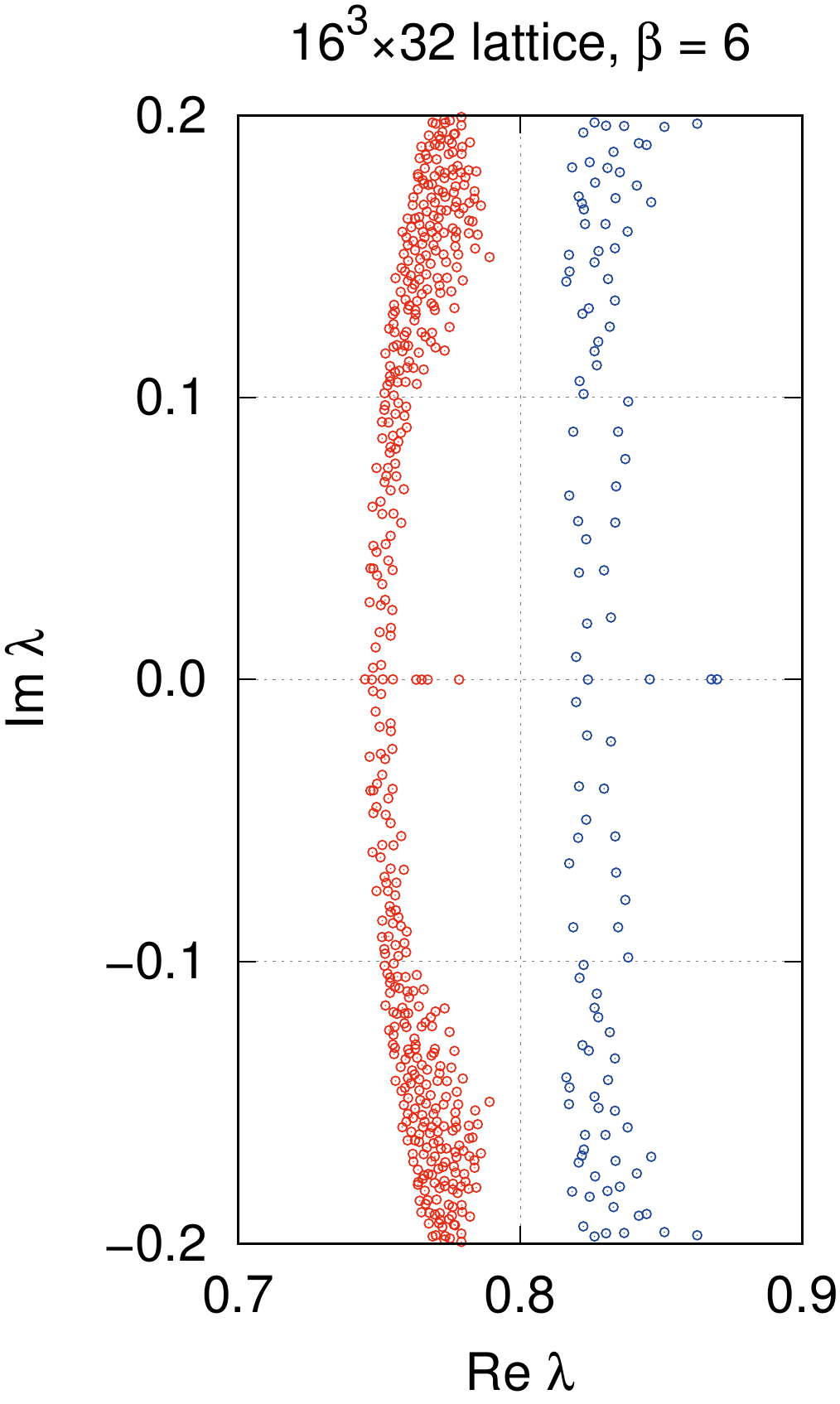}
\par\end{centering}
}
\par\end{centering}
\caption{Comparison of the staggered Wilson (red) and Wilson Dirac spectrum
(blue). \label{fig:ComparisionPhysBranch}}
\end{figure}

Let us now consider the case of (unsmeared) quenched QCD background
fields. The calculation of eigenvalue spectra on large lattices is
numerically and computationally challenging. The Dirac operator is
represented by a large, nonsymmetric, complex-valued matrix of size
$\left(N_{\mathsf{c}}N_{\mathsf{d}}N_{\mathsf{s}}^{3}N_{\mathsf{t}}\right)^{2}$,
where for quantum chromodynamics the number of colors is $N_{\mathsf{c}}=3$
and the number of spinor components is given by $N_{\mathsf{d}}=4$
for Wilson and by $N_{\mathsf{d}}=1$ for staggered Wilson fermions.
In particular, for larger lattices such as a $N_{s}\times N_{t}=16^{3}\times32$
lattice, we have more than $1.5\times10^{6}$ eigenvalues for the
Wilson Dirac operator and more than $5\times10^{5}$ for the staggered
Wilson Dirac operator. As a result, we can compute complete eigenvalue
spectra only for small lattice sizes. For larger lattices, we have
to restrict ourselves to subsets of the spectrum. Of particular interest
is here the physically relevant low-lying part around the physical
branch.

\subsection{Implementation}

To determine the eigenvalues of interest, our numerical implementation
works as follows. For small lattices we compute complete eigenvalue
spectra using the \textsc{lapack} \citep{LAPACK} library. The library
function \texttt{zgeev} computes all eigenvalues of a complex nonsymmetric
matrix in double precision using the $QR$-algorithm \citep{francis1961qr,francis1962qr,kublanovskaya1962some}.
As the function requires the input matrix to be constructed explicitly
(i.e.\ densely) in memory, its use for larger lattices is limited
by quickly growing memory requirements. As a consequence we use this
method to compute complete spectra for lattices of up to a size of
$6^{4}$ for the Wilson Dirac operator and up to a size of $10^{4}$
in the case of the staggered Wilson Dirac operator.

For larger lattices we use \textsc{arpack} \citep{ARPACK}, a library
to solve large-scale eigenvalue problems involving sparse and structured
matrices. It uses the implicitly restarted Arnoldi method \citep{IRAM}
to compute the low-lying part of the eigenvalue spectrum. Here low-lying
refers to eigenvalues with smallest absolute values. The implicitly
restarted Arnoldi method is an iterative method and is not suitable
for the computation of complete spectra, hence our choice of \textsc{lapack}
for smaller lattices. It has nontrivial convergence behavior, in particular
choosing the number of Lanczos basis vectors has significant impact
on the performance. The \texttt{znaupd} function deals with the case
of a complex nonsymmetric matrix in double precision. By choosing
the bare mass parameter $m$ appropriately, one can shift the part
of the spectrum of interest to the origin. With our setup we can compute
typically a few hundred eigenvalues for lattice sizes of up to $16^{3}\times32$
in quenched QCD background fields.

In the following, all reported numerical results are for periodic
boundary conditions, use the canonical choice $r=1$ and are shifted
back to $m=0$. All numerical calculations are done in double precision
as the use of single precision resulted in erroneous eigenvalues due
to accumulated rounding errors.

\subsection{Eigenvalue spectra}

We now begin our discussion of the eigenvalue spectra by comparing
the free-field case with a $\beta=6$ quenched QCD background field
on small lattices. In Fig.~\ref{fig:FullSpecBeta6}, we can find
two typical examples for the spectrum for a $8^{4}$ and a $10^{4}$
lattice. Compared to the free-field spectrum in the same figure, we
can see how the whole spectrum is contracted noticeably and the gap
is narrow. In Ref.~\citep{deForcrand:2011ak}, this degradation of
the spectrum was attributed to the four-hop terms in the staggered
Wilson action, which raise gauge fluctuations to the fourth power.
The physical branch and the doubler branch get closer as the spectrum
collapses into a vertical stripe and one would expect a reduced computational
efficiency of staggered Wilson fermions.

For $\beta=5.8$ the spectrum degrades further. In Fig.~\ref{fig:FullSpecBeta58},
we can find two exemplary eigenvalue spectra. We see that the branches
become even more diffuse and the gap is very narrow. As a result,
we expect a further decreased performance compared to the $\beta=6$
case.

The situation changes when we move to larger lattices. In Fig.~\ref{fig:CloseUpPhysBranch},
we can find a close-up of the physical branch over an increasing range
of lattice sizes from $8^{4}$ to $12^{3}\times16$ at $\beta=6$.
We can observe how for larger lattices the physical branch becomes
increasingly sharp. Finally, in Fig.~\ref{fig:ComparisionPhysBranch}
we can find a direct comparison of the physical branch of both the
Wilson and staggered Wilson Dirac operator up to the largest lattices
we can access, i.e.\ $16^{3}\times32$. Here we can see very clearly
how the spectrum improves from small to large lattices. While on small
lattices the branches are diffuse and not well-separated, on larger
lattices the situation improves remarkably.

\section{Conclusions \label{sec:EV-Conclusions}}

Our observations give a possible explanation for the modest speedup
factors for staggered overlap fermions observed in Refs.~\citep{deForcrand:2011ak,deForcrand:2012bm}
compared to the large speedup factors for staggered Wilson fermions
found in Refs.~\citep{Adams:2013tya,ZielinskiLat14}. While the former
study was carried out on small lattices up to a size of $12^{4}$,
the latter one was done on lattices sizes of up to a size of $20^{3}\times40$.

As we confirmed, on small lattices the eigenvalue spectrum has a small
gap and is lacking a clear separation of the branches. Here one can
often find eigenvalues close to the center of the ``belly'' of the
spectrum, making the overlap construction computationally more expensive.
The situation quickly improves on both the infrared and ultraviolet
part of the spectrum when increasing the lattice size, see Fig.~\ref{fig:CloseUpPhysBranch}.
While this also holds for Wilson fermions, it appears to be more pronounced
for the staggered Wilson Dirac operator, which is likely linked to
the presence of four-hop terms in the staggered Wilson term.

That the eigenvalue spectrum of the staggered Wilson Dirac operator
is well-behaved on larger lattices can also be seen from the condition
number. The diffuse spectrum could in principle give rise to almost
vanishing eigenvalues when the pion mass is small. However, as we
observed the condition number in our numerical studies in Sec.\ \ref{sec:CE-Numerical-study}
to be typically a factor of $\mathcal{O}\left(4\right)$ smaller compared
to Wilson fermions, we conclude that these pathological cases are
rare.

Finally the smaller additive mass renormalization of staggered Wilson
fermions is suggesting improved chiral properties compared to Wilson
fermions, which are expected to have a positive impact on the computational
efficiency.

As a consequence, we believe that the speedup factors reported in
Refs.~\citep{deForcrand:2011ak,deForcrand:2012bm} are not representative
for the actual performance gains achievable with staggered overlap
fermions on larger lattices. We would expect a higher computational
efficiency in cases where the staggered Wilson eigenvalue spectrum
shows a wide gap and well-separated, sharp branches. As our study
suggests, this can be (at least partially) achieved by moving to larger
lattices, although only a comprehensive numerical study can clarify
this point in the end. Alternatively, as suggested in Ref.~\citep{Durr:2013gp},
smearing also significantly improves the eigenvalue spectrum and,
thus, is expected to result in an improved computational performance.

\chapter{Staggered domain wall fermions \label{chap:Staggered-domain-wall}}

Besides staggered overlap fermions, the staggered Wilson kernel allows
the construction of a related chiral lattice formulation, namely staggered
domain wall fermions as proposed by Adams in Refs.~\citep{Adams:2009eb,Adams:2010gx}.
In the following, we investigate spectral properties and quantify
chiral symmetry violations of various formulations of staggered and
usual domain wall fermions. We do this in the free-field case, on
quenched thermalized background fields in the setting of the Schwinger
model and on smooth topological configurations. Moreover, we present
first results for four-dimensional quantum chromodynamics. We closely
follow the discussion from our original\footnote{Discussion based on and figures reprinted with permission from \sloppy
\bibentry{Hoelbling:2016qfv}. Copyright 2016 by the American Physical
Society.} and subsequent\footnote{Discussion based on and figures reprinted from \sloppy \bibentry{Hoelbling:2016dug}.
Copyright owned by the authors under the terms of the Creative Commons
Attribution-NonCommercial-NoDerivatives 4.0 International License
(CC BY-NC-ND 4.0).} reports given in Refs.~\citep{Hoelbling:2016qfv,ZielinskiDPG16,Hoelbling:2016dug},
where we previously presented our results.

\section{Introduction}

For our understanding of the low-energy dynamics of quantum chromodynamics,
such as hadron phenomenology, chiral symmetry plays an essential role.
While the proper implementation of chiral symmetry on the lattice
turned out to be a notorious problem, it was eventually overcome with
the overlap construction as discussed in Chapter\ \ref{chap:Overlap-fermions}.
Overlap fermions obey an exact chiral symmetry, thus evading the Nielsen-Ninomiya
theorem (cf.\ Subsec.\ \ref{subsec:GT-Nielsen-Ninomiya}). From
a theoretical perspective overlap fermions have many desirable properties,
but their practical use is limited due to the fact that they typically
require a factor of $\mathcal{O}\left(10\text{--}100\right)$ more
computational resources compared to Wilson fermions. Moreover, even
at moderate lattice spacings tunneling between topological sectors
is strongly suppressed \citep{Egri:2005cx,Fukaya:2006vs,Cundy:2008zc,Cundy:2011fz},
so that adequate coverage of the different topological sectors becomes
a concern.

An alternative to overlap fermions is given by domain wall fermions
\citep{Kaplan:1992bt,Shamir:1993zy,Furman:1994ky}, which implement
an approximate chiral symmetry in $d$ dimensions by means of a theory
of massive coupled fermions in $d+1$ dimensions ($d=2,4$). The extent
of this extra dimension controls the degree of chiral symmetry violations
and in the limit of an infinite extent one recovers the overlap operator
with its exact chiral symmetry. The case of a finite extent can be
interpreted as a truncation of overlap fermions.

Implementations of domain wall fermions can be easily parallelized
and bear the potential of reduced computational costs. While they
only have an approximate chiral symmetry, violations of chiral symmetry
are well-controlled. Based on theoretical grounds, these violations
are expected to be exponentially suppressed \citep{Shamir:1993zy,Vranas:1997da,Aoki:1997xg,Kikukawa:1997tf}.
In practice, however, this suppression typically still requires large
extents of the extra dimension \citep{Chen:1998xw,Blum:1998ud,Fleming:1999eq,Hernandez:2000iw,Aoki:2000pc}.
At the same time, these violations allow for an easier tunneling between
different topological sectors. 

Typically domain wall fermions are formulated with a Wilson-like kernel
operator. However, recently Adams laid the foundations \citep{Adams:2009eb,Adams:2010gx}
for the use of staggered kernels in the domain wall construction.
This was enabled by the introduction of a flavored mass term as described
in Sec.\ \ref{sec:SW-Adams-mass-term}. While in previous studies
the properties of the staggered Wilson kernel \citep{Misumi:2011su,Misumi:2012sp,Nakano:2012wa,deForcrand:2012bm,Durr:2013gp,Adams:2013tya}
and the staggered overlap construction \citep{Adams:2010gx,deForcrand:2011ak,Adams:2013lpa}
were investigated, the staggered domain wall fermion formulation was
largely ignored in the literature. For this reason, we investigate
Adams' proposal in this chapter. We give explicit constructions of
these lattice fermions in different variants, study their spectral
properties and compare the degree of chiral symmetry breaking with
the one of usual domain wall fermions in the setting of the Schwinger
model \citep{Schwinger:1962tp}.

The Schwinger model, i.e.\ quantum electrodynamics (QED) in two dimensions,
serves us a toy-model for quantum chromodynamics (QCD). Like quantum
chromodynamics itself, it shows confinement and has topological structure.
At the same time its numerical treatment is simpler and allows the
computation of complete eigenvalue spectra of lattice Dirac operators
on nontrivial background fields. Moreover, the study of the Schwinger
model is of interest in its own right as it is relevant for the description
of conducting electrons in metals in the low-energy regime, see e.g.\ Ref.~\citep{tsvelik2005quantum}.

This chapter is organized as follows. In Sec.\ \ref{sec:DW-kernel-operators},
we give a short overview of the kernel operators we use in our study.
In Sec.\ \ref{sec:DW-Domain-wall-fermions}, we discuss the construction
of (staggered) domain wall fermions and their variants. In Sec.\ \ref{sec:DW-Effective-operator},
we introduce effective low-energy Dirac operators and discuss their
relations to the overlap formalism. In Sec.\ \ref{sec:DW-Setting},
we introduce the setting of our numerical calculations. In Sec.\ \ref{sec:DW-Numerical-results},
we discuss the numerical results in detail. In Sec.\ \ref{sec:DW-Quenched-QCD},
we present first results for four-dimensional quantum chromodynamics.
In Sec.\ \ref{sec:DW-Optimal-weights}, we provide some exemplary
weight factors for the optimal domain wall construction and finally
in Sec.\ \ref{sec:DW-Conclusions}, we end with some concluding remarks.

\section{Kernel operators \label{sec:DW-kernel-operators}}

Throughout this chapter we consider domain wall fermions with two
different kernel operators, i.e.\ Wilson and staggered Wilson fermions.
In the following, we adopt the notation previously used in Chapters\ \ref{chap:GaugeTheories},
\ref{chap:Staggered-Wilson-fermions} and \ref{chap:Overlap-fermions}.
In the case of $d=2$ dimensions the $\gamma_{\mu}\in\mathbb{C}^{2\times2}$
matrices ($\mu=1,2$) correspond to a representation of the Dirac
algebra $\left\{ \gamma_{\mu},\gamma_{\nu}\right\} =2\delta_{\mu\nu}\One$
and we denote the chirality matrix as $\gamma_{3}$. At the same time,
the $\xi_{\mu}\in\mathbb{C}^{2\times2}$ matrices are a representation
of the Dirac algebra in flavor space.

\paragraph{Wilson fermions.}

We introduced and discussed the Wilson Dirac operator in Subsec.\ \ref{subsec:GT-Wilson-fermions}.
The Dirac operator is defined in Eq.~\eqref{eq:DefWilsonDiracOp},
where we consider the case of two space-time dimensions for our numerical
studies.

\paragraph{Staggered Wilson fermions.}

The second kernel operator of interest is Adams' staggered Wilson
Dirac operator. In Sec.\ \ref{sec:SW-Adams-mass-term} and Sec.\ \ref{sec:SW-Generalized-mass-terms},
we discussed its construction and properties in detail. In the numerical
part of this chapter we use the formulation in $d=2$ dimensions as
defined in Eq.~\eqref{eq:StagWilsonDiracTwoDim}. We note that in
principle also the use of the other flavored mass terms in Chapter\ \ref{chap:Staggered-Wilson-fermions}
can give rise to suitable kernel operators.

\section{Domain wall fermions \label{sec:DW-Domain-wall-fermions}}

Domain wall fermion were originally proposed by Kaplan in Ref.~\citep{Kaplan:1992bt}
and subsequently refined by Shamir and Furman in Refs.~\citep{Shamir:1993zy,Furman:1994ky}.
By means of a $\left(d+1\right)$-dimensional theory, the domain wall
formulation realizes approximately massless fermions in $d$ dimensions.
Alternatively, one can interpret the construction as a tower of $N_{s}$
fermions in $d$ dimensions with a specific flavor structure.

We begin by introducing the $\left(d+1\right)$-dimensional bulk operators,
where we fix the lattice spacing to $a=1$ in the first $d$ dimensions
and the (staggered) Wilson parameter to $r=1$. Although for the numerical
work we are interested in the $d=2$ setting, we keep the following
discussion general and consider the case of even $d$ dimensions,
where $\gamma_{d+1}$ refers to the corresponding chirality matrix
in that dimension.

\subsection{Standard construction}

We begin with the originally proposed construction. We define
\begin{equation}
D_{\mathsf{w}}^{\pm}=a_{d+1}D_{\mathsf{w}}\left(-M_{0}\right)\pm\One
\end{equation}
and explicitly keep the lattice spacing $a_{d+1}$ in the extra dimension.
The mass parameter $M_{0}$ is commonly referred to as the domain
wall height and has to be suitably chosen to give rise to a one flavor
theory. For the free-field case, its values are restricted to the
range $M_{0}\in\left(0,2r\right)$.

We can write down the Dirac operator, which reads
\begin{equation}
\overline{\Psi}D_{\mathsf{dw}}\Psi=\sum_{s=1}^{N_{s}}\overline{\Psi}_{s}\left[D_{\mathsf{w}}^{+}\Psi_{s}-P_{-}\Psi_{s+1}-P_{+}\Psi_{s-1}\right]
\end{equation}
with chiral projectors $P_{\pm}=\left(\One\pm\gamma_{d+1}\right)/2$
and $\left(d+1\right)$-dimensional fermion fields $\overline{\Psi}$,
$\Psi$. Throughout this chapter, the index $s=1,\dots,N_{s}$ refers
to the additional spatial coordinate (or equivalently to the flavor).
Along the extra spatial coordinate, the gauge links are taken to be
the unit matrix. Furthermore, for finite extents of the $s$-coordinate
we must impose boundary conditions, which we take to be
\begin{equation}
\begin{aligned}P_{+}\left(\Psi_{0}+m\Psi_{N_{s}}\right) & =0,\\
P_{-}\left(\Psi_{N_{s}+1}+m\Psi_{1}\right) & =0.
\end{aligned}
\end{equation}
Here the parameter $m$ is related to the bare fermion mass, see Eq.~\eqref{eq:InducedMass}.
In the special case of $m=0$ the condition reduces to Dirichlet type
boundary conditions, while for $m=\pm1$ we find (anti\nobreakdash-)periodic
boundary conditions. Writing the Dirac operator explicitly in the
extra dimension, we find
\begin{equation}
D_{\mathsf{dw}}=\left(\begin{array}{ccccc}
D_{\mathsf{w}}^{+} & -P_{-} &  &  & mP_{+}\\
-P_{+} & D_{\mathsf{w}}^{+} & -P_{-}\\
 & \ddots & \ddots & \ddots\\
 &  & -P_{+} & D_{\mathsf{w}}^{+} & -P_{-}\\
mP_{-} &  &  & -P_{+} & D_{\mathsf{w}}^{+}
\end{array}\right).\label{eq:DefDdw}
\end{equation}
The $d$-dimensional fermion fields can be defined from the boundary
as
\begin{equation}
\begin{aligned}q & =P_{+}\Psi_{N_{s}}+P_{-}\Psi_{1},\\
\overline{q} & =\overline{\Psi}_{1}P_{+}+\overline{\Psi}_{N_{s}}P_{-}.
\end{aligned}
\label{eq:DefLightFermions}
\end{equation}
Now let us also introduce the reflection operator in the extra dimension
\begin{equation}
R=\left(\begin{array}{ccc}
 &  & \One\\
 & \iddots\\
\One
\end{array}\right).
\end{equation}
We can easily verify that $D_{\mathsf{dw}}$ is $R\gamma_{d+1}$ Hermitian,
that is
\begin{equation}
D_{\mathsf{dw}}^{\dagger}=R\gamma_{d+1}\cdot D_{\mathsf{dw}}\cdot R\gamma_{d+1}.\label{eq:RGHermiticity}
\end{equation}
This property implies the reality of the fermion determinant, i.e.\ $\det D_{\mathsf{dw}}\in\mathbb{R}$.

Besides this standard formulation, several variants of domain wall
fermions have been proposed after the original proposal.

\subsection{Boriçi's construction}

Among them we can find Boriçi's construction \citep{Borici:1999zw},
which follows from the standard construction by the replacements
\begin{equation}
\begin{aligned}P_{+}\Psi_{s-1} & \to-D_{\mathsf{w}}^{-}P_{+}\Psi_{s-1},\\
P_{-}\Psi_{s+1} & \to-D_{\mathsf{w}}^{-}P_{-}\Psi_{s+1}
\end{aligned}
\end{equation}
and is an $\mathcal{O}\left(a_{d+1}\right)$ modification. After this
replacement, the Dirac operator takes the form
\begin{equation}
\overline{\Psi}D_{\mathsf{dw}}\Psi=\sum_{s=1}^{N_{s}}\overline{\Psi}_{s}\left[D_{\mathsf{w}}^{+}\Psi_{s}+D_{\mathsf{w}}^{-}P_{-}\Psi_{s+1}+D_{\mathsf{w}}^{-}P_{+}\Psi_{s-1}\right]
\end{equation}
or explicitly
\begin{equation}
D_{\mathsf{dw}}=\left(\begin{array}{ccccc}
D_{\mathsf{w}}^{+} & D_{\mathsf{w}}^{-}P_{-} &  &  & -mD_{\mathsf{w}}^{-}P_{+}\\
D_{\mathsf{w}}^{-}P_{+} & D_{\mathsf{w}}^{+} & D_{\mathsf{w}}^{-}P_{-}\\
 & \ddots & \ddots & \ddots\\
 &  & D_{\mathsf{w}}^{-}P_{+} & D_{\mathsf{w}}^{+} & D_{\mathsf{w}}^{-}P_{-}\\
-mD_{\mathsf{w}}^{-}P_{-} &  &  & D_{\mathsf{w}}^{-}P_{+} & D_{\mathsf{w}}^{+}
\end{array}\right).\label{eq:DefDdwBorici}
\end{equation}
Moreover, the generalization of Eq.~\eqref{eq:DefLightFermions}
reads
\begin{equation}
\begin{aligned}q & =P_{+}\Psi_{N_{s}}+P_{-}\Psi_{1},\\
\overline{q} & =-\overline{\Psi}_{1}D_{\mathsf{w}}^{-}P_{+}-\overline{\Psi}_{N_{s}}D_{\mathsf{w}}^{-}P_{-}
\end{aligned}
\end{equation}
and the equivalent of Eq.~\eqref{eq:RGHermiticity} takes the form
\begin{equation}
\begin{aligned}\left(\mathcal{D}^{-1}D_{\mathsf{dw}}\right)^{\dagger} & =R\gamma_{d+1}\cdot\left(\mathcal{D}^{-1}D_{\mathsf{dw}}\right)\cdot R\gamma_{d+1},\\
\mathcal{D} & =\One_{N_{s}}\otimes D_{\mathsf{w}}^{-}
\end{aligned}
\label{eq:RGHermiticityBorici}
\end{equation}
(see Ref.~\citep{Brower:2004xi}). Boriçi referred to this formulation
as \textit{truncated overlap fermions} due to the fact that the corresponding
$d$-dimensional effective operator, as discussed in Sec.\ \ref{sec:DW-Effective-operator},
corresponds to the polar decomposition approximation \citep{Neuberger:1998my,higham1994matrix}
of Neuberger's overlap operator of order $N_{s}/2$ (where $N_{s}$
is even).

\subsection{Optimal construction \label{subsec:DW-Optimal-construction}}

Another modification of domain wall fermions is the optimal construction
as proposed by Chiu \citep{Chiu:2002ir,Chiu:2002kj}. Here one modifies
$D_{\mathsf{dw}}$ so that the effective Dirac operator is formulated
via Zolotarev's optimal rational function approximation of the $\sign$
function \citep{zolotarev1877application,Akhiezer1990,achieser2013theory},
see also Refs.~\citep{vandenEshof:2002ms,Chiu:2002eh}. We now summarize
the construction given in Ref.~\citep{Chiu:2002ir}.

As a starting point we use Boriçi's formulation of domain wall fermions.
We generalize the Dirac operator by introducing weight factors
\begin{equation}
\overline{\Psi}D_{\mathsf{dw}}\Psi=\sum_{s=1}^{N_{s}}\overline{\Psi}_{s}\left[D_{\mathsf{w}}^{+}\left(s\right)\Psi_{s}+D_{\mathsf{w}}^{-}\left(s\right)P_{-}\Psi_{s+1}+D_{\mathsf{w}}^{-}\left(s\right)P_{+}\Psi_{s-1}\right]
\end{equation}
via
\begin{equation}
D_{\mathsf{w}}^{\pm}\left(s\right)=a_{d+1}\omega_{s}D_{\mathsf{w}}\left(-M_{0}\right)\pm\One.
\end{equation}
The weight factors $\omega_{s}$ take the values
\begin{equation}
\omega_{s}=\frac{1}{\lambda_{\mathsf{min}}}\sqrt{1-\kappa^{\prime2}\sn\left(v_{s},\kappa^{\prime}\right)}\label{eq:OptWeights}
\end{equation}
with $\sn\left(v_{s},\kappa^{\prime}\right)$ being the respective
Jacobi elliptic function with the argument $v_{s}$ and modulus $\kappa^{\prime}$.
The modulus is taken as
\begin{equation}
\kappa^{\prime}=\sqrt{1-\lambda_{\mathsf{min}}^{2}/\lambda_{\mathsf{max}}^{2}}
\end{equation}
and $\lambda_{\mathsf{min}}^{2}$ ($\lambda_{\mathsf{max}}^{2}$)
denotes the smallest (largest) eigenvalue of $H_{\mathsf{w}}^{2}$
with
\begin{equation}
H_{\mathsf{w}}=\gamma_{d+1}D_{\mathsf{w}}\left(-M_{0}\right).
\end{equation}
The argument $v_{s}$ reads
\begin{equation}
v_{s}=\left(-1\right)^{s-1}M\sn^{-1}\left(\sqrt{\frac{1+3\lambda}{\left(1+\lambda\right)^{3}}},\sqrt{1-\lambda^{2}}\right)+\left\lfloor \frac{s}{2}\right\rfloor \frac{2K^{\prime}}{N_{s}},
\end{equation}
where
\begin{align}
\lambda & =\prod_{\ell=1}^{N_{s}}\frac{\Theta^{2}\left(2\ell K^{\prime}/N_{s},\kappa^{\prime}\right)}{\Theta^{2}\left(\left(2\ell-1\right)K^{\prime}/N_{s},\kappa^{\prime}\right)},\\
M & =\prod_{\ell=1}^{\left\lfloor N_{s}/2\right\rfloor }\frac{\sn^{2}\left(\left(2\ell-1\right)K^{\prime}/N_{s},\kappa^{\prime}\right)}{\sn^{2}\left(2\ell K^{\prime}/N_{s},\kappa^{\prime}\right)}.
\end{align}
Here $\left\lfloor \cdot\right\rfloor $ denotes the floor function,
$K^{\prime}=K\left(\kappa^{\prime}\right)$ is the complete elliptic
integral of the first kind with
\begin{equation}
K\left(k\right)=\intop_{0}^{\pi/2}\frac{\mathrm{d}\theta}{\sqrt{1-k^{2}\sin^{2}\theta}}.
\end{equation}
We also introduced the elliptic Theta function via
\begin{equation}
\Theta\left(w,k\right)=\vartheta_{4}\left(\frac{\pi w}{2K},k\right)
\end{equation}
with $K=K\left(k\right)$ and elliptic theta functions $\vartheta_{i}$,
see e.g.\ Ref.~\citep{abramowitz2012handbook}. To allow for the
verification of implementations of these so-called \textit{optimal
domain wall fermions}, we provide some reference values for the weight
factors $\omega_{s}$ in Sec.\ \ref{sec:DW-Optimal-weights}.

In the optimal construction, Eq.~\eqref{eq:DefLightFermions} is
replaced by
\begin{equation}
\begin{aligned}q & =P_{+}\Psi_{N_{s}}+P_{-}\Psi_{1},\\
\overline{q} & =-\overline{\Psi}_{1}D_{\mathsf{w}}^{-}\left(1\right)P_{+}-\overline{\Psi}_{N_{s}}D_{\mathsf{w}}^{-}\left(N_{s}\right)P_{-}
\end{aligned}
\end{equation}
and Eq.~\eqref{eq:RGHermiticity} again takes the form of Eq.~\eqref{eq:RGHermiticityBorici},
but now with
\begin{equation}
\mathcal{D}=\diag\left[D_{\mathsf{w}}^{-}\left(1\right),\dots,D_{\mathsf{w}}^{-}\left(N_{s}\right)\right]
\end{equation}
as pointed out in Ref.~\citep{Brower:2004xi}. Optimal domain wall
fermions are one of the most popular domain wall fermion formulations
and were used in numerous studies. In Refs.~\citep{Chen:2011qy,Hsieh:2011qx,Chiu:2011dz,Chiu:2011bm,Chen:2012jya,Chiu:2012jm}
some of these works can be found.

Besides the original optimal domain wall fermion formulation, there
is also a modified version \citep{Chiu:2015sea} which is reflection-symmetric
along the extra dimension. Finally, we note that all of the preceding
domain wall fermion variants can be interpreted as special cases of
Möbius domain wall fermions \citep{Brower:2004xi,Brower:2005qw,Brower:2009sb}.

\subsection{Staggered formulations}

As recently clarified by Adams, it is possible to employ a staggered
kernel for the formulation of domain wall fermions \citep{Adams:2009eb,Adams:2010gx}.
Following this idea, the Dirac operator takes the form
\begin{equation}
\overline{\Upsilon}D_{\mathsf{sdw}}\Upsilon=\sum_{s=1}^{N_{s}}\overline{\Upsilon}_{s}\left[D_{\mathsf{sw}}^{+}\Upsilon_{s}-P_{-}\Upsilon_{s+1}-P_{+}\Upsilon_{s-1}\right],
\end{equation}
where $\Upsilon$ refers to the $\left(d+1\right)$-dimensional staggered
fermion field. Similarly to the Wilson case, we let
\begin{equation}
D_{\mathsf{sw}}^{\pm}=a_{d+1}D_{\mathsf{sw}}\left(-M_{0}\right)\pm\One
\end{equation}
with $D_{\mathsf{sw}}$ being the staggered Wilson Dirac operator.
The chiral projectors are given by $P_{\pm}=\left(\One\pm\epsilon\right)/2$,
where $\epsilon$ is defined in Eq.~\eqref{eq:EpsilonDDim} and we
note that $\epsilon^{2}=\One$. Recall that $\epsilon\sim\gamma_{d+1}\otimes\xi_{d+1}$,
which reduces to $\epsilon\sim\gamma_{d+1}\otimes\One$ on the physical
species. The $R\gamma_{d+1}$ Hermiticity of $D_{\mathsf{dw}}$ generalizes
to a $R\epsilon$ Hermiticity of $D_{\mathsf{sdw}}$. With our sign
convention $D_{\mathsf{sdw}}$ is in full analogy with $D_{\mathsf{dw}}$,
while a slightly different convention was used in the original proposal
in Ref.~\citep{Adams:2010gx}.

In Ref.~\citep{Adams:2010gx}, a replacement rule was specified,
which can be written in our general $d$-dimensional setting as
\begin{equation}
\gamma_{d+1}\to\epsilon,\qquad D_{\mathsf{w}}\to D_{\mathsf{sw}}\label{eq:ReplacementRule}
\end{equation}
and describes how $D_{\mathsf{sdw}}$ can be constructed from $D_{\mathsf{dw}}$.
Using Eq.~\eqref{eq:ReplacementRule}, it is possible for us to generalize
also Boriçi's and Chiu's work to the case of a staggered Wilson kernel.
This allows us to write down the Dirac operator for what we are going
to denote as \textit{truncated staggered domain wall fermions}\footnote{We chose the name ``truncated staggered domain wall fermions'' in
Refs.~\citep{Hoelbling:2016qfv,Hoelbling:2016dug} rather than ``truncated
staggered overlap fermions'' in order to emphasize the fact that
we are dealing with a $\left(d+1\right)$-dimensional domain wall
fermion formulation, which is motivated by a truncation of the staggered
overlap operator.} and which reads
\begin{equation}
\overline{\Upsilon}D_{\mathsf{sdw}}\Upsilon=\sum_{s=1}^{N_{s}}\overline{\Upsilon}_{s}\left[D_{\mathsf{sw}}^{+}\Upsilon_{s}+D_{\mathsf{sw}}^{-}P_{-}\Upsilon_{s+1}+D_{\mathsf{sw}}^{-}P_{+}\Upsilon_{s-1}\right].
\end{equation}
Moreover, the Dirac operator of \textit{optimal staggered domain wall
fermions}, a generalization of Chiu's construction, takes the form
\begin{equation}
\overline{\Upsilon}D_{\mathsf{sdw}}\Upsilon=\sum_{s=1}^{N_{s}}\overline{\Upsilon}_{s}\left[D_{\mathsf{sw}}^{+}\left(s\right)\Upsilon_{s}+D_{\mathsf{sw}}^{-}\left(s\right)P_{-}\Upsilon_{s+1}+D_{\mathsf{sw}}^{-}\left(s\right)P_{+}\Upsilon_{s-1}\right].
\end{equation}
Here we let $D_{\mathsf{sw}}^{\pm}\left(s\right)=a_{d+1}\omega_{s}D_{\mathsf{sw}}\left(-M_{0}\right)\pm\One$
and the weight factors $\omega_{s}$ are given by Eq.~\eqref{eq:OptWeights}
for the kernel $H_{\mathsf{sw}}=\epsilon D_{\mathsf{sw}}\left(-M_{0}\right)$.

\section{Effective Dirac operator \label{sec:DW-Effective-operator}}

We now discuss the effective low-energy $d$-dimensional Dirac operator
derived in Refs.~\citep{Neuberger:1997bg,Kikukawa:1999sy,Kikukawa:1999dk,Borici:1999zw}
(cf.\ Refs.~\citep{Shamir:1998ww,Edwards:2000qv}). This will shine
some light on the relation between the light $d$-dimensional $q$,
$\overline{q}$ fields at the boundary and the $\left(d+1\right)$-dimensional
theory. We follow Refs.~\citep{Kikukawa:1999sy,Borici:1999zw,Edwards:2000qv}
in our brief discussion of the construction.

\subsection{Derivation}

After integrating out the $N_{s}-1$ heavy modes in the theory, one
can define an effective low-energy $d$-dimensional action
\begin{equation}
S_{\mathsf{eff}}=\sum_{x}\overline{q}\left(x\right)D_{\mathsf{eff}}\,q\left(x\right),
\end{equation}
where the effective Dirac operator is defined by means of the propagator
of light modes
\begin{equation}
D_{\mathsf{eff}}^{-1}\left(x,y\right)=\left\langle q\left(x\right)\overline{q}\left(y\right)\right\rangle .
\end{equation}
In the $\left(d+1\right)$-dimensional theory we find one light and
$N_{s}-1$ heavy Dirac fermions (for suitable choices of $M_{0}$).

If we take the chiral limit, where at fixed bare coupling $\beta$
we take the extent of the extra dimension to infinity $N_{s}\to\infty$,
the contribution from the heavy modes diverges. This bulk contribution
can be canceled by the introduction of pseudofermionic fields, where
the pseudofermion action is typically taken to be the fermion action
with $m=1$.

Now let us define the Hermitian operators
\begin{equation}
H_{\mathsf{w}}=\gamma_{d+1}D_{\mathsf{w}}\left(-M_{0}\right),\qquad H_{\mathsf{m}}=\gamma_{d+1}D_{\mathsf{m}}\left(-M_{0}\right),
\end{equation}
where for standard domain wall fermions the kernel operator reads
\begin{equation}
D_{\mathsf{m}}\left(-M_{0}\right)=\frac{D_{\mathsf{w}}\left(-M_{0}\right)}{2\cdot\One+a_{d+1}D_{\mathsf{w}}\left(-M_{0}\right)}.
\end{equation}
The transfer matrix in the extra dimension takes the form
\begin{equation}
T=\frac{T_{-}}{T_{+}},\qquad T_{\pm}=\One\pm a_{d+1}H,
\end{equation}
where the Hermitian operator $H$ depends on the construction given
by
\begin{equation}
H=\begin{cases}
H_{\mathsf{m}} & \textrm{for standard construction},\\
H_{\mathsf{w}} & \textrm{for Boriçi's construction}.
\end{cases}\label{eq:DefH}
\end{equation}
One can derive the explicit form \citep{Kikukawa:1999sy} of the effective
Dirac operator to find
\begin{equation}
D_{\mathsf{eff}}=\frac{1+m}{2}\One+\frac{1-m}{2}\gamma_{d+1}\frac{T_{+}^{N_{s}}-T_{-}^{N_{s}}}{T_{+}^{N_{s}}+T_{-}^{N_{s}}}.\label{eq:DefDeff}
\end{equation}
Rewriting Eq.~\eqref{eq:DefDeff} as
\begin{equation}
D_{\mathsf{eff}}=\left(1-m\right)\left[D_{\mathsf{eff}}\left(0\right)+\frac{m}{1-m}\right],\label{eq:InducedMass}
\end{equation}
where $D_{\mathsf{eff}}\left(0\right)$ refers to the effective operator
$D_{\mathsf{eff}}$ at $m=0$, we find a bare mass of $m/\left(1-m\right)$
for a given choice of the parameter $m$, see Ref.~\citep{Boyle:2016imm}.

The effective operator in Eq.~\eqref{eq:DefDeff} can also be written
in the compact form
\begin{equation}
D_{\mathsf{eff}}=\left(\mathcal{P}^{\intercal}D_{1}^{-1}D_{m}\mathcal{P}\right)_{1,1}\label{eq:ProjMethod}
\end{equation}
as shown in Refs.~\citep{Borici:1999zw,Edwards:2000qv}, where the
matrix $\mathcal{P}$ reads
\begin{equation}
\mathcal{P}=\left(\begin{array}{ccccc}
P_{-} & P_{+}\\
 & P_{-} & P_{+}\\
 &  & \ddots & \ddots\\
 &  &  & P_{-} & P_{+}\\
P_{+} &  &  &  & P_{-}
\end{array}\right)
\end{equation}
and $\mathcal{P}^{-1}=\mathcal{P}^{\intercal}$. To simplify the notation
we introduced $D_{m}\equiv D_{\mathsf{dw}}\left(m\right)$ and the
$\left(1,1\right)$-index refers to the corresponding $s$-block of
the matrix product.

For Chiu's optimal domain wall fermions the derivation follows Boriçi's
case, where the additional weight factors $\omega_{s}$ have to be
taken into account \citep{Chiu:2002ir}. By construction, we find
that in this case the $\sign$-function approximation in the effective
Dirac operator equals Zolotarev's optimal rational function approximation.

\subsection{The $N_{s}\to\infty$ limit}

For the following discussion we restrict ourselves to the $m=0$ case.
First, we remark that Eq.~\eqref{eq:DefDeff} can be written as
\begin{equation}
\frac{T_{+}^{N_{s}}-T_{-}^{N_{s}}}{T_{+}^{N_{s}}+T_{-}^{N_{s}}}=\epsilon_{N_{s}/2}\left(a_{d+1}H\right),
\end{equation}
where $\epsilon_{N_{s}/2}$ is Neuberger’s polar decomposition approximation
\citep{Neuberger:1998my,higham1994matrix} of the $\sign$-function.
One can easily see that in the $N_{s}\to\infty$ limit we recover
the overlap Dirac operator,
\begin{align}
D_{\mathsf{ov}} & =\lim_{N_{s}\to\infty}D_{\mathsf{eff}}\nonumber \\
 & =\frac{1}{2}\One+\frac{1}{2}\gamma_{d+1}\sign H\nonumber \\
 & =\frac{1}{2}\left[\One+D_{-M_{0}}\left(D_{-M_{0}}^{\dagger}D_{-M_{0}}\right)^{-\frac{1}{2}}\right],\label{eq:DefDov}
\end{align}
with $H$ given in Eq.~\eqref{eq:DefH}, using the shorthand notation
$D_{-M_{0}}=D\left(-M_{0}\right)$ and
\begin{equation}
D=\begin{cases}
D_{\mathsf{m}} & \textrm{for standard construction},\\
D_{\mathsf{w}} & \textrm{for Boriçi's/Chiu's construction}.
\end{cases}\label{eq:DefD}
\end{equation}
The Ginsparg-Wilson equation
\begin{equation}
\left\{ \gamma_{d+1},D_{\mathsf{ov}}\right\} =2D_{\mathsf{ov}}\gamma_{d+1}D_{\mathsf{ov}}\label{eq:GWR}
\end{equation}
can be shown to hold for $D_{\mathsf{ov}}$, thus implementing an
exact chiral symmetry and implying the normality of the operator.
We note that compared to the discussion in Subsec.\ \ref{subsec:OV-Ginsparg-Wilson-relation},
we find an additional factor of two in the Ginsparg-Wilson relation
due to a different normalization of the overlap operator.

If we now compare Eq.~\eqref{eq:DefDov} with the standard definition
\begin{equation}
D_{\mathsf{ov}}=\rho\left[\One+D_{-\rho}\left(D_{-\rho}^{\dagger}D_{-\rho}\right)^{-\frac{1}{2}}\right]
\end{equation}
of the overlap operator and we use the relation for the effective
negative mass parameter
\begin{equation}
\rho=\begin{cases}
M_{0}-\frac{a_{d+1}}{2}M_{0}^{2} & \textrm{for standard construction},\\
M_{0} & \textrm{for Boriçi's/Chiu's construction},
\end{cases}\label{eq:Defrho}
\end{equation}
this would result in a restriction on the domain wall height $M_{0}$
from $\rho=1/2$. This limitation can be overcome by rescaling $D_{\mathsf{eff}}$
by a factor $\varrho=2\rho$, so that in the free-field case the low-lying
eigenvalues of the kernel operator remain invariant (up to discretization
effects) when the effective operator projection is applied. This can
also be seen in Fig.~\ref{fig:free-dov}, which we later discuss
in Sec.\ \ref{subsec:Free-field-case}. In all numerical investigations
of this chapter this rescaling is taken into account.

\subsection{Approximate sign functions}

\begin{figure}[t]
\begin{centering}
\subfloat[$N_{s}=2$]{\includegraphics[width=0.49\textwidth]{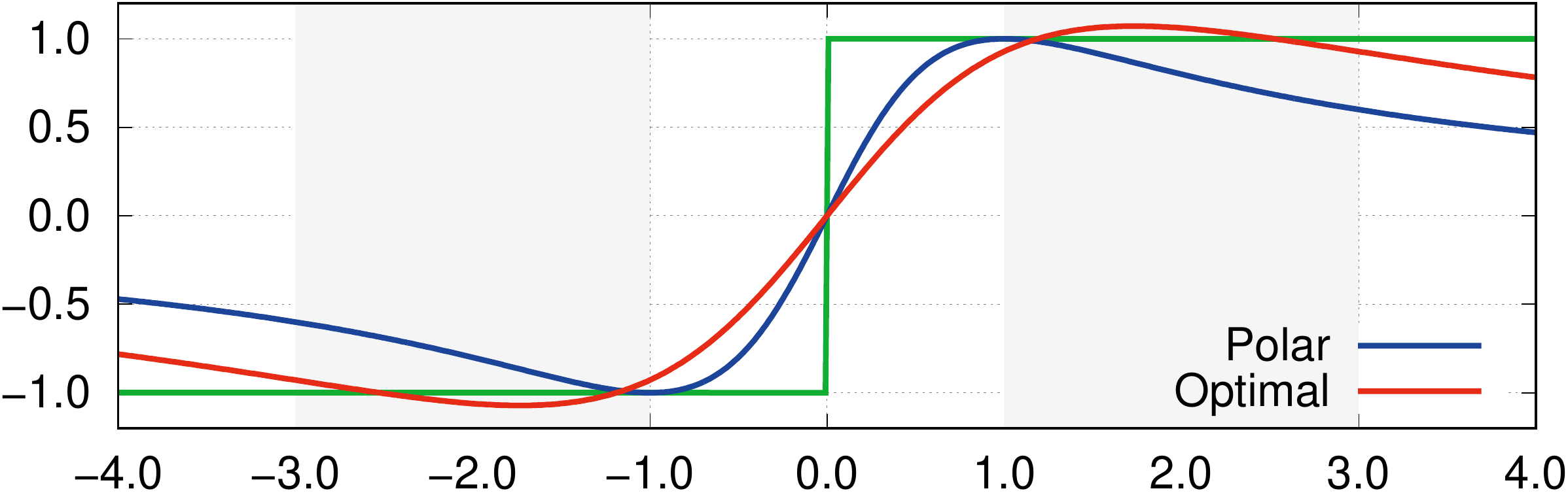}

}\hfill{}\subfloat[$N_{s}=6$]{\includegraphics[width=0.49\textwidth]{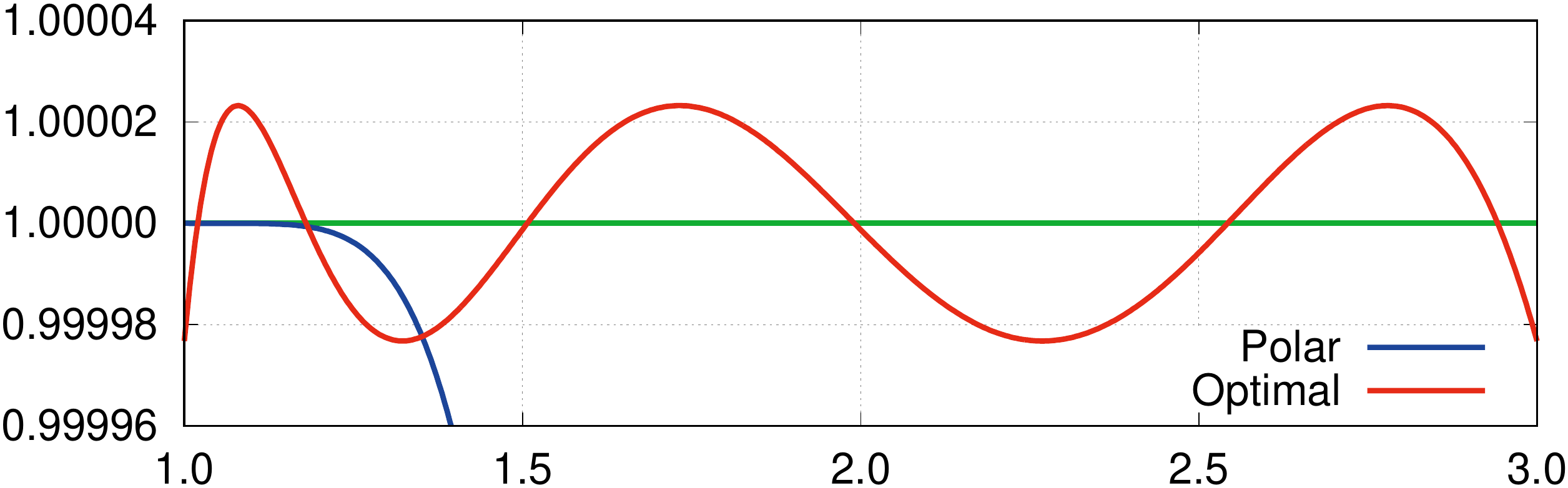}

}
\par\end{centering}
\caption{Approximations of the $\protect\sign$-function by $r\left(z\right)$.
The optimal construction is illustrated for the case of $\lambda_{\mathsf{min}}=1$
and $\lambda_{\mathsf{max}}=3$. \label{fig:sign-approximations} }
\end{figure}

The effective operator, as previously derived, is an approximation
to the overlap operator. Explicitly, the $\sign$-function is approximated
by the rational function
\begin{equation}
r\left(z\right)=\frac{\Pi_{+}\left(z\right)-\Pi_{-}\left(z\right)}{\Pi_{+}\left(z\right)+\Pi_{-}\left(z\right)}\label{eq:SignApproximation}
\end{equation}
with
\begin{equation}
\Pi_{\pm}\left(z\right)=\begin{cases}
\left(1\pm z\right)^{N_{s}} & \textrm{for standard/Boriçi's construction},\\
\prod_{s}\left(1\pm\omega_{s}z\right) & \textrm{for Chiu's construction}.
\end{cases}\label{eq:SignFuncApprox}
\end{equation}
As expected, we can verify that $r\left(z\right)\to\sign\left(z\right)$
for $N_{s}\to\infty$. We note that for the standard construction
and Boriçi's construction the approximation is identical, but is applied
to different kernel operators, namely $H_{\mathsf{m}}$ or $H_{\mathsf{w}}$
respectively. To illustrate the approximations in Eq.~\eqref{eq:SignFuncApprox},
we compare the polar decomposition approximation in Boriçi's formulation
with the optimal approximation in Chiu's construction in Fig.~\ref{fig:sign-approximations}.
As we can see, the approximation error over the construction range
is significantly lower for the optimal construction. We recall that
the coefficients $\omega_{s}$ are related to Zolotarev's coefficients,
cf.\ Refs.~\citep{zolotarev1877application,vandenEshof:2002ms,Chiu:2002eh}.

\minisec{Staggered formulations}

The discussion in this section naturally generalizes to the case of
a staggered Wilson kernel, when the replacement rule in Eq.~\eqref{eq:ReplacementRule}
is applied. In particular, the effective and overlap operators follow
from this replacement and we can interpret staggered domain wall fermions
as a truncation of staggered overlap fermions \citep{Adams:2010gx}.

\section{Setting \label{sec:DW-Setting}}

We now move on to the numerical part of the study and begin by discussing
our setting. In particular, we elaborate on our method to quantify
chiral symmetry violations to benchmark the various domain wall fermion
constructions.

\subsection{Setting the domain wall height}

To give rise to a one-flavor theory, in the free-field case the domain
wall height $M_{0}$ has to be chosen in the range $0<M_{0}<2r$.
When moving to a nontrivial background field, this interval generally
shifts and contracts. In less rigor terms, the parameter $M_{0}$
has to be chosen so that the origin is shifted inside the leftmost
``belly'' of the eigenvalue spectrum of the kernel operator. Although
even in the free-field case there is no optimal choice for $M_{0}$
as discussed in Ref.~\citep{Durr:2005an}, one typically uses the
canonical choice $M_{0}=1$ for the free field, which is at the center
of the eigenvalue belly. More rigorously speaking, valid choices of
$M_{0}$ are restricted by the position of the mobility edge \citep{Golterman:2003qe,Golterman:2003ui,Golterman:2004cy,Golterman:2005fe}.

In the Schwinger model, for reasonable choices of the inverse coupling
$\beta$ the eigenvalue spectrum is not distorted too far away from
the free-field spectrum. As a consequence, valid choices for $M_{0}$
remain close to the free-field case. While in four-dimensional quantum
chromodynamics one often uses $M_{0}=1.8$ (see e.g.\ Ref.~\citep{Blum:2000kn}),
in our present case $M_{0}=1$ remains a sensible and simple choice,
which we consequently use for all our numerical work.

\subsection{Effective mass \label{subsec:DW-Effective-mass}}

Commonly employed measures in the literature to quantify the degree
of chiral symmetry breaking of domain wall fermions are the residual
mass $m_{\mathsf{res}}$ \citep{Furman:1994ky,Fleming:1999eq,Blum:2000kn,Blum:2001sr}
and the effective mass $m_{\mathsf{eff}}$ \citep{Gadiyak:2000kz,Jung:2000fh,Jung:2000ys}.
While the residual mass makes use of the explicit mass dependence
in the chiral Ward-Takahashi identities, the effective mass is defined
as the lowest eigenvalue (by magnitude) of the Hermitian operator
in a background field with nontrivial topology. Although these measures
are not identical, they usually agree within a factor of order $\mathcal{O}\left(1\right)$.
Consequently both define valid measures to quantify the chirality
of a given lattice fermion formulation.

Because of its conceptual simplicity, we use the effective mass $m_{\mathsf{eff}}$
as one of our measures for the numerical studies. To give a precise
definition, we consider a given Dirac operator $D$ on a topologically
nontrivial background configuration with periodic boundary conditions
and define the effective mass $m_{\mathsf{eff}}$ as the lowest eigenvalue
of the Hermitian kernel operator $H$. Explicitly, we let
\begin{equation}
m_{\mathsf{eff}}=\min_{\lambda\in\spec H}\left|\lambda\right|=\min_{\Lambda\in\spec D^{\dagger}D}\sqrt{\Lambda},
\end{equation}
where we used the fact that $H^{2}=D^{\dagger}D$. For a normal operator
$D$, this definition simplifies to $m_{\mathsf{eff}}=\min_{\lambda\in\spec D}\left|\lambda\right|$.
For a non-normal operator, however, the eigenvalues of $D$ and $H$
are not directly related.

For gauge configurations with topological charge $Q\neq0$, the existence
of zero modes of $H$ is guaranteed in the continuum by the Atiyah-Singer
index theorem \citep{AtiyahI,AtiyahIII,AtiyahIV,AtiyahV}. Similarly,
the presence of zero modes of the overlap operator as defined in Eq.~\eqref{eq:DefDov}
can be shown as well \citep{Ginsparg:1981bj,Hasenfratz:1998ri,Luscher:1998pqa}.
For the effective Dirac operators these zero modes are approximate,
but become exact in the $N_{s}\to\infty$ limit. The absolute value
of the eigenvalues of these approximate zero modes can then serve
as a measure of chirality.

In the context of a gauge theory, the Atiyah-Singer index theorem
takes the following form. Let $n_{\mp}$ refer to the number of left-handed
and right-handed zero modes, then their difference is related to the
topological charge by
\begin{equation}
n_{-}-n_{+}=\left(-1\right)^{d/2}Q,
\end{equation}
see Ref.~\citep{Adams:2009eb}. In Eq.~\eqref{eq:DefQ}, we will
give a precise definition of $Q$. We note that one can show the Vanishing
Theorem \citep{Kiskis:1977vh,Nielsen:1977aw,Ansourian:1977qe} in
our two-dimensional setting. The theorem states that on gauge configurations
with $Q\neq0$, either $n_{-}$ or $n_{+}$ vanishes, i.e.\ $n_{-}$
and $n_{+}$ are not simultaneously nonzero.

\subsection{Normality and Ginsparg-Wilson relation}

The continuum Dirac operator, the naïve and overlap lattice Dirac
operators are all normal operators. At the same time, the effective
Dirac operators are not normal. This makes operator normality an interesting
aspect in the context of chirality, especially as it has been shown
that chiral properties imply the normality of the Dirac operator \citep{Kerler:1999dk},
see also Refs.~\citep{Hip:2001hc,Hip:2001mh}.

By definition a normal operator $D$ satisfies $\left[D,D^{\dagger}\right]=0$.
Deviation from normality can be quantified by the norm of this commutator
and we define
\begin{equation}
\Delta_{\mathsf{N}}=\left\Vert \left[D,D^{\dagger}\right]\right\Vert _{\infty},
\end{equation}
where $\left\Vert \cdot\right\Vert _{\infty}$ is the matrix norm
induced by the $L_{\infty}$-norm. As previously discussed, the normality
measure $\Delta_{\mathsf{N}}$ vanishes for the overlap operator and,
thus, for the effective Dirac operators in the limit $N_{s}\to\infty$.

Besides normality, violations of the Ginsparg-Wilson relation as given
in Eq.~\eqref{eq:GWR} are a natural measure for chirality of a lattice
fermion formulation. We let
\begin{equation}
\Delta_{\mathsf{GW}}=\left\Vert \left\{ \gamma_{3},D\right\} -\rho^{-1}D\gamma_{3}D\right\Vert _{\infty}\label{eq:DefDeltaGWR}
\end{equation}
and find that for the effective operators $\Delta_{\mathsf{GW}}\to0$
for $N_{s}\to\infty$. For a staggered Wilson kernel, the chirality
matrix $\gamma_{3}$ has to be replaced by the staggered $\epsilon$
in the definition of $\Delta_{\mathsf{GW}}$. We note that the additional
factor of $1/\rho$ in Eq.~\eqref{eq:DefDeltaGWR} is due to the
different scales of the effective Dirac operators. The measures $\Delta_{\mathsf{N}}$
and $\Delta_{\mathsf{GW}}$ were previously considered in Refs.~\citep{Durr:2005an,Durr:2010ch}
and together with $m_{\mathsf{eff}}$ serve here as a measure for
chiral symmetry violations.

\subsection{Topological charge}

To determine the topological charge of a gauge configuration, we use
the index formula for the standard overlap operator,
\begin{equation}
Q=\frac{1}{2}\Tr\left(H_{\mathsf{w}}/\sqrt{H_{\mathsf{w}}^{2}}\right),
\end{equation}
and its staggered equivalent,
\begin{equation}
Q=\frac{1}{2}\Tr\left(H_{\mathsf{sw}}/\sqrt{H_{\mathsf{sw}}^{2}}\right),\label{eq:DefQ}
\end{equation}
with $H_{\mathsf{sw}}=\epsilon D_{\mathsf{sw}}\left(-M_{0}\right)$
as discussed in Ref.~\citep{Adams:2009eb}. We numerically verified
that these two definition yield the same topological charge on a set
of sample configurations. This observation is in agreement with analytical
results \citep{Adams:2013lpa} and other numerical studies \citep{deForcrand:2012bm,Azcoiti:2014pfa}.

\section{Numerical results \label{sec:DW-Numerical-results}}

For the numerical part of this work, we calculate complete eigenvalue
spectra and evaluate our measures of chirality for the previously
discussed Dirac operators. We do this on various gauge configurations
in the Schwinger model and cover the free-field case, thermalized
configurations and the smooth topological configurations discussed
in Ref.~\citep{Smit:1986fn}.

Throughout this section, we set the lattice spacings in all dimensions
to $a=a_{d+1}=1$ and fix the (staggered) Wilson parameter to $r=1$.
For the extra dimension we consider extents within the range $2\leq N_{s}\leq8$.
In order to allow the calculation of the effective mass $m_{\mathsf{eff}}$
as discussed in Sec.\ \ref{subsec:DW-Effective-mass}, we impose
periodic boundary conditions in all dimensions.

We determine complete eigenvalue spectra using the \textsc{lapack}
\citep{LAPACK} library, while extremal eigenvalues, such as the effective
mass, are computed with \textsc{arpack} \citep{ARPACK}. For all numerical
calculations we use double precision. In the figures, the standard
construction is abbreviated with ``std'', Boriçi's construction
with ``Bor'' and Chiu's optimal construction with ``opt''. When
discussing overlap operators, the construction with kernel $H_{\mathsf{m}}$
is referred to as ``DW'', Neuberger's overlap with kernel $H_{\mathsf{w}}$
is abbreviated as ``Neub'' and Adams' staggered overlap with kernel
$H_{\mathsf{sw}}$ is denoted as ``Adams''.

\subsection{Free-field case \label{subsec:Free-field-case}}

In our discussion we begin with the simplest case of the free field.
Periodicity allows a Fourier transformation of the kernel operators
and simplifies the computation of the eigenvalue spectra, cf.\ Chapter\ \ref{chap:Eigenvalue-spectra}.
In the momentum representation, the Wilson kernel takes the form of
the following $2\times2$ linear map
\begin{equation}
D_{\mathsf{w}}=\left(m_{\mathsf{f}}+2r\right)\One+\ii\sum_{\mu}\gamma_{\mu}\sin p_{\mu}-r\sum_{\nu}\cos p_{\nu}\One,
\end{equation}
where $p_{\mu}=2\pi n_{\mu}/N_{\mu}$ with $n_{\mu}=0,1,\dots,N_{\mu}-1$
and $N_{\mu}$ denotes the number of slices in $\mu$-direction. The
staggered Wilson kernel is represented by the $4\times4$ linear map
\begin{equation}
D_{\mathsf{sw}}=m_{\mathsf{f}}\left(\One\otimes\One\right)+\ii\sum_{\mu}\sin p_{\mu}\left(\gamma_{\mu}\otimes\One\right)+r\One\otimes\left(\One+\xi_{3}\prod_{\nu}\cos p_{\nu}\right)
\end{equation}
with $n_{\mu}=0,1,\dots,N_{\mu}/2-1$, cf.\ Ref.~\citep{deForcrand:2012bm}.

For the three-dimensional bulk operator, the extra dimension is lacking
periodicity and we leave the corresponding coordinate in the position-space
representation. By employing a momentum-space representation for the
kernel operators, we reduce the dimensionality of the problem and
avoid numerical instabilities otherwise encountered in the free-field
case. All free-field results discussed in the following are for the
case of a $N_{s}\times N_{t}=20\times20$ lattice.

\paragraph{Kernel operators.}

\begin{figure}[t]
\begin{centering}
\includegraphics[width=0.75\textwidth]{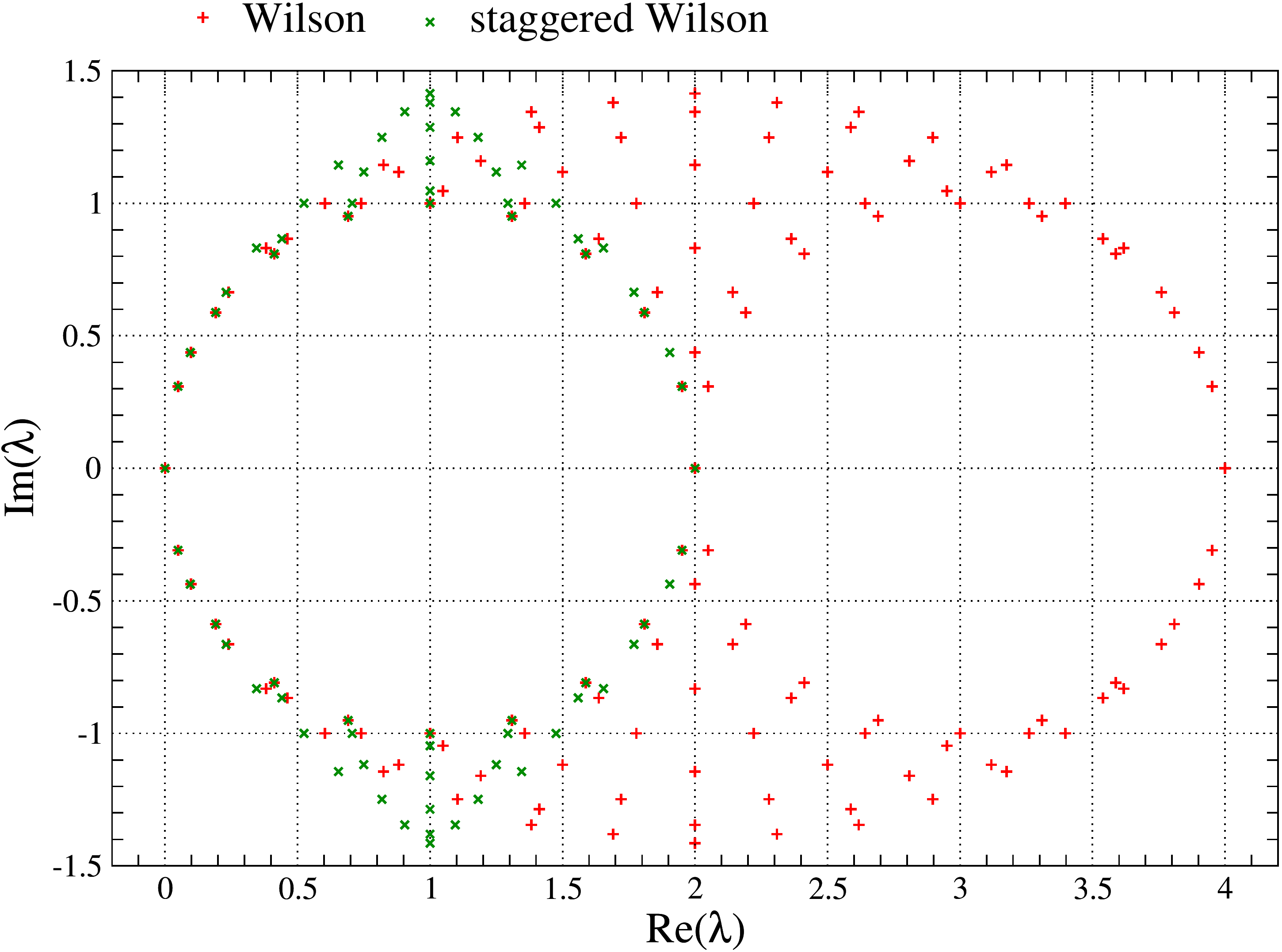} 
\par\end{centering}
\caption{Spectrum of the kernel operators in the free field case. \label{fig:free-kernel} }
\end{figure}

In Fig.~\ref{fig:free-kernel}, we show the free-field eigenvalue
spectra of the Wilson and staggered Wilson Dirac operator. While in
four dimensions Wilson fermions have four doubler branches, in our
setting of the two-dimensional Schwinger model only two doubler branches
are present. Adams' staggered Wilson term, however, always splits
the tastes into two groups with positive and negative flavor-chirality.
Compared to Wilson fermions, we observe how the spectrum of the staggered
Wilson kernel resembles the Ginsparg-Wilson circle much more closely.
On sufficiently smooth configurations one can then hope for lower
computational costs for the construction of chiral fermion formulations.

We note that the Wilson and staggered Wilson eigenvalue spectrum differs
mostly in their ultraviolet parts and very little in the physically
relevant low-lying part. Nevertheless, these discretization effects
alter the properties and shape of the bulk, effective and overlap
operators and in general have an impact on computational efficiency
and chiral properties.

\paragraph{Bulk operators.}

\begin{figure}[t]
\begin{centering}
\subfloat[Standard construction]{\includegraphics[width=0.32\textwidth]{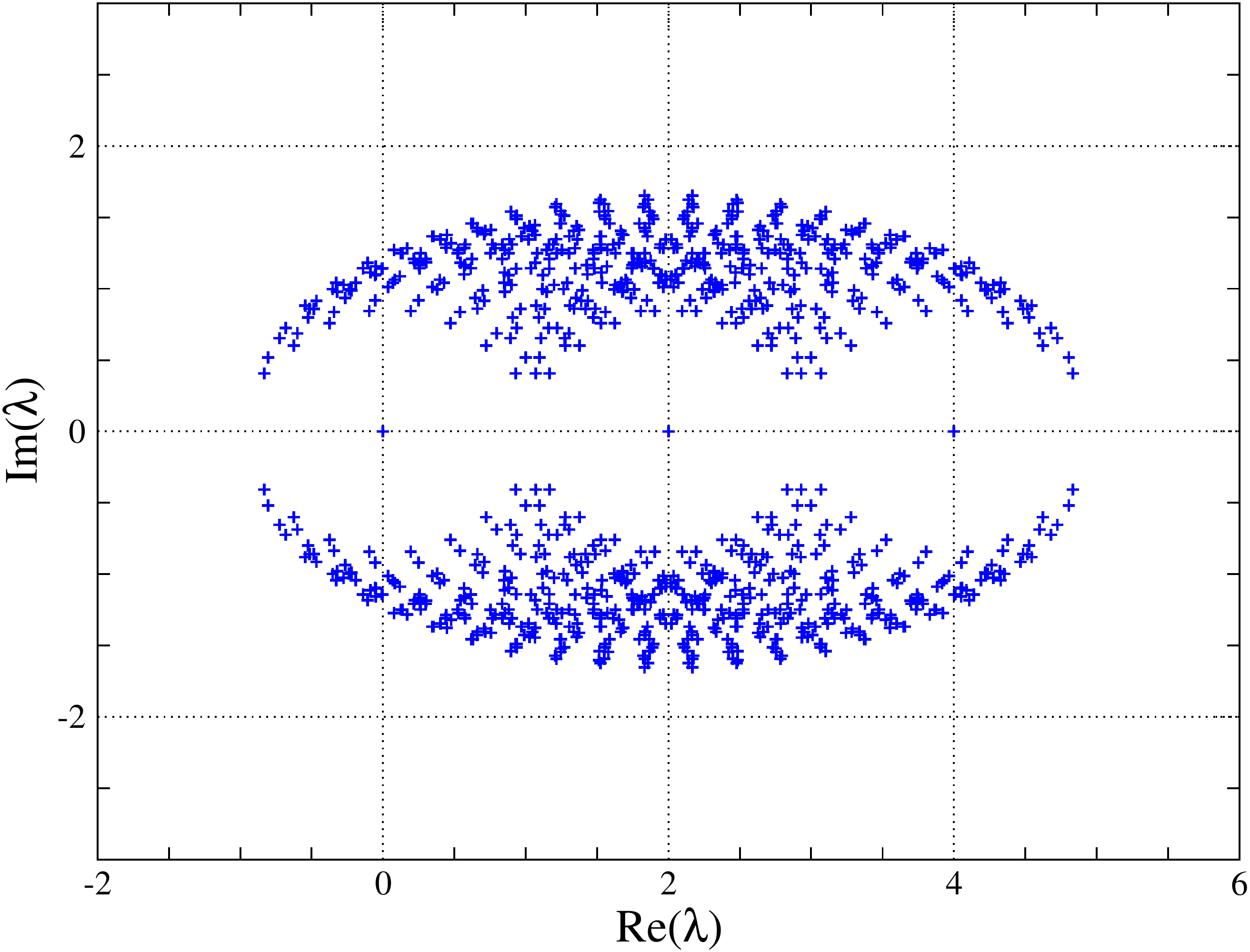}

}\hfill{}\subfloat[Boriçi's construction]{\includegraphics[width=0.32\textwidth]{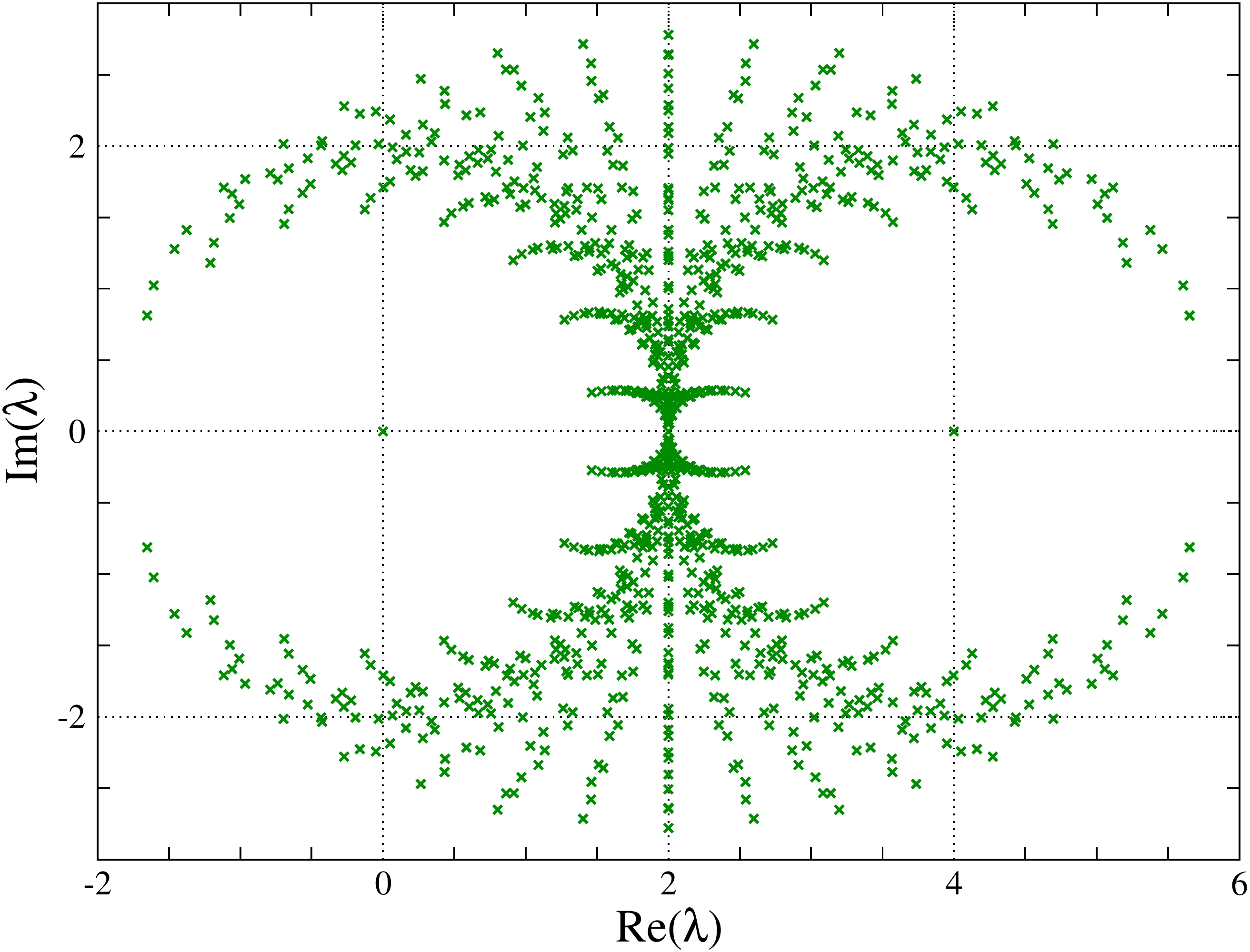}

}\hfill{}\subfloat[Optimal construction]{\includegraphics[width=0.32\textwidth]{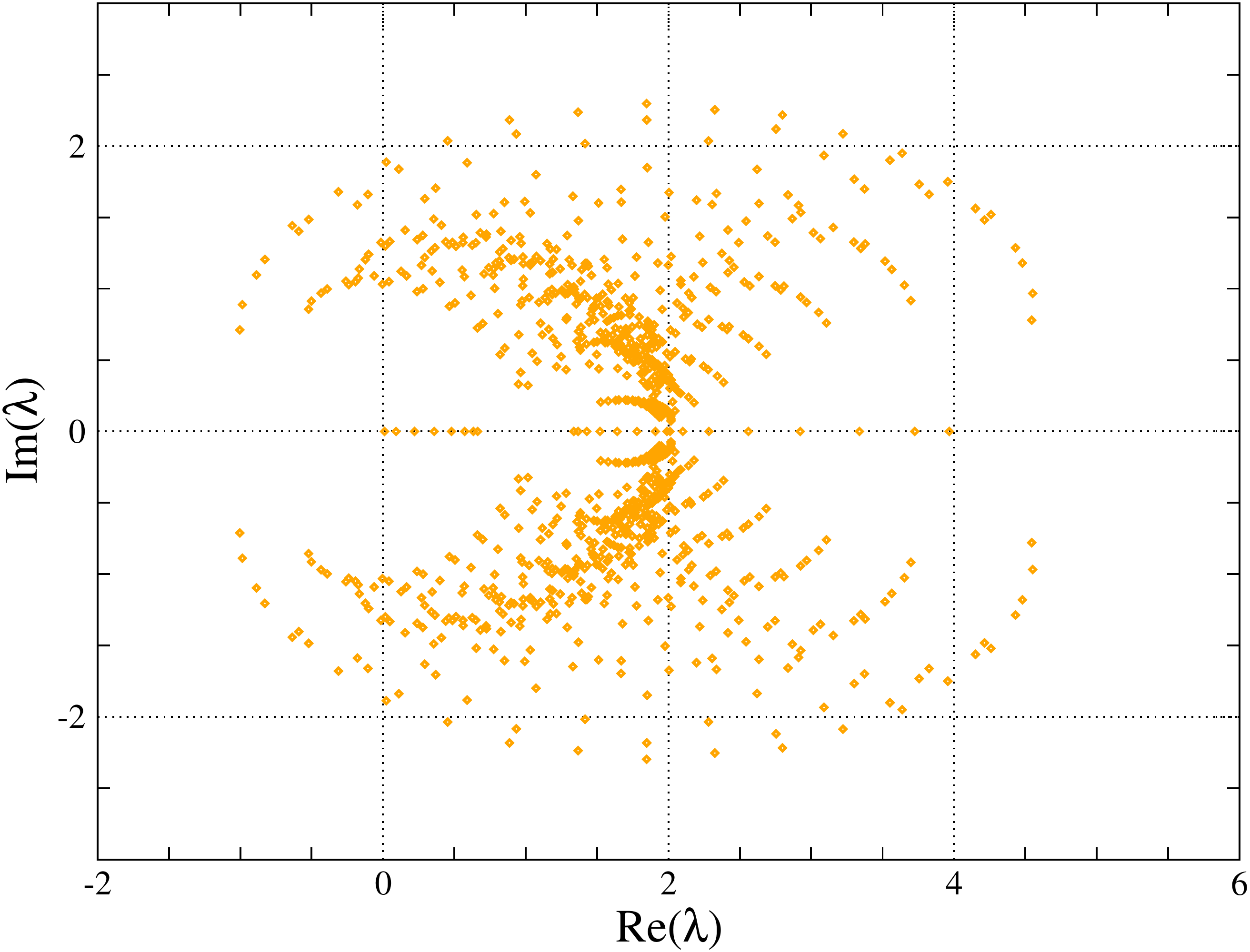}

}
\par\end{centering}
\caption{Spectrum of $D_{\mathsf{dw}}$ with Wilson kernel for $m=0$ at $N_{s}=8$
in the free field case. \label{fig:free-wilson-bulk}}
\end{figure}
\begin{figure}[t]
\begin{centering}
\subfloat[Standard construction]{\includegraphics[width=0.32\textwidth]{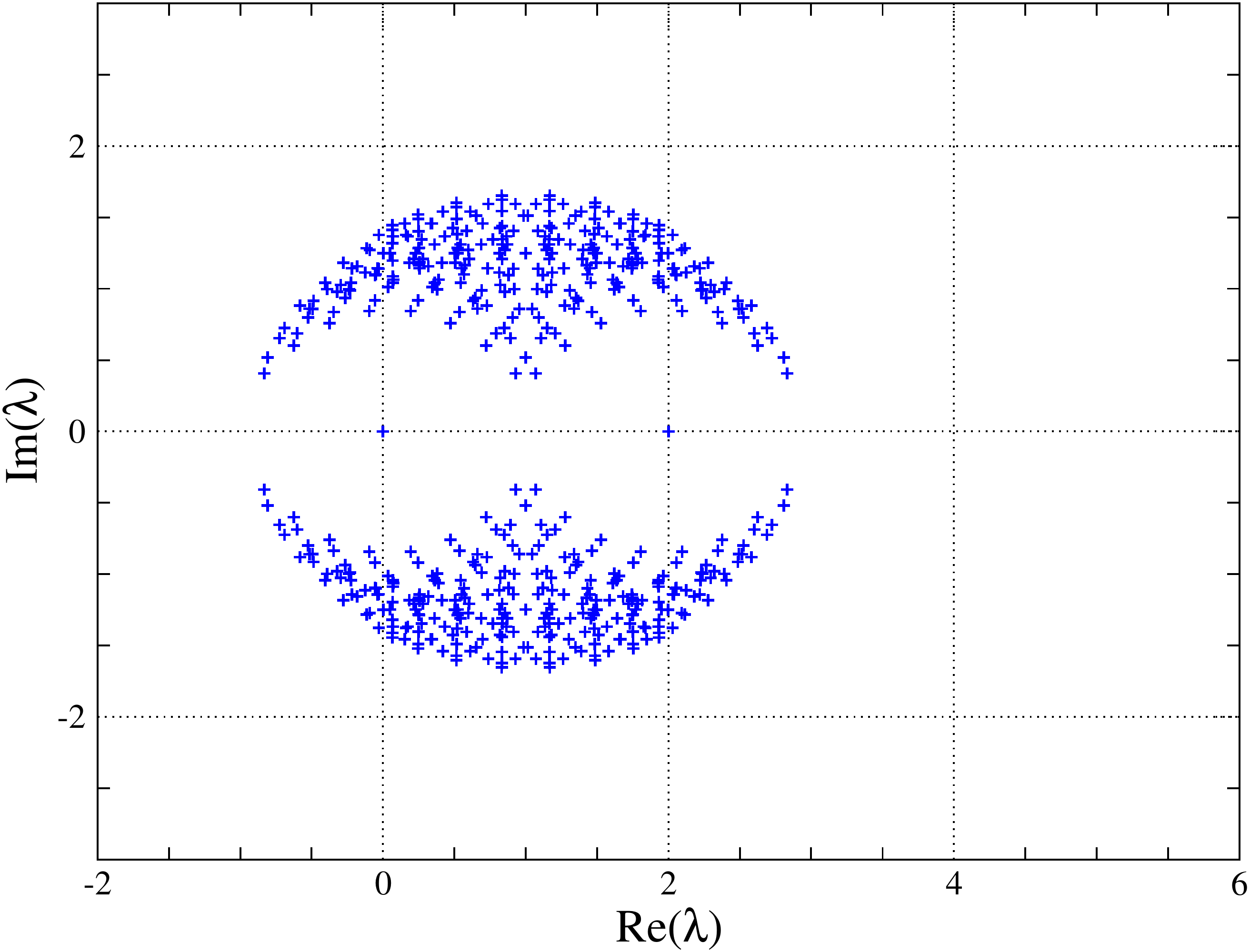}

}\hfill{}\subfloat[Boriçi's construction]{\includegraphics[width=0.32\textwidth]{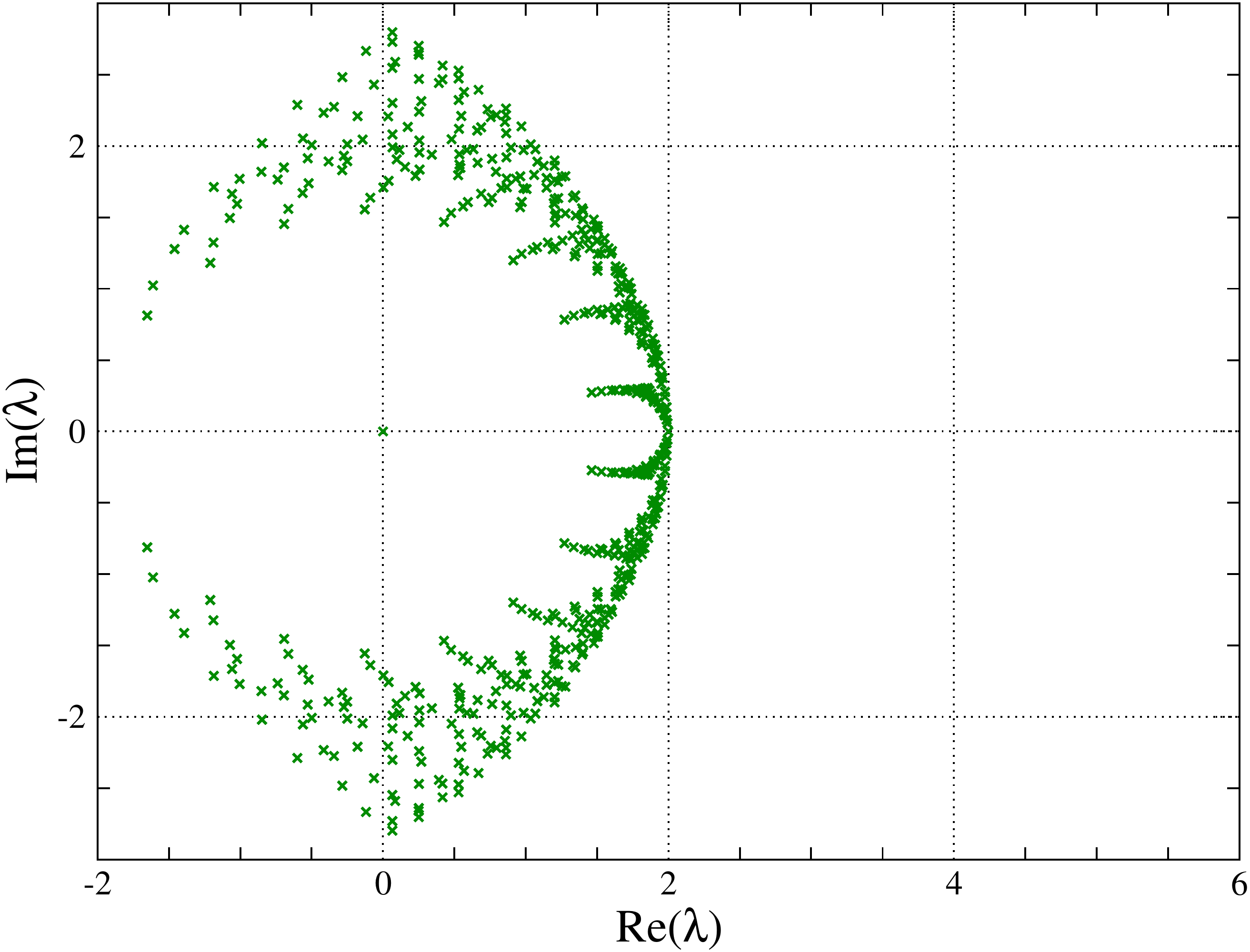}

}\hfill{}\subfloat[Optimal construction]{\includegraphics[width=0.32\textwidth]{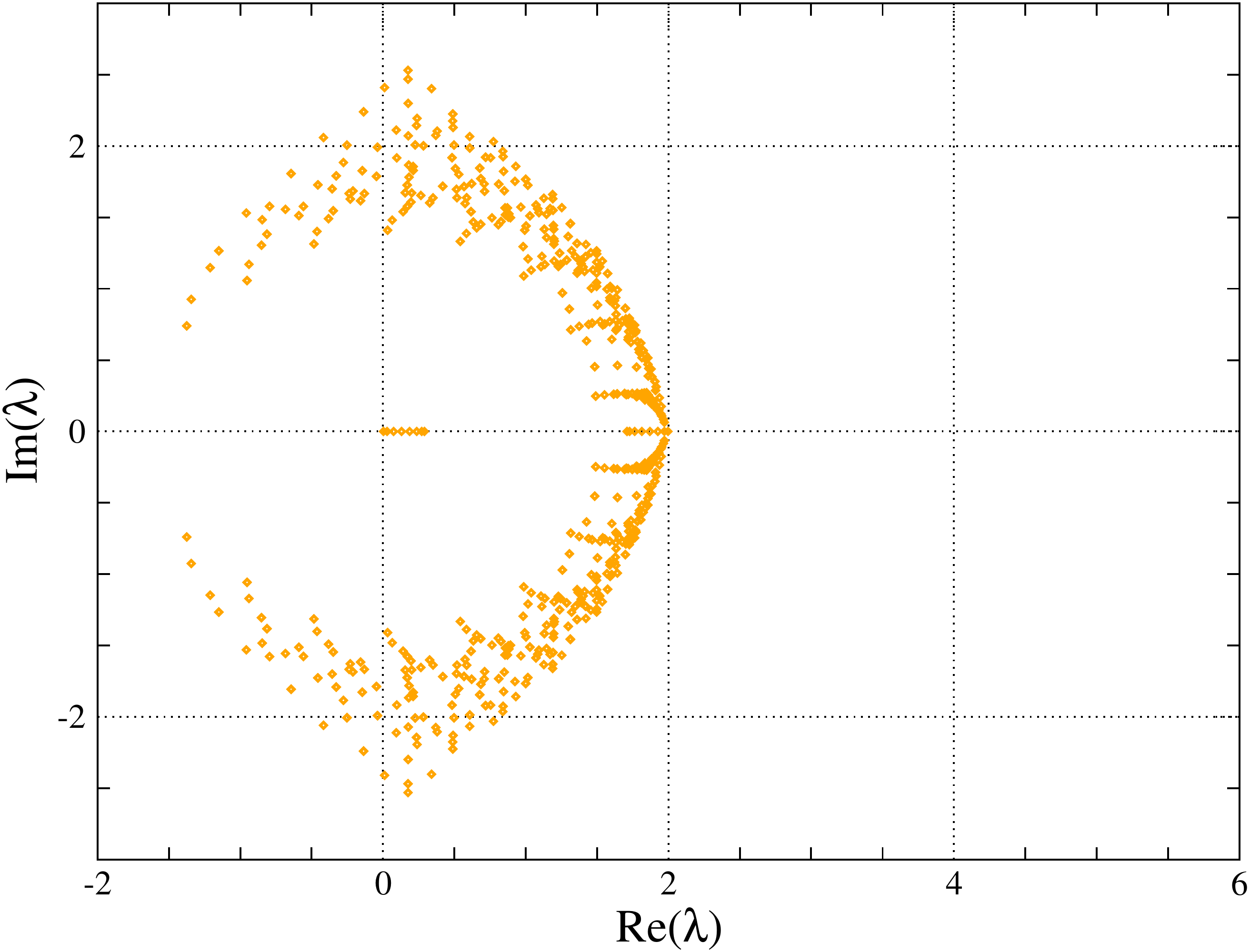}

}
\par\end{centering}
\caption{Spectrum of $D_{\mathsf{sdw}}$ with staggered Wilson kernel for $m=0$
at $N_{s}=8$ in the free field case. \label{fig:free-stw-bulk}}
\end{figure}
\begin{figure}[t]
\begin{centering}
\hfill{}\subfloat[Periodic case ($m=-1$)]{\includegraphics[width=0.4\textwidth]{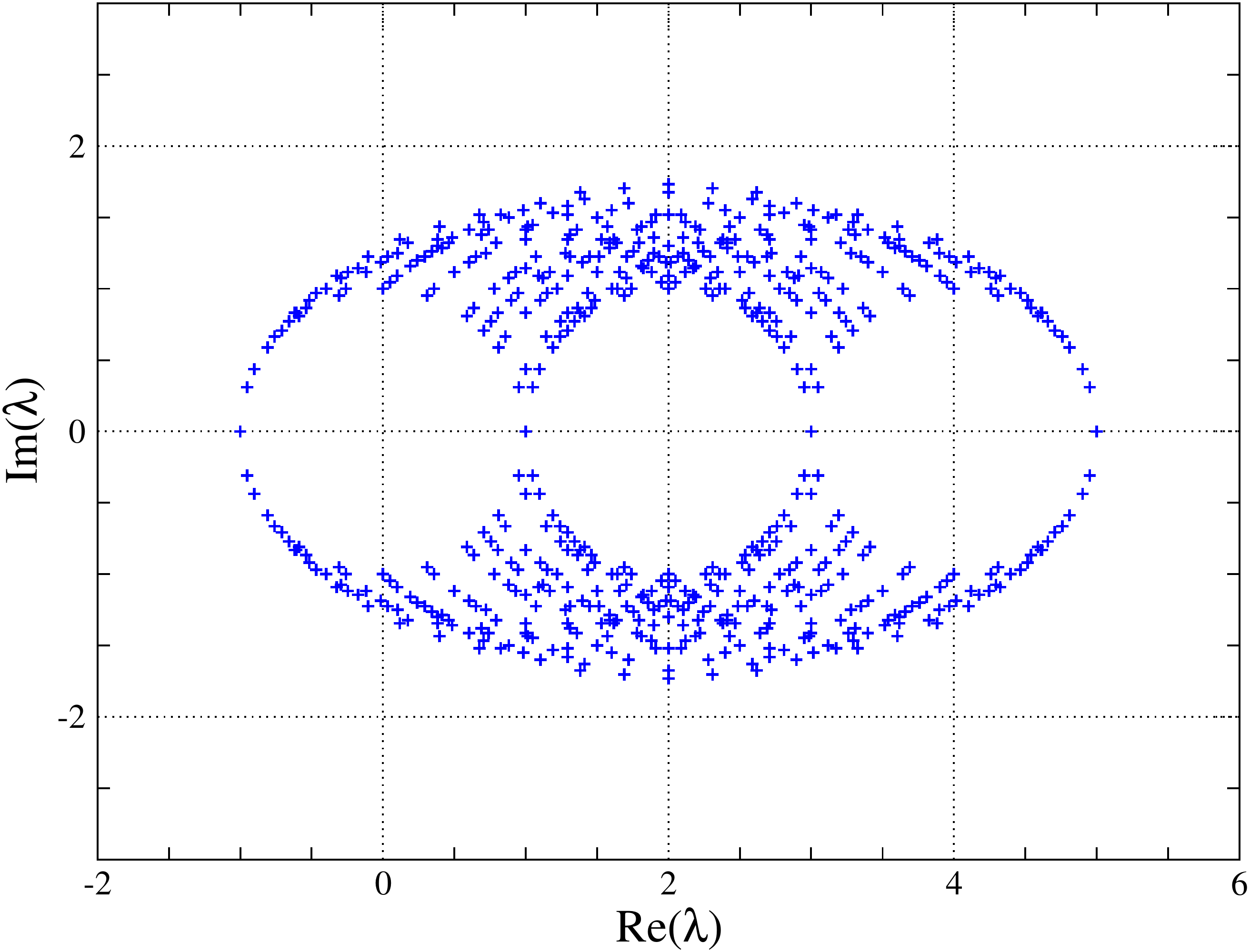}

}\hfill{}\subfloat[Antiperiodic case ($m=1$)]{\includegraphics[width=0.4\textwidth]{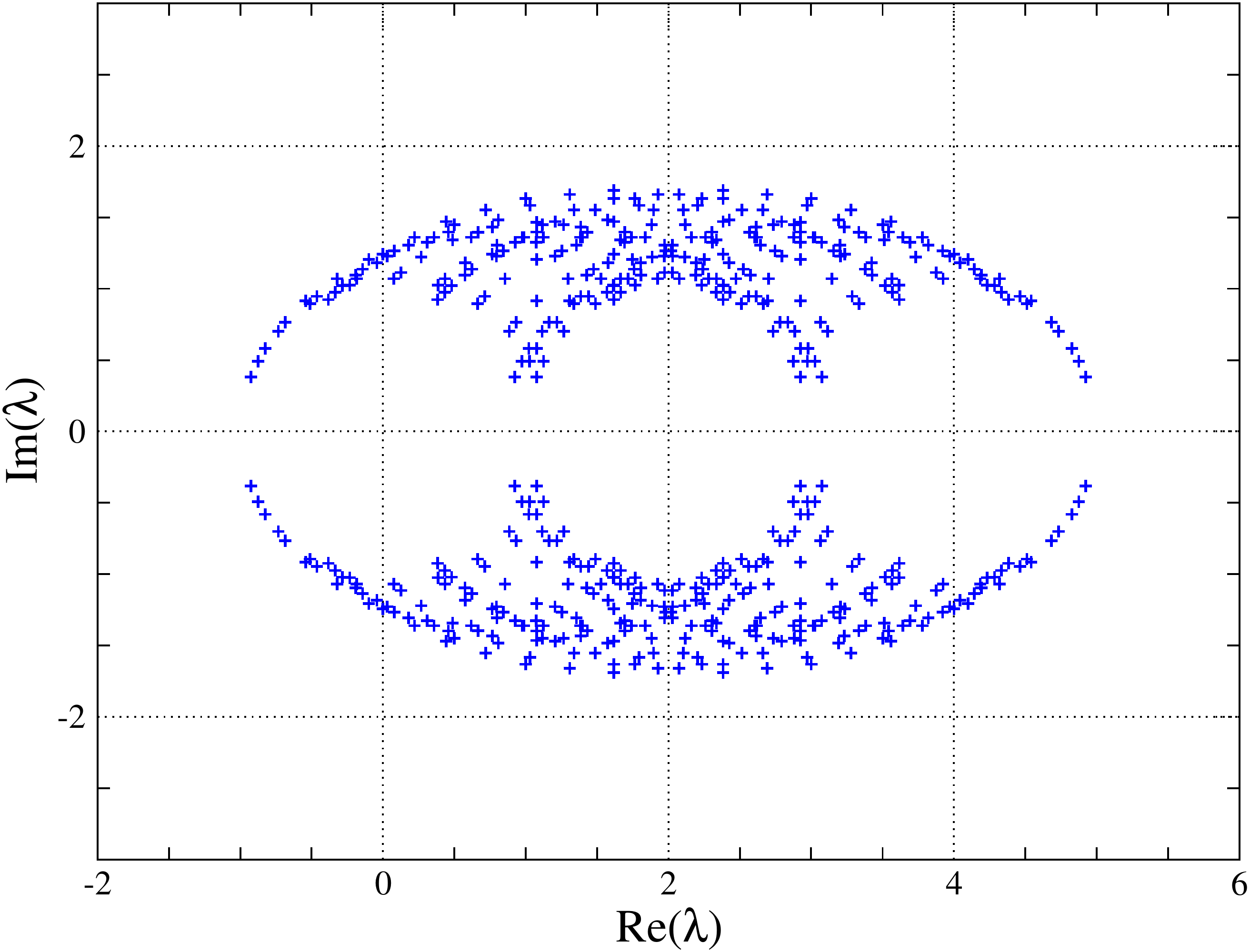}

}\hfill{} 
\par\end{centering}
\caption{Spectrum of $D_{\mathsf{dw}}$ with Wilson kernel in the standard
construction at $N_{s}=8$ in the free field case with different boundary
conditions (cf.\ Fig.~\ref{fig:free-wilson-bulk}) \label{fig:free-wilson-bulk-bcs}}
\end{figure}

Moving on to the $\left(2+1\right)$-dimensional bulk operators, in
Figs.~\ref{fig:free-wilson-bulk} and \ref{fig:free-stw-bulk} we
show the eigenvalue spectra in the standard, Boriçi's and Chiu's construction.
To illustrate the effect of changing boundary conditions in the extra
dimension, we show in Fig.~\ref{fig:free-wilson-bulk-bcs} the case
of periodic ($m=-1$) and antiperiodic ($m=1$) boundary conditions,
which can be compared to the Dirichlet ($m=0)$ case.

In the spectrum of the bulk operator with the standard construction,
we find (two) three doubler branches for the case of a (staggered)
Wilson kernel. For Boriçi's construction the number of branches is
reduced by one. In both cases we find $2N_{s}$ exact zero modes in
the Dirichlet case \citep{Shamir:1993zy,Gadiyak:2000kz}, which disappear
for non-vanishing values of $m$. For Chiu's optimal construction
we only find two approximate zero modes and observe that the real
eigenvalues are spread along the real axis. In the case of Boriçi's
and the optimal construction with a Wilson kernel, we notice how the
spectrum of the bulk operators lost its resemblance to the three-dimensional
Wilson operator.

\paragraph{Effective operators.}

\begin{figure}[t]
\begin{centering}
\subfloat[Wilson kernel]{\includegraphics[width=0.45\textwidth]{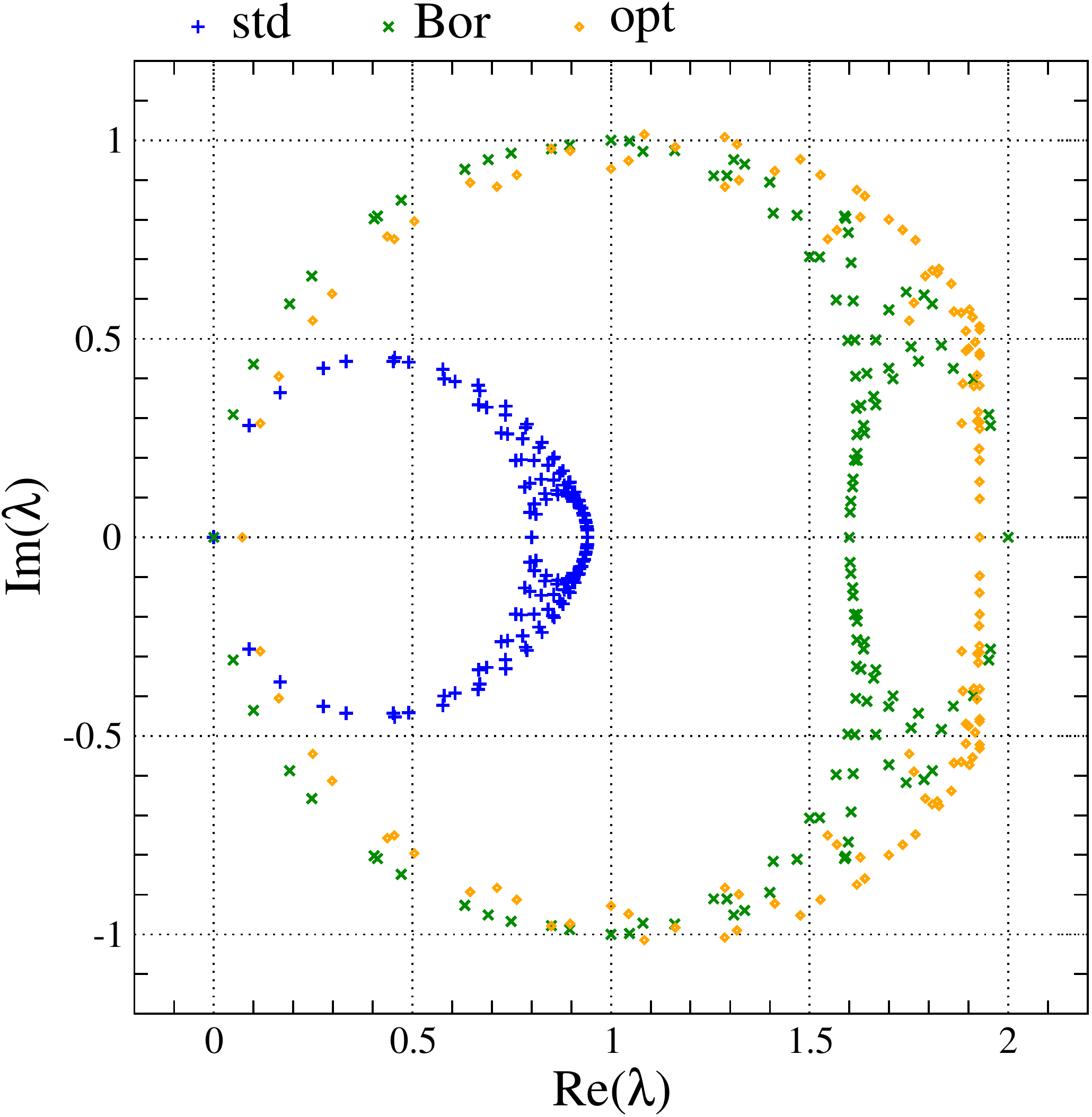}

}\hfill{}\subfloat[Staggered Wilson kernel]{\includegraphics[width=0.45\textwidth]{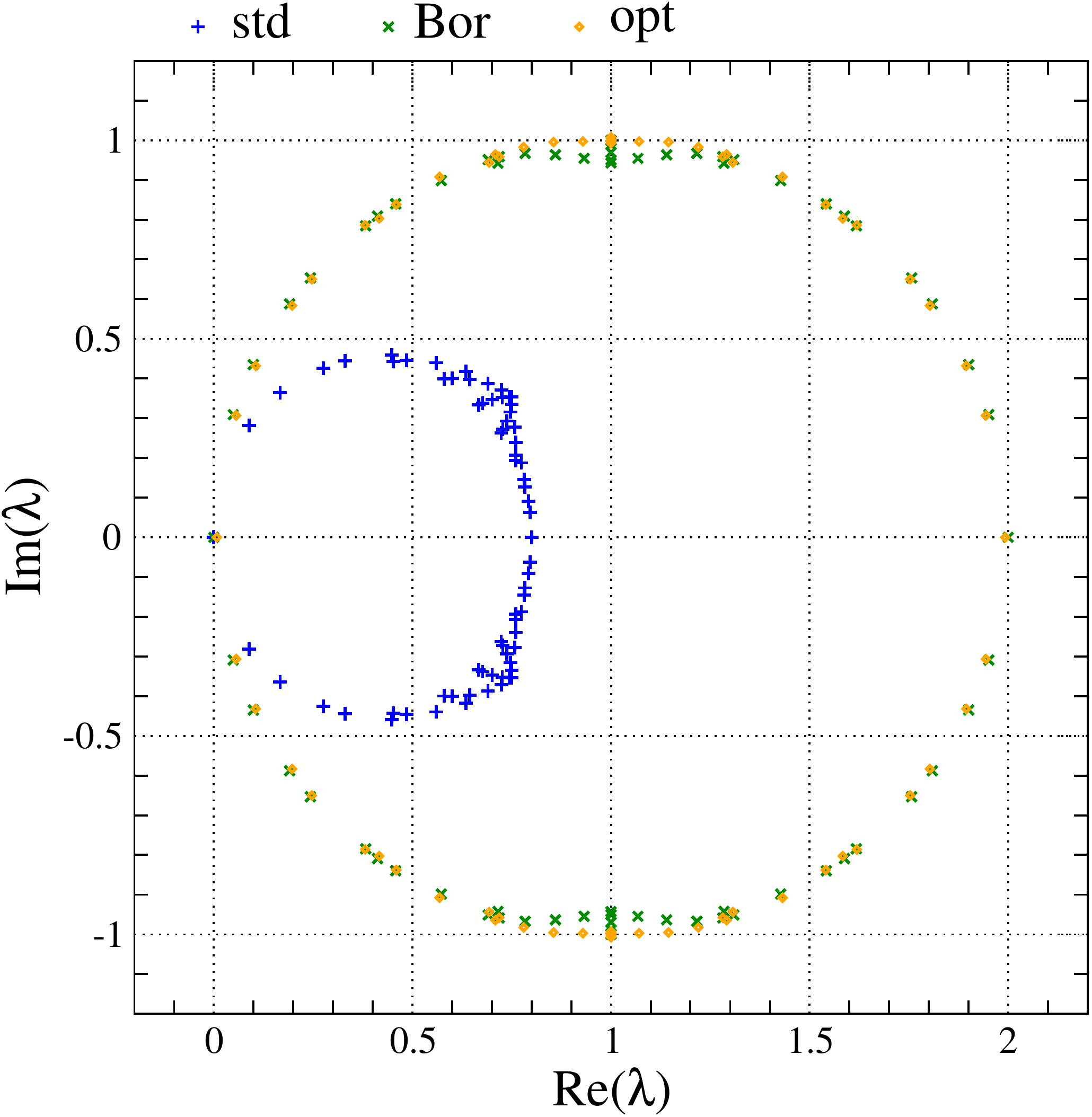}

}
\par\end{centering}
\caption{Spectrum of $\varrho D_{\mathsf{eff}}$ for the standard (std), Boriçi
(Bor) and optimal (opt) construction at $N_{s}=2$. \label{fig:free-deff-Ns2}}
\end{figure}
\begin{figure}[t]
\begin{centering}
\subfloat[Wilson kernel]{\includegraphics[width=0.45\textwidth]{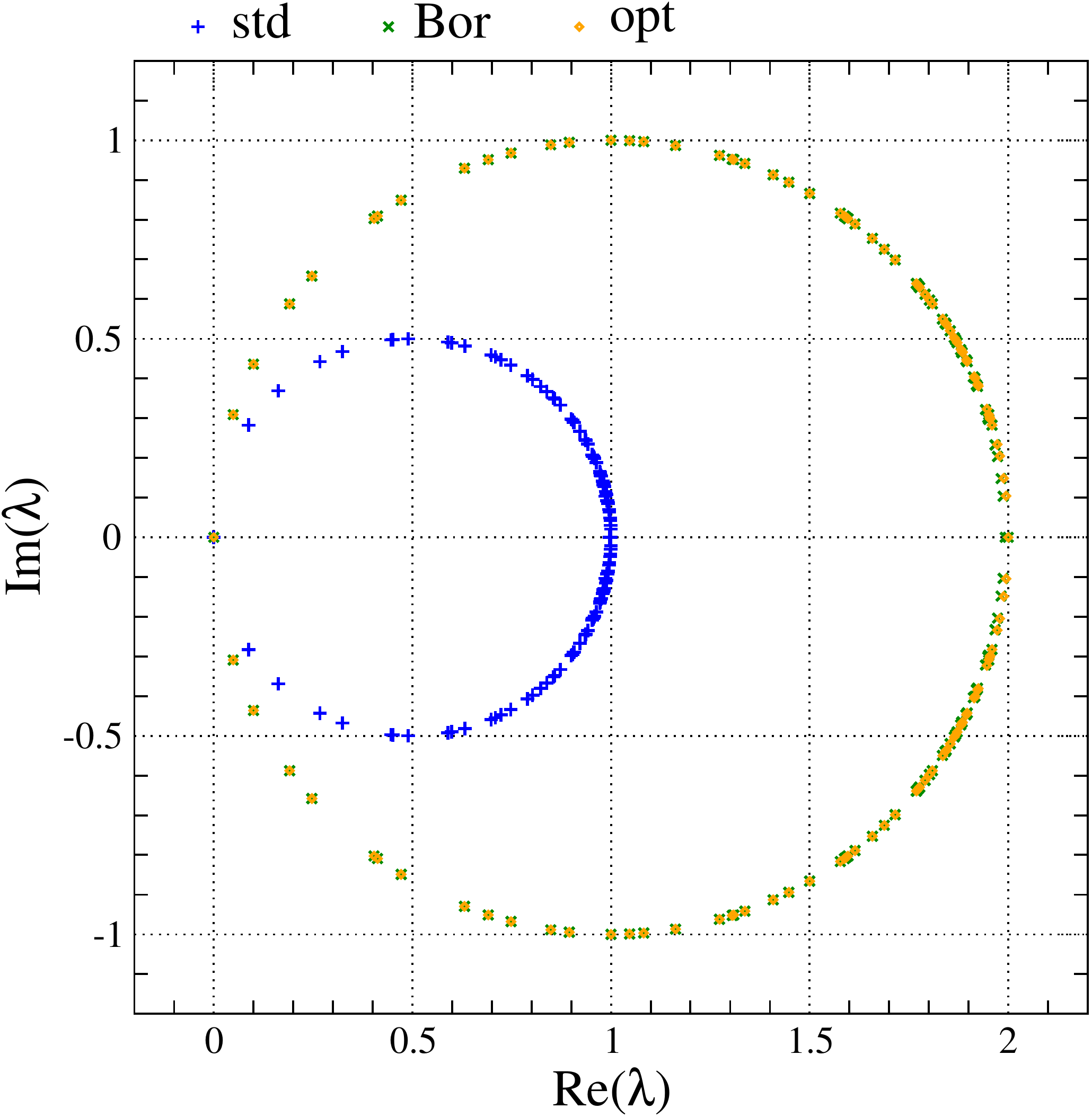}

}\hfill{}\subfloat[Staggered Wilson kernel]{\includegraphics[width=0.45\textwidth]{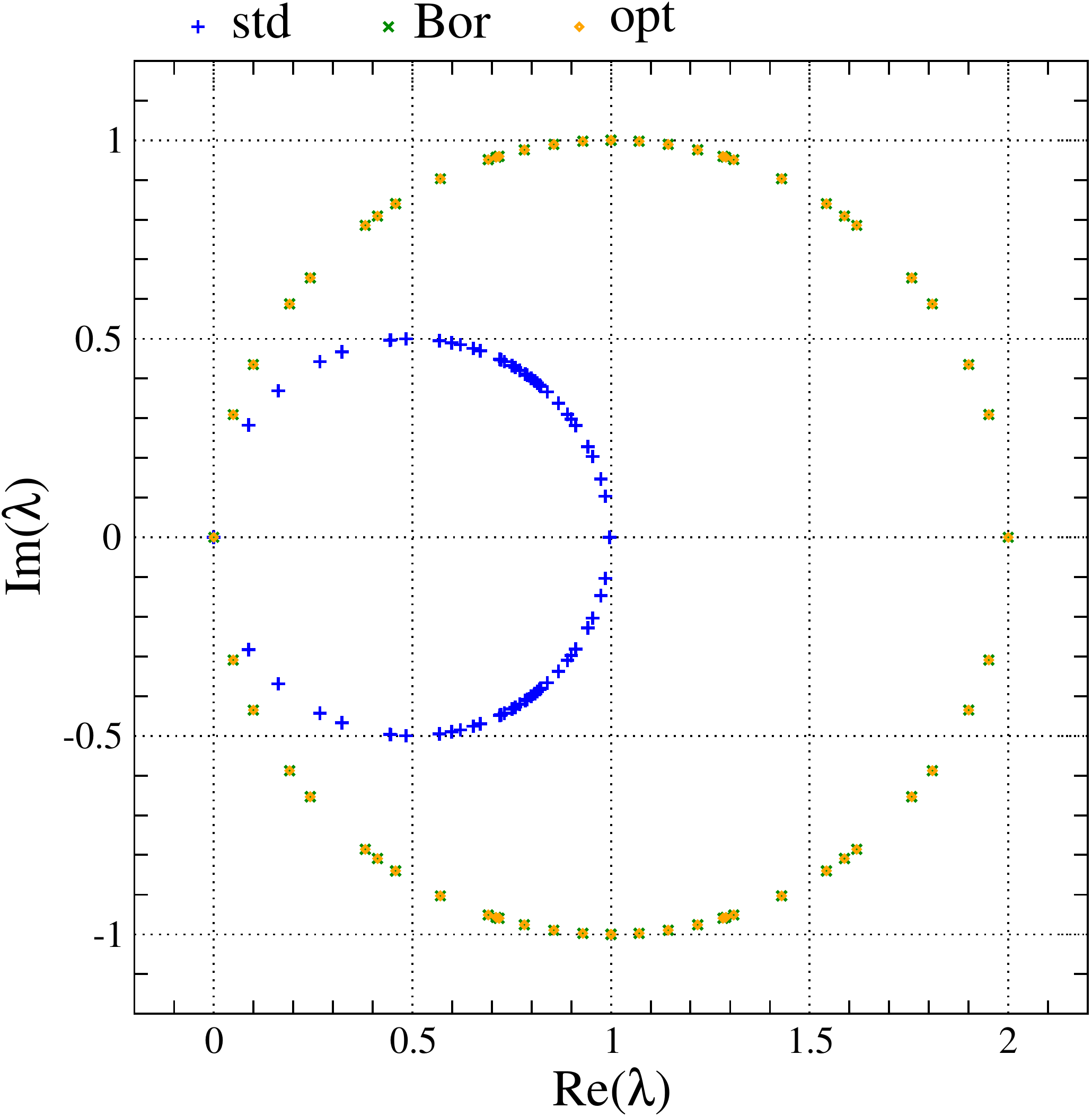}

}
\par\end{centering}
\caption{Spectrum of $\varrho D_{\mathsf{eff}}$ for the standard (std), Boriçi
(Bor) and optimal (opt) construction at $N_{s}=8$. \label{fig:free-deff-Ns8}}
\end{figure}

In Figs.~\ref{fig:free-deff-Ns2} and \ref{fig:free-deff-Ns8}, we
can find the eigenvalue spectra of the effective Dirac operators $\varrho D_{\mathsf{eff}}$
with different kernel operators as defined in Eq.~\eqref{eq:DefDeff}.
The spectra resemble the Ginsparg-Wilson circle and rapidly converge
towards it for increasing values of $N_{s}$. As the free field is
smooth, this fast convergence is not unexpected. Already for small
values of $N_{s}$ (such as $N_{s}=8$), it is almost impossible to
visually distinguish the spectrum from the one of the associated overlap
operator. We also point out the rapid convergence of Boriçi's and
Chiu's construction, in particular with a staggered Wilson kernel.

By construction, the optimal construction shows significantly improved
chiral properties. For a given spectral range $I=\left[\lambda_{\mathsf{min}},\lambda_{\mathsf{max}}\right]$
of the Hermitian kernel operator, the optimal rational function approximation
$r_{\mathsf{opt}}\left(z\right)$ minimizes the maximal deviation
from the $\sign$ function
\begin{equation}
\delta_{\mathsf{max}}=\max_{z\in-I\cup I}\left|\sign\left(z\right)-r_{\mathsf{opt}}\left(z\right)\right|=1\mp r_{\mathsf{opt}}\left(\pm\lambda_{\mathsf{min}}\right).\label{eq:DeltaMax}
\end{equation}
For a normal operator, such as in the free-field case, the eigenvalues
are confined in a tube around the Ginsparg-Wilson circle with radius
$\delta_{\mathsf{max}}$. As Zolotarev's optimal approximation to
the $\sign$-function minimizes the $\left\Vert \cdot\right\Vert _{\infty}$-norm,
we find a point of maximal deviation at both $\lambda_{\mathsf{min}}$
and $\lambda_{\mathsf{max}}$, resulting in the absence of an exact
zero mode. However, this approximate zero mode is decreasing rapidly
in magnitude with increasing values of $N_{s}$ due to the rapid convergence
of the underlying $\sign$-function approximation.

\paragraph{Overlap operators.}

\begin{figure}[t]
\begin{centering}
\subfloat[Wilson kernel]{\includegraphics[width=0.45\textwidth]{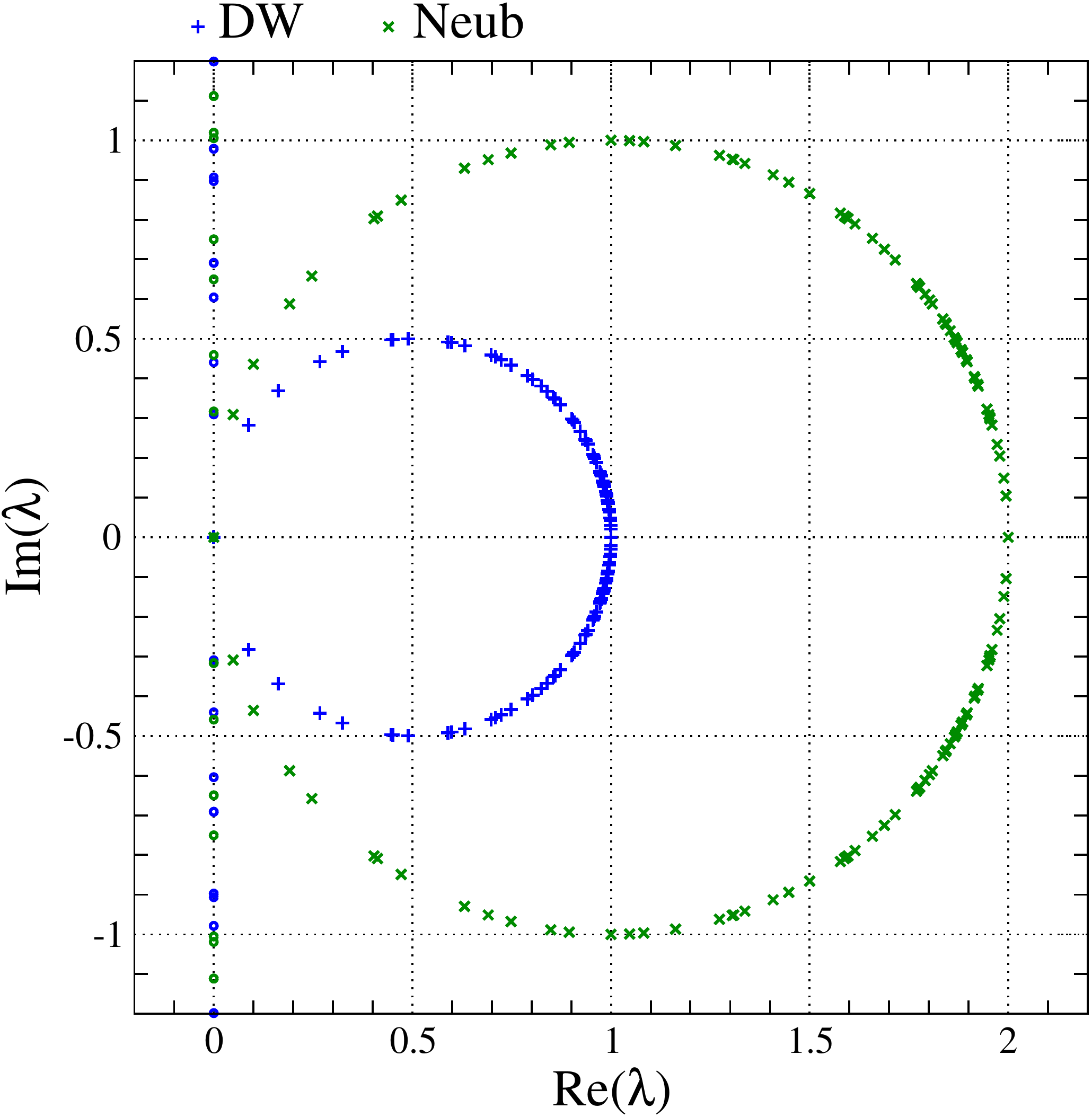}

}\hfill{}\subfloat[Staggered Wilson kernel]{\includegraphics[width=0.45\textwidth]{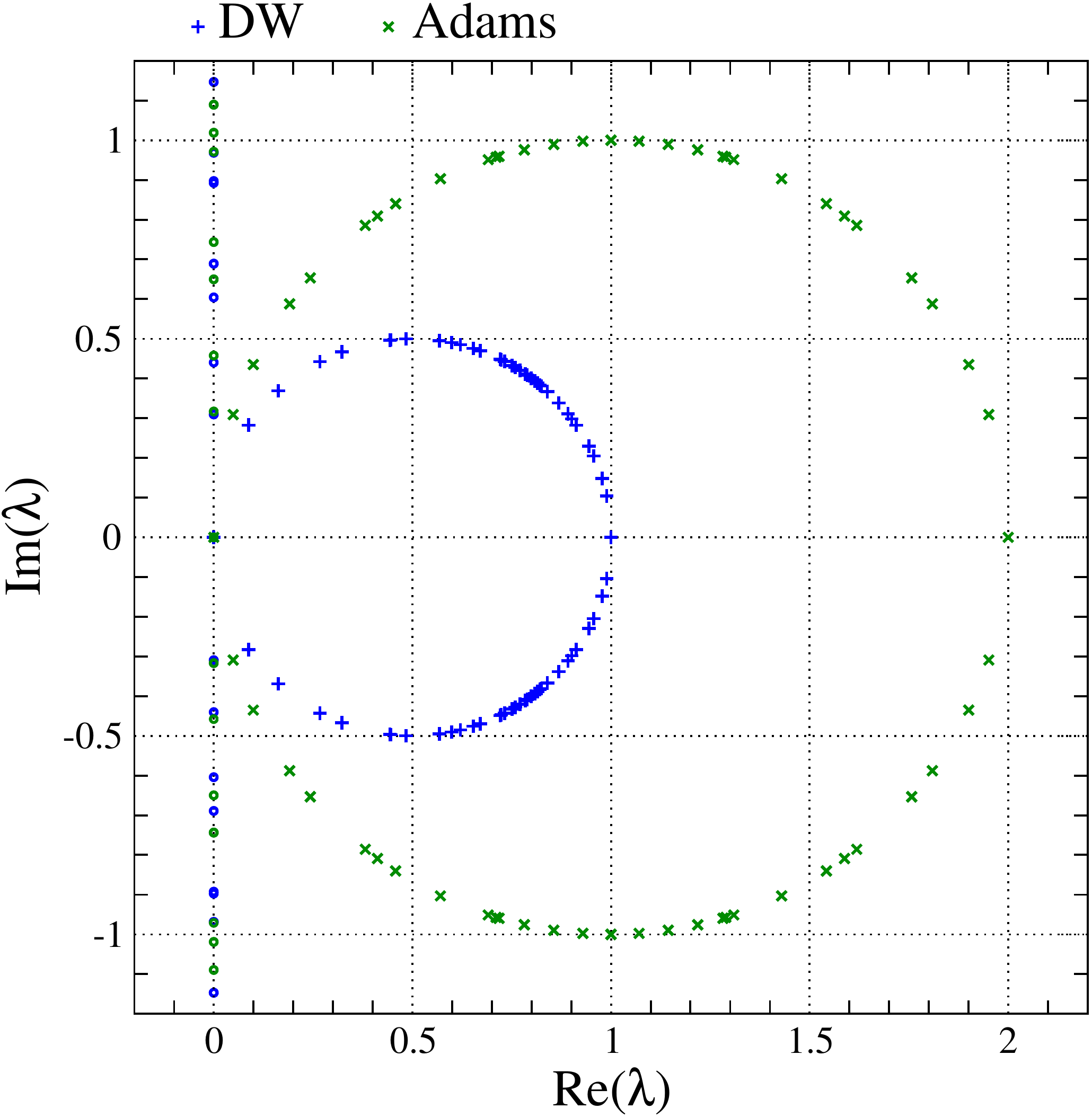}

}
\par\end{centering}
\caption{Spectrum of $\varrho D_{\mathsf{ov}}$ with its stereographic projection
for the domain wall (DW) kernel and the standard overlap (Neub/Adams)
kernel. \label{fig:free-dov} }
\end{figure}

Taking the $N_{s}\to\infty$ limit of the effective Dirac operators,
we find an overlap operator. The Hermitian kernel operator $H$ of
the overlap operator as given in Eq.~\eqref{eq:DefH} depends on
the choice of the domain wall construction. In Fig.~\ref{fig:free-dov},
we show the eigenvalue spectrum of $\varrho D_{\mathsf{ov}}$ together
with the stereographic projection $\pi$ of the eigenvalues, given
by the map
\begin{equation}
\pi\left(\lambda\right)=\frac{\lambda}{1-\frac{1}{\varrho}\lambda}.
\end{equation}
The result is the projection of the eigenvalues onto the imaginary
axis, where the physical low-lying eigenvalues line up as required
due to the introduction of the scale factor in Eq.~\eqref{eq:Defrho}.

For Adams' overlap operator the eigenvalue spectrum is highly symmetric.
By construction, the effective operators in Boriçi's and the optimal
construction converge, depending on the kernel operator, towards Neuberger's
and Adams' overlap operator for $N_{s}\to\infty$. For the standard
construction we find an overlap operator with a modified kernel operator.

\subsection{$\protect\Uone$ gauge field case}

After the discussion of the free-field case, we now consider nontrivial
gauge configurations. The Schwinger model is an Abelian gauge theory
with symmetry group $\Uone$ and in the following we consider quenched
thermalized gauge configurations with the setup of Refs.~\citep{Durr:2003xs,Durr:2004ta}.
Although we restrict ourselves to quenched configurations, it is in
principle possible to reweight to an unquenched ensemble with arbitrary
mass \citep{Giusti:2001cn,Durr:2003xs,Durr:2004ta,Durr:2006ze}. We
begin our discussion with the analysis of the eigenvalue spectra on
a few selected $20^{2}$ configurations at an inverse coupling of
$\beta=5$.

\paragraph{Kernel operators.}

\begin{figure}[t]
\begin{centering}
\includegraphics[width=0.75\textwidth]{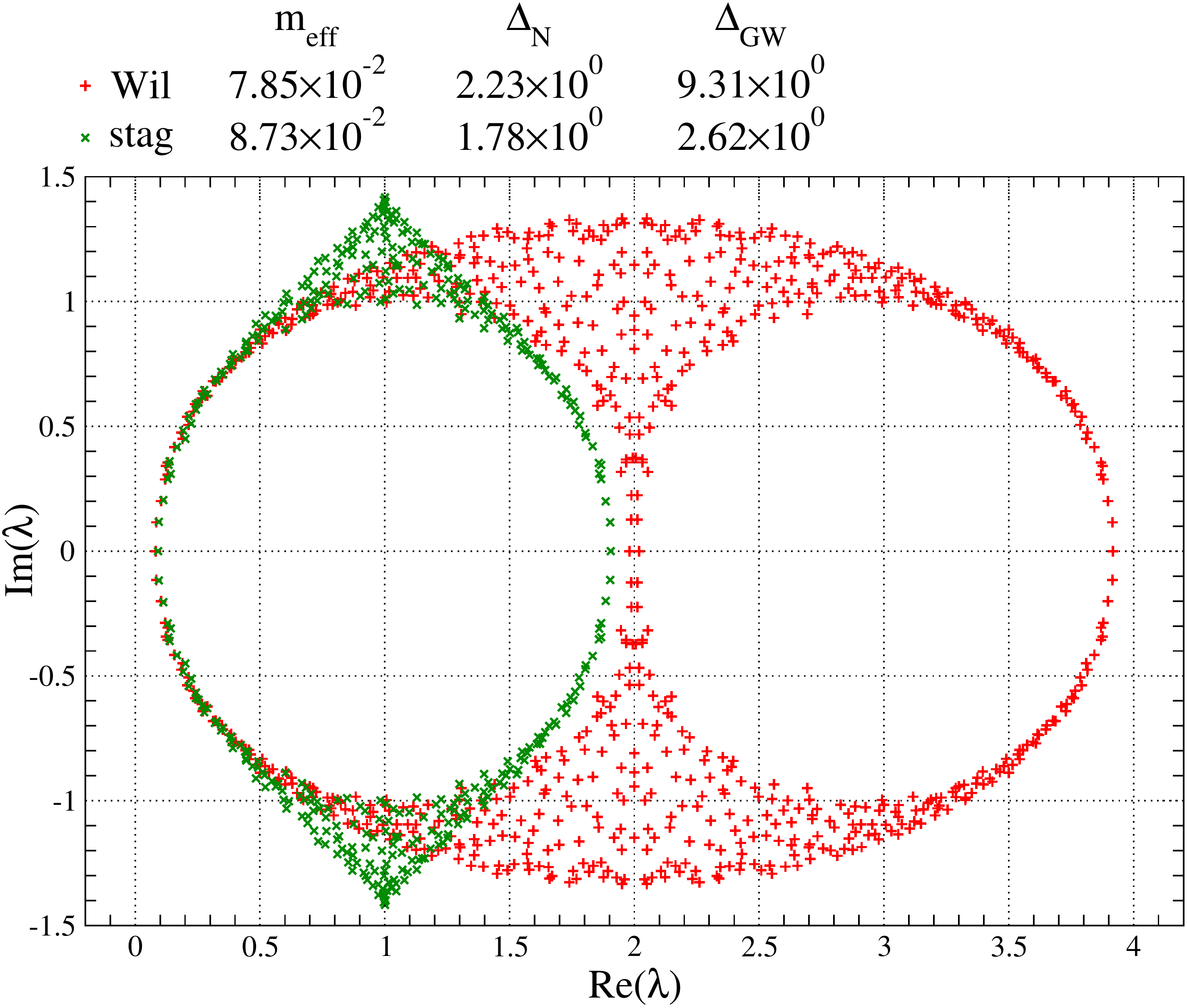} 
\par\end{centering}
\caption{Spectrum of kernel operators in a $\protect\Uone$ background field.
\label{fig:gauge-kernel}}
\end{figure}

An example of a kernel eigenvalue spectrum on a gauge configuration
with topological charge $Q=1$ can be found in Fig.~\ref{fig:gauge-kernel}.
Compared to the case of four-dimensional quantum chromodynamics, the
branches are much sharper and better separated, cf.\ Chapter\ \ref{chap:Eigenvalue-spectra}
and Refs.~\citep{deForcrand:2011ak,deForcrand:2012bm,Durr:2013gp,Adams:2013tya,ZielinskiLat14,ZielinskiICCP9}.

\paragraph{Bulk operators.}

\begin{figure}[t]
\begin{centering}
\subfloat[Standard construction]{\includegraphics[width=0.32\textwidth]{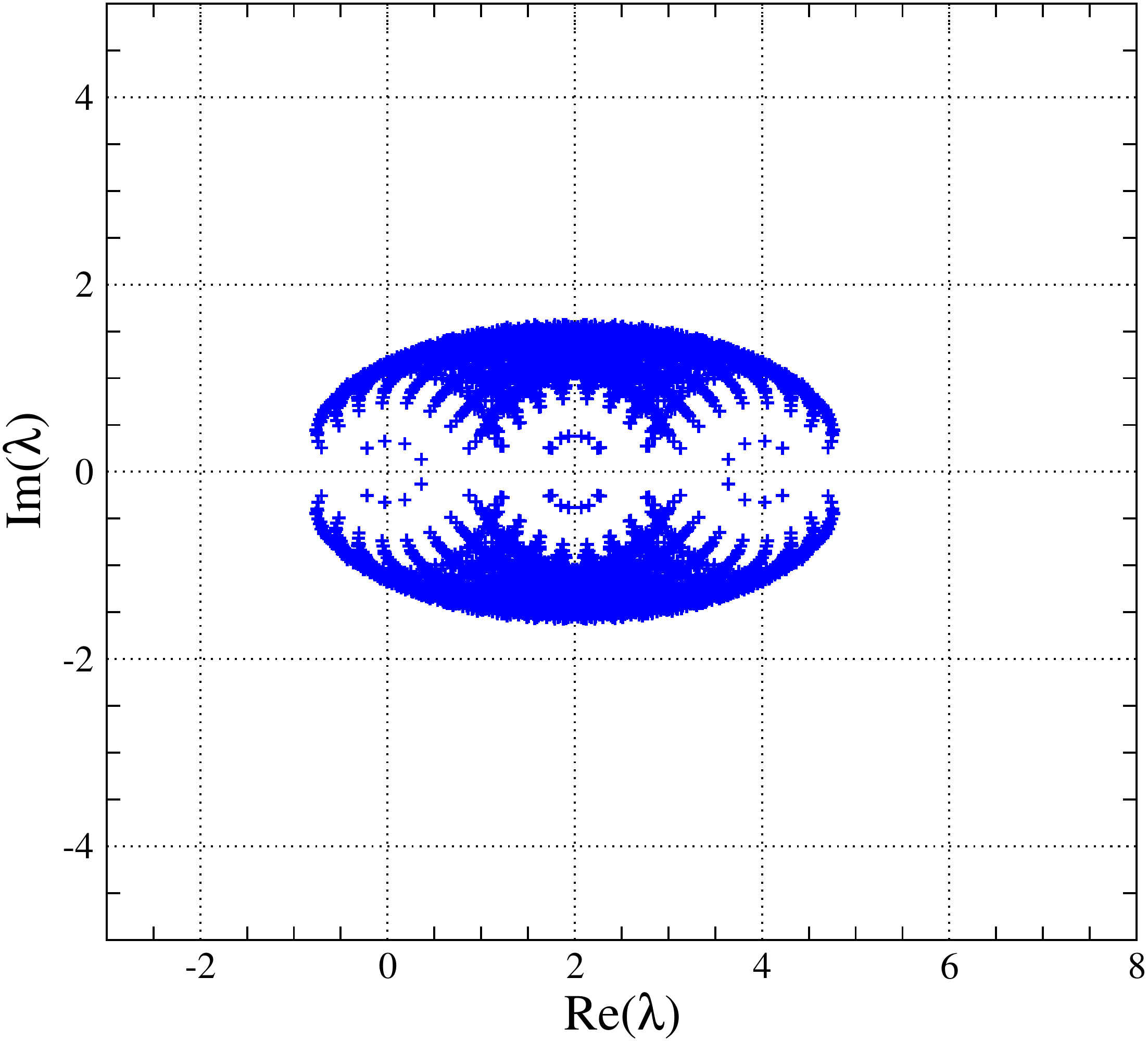}

}\hfill{}\subfloat[Boriçi's construction]{\includegraphics[width=0.32\textwidth]{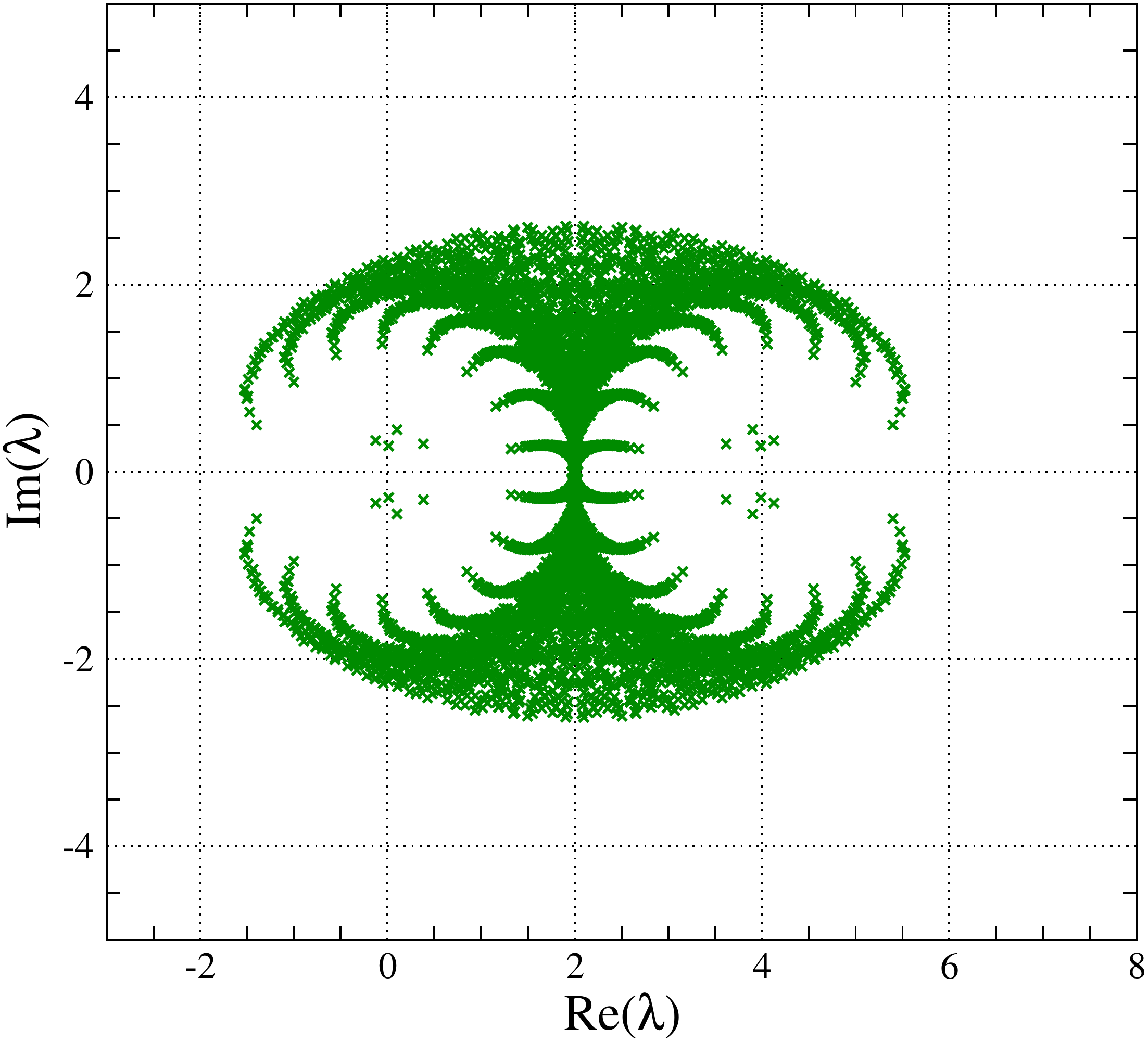}

}\hfill{}\subfloat[Optimal construction]{\includegraphics[width=0.32\textwidth]{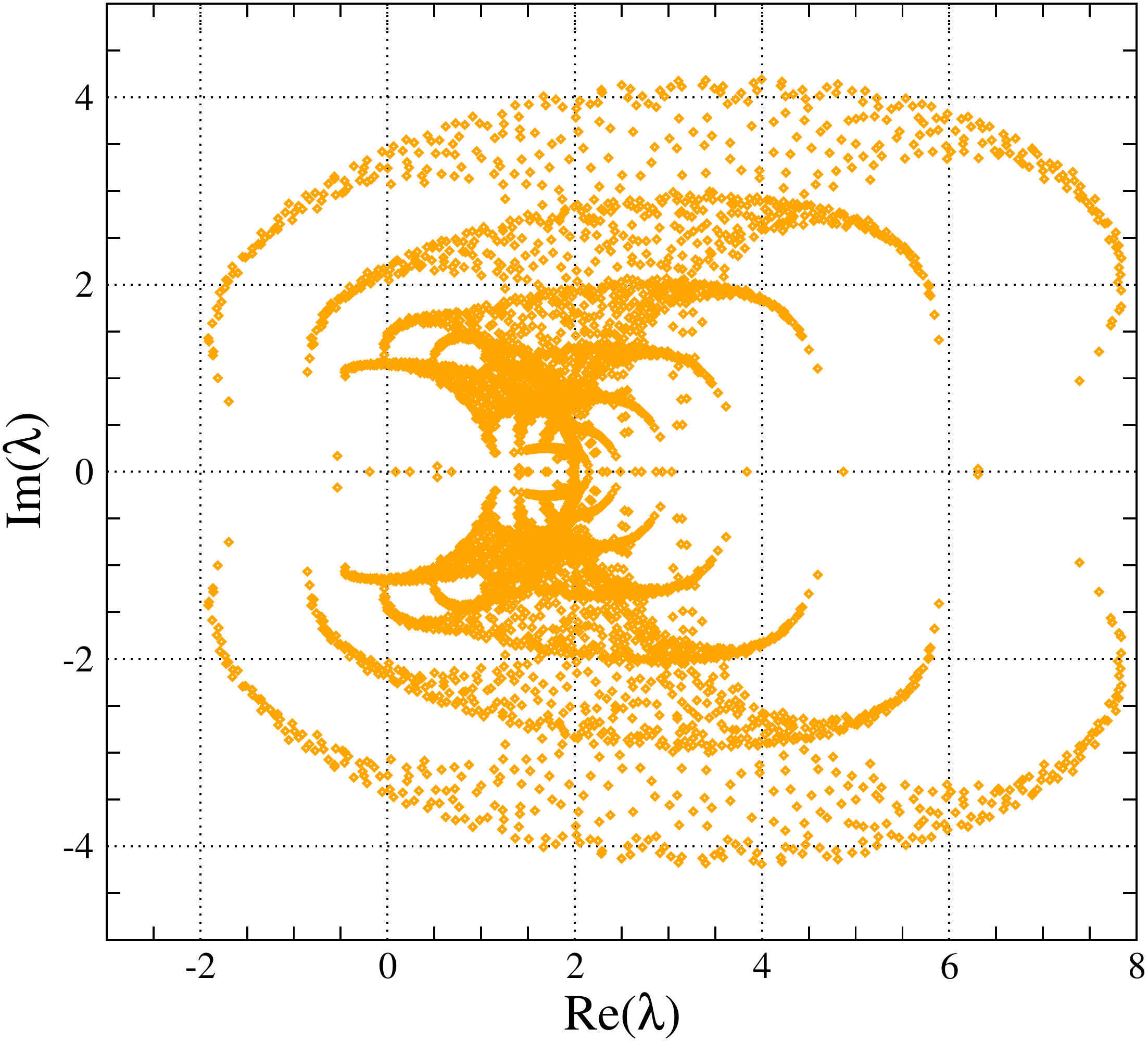}

}
\par\end{centering}
\caption{Spectrum of $D_{\mathsf{dw}}$ with Wilson kernel for $m=0$ at $N_{s}=8$.
\label{fig:gauge-wilson-dbulk}}
\end{figure}
\begin{figure}[t]
\begin{centering}
\subfloat[Standard construction]{\includegraphics[width=0.32\textwidth]{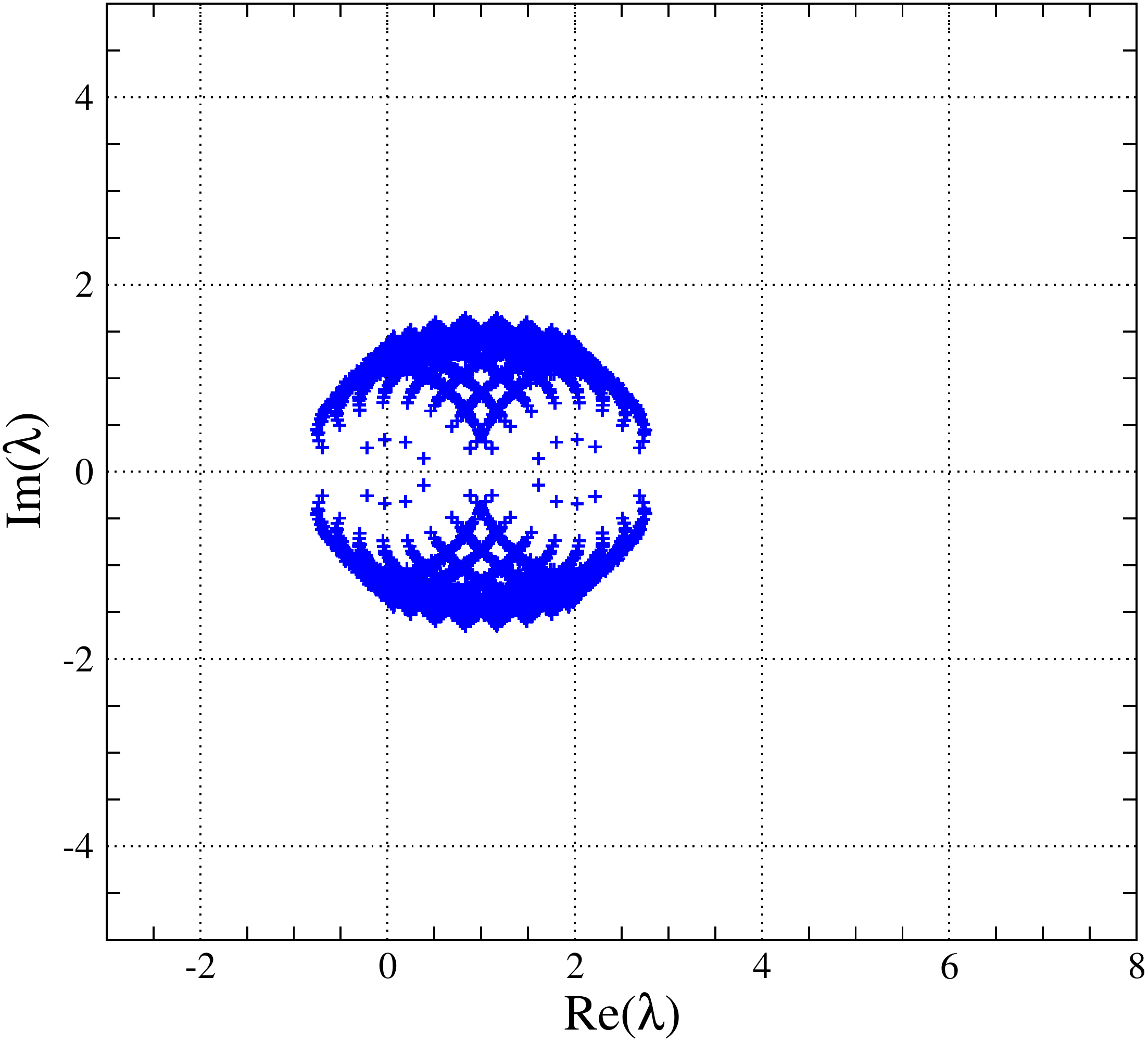}

}\hfill{}\subfloat[Boriçi's construction]{\includegraphics[width=0.32\textwidth]{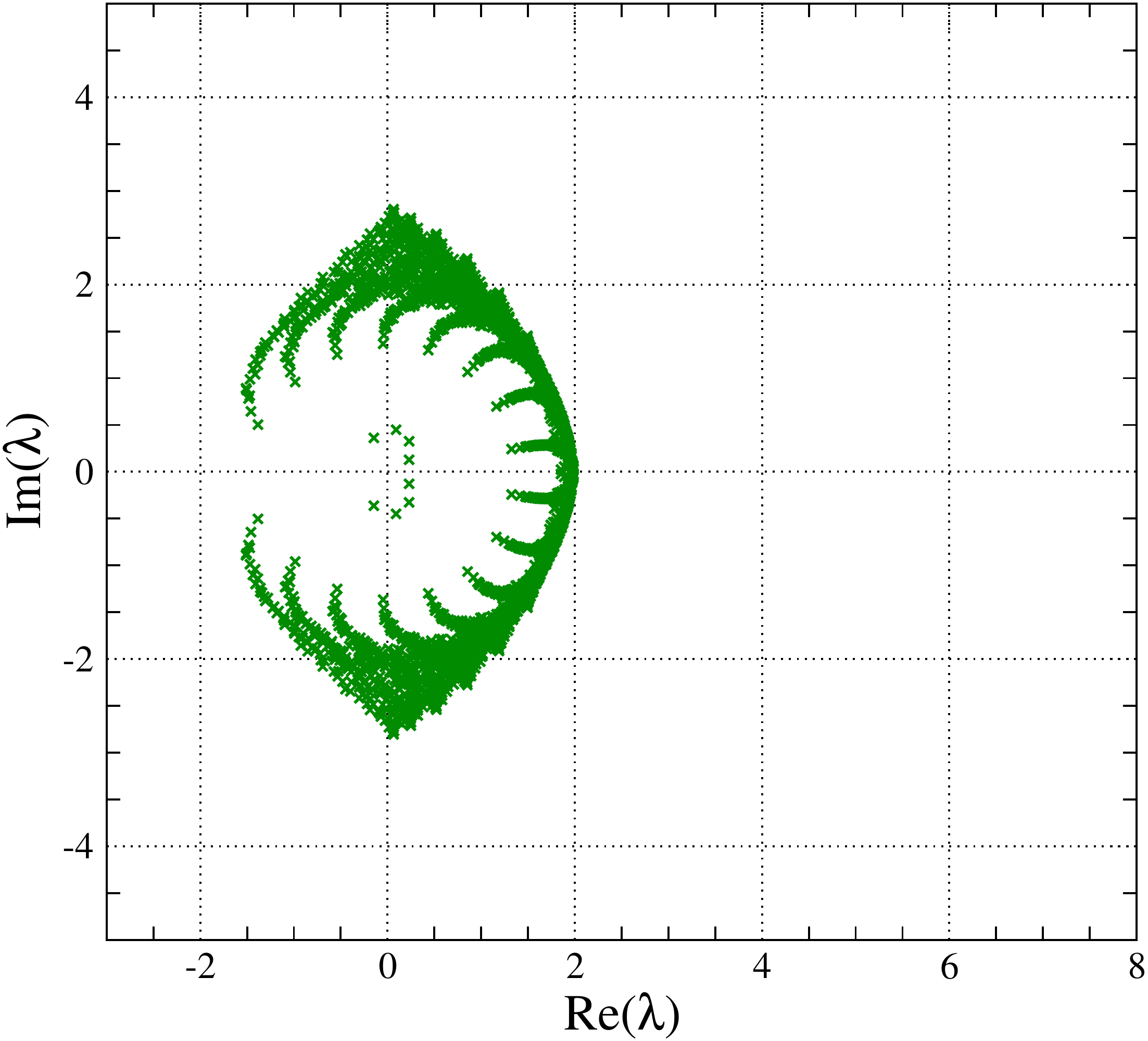}

}\hfill{}\subfloat[Optimal construction]{\includegraphics[width=0.32\textwidth]{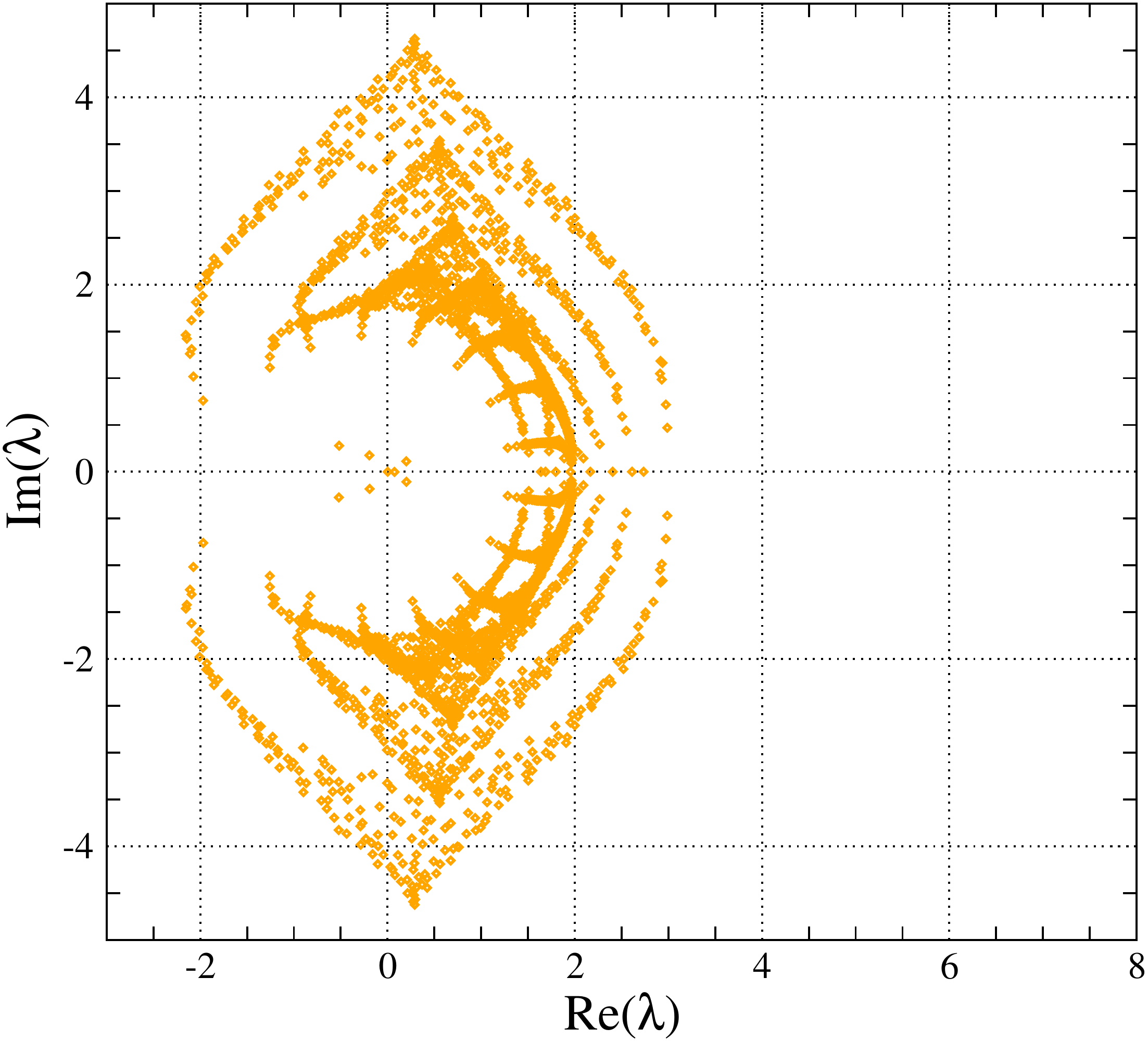}

}
\par\end{centering}
\caption{Spectrum of $D_{\mathsf{sdw}}$ with staggered Wilson kernel for $m=0$
at $N_{s}=8$. \label{fig:gauge-stw-dbulk}}
\end{figure}

For the eigenvalue spectra of the bulk operators in Figs.~\ref{fig:gauge-wilson-dbulk}
and \ref{fig:gauge-stw-dbulk} we use the same gauge background field
as in Fig.~\ref{fig:gauge-kernel}. Being a topologically nontrivial
configuration, the effective operator has $\left|Q\right|$ approximate
zero modes which become exact in the limit $N_{s}\to\infty$. For
the bulk operators, there are $N_{s}\cdot\left|Q\right|$ eigenvalues
in the neighborhood of the origin. It is an interesting observation
to see how the eigenvalue spectrum is distorted in the optimal construction
in order to improve chiral properties.

\paragraph{Effective operators.}

\begin{figure}[t]
\begin{centering}
\subfloat[Wilson kernel]{\includegraphics[width=0.45\textwidth]{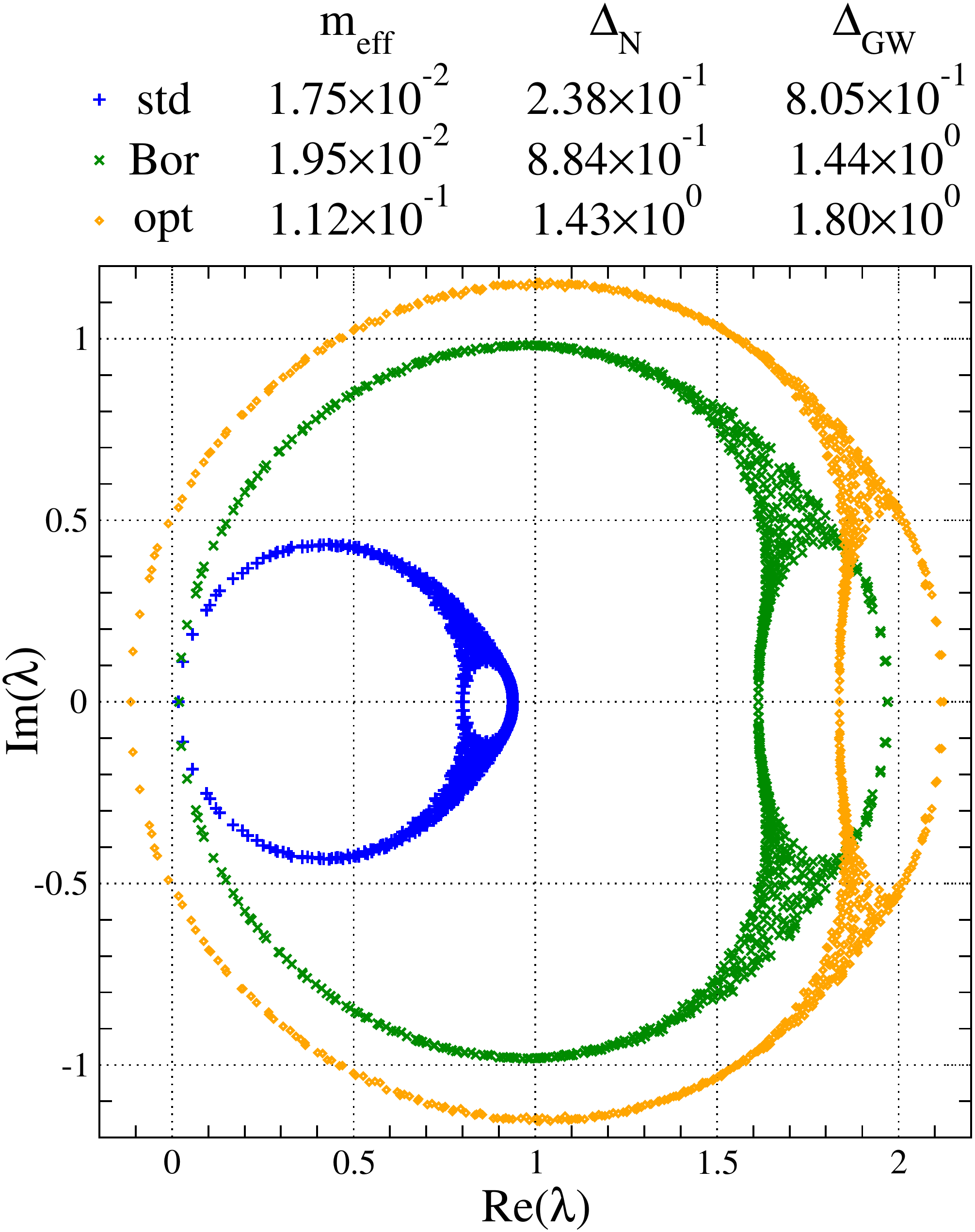}

}\hfill{}\subfloat[Staggered Wilson kernel]{\includegraphics[width=0.45\textwidth]{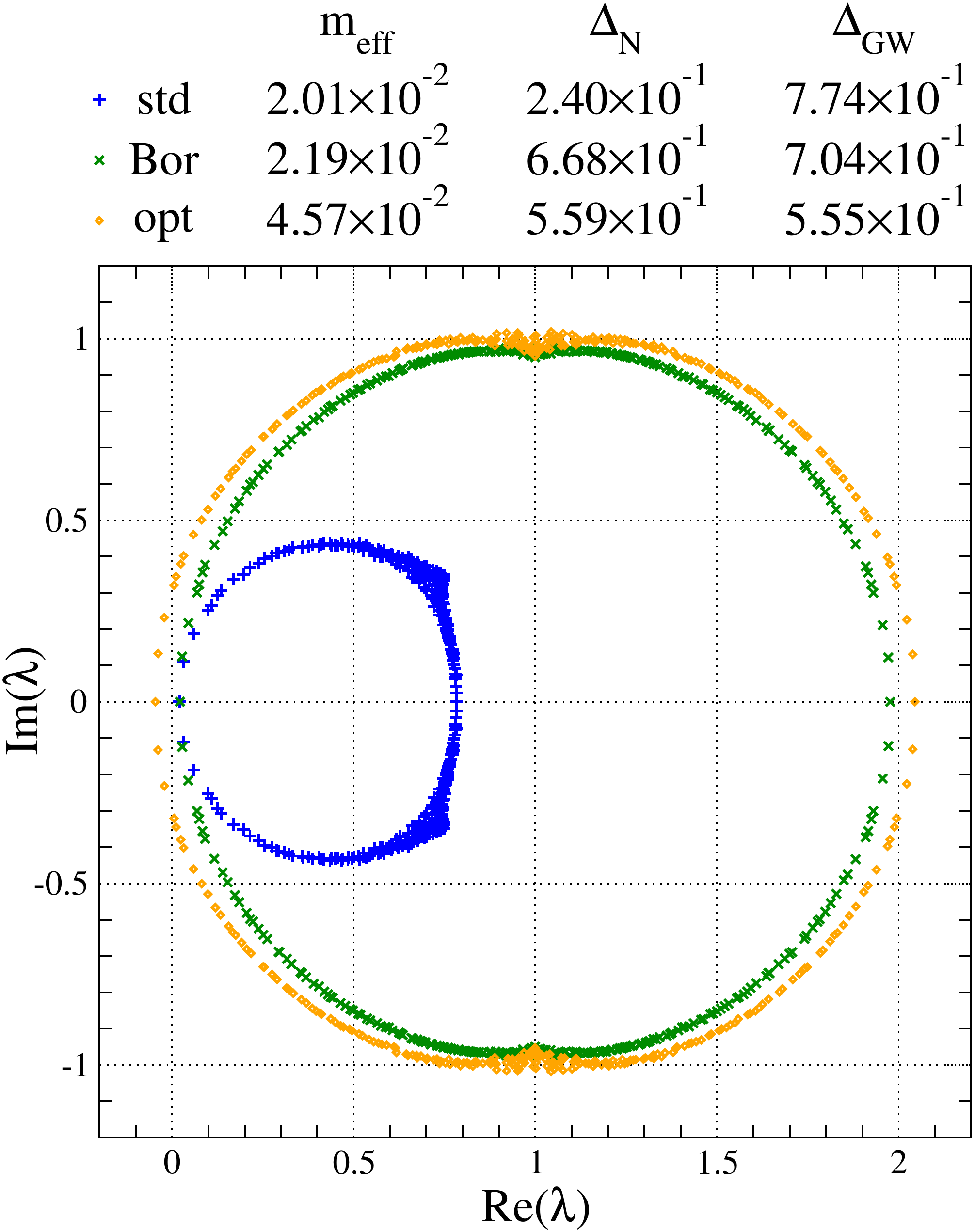}

}
\par\end{centering}
\caption{Spectrum of $\varrho D_{\mathsf{eff}}$ for the standard (std), Boriçi
(Bor) and optimal (opt) construction at $N_{s}=2$. \label{fig:gauge-deff-Ns2}}
\end{figure}
\begin{figure}[t]
\begin{centering}
\subfloat[Wilson kernel]{\includegraphics[width=0.45\textwidth]{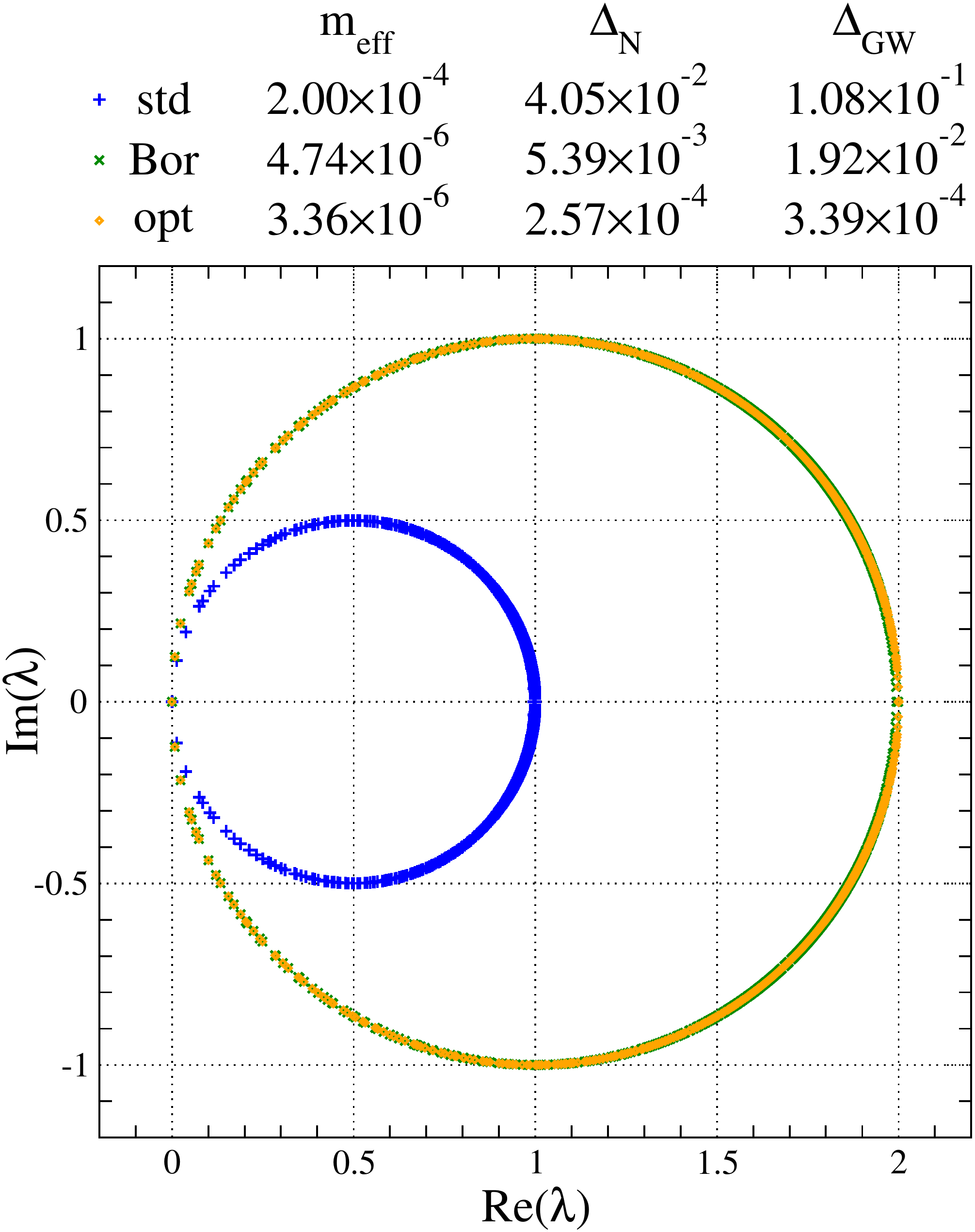}

}\hfill{}\subfloat[Staggered Wilson kernel \label{fig:gauge-deff-Ns8-b}]{\includegraphics[width=0.45\textwidth]{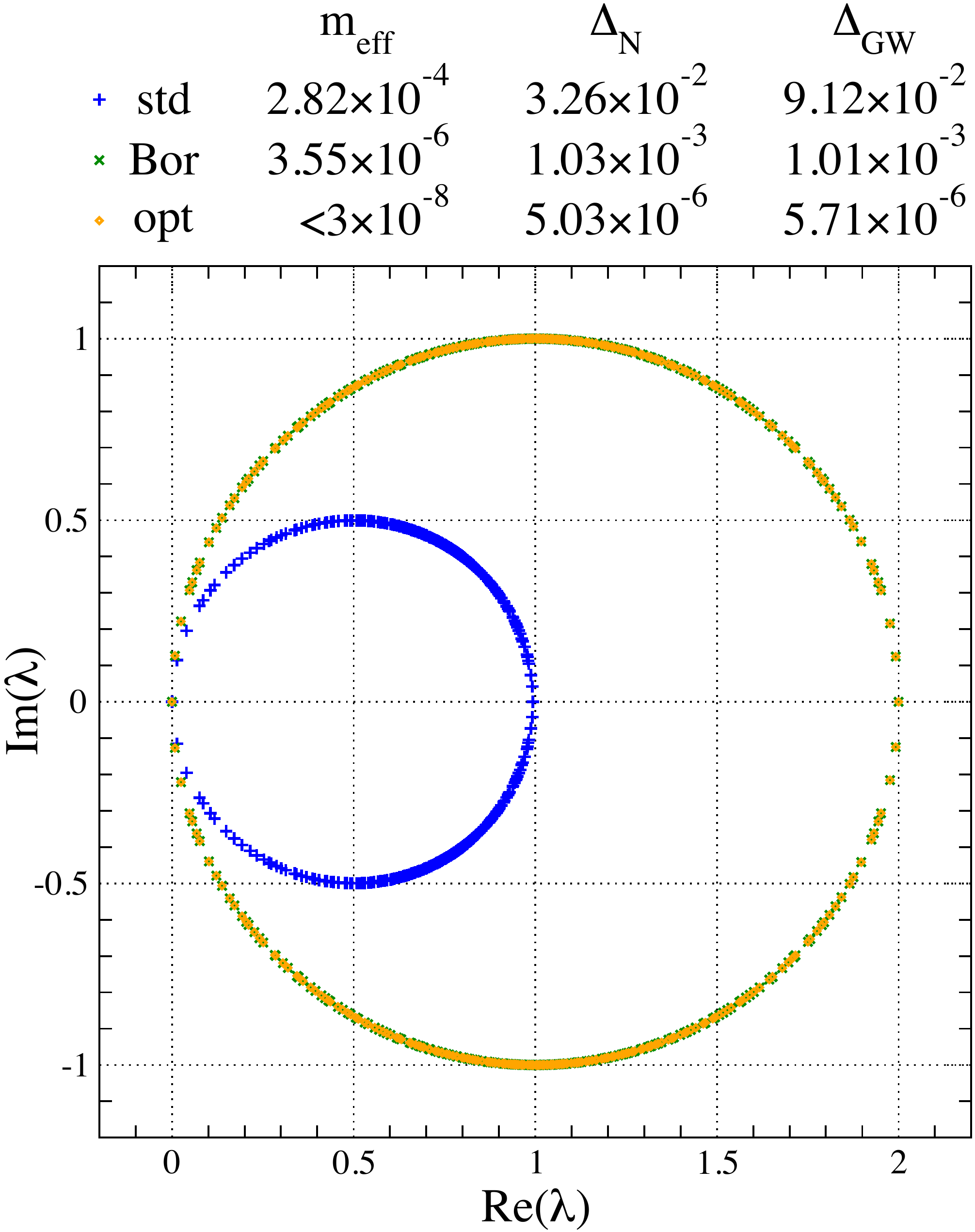}

}
\par\end{centering}
\caption{Spectrum of $\varrho D_{\mathsf{eff}}$ for the standard (std), Boriçi
(Bor) and optimal (opt) construction at $N_{s}=8$. \label{fig:gauge-deff-Ns8}}
\end{figure}

Again using the same gauge configuration, in Figs.~\ref{fig:gauge-deff-Ns2}
and \ref{fig:gauge-deff-Ns8} we find the eigenvalue spectra of the
effective operators. Here Boriçi's construction shows superior chiral
properties compared to the standard construction for $N_{s}\geq4$
with respect to our measures $m_{\mathsf{eff}}$, $\Delta_{\mathsf{N}}$
and $\Delta_{\mathsf{GW}}$.

The optimal construction further improves upon that and shows an ever
lower degree of chiral symmetry violations. In the case of Fig.~\ref{fig:gauge-deff-Ns8-b},
the effective mass $m_{\mathsf{eff}}$ is already so small that it
is in the order of the round-off error and, thus, we are only able
to quote an upper bound. In this nontrivial background field the kernel
operators lose their normality and the maximum deviation $\delta_{\mathsf{max}}$
of eigenvalues from the Ginsparg-Wilson circle in the optimal construction
does not have to be saturated. We also note that a smaller $\delta_{\mathsf{max}}$
does not necessarily imply a smaller value of $\Delta_{\mathsf{GW}}$,
but that also the distribution of eigenvalues of $H$ is important.
There is, however, a strong correlation between larger $N_{s}$ and
smaller values of $\Delta_{\mathsf{GW}}$, where the violations of
the Ginsparg-Wilson relation rapidly tend to zero for $N_{s}\to\infty$.

Chiu's construction is only in a very particular sense optimal, namely
the minimization of the deviation $\delta_{\mathsf{max}}$ from the
$\sign$-function as defined in Eq.~\eqref{eq:DeltaMax}. With respect
to other criteria, such as the number of iterations required for solving
a linear system, optimal domain wall fermions are typically not optimal
\citep{Brower:2005qw}. In theory, it is possible to formulate variants
of domain wall fermions which are optimal with respect to other measures
such as $\Delta_{\mathsf{GW}}$. However, such a construction would
likely require more knowledge about the eigenvalue spectrum of $H$.

For the standard construction $m_{\mathsf{eff}}$, $\Delta_{\mathsf{N}}$
and $\Delta_{\mathsf{GW}}$ have typically the same order of magnitude
for both a Wilson and a staggered Wilson kernel. On the other hand,
for Boriçi's and the optimal construction a staggered Wilson kernel
results in reduced chiral symmetry violations compared to the Wilson
kernel.

An interesting, although rather artificial case is $N_{s}=2$. Here
we observe that the performance of all constructions is comparable
in the case of a staggered Wilson kernel, while for the case of a
Wilson kernel the standard construction performs better than Boriçi's,
which itself clearly outperforms the optimal construction.

\paragraph{Overlap operators.}

\begin{figure}[t]
\begin{centering}
\subfloat[Wilson kernel]{\includegraphics[width=0.45\textwidth]{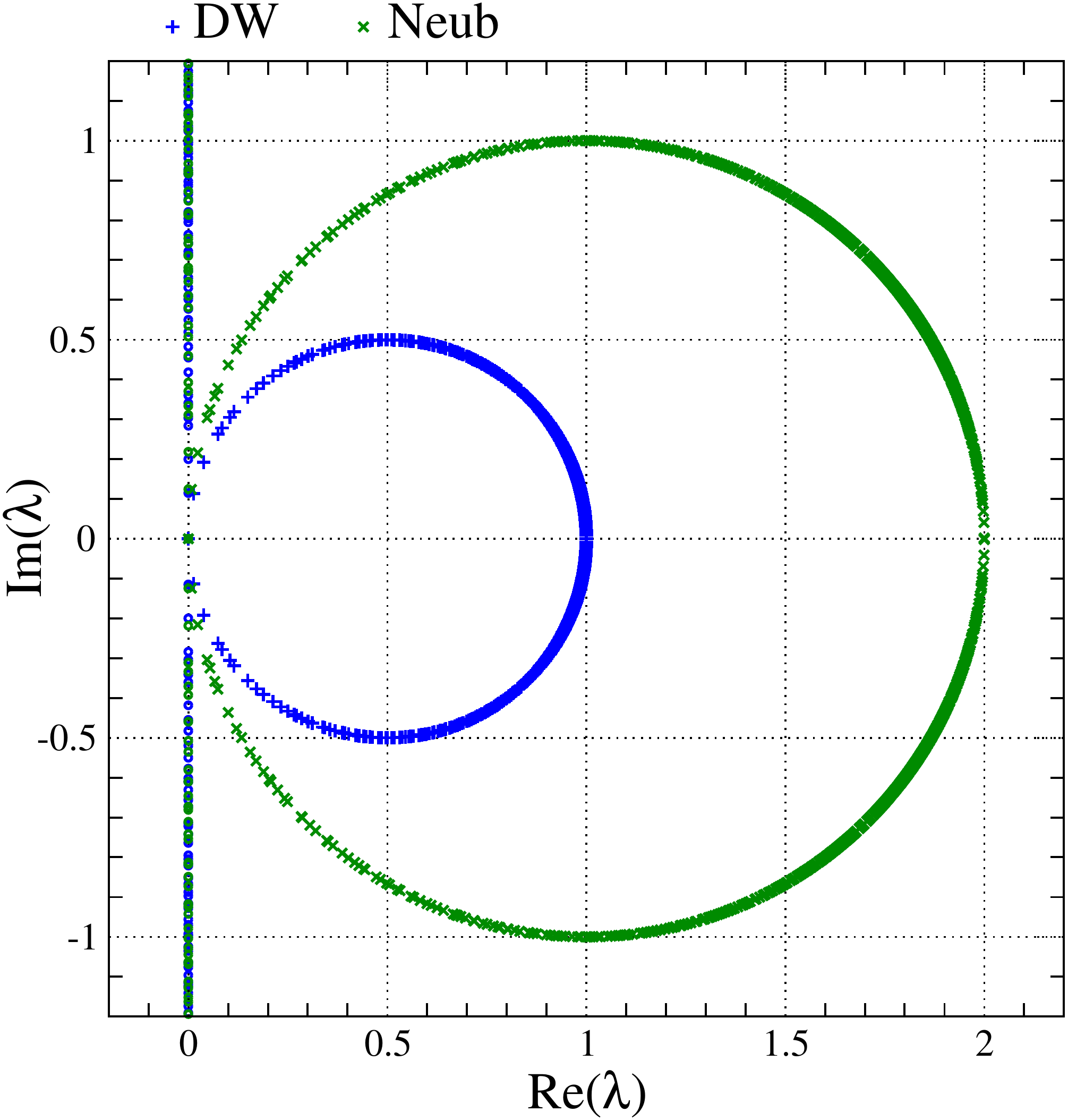}

}\hfill{}\subfloat[Staggered Wilson kernel]{\includegraphics[width=0.45\textwidth]{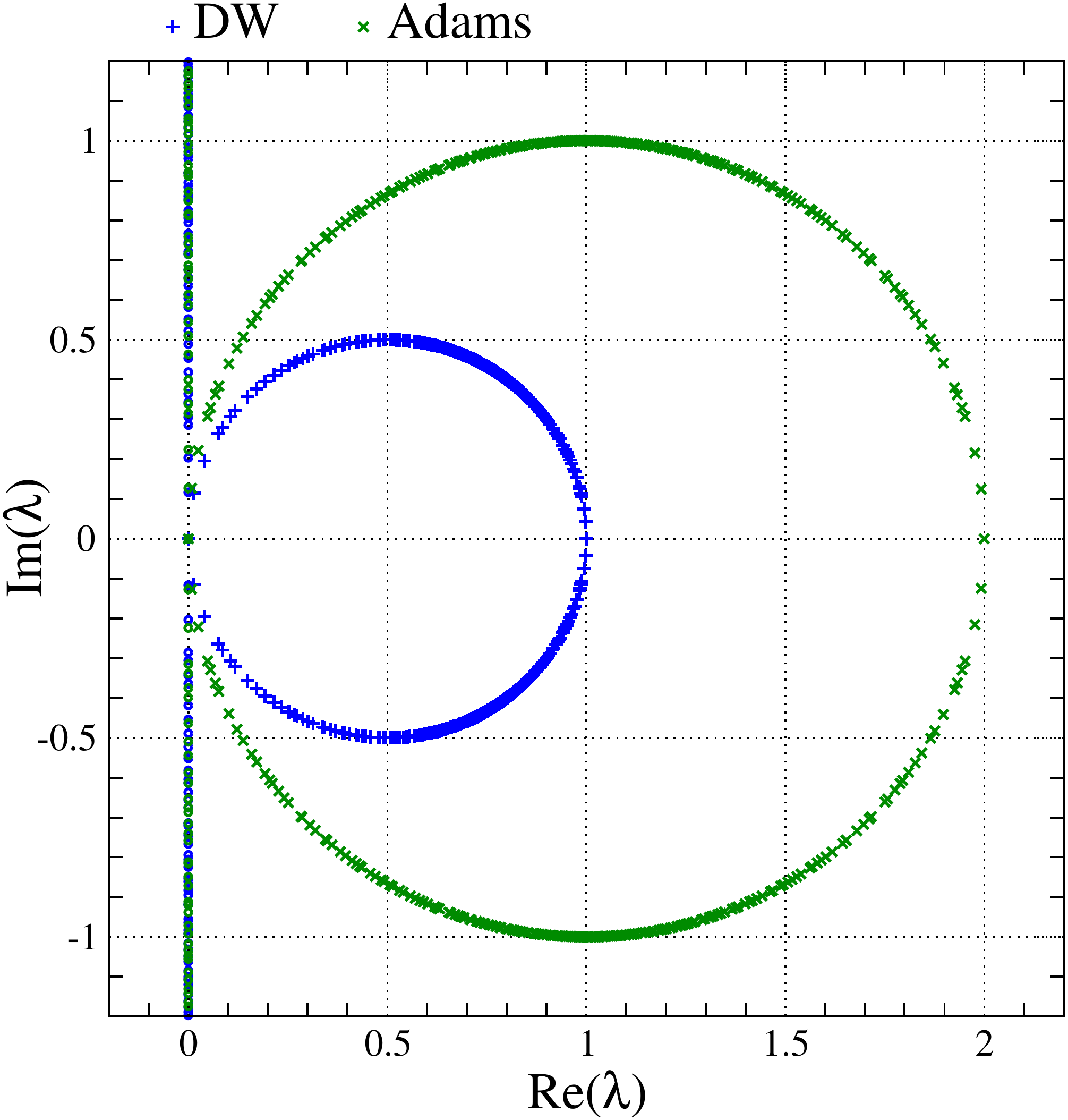}

}
\par\end{centering}
\caption{Spectrum of $\varrho D_{\mathsf{ov}}$ with its stereographic projection
for the domain wall (DW) kernel and the standard overlap (Neub/Adams)
kernel. \label{fig:gauge-dov}}
\end{figure}

In Fig.~\ref{fig:gauge-dov}, we again show the spectrum of the associated
overlap operators together with a stereographic projection of their
eigenvalues. We omit the chirality measures $m_{\mathsf{eff}}$, $\Delta_{\mathsf{N}}$
and $\Delta_{\mathsf{GW}}$ in the figure labels as they vanish up
to round-off errors.

\subsection{Smearing}

\begin{figure}[t]
\begin{centering}
\subfloat[No smearing]{\includegraphics[width=0.45\textwidth]{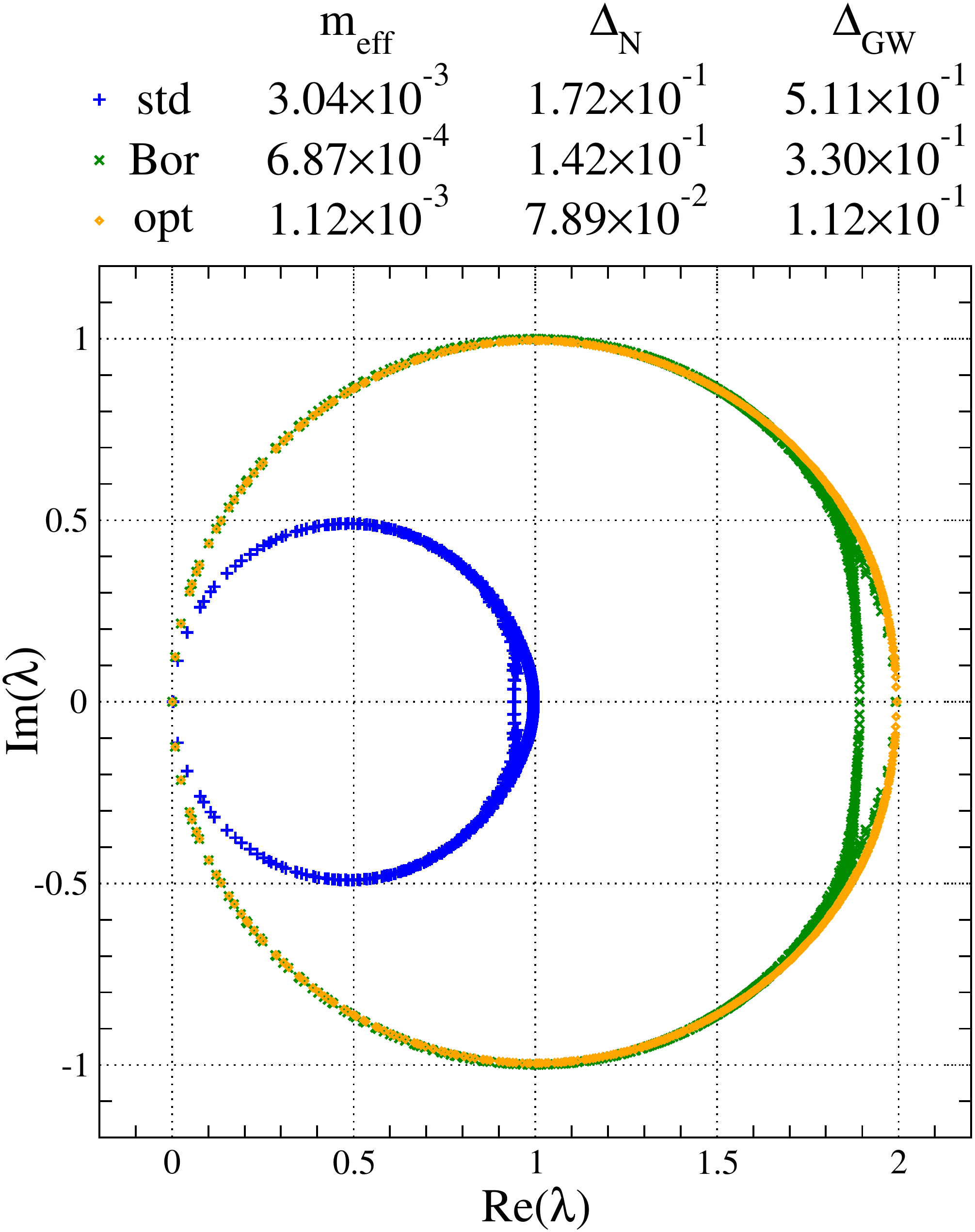}

}\hfill{}\subfloat[Three smearing iterations]{\includegraphics[width=0.45\textwidth]{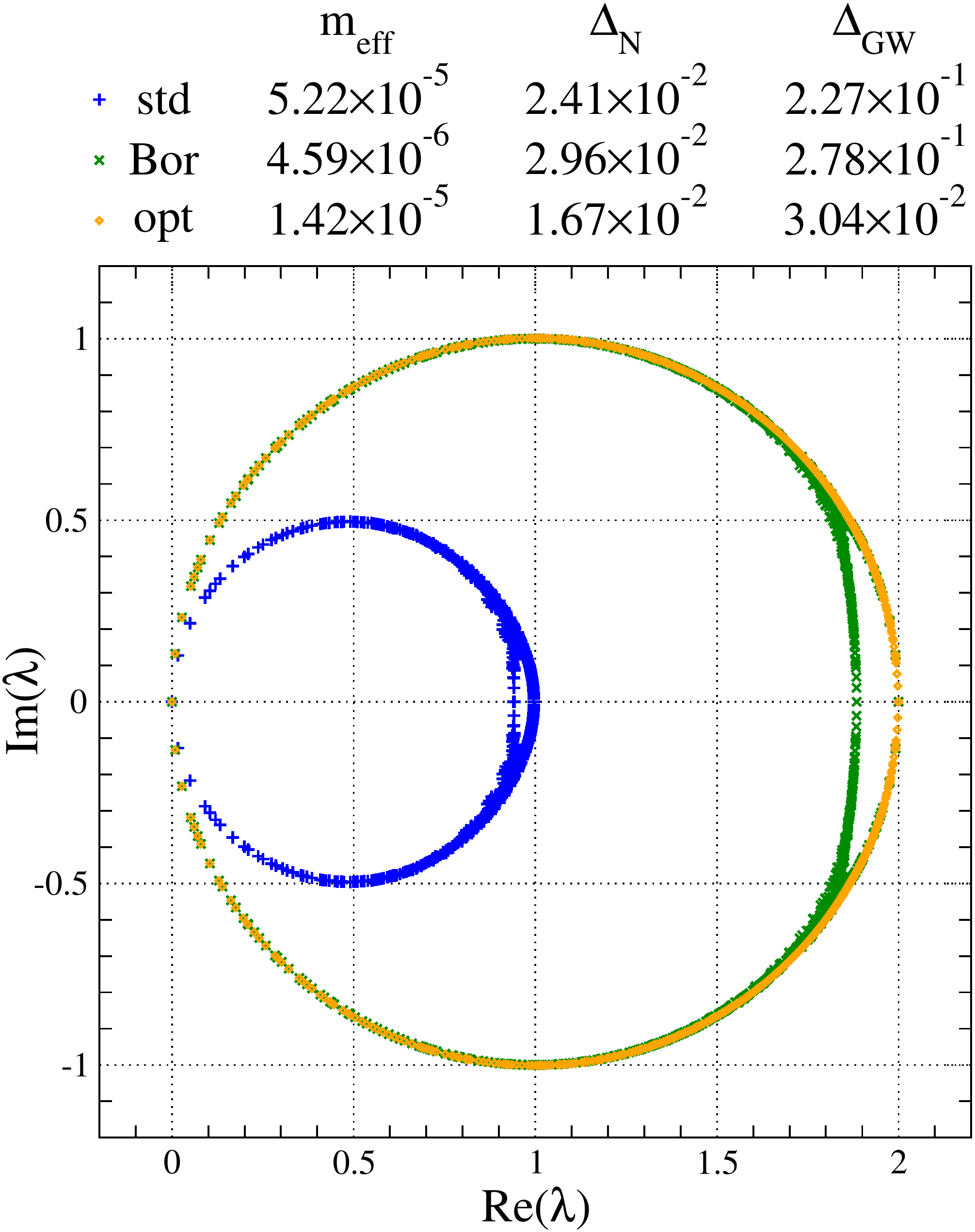}

}
\par\end{centering}
\caption{Spectrum of $\varrho D_{\mathsf{eff}}$ with Wilson kernel for the
standard (std), Boriçi (Bor) and optimal (opt) construction at $N_{s}=4$
on a smeared configuration with $\alpha=0.5$.\label{fig:gauge-wilson-smearing}}
\end{figure}
\begin{figure}[t]
\begin{centering}
\subfloat[No smearing]{\includegraphics[width=0.45\textwidth]{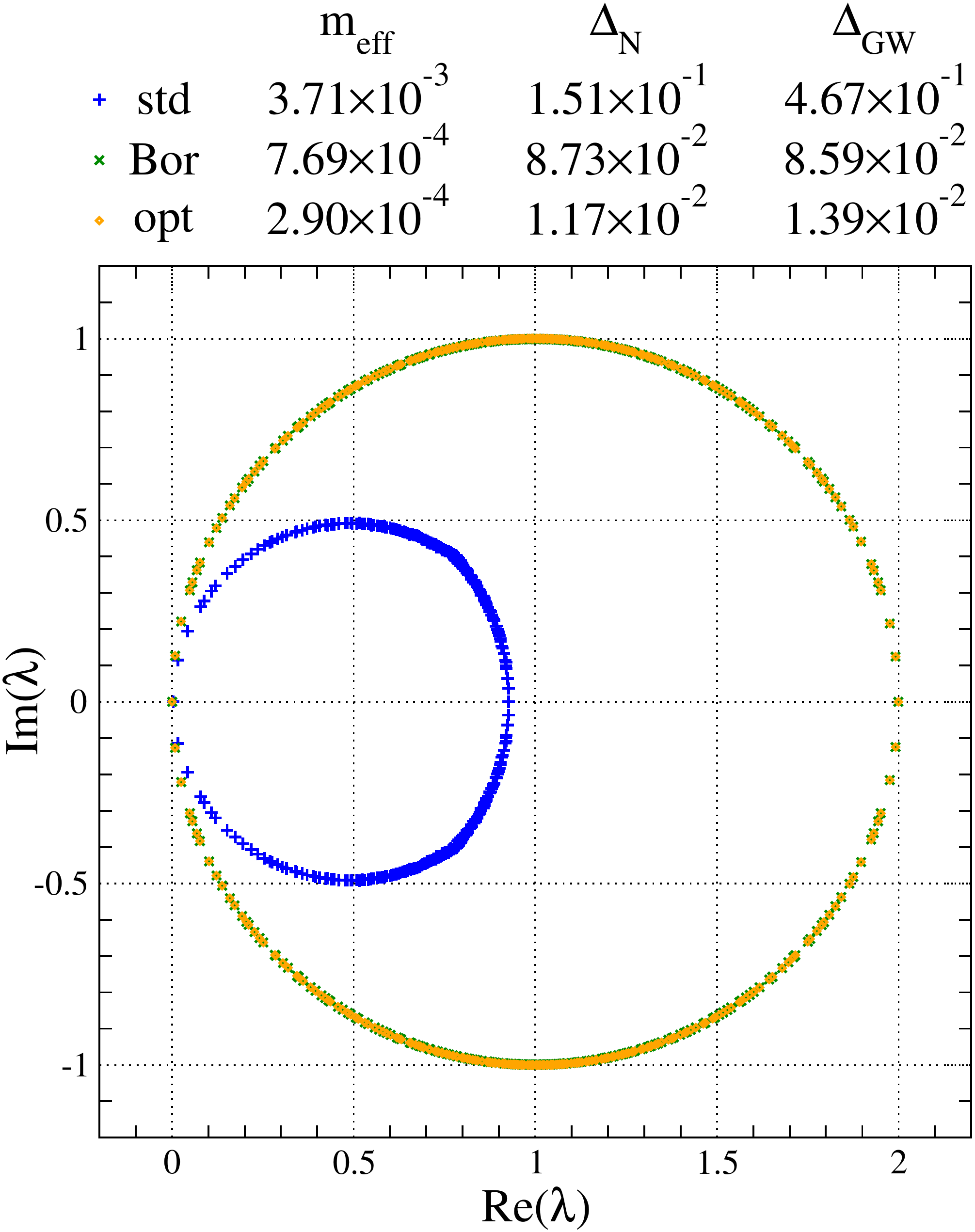}

}\hfill{}\subfloat[Three smearing iterations]{\includegraphics[width=0.45\textwidth]{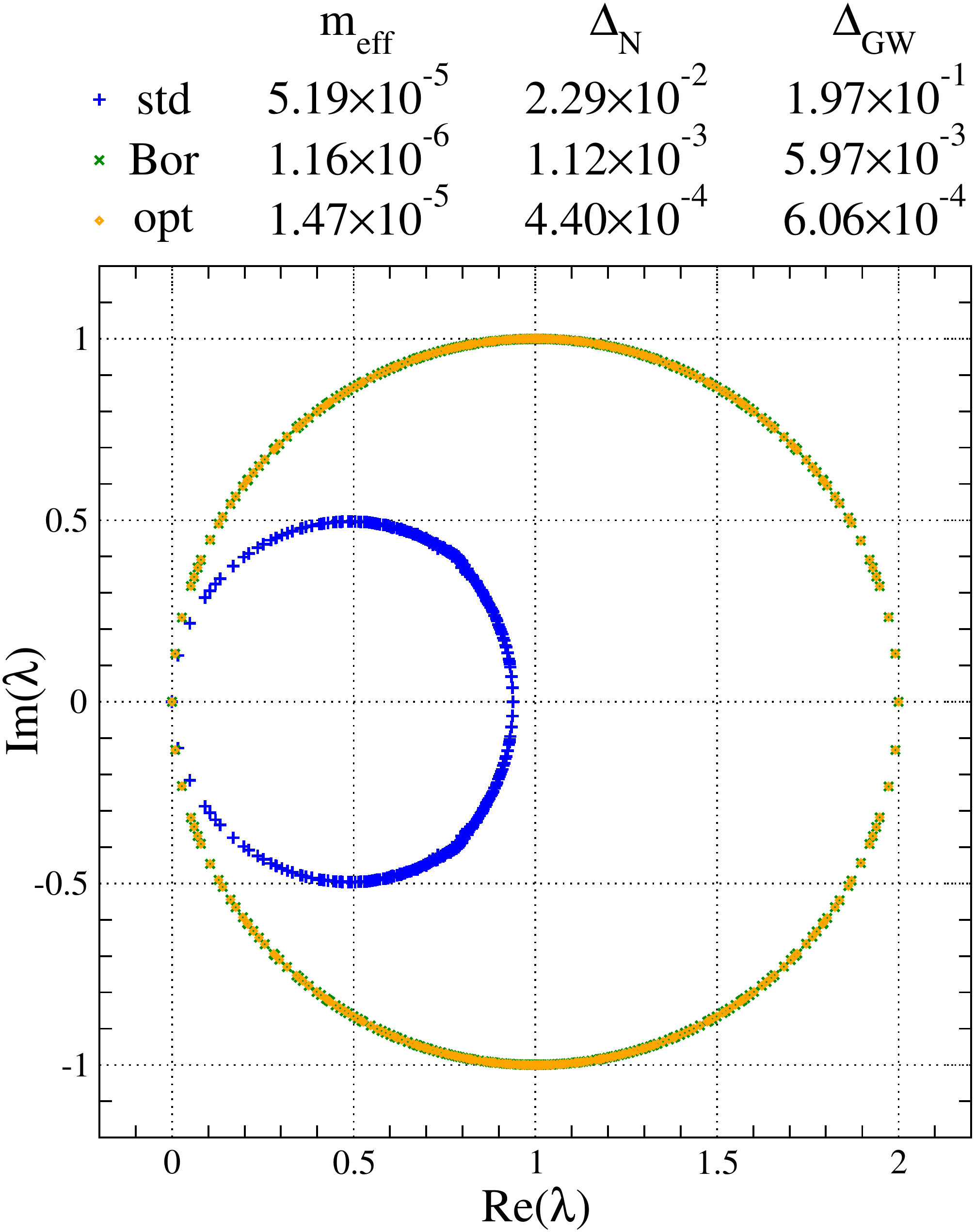}

}
\par\end{centering}
\caption{Spectrum of $\varrho D_{\mathsf{eff}}$ with staggered Wilson kernel
for the standard (std), Boriçi (Bor) and optimal (opt) construction
at $N_{s}=4$ on a smeared configuration with $\alpha=0.5$.\label{fig:gauge-stw-smearing}}
\end{figure}

Smearing is a common technique in the simulations of lattice gauge
theories. It is expected to be especially beneficial for the staggered
Wilson kernel (in particular in $3+1$ dimensions) as discussed in
Ref.~\citep{Durr:2013gp}. For this reason we consider gauge fields
with three-step \textsc{ape} smearing \citep{Falcioni:1984ei,Albanese:1987ds}.
For the smearing parameter, we choose the maximal value $\alpha=0.5$
in two dimensions within the perturbatively reasonable range~\citep{Capitani:2006ni}.

To test the effects of smearing on the violations of chiral symmetry,
in Figs.~\ref{fig:gauge-wilson-smearing} and \ref{fig:gauge-stw-smearing}
we directly compare the effective operators on an unsmeared and smeared
version of a gauge configuration at $N_{s}=4$. We observe that after
three smearing steps, chiral symmetry violations as measured by $m_{\mathsf{eff}}$,
$\Delta_{\mathsf{N}}$ and $\Delta_{\mathsf{GW}}$ are reduced significantly.
In some cases we observe a reduction by up to two order of magnitude
with the exception of the effective mass for the standard construction
with a Wilson kernel.

\subsection{Topology \label{subsec:DW-Topology}}

\begin{figure}[t]
\subfloat[Configuration with $Q=0$]{\includegraphics[width=0.32\textwidth]{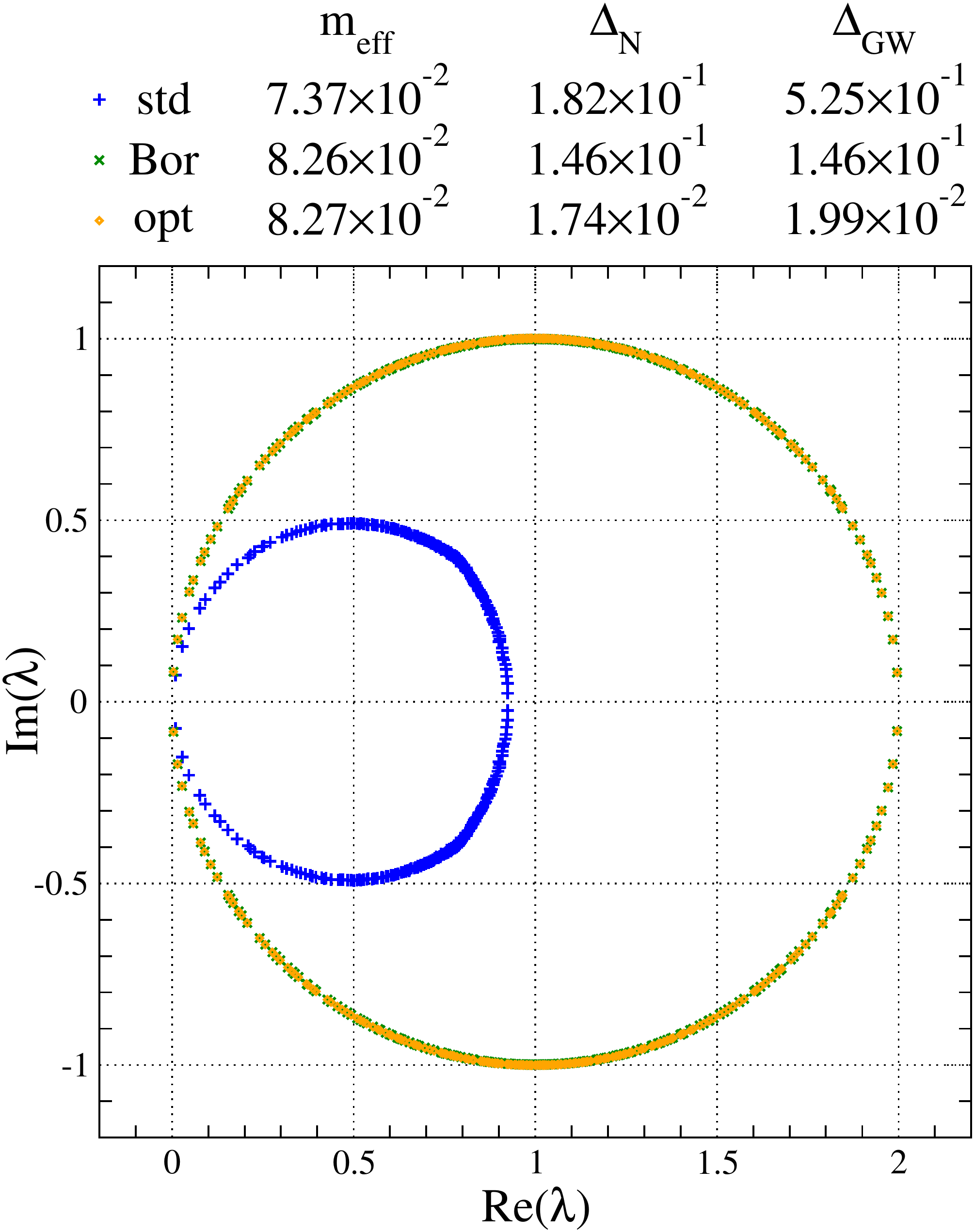}

}\hfill{}\subfloat[Configuration with $Q=1$]{\includegraphics[width=0.32\textwidth]{figures/domain_wall/gauge_field/schw_20_20_5_1009_0.65_0/stw_deff_4}

}\hfill{}\subfloat[Configuration with $Q=3$]{\includegraphics[width=0.32\textwidth]{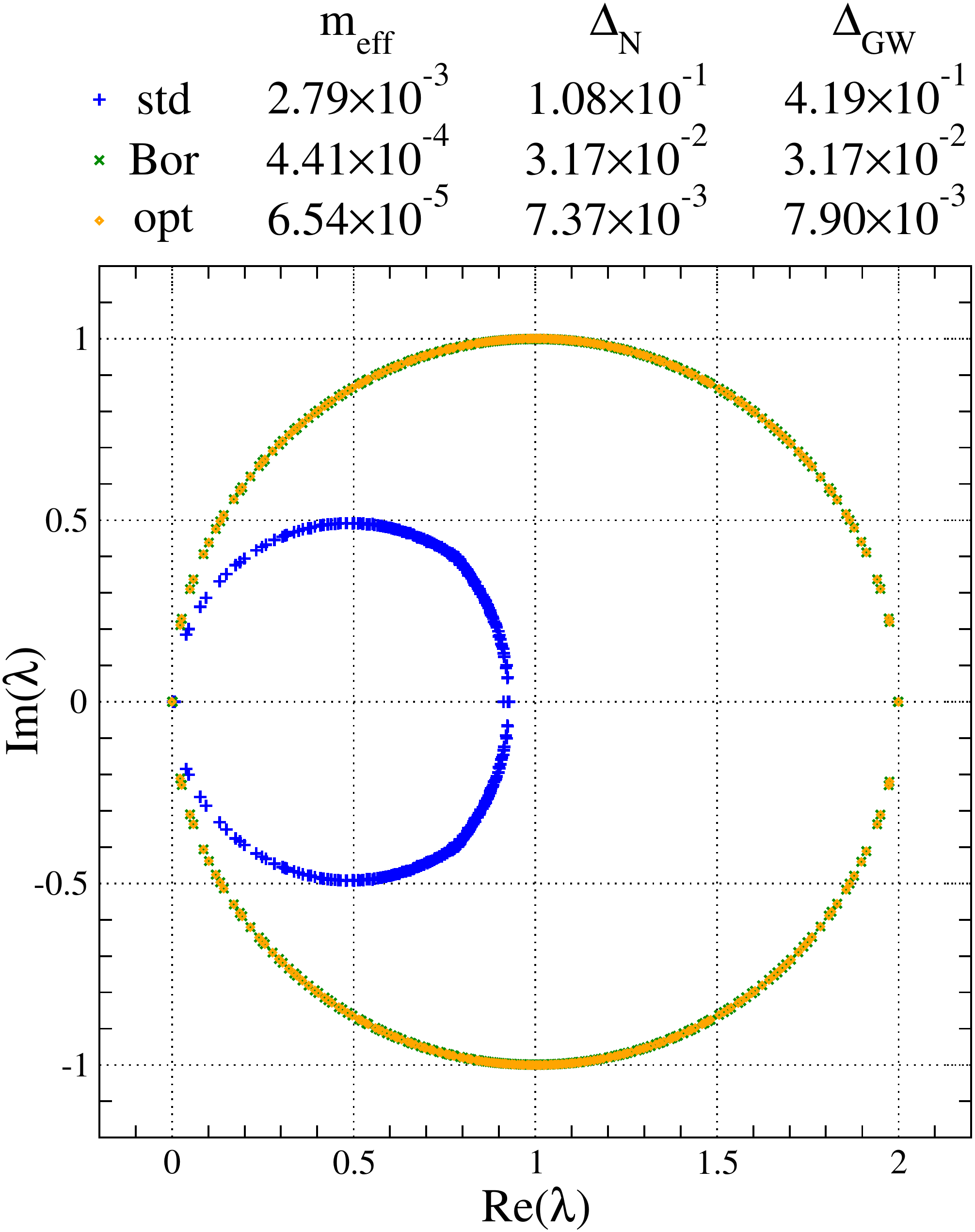}

}

\caption{Spectrum of $\varrho D_{\mathsf{eff}}$ with staggered Wilson kernel
for the standard (std), Boriçi (Bor) and optimal (opt) construction
at $N_{s}=4$ on configurations with various $Q$. \label{fig:gauge-stw-topology}}
\end{figure}
\begin{figure}[!h]
\begin{centering}
\subfloat[Wilson kernel]{\includegraphics[height=0.45\textwidth]{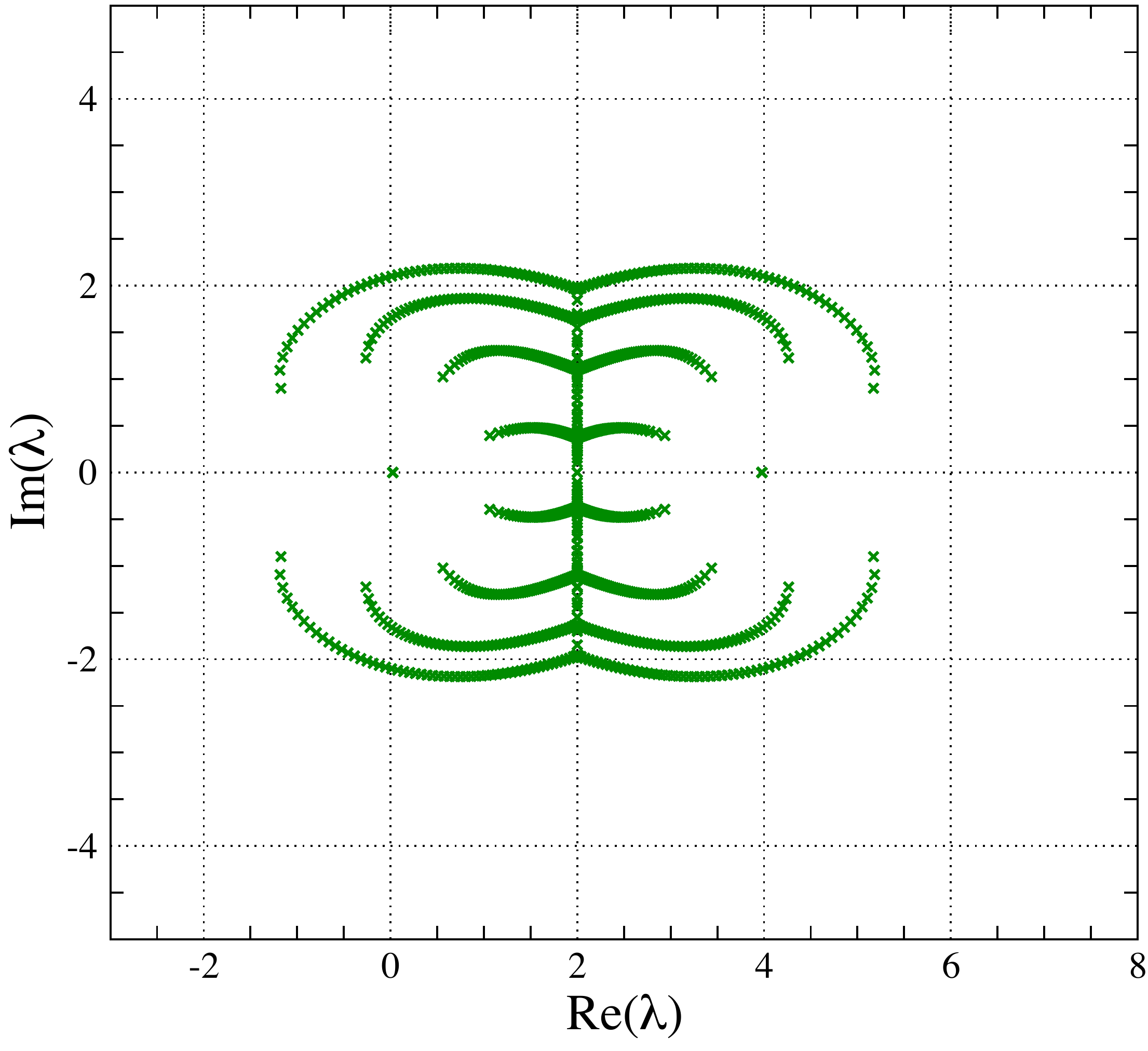}

}\hfill{}\subfloat[Staggered Wilson kernel]{\includegraphics[height=0.45\textwidth]{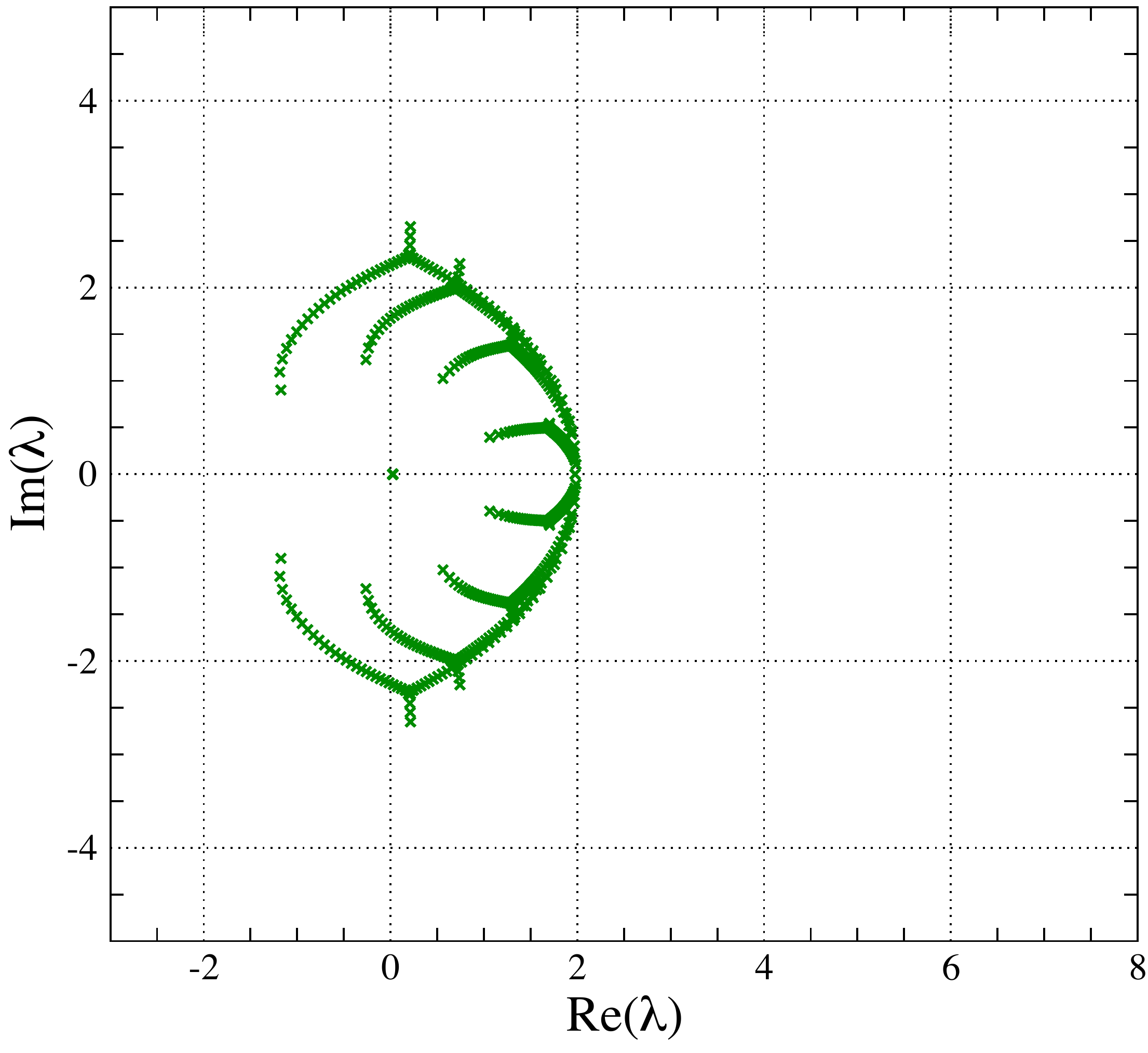}

}
\par\end{centering}
\caption{Spectrum of $D_{\mathsf{dw}}$ in Boriçi's construction at $N_{s}=4$
on a smooth topological configuration with $Q=3$. \label{fig:topological-configs}}
\end{figure}

On topologically nontrivial configurations with $Q\neq0$, the Atiyah-Singer
index theorem ensures that the continuum Dirac operator has exact
zero modes. On the lattice the same can be shown for the overlap operator,
while for the effective Dirac operators we only find approximate zero
modes as illustrated in Fig.~\ref{fig:gauge-stw-topology}. These
modes come with a multiplicity of $\left|Q\right|$ as expected from
the Vanishing Theorem.

By using the technique outlined in Ref.~\citep{Smit:1986fn} for
constructing a smooth configuration for any given topological charge
$Q$, we can study topological aspects also in a more abstract setting.
The gauge configurations are maximally smooth in the sense that they
are invariant under the application of the \textsc{ape} smearing prescription.
We measure $m_{\mathsf{eff}}$, $\Delta_{\mathsf{N}}$ and $\Delta_{\mathsf{GW}}$
for the effective Dirac operators on a wide range of smooth topological
configurations with various $Q$. Our measures of chirality are identical
for the configurations with $\pm Q$ and only depend on $\left|Q\right|$.

Due to the smoothness of the configurations, the measures $m_{\mathsf{eff}}$,
$\Delta_{\mathsf{N}}$ and $\Delta_{\mathsf{GW}}$ are all small compared
to the typical values one measures on thermalized configurations at
reasonable values of $\beta$. We also observe that with increasing
values of $\left|Q\right|$ the chiral symmetry violations become
larger. Finally, we show two eigenvalue spectra of bulk operators
on a $20^{2}$ lattice in Fig.~\ref{fig:topological-configs}, revealing
a very clear structure of the spectrum on a topologically nontrivial
configuration.

\subsection{Spectral flow}

\begin{figure}[t]
\subfloat[Unsmeared configuration]{\includegraphics[width=0.32\textwidth]{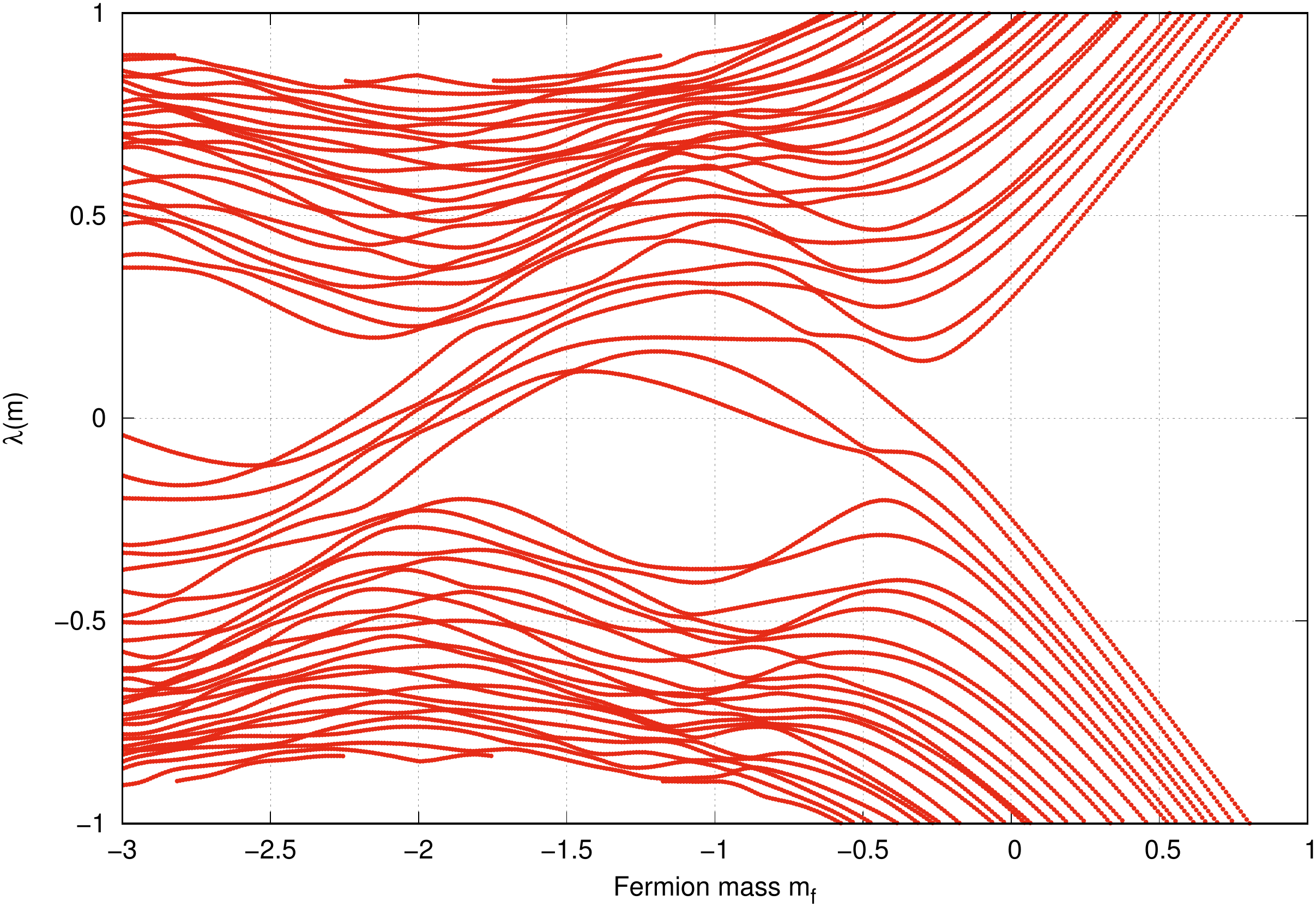}

}\hfill{}\subfloat[Smeared configuration]{\includegraphics[width=0.32\textwidth]{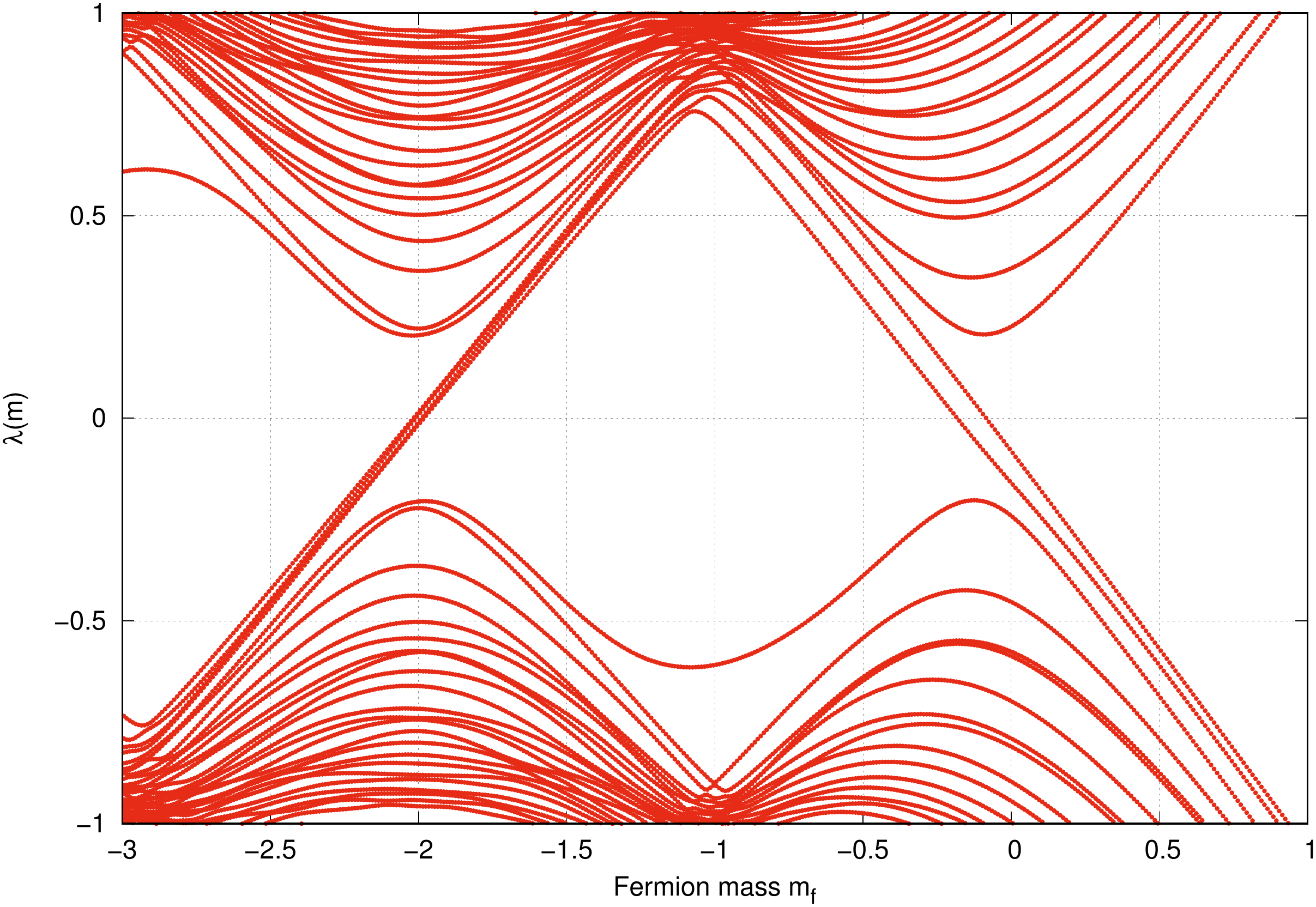}

}\hfill{}\subfloat[Topological configuration]{\includegraphics[width=0.32\textwidth]{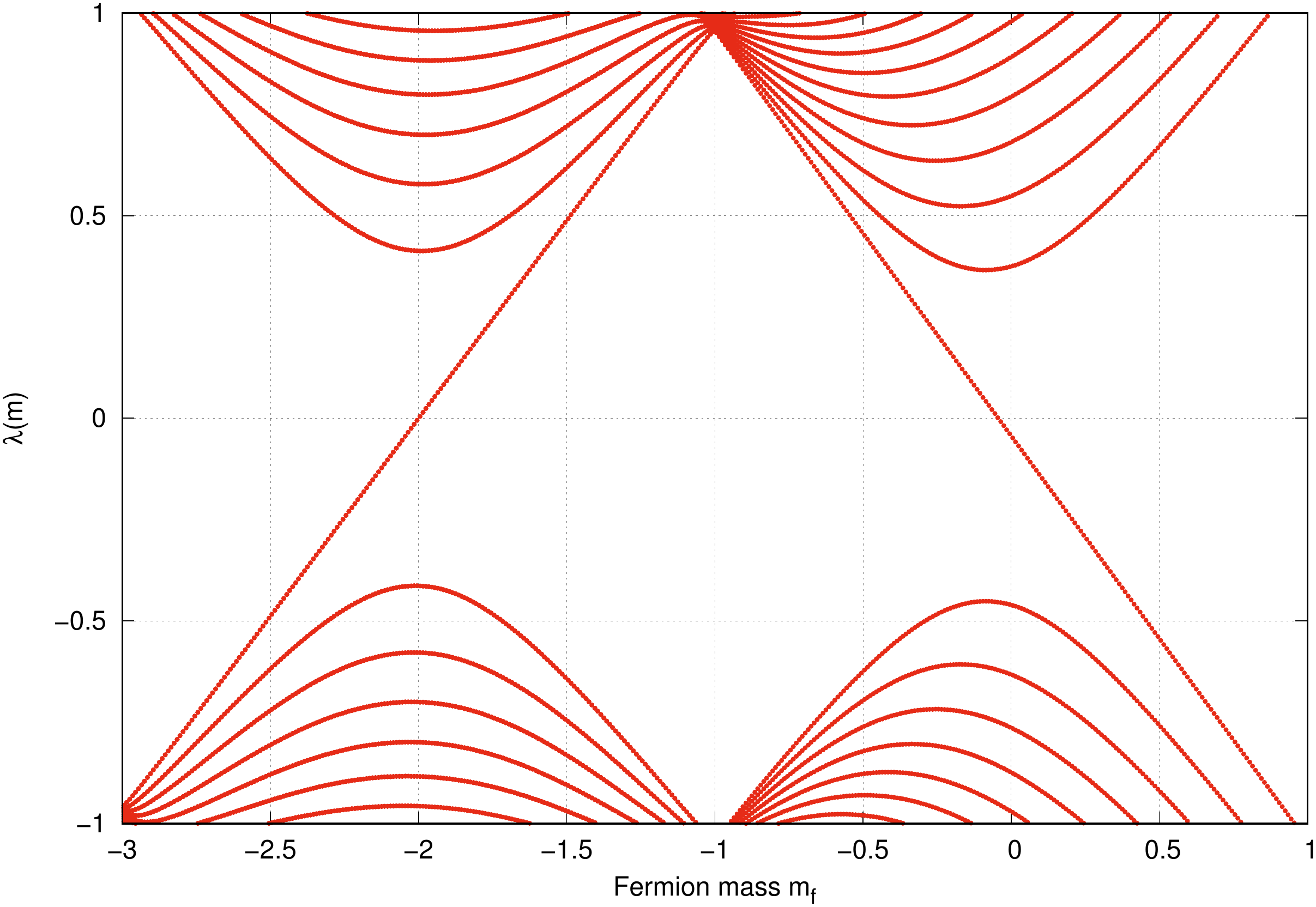}

}

\caption{Spectral flow of the Wilson kernel for configurations with $Q=2$.
\label{fig:spectral-flow-wilson}}
\end{figure}
\begin{figure}[t]
\subfloat[Unsmeared configuration]{\includegraphics[width=0.32\textwidth]{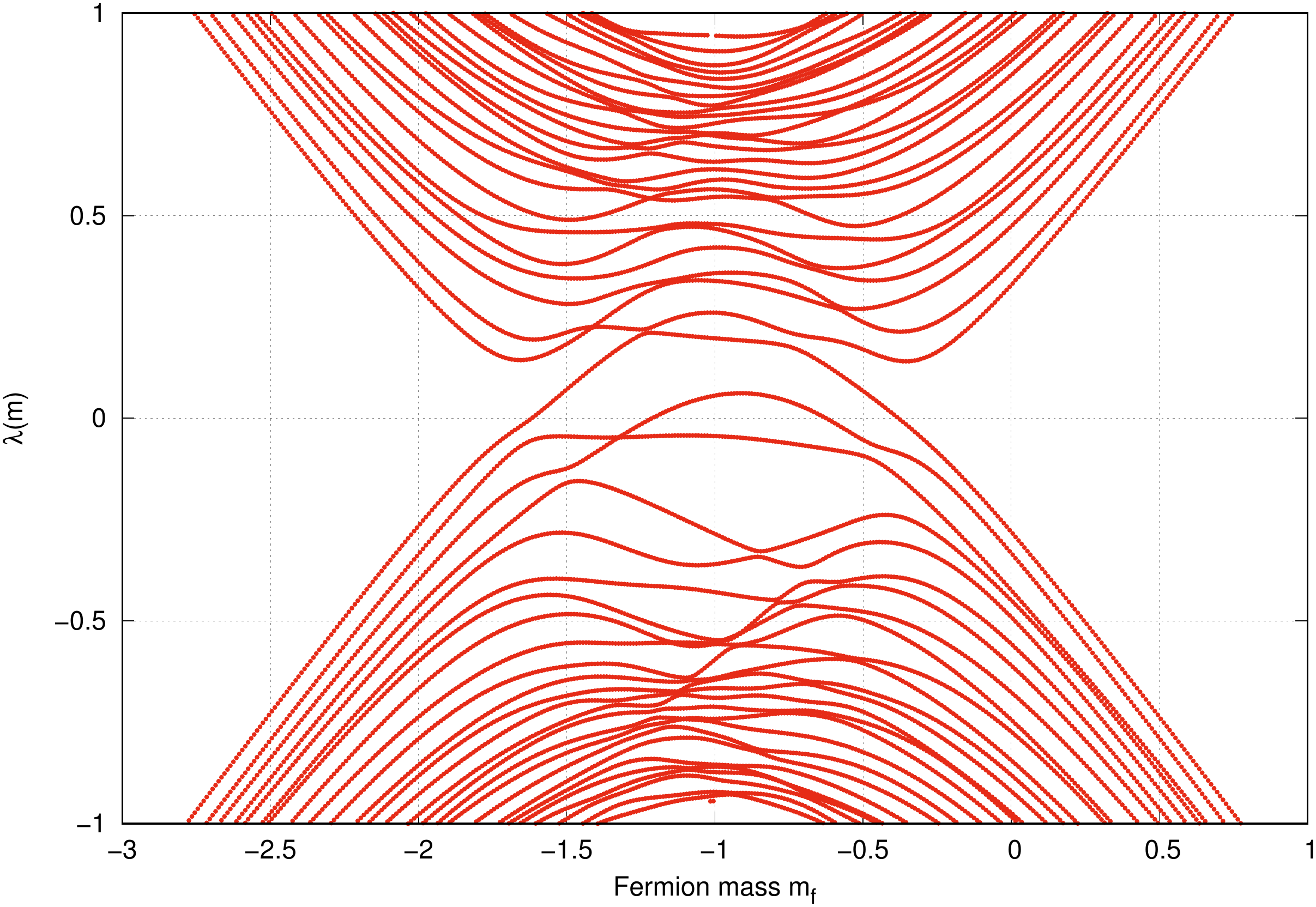}

}\hfill{}\subfloat[Smeared configuration]{\includegraphics[width=0.32\textwidth]{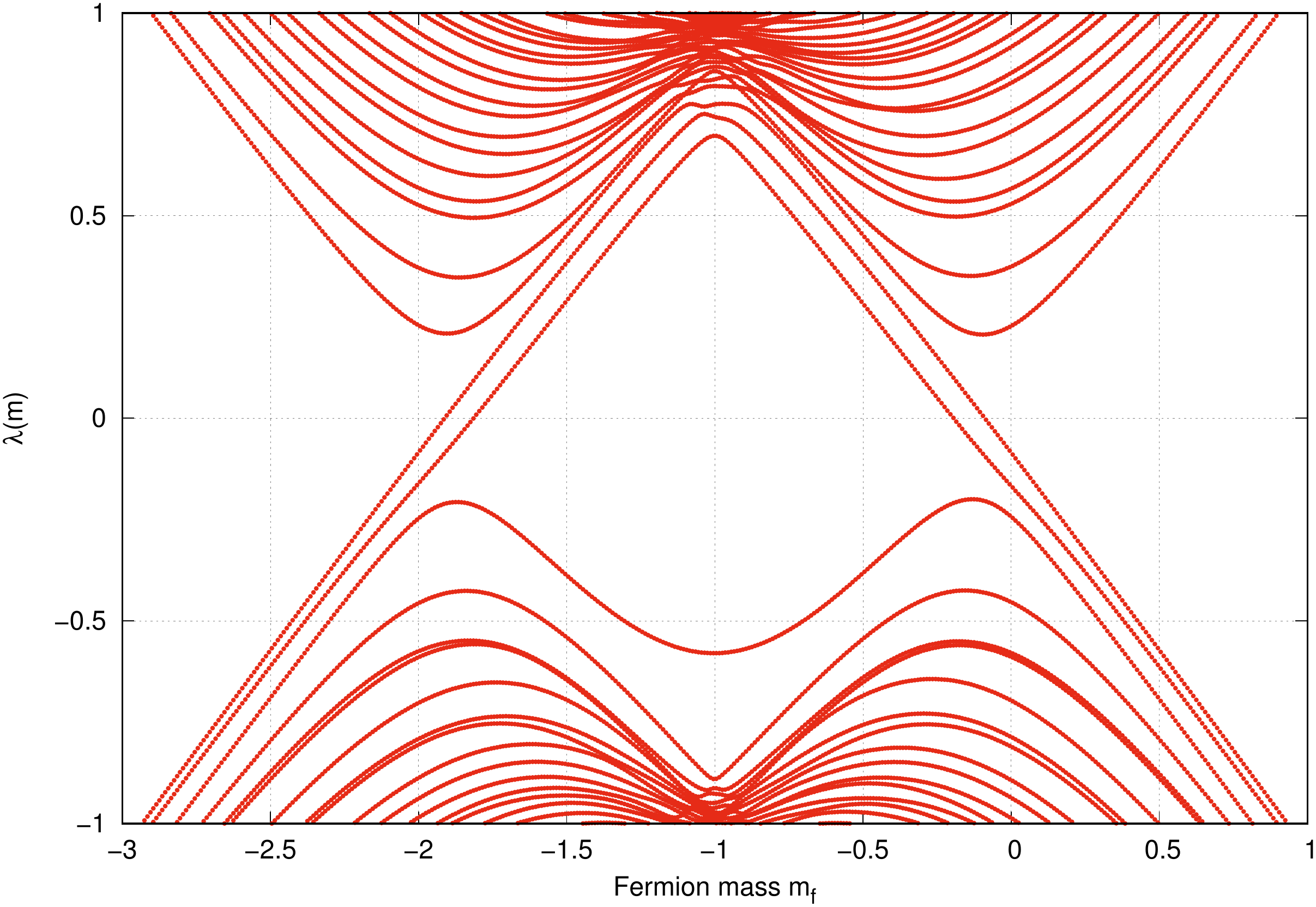}

}\hfill{}\subfloat[Topological configuration]{\includegraphics[width=0.32\textwidth]{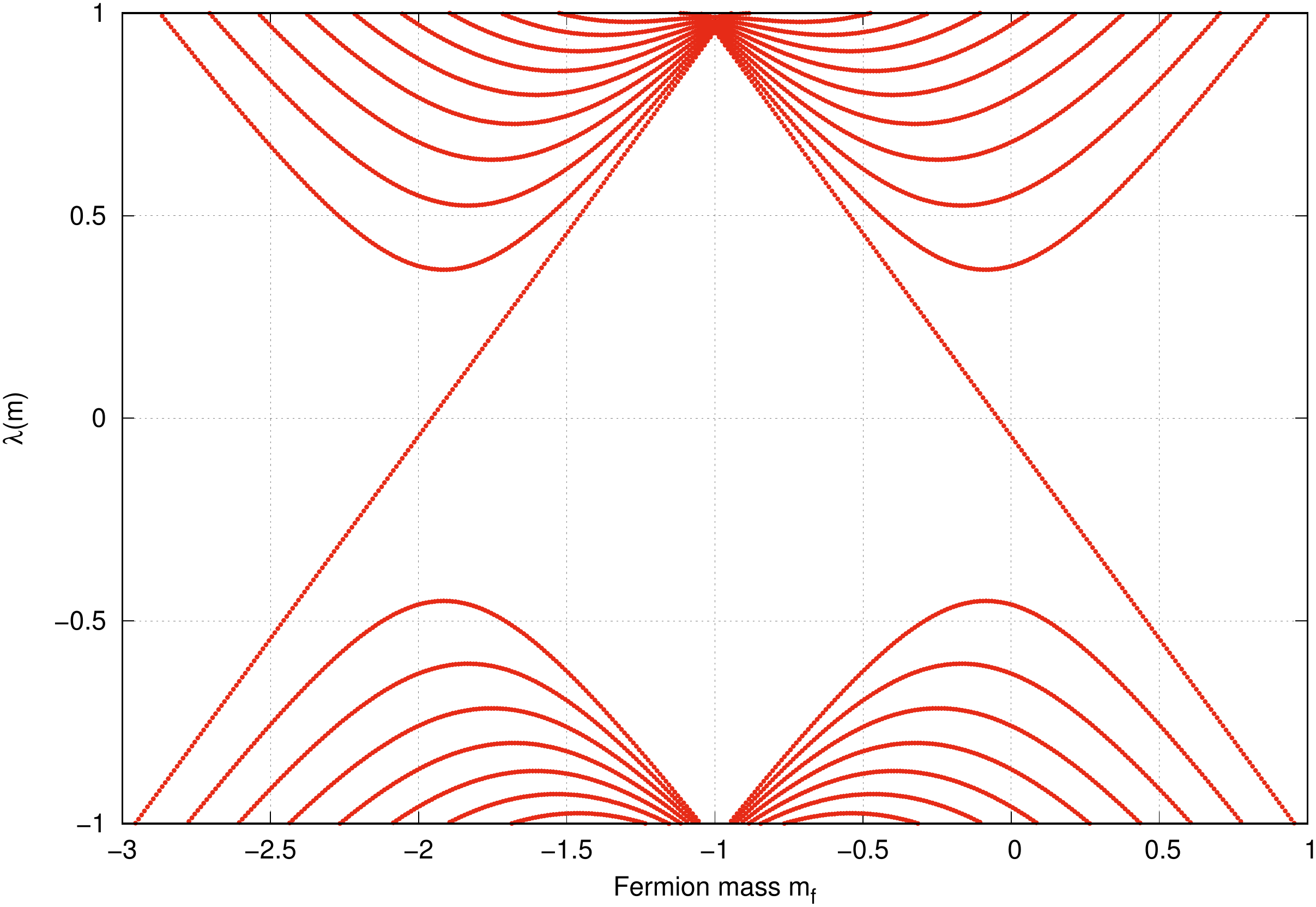}

}

\caption{Spectral flow of the staggered Wilson kernel for configurations with
$Q=2$. \label{fig:spectral-flow-stw}}
\end{figure}

Besides the study of the eigenvalue spectra, topological aspects—such
as the index theorem—can be studied on the lattice also with the help
of spectral flows \citep{Itoh:1987iy,Narayanan:1994gw}.

Commonly used with a Wilson kernel, one studies the eigenvalues $\lambda\left(m_{\mathsf{f}}\right)$
of the Hermitian operator $H_{\mathsf{w}}\left(m_{\mathsf{f}}\right)$
as a function of $m_{\mathsf{f}}$. One finds a direct one-to-one
correspondence between the eigenvalue crossings of $H_{\mathsf{w}}\left(m_{\mathsf{f}}\right)$
and the real eigenvalues of $D_{\mathsf{w}}\left(m_{\mathsf{f}}\right)$.
Moreover, the low-lying real eigenvalues of the Wilson Dirac operator
$D_{\mathsf{w}}\left(m_{\mathsf{f}}\right)$ correspond to the would-be
zero modes \citep{Smit:1986fn}. For each respective mode, the slope
of the eigenvalue crossings at small values of $m_{\mathsf{f}}$ equals
minus its chirality. By studying the eigenvalue crossings of the Hermitian
operator in a background field we can, thus, determine the respective
topological charge.

Although originally spectral flows of Wilson-like kernels were studied,
Adams' generalized the method to staggered kernels \citep{Adams:2009eb}
allowing the study of topological aspects within the staggered framework
\citep{deForcrand:2011ak,Follana:2011kh,deForcrand:2012bm,Azcoiti:2014pfa}.
While in the literature these spectral flows have been investigated
before, here we want to show the effectiveness of smearing when studying
topological aspects on the lattice.

To this end we show an example for the flow $\lambda\left(m_{\mathsf{f}}\right)$
of the lowest $50$ eigenvalues for the Wilson kernel $H_{\mathsf{w}}\left(m_{\mathsf{f}}\right)$
and the staggered Wilson kernel $H_{\mathsf{sw}}\left(m_{\mathsf{f}}\right)$
in Figs.~\ref{fig:spectral-flow-wilson} and \ref{fig:spectral-flow-stw}.
The $12^{2}$ gauge configuration under consideration was generated
at $\beta=1.8$ and has a topological charge of $Q=2$. We compare
the eigenvalue flow on both the unsmeared and the three-step \textsc{ape}
smeared ($\alpha=0.5$) version of the configuration and also added
the smooth $Q=2$ topological configuration described in Sec.\ \ref{subsec:DW-Topology}.
While the determination of the topological charge $Q$ via the spectral
flow is not easily possible for the rough and unsmeared configuration,
the use of smearing allows an unambiguous identification of the topological
sector. In the case of the smooth topological configuration, the two
eigenvalue crossings lie essentially on top of each other. The direct
comparison between the figures shows the effectiveness of smearing
for the study of topological aspects on the lattice.

\subsection{Approaching the continuum}

\begin{table}[t]
\begin{centering}
\begin{tabular*}{1\textwidth}{@{\extracolsep{\fill}}cccccc}
\toprule 
Kernel & Construction & $N_{s}$ & $m_{\mathsf{eff}}$ & $\Delta_{\mathsf{N}}$ & $\Delta_{\mathsf{GW}}$\tabularnewline
\midrule
 &  & 2 & $5.47\times10^{-3}$ & $1.53\times10^{-1}$ & $6.72\times10^{-1}$\tabularnewline
\cmidrule{3-6} 
 & Standard & 4 & $5.90\times10^{-4}$ & $6.87\times10^{-2}$ & $3.10\times10^{-1}$\tabularnewline
\cmidrule{3-6} 
 &  & 6 & $1.00\times10^{-4}$ & $2.30\times10^{-2}$ & $1.02\times10^{-1}$\tabularnewline
\cmidrule{2-6} 
 &  & 2 & $5.73\times10^{-3}$ & $3.71\times10^{-1}$ & $1.16\times10^{0\phantom{-}}$\tabularnewline
\cmidrule{3-6} 
Wilson & Boriçi & 4 & $8.56\times10^{-5}$ & $6.64\times10^{-2}$ & $3.00\times10^{-1}$\tabularnewline
\cmidrule{3-6} 
 &  & 6 & $5.42\times10^{-6}$ & $1.57\times10^{-2}$ & $7.60\times10^{-2}$\tabularnewline
\cmidrule{2-6} 
 &  & 2 & $8.30\times10^{-3}$ & $7.51\times10^{-1}$ & $1.16\times10^{0\phantom{-}}$\tabularnewline
\cmidrule{3-6} 
 & Optimal & 4 & $2.11\times10^{-3}$ & $3.59\times10^{-2}$ & $4.78\times10^{-2}$\tabularnewline
\cmidrule{3-6} 
 &  & 6 & $8.71\times10^{-6}$ & $1.21\times10^{-3}$ & $1.74\times10^{-3}$\tabularnewline
\midrule
 &  & 2 & $6.22\times10^{-3}$ & $1.48\times10^{-1}$ & $6.49\times10^{-1}$\tabularnewline
\cmidrule{3-6} 
 & Standard & 4 & $7.18\times10^{-4}$ & $5.88\times10^{-2}$ & $2.88\times10^{-1}$\tabularnewline
\cmidrule{3-6} 
 &  & 6 & $1.34\times10^{-4}$ & $2.01\times10^{-2}$ & $9.72\times10^{-2}$\tabularnewline
\cmidrule{2-6} 
 &  & 2 & $6.38\times10^{-3}$ & $2.02\times10^{-1}$ & $2.92\times10^{-1}$\tabularnewline
\cmidrule{3-6} 
Staggered Wilson & Boriçi & 4 & $5.13\times10^{-5}$ & $5.45\times10^{-3}$ & $8.68\times10^{-3}$\tabularnewline
\cmidrule{3-6} 
 &  & 6 & $5.91\times10^{-7}$ & $1.53\times10^{-4}$ & $2.69\times10^{-4}$\tabularnewline
\cmidrule{2-6} 
 &  & 2 & $1.72\times10^{-2}$ & $2.32\times10^{-1}$ & $2.66\times10^{-1}$\tabularnewline
\cmidrule{3-6} 
 & Optimal & 4 & $3.35\times10^{-5}$ & $2.23\times10^{-3}$ & $2.63\times10^{-3}$\tabularnewline
\cmidrule{3-6} 
 &  & 6 & $5.02\times10^{-8}$ & $2.01\times10^{-5}$ & $2.36\times10^{-5}$\tabularnewline
\bottomrule
\end{tabular*}
\par\end{centering}
\caption{Median values for the various measures of chirality on unsmeared $32^{2}$
configurations generated at $\beta=12.8$. For $m_{\mathsf{eff}}$
only configurations with $Q\protect\neq0$ are considered. \label{tab:cont-limit-values}}
\end{table}
\begin{figure}[t]
\begin{centering}
\subfloat[Wilson kernel]{\includegraphics[width=0.49\textwidth]{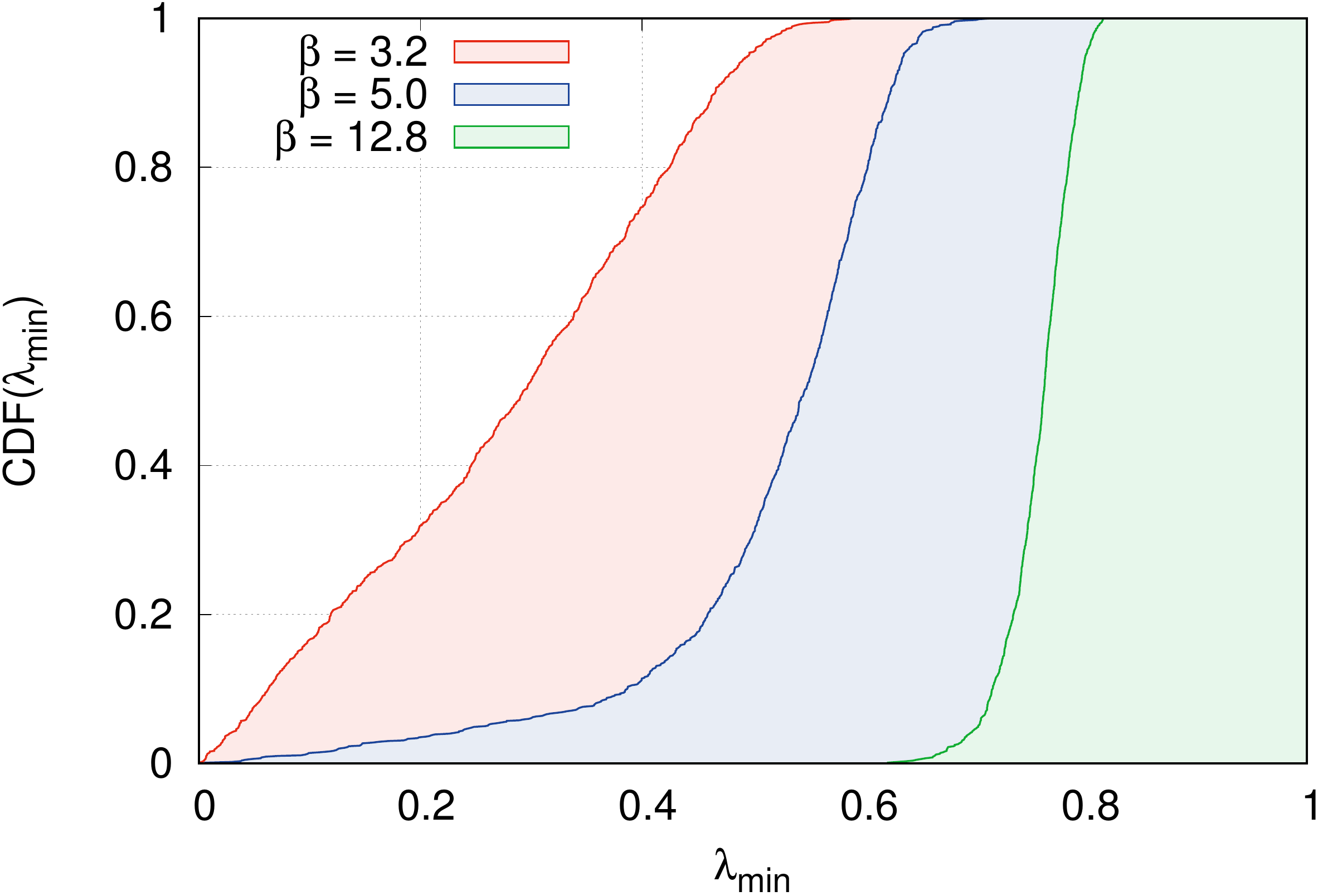}

}\hfill{}\subfloat[Staggered Wilson kernel]{\includegraphics[width=0.49\textwidth]{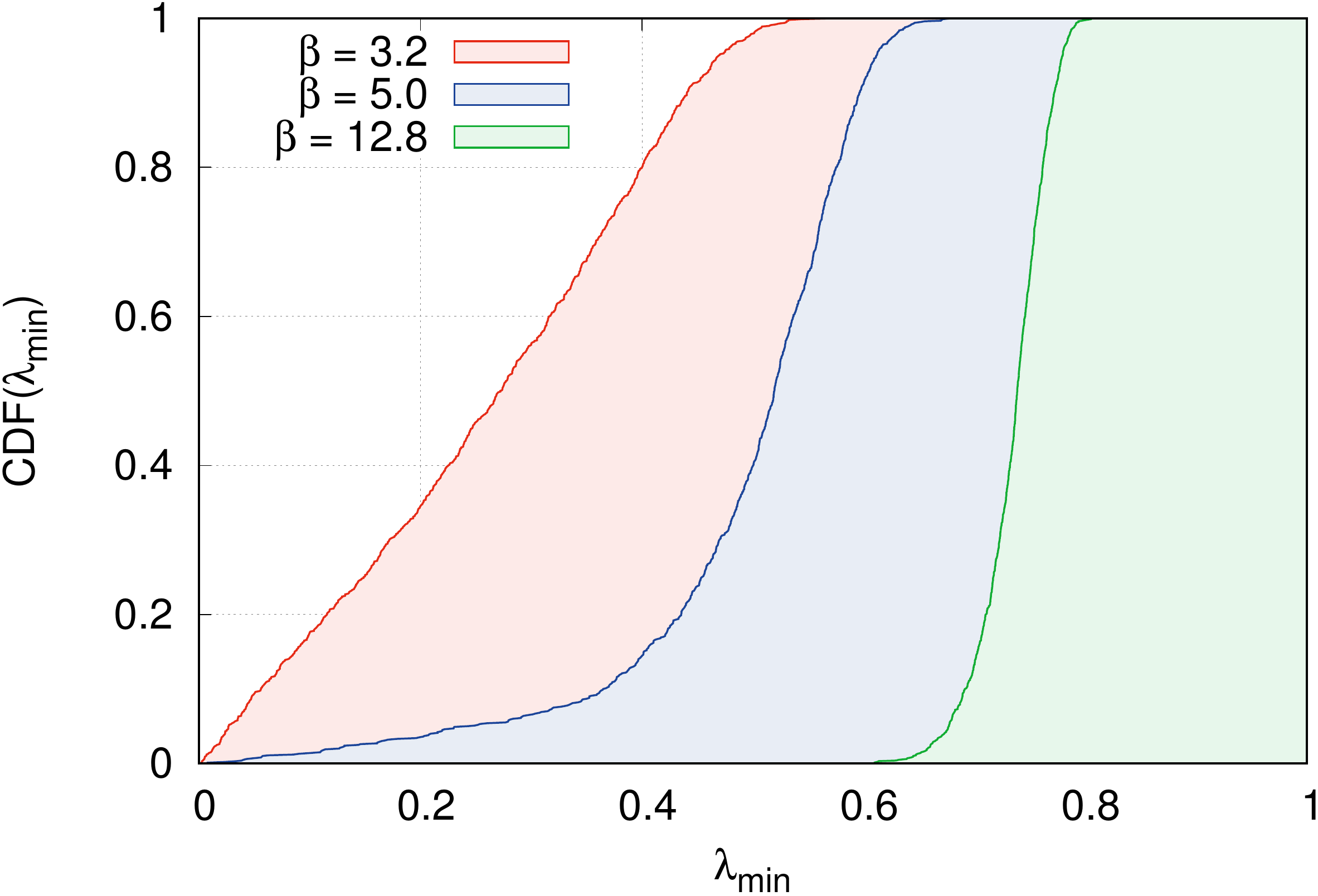}

}
\par\end{centering}
\caption{Cumulative distribution function $\mathsf{CDF}$ of $\lambda_{\mathsf{min}}$
for the Hermitian kernel operators. \label{fig:cdf-lmin}}
\end{figure}

To evaluate the chiral properties of the various effective Dirac operators
when approaching the continuum, we evaluated them on seven ensembles
with $1000$ configurations each. While keeping the physical volume
fixed, we consider the following ensembles: $8^{2}$ at $\beta=0.8$,
$12^{2}$ at $\beta=1.8$, $16^{2}$ at $\beta=3.2$, $20^{2}$ at
$\beta=5.0$, $24^{2}$ at $\beta=7.2$, $28^{2}$ at $\beta=9.8$
and $32^{2}$ at $\beta=12.8$. Besides the unsmeared configurations,
we also consider the smeared versions, resulting in $N=14\,000$ gauge
configurations in total.

While we are interested in the comparison of the different formulations
on finer lattices, we do not attempt a rigorous continuum limit analysis.
The chirality measures on our finest lattice at $\beta=12.8$ serve
us as an indicator for the relative performance of the different domain
wall formulations. In Table\ \ref{tab:cont-limit-values}, we quote
the median values for the parameter range $N_{s}\in\left\{ 2,4,6\right\} $.
For $N_{s}\geq8$ some values already become so small, that they are
comparable with the round-off error and, thus, cannot be considered
reliable. For the standard construction we find that the chiral symmetry
violations at large values of $\beta$ are comparable for the cases
of a Wilson and a staggered Wilson kernel. At the same time, in the
case of Boriçi's and the optimal construction we often find differences
between the two kernel operators of one to two orders of magnitude.

In practical applications the spectral bounds $\lambda_{\mathsf{min}}$
and $\lambda_{\mathsf{max}}$ of optimal domain wall fermions are
set on a per ensemble basis. As especially for small $\beta$ and
large ensembles one frequently observes values of $\lambda_{\mathsf{min}}$
close to zero, one typically projects out the lowest lying eigenvalues
and treats them exactly in order to arrive at a kernel operator with
a narrower spectrum. As in our study we want to compare the ``baseline
performance'' of the various domain wall fermion formulations, we
decided to not add such an additional algorithmic component mixing
into the results of our benchmarks. However, as we also investigate
large ensembles at small $\beta$, the observed ratios of $\lambda_{\min}^{2}/\lambda_{\min}^{2}$
can be very small in these cases, rendering the optimal $L_{\infty}$-approximation
of the $\sign$-function inadequate and of no practical use. To mitigate
this phenomenon, we decided to construct the $\sign$-function approximation
on a per-configuration rather than a per-ensemble basis. This means
that a small eigenvalue on a given configuration does not render the
$\sign$-function approximation inadequate for the whole ensemble,
but only on that particular configuration. On the other hand this
implies that in our benchmarks the performance of optimal domain wall
fermions tend to be overestimated and can only be considered indicative
of the performance of Chiu's formulation of domain wall fermions.

\begin{figure}[t]
\begin{centering}
\subfloat[Without smearing]{\includegraphics[width=0.49\textwidth]{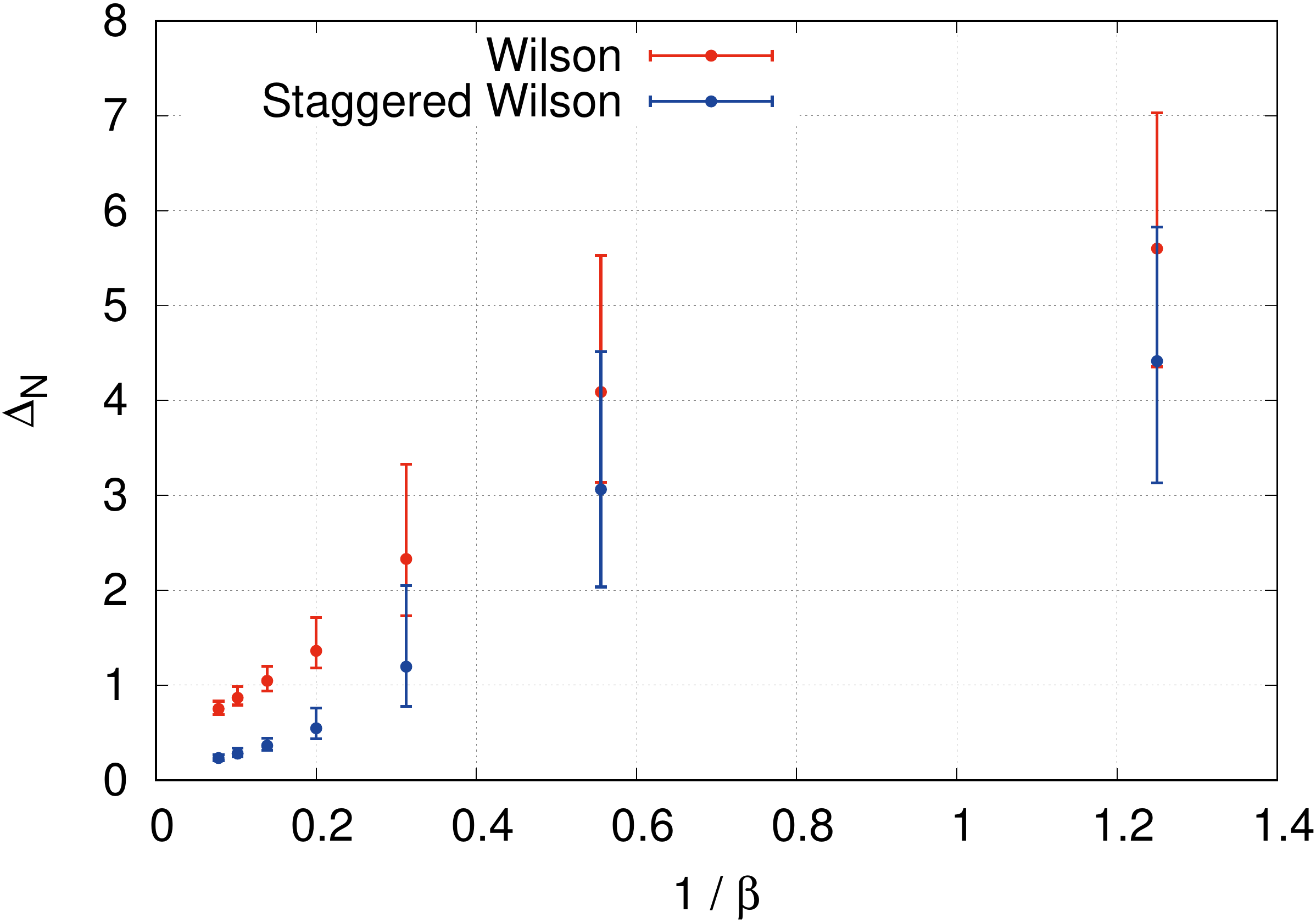}

}\hfill{}\subfloat[With smearing]{\includegraphics[width=0.49\textwidth]{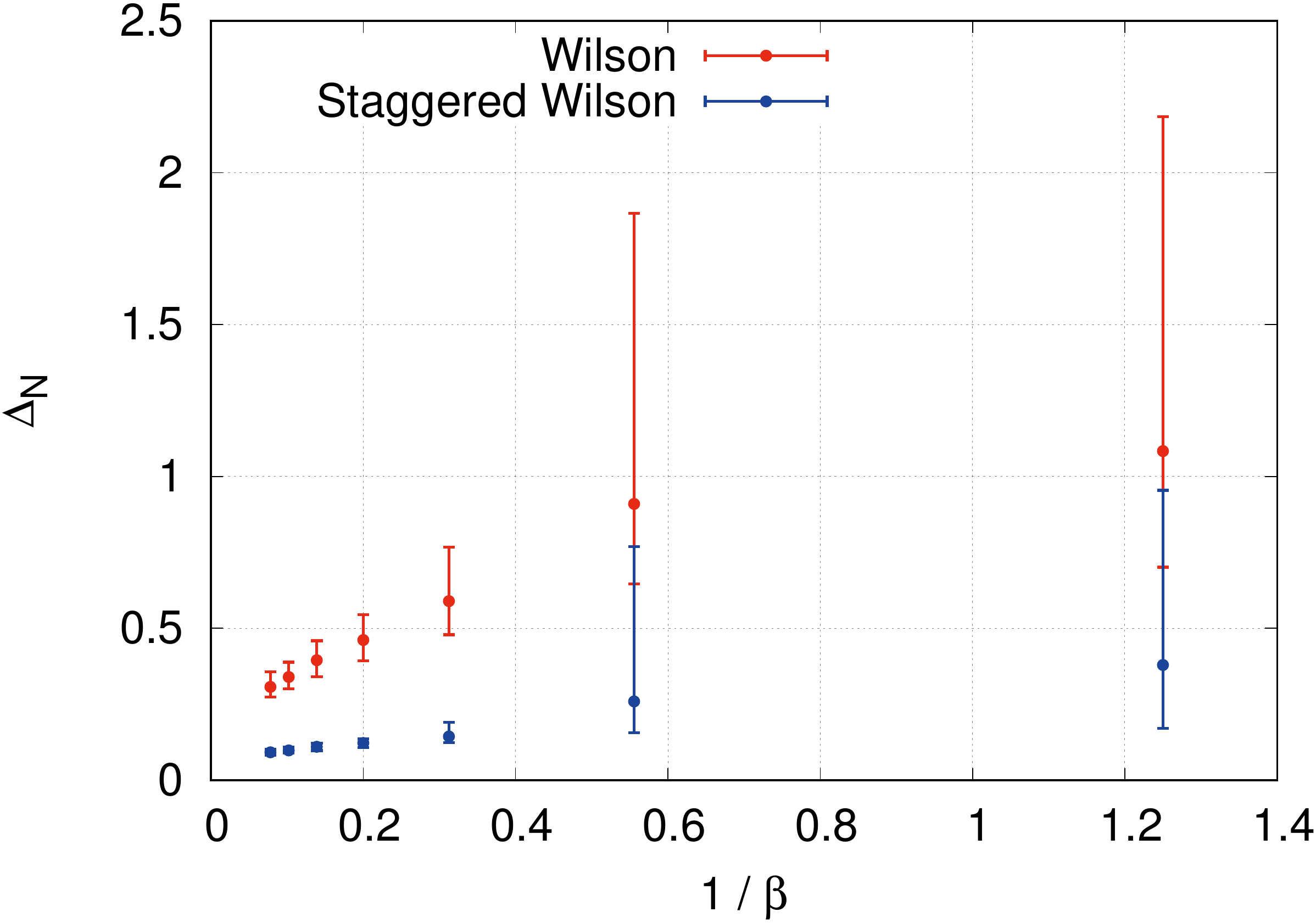}

}
\par\end{centering}
\caption{Violation of normality $\Delta_{\mathsf{N}}$ as a function of $1/\beta$
for $\varrho D_{\mathsf{eff}}$ in the optimal construction at $N_{s}=2$
(this figure is not part of Ref.~\citep{Hoelbling:2016qfv}). \label{fig:cont-limit-norm}}
\end{figure}
\begin{figure}[t]
\begin{centering}
\subfloat[Without smearing]{\includegraphics[width=0.49\textwidth]{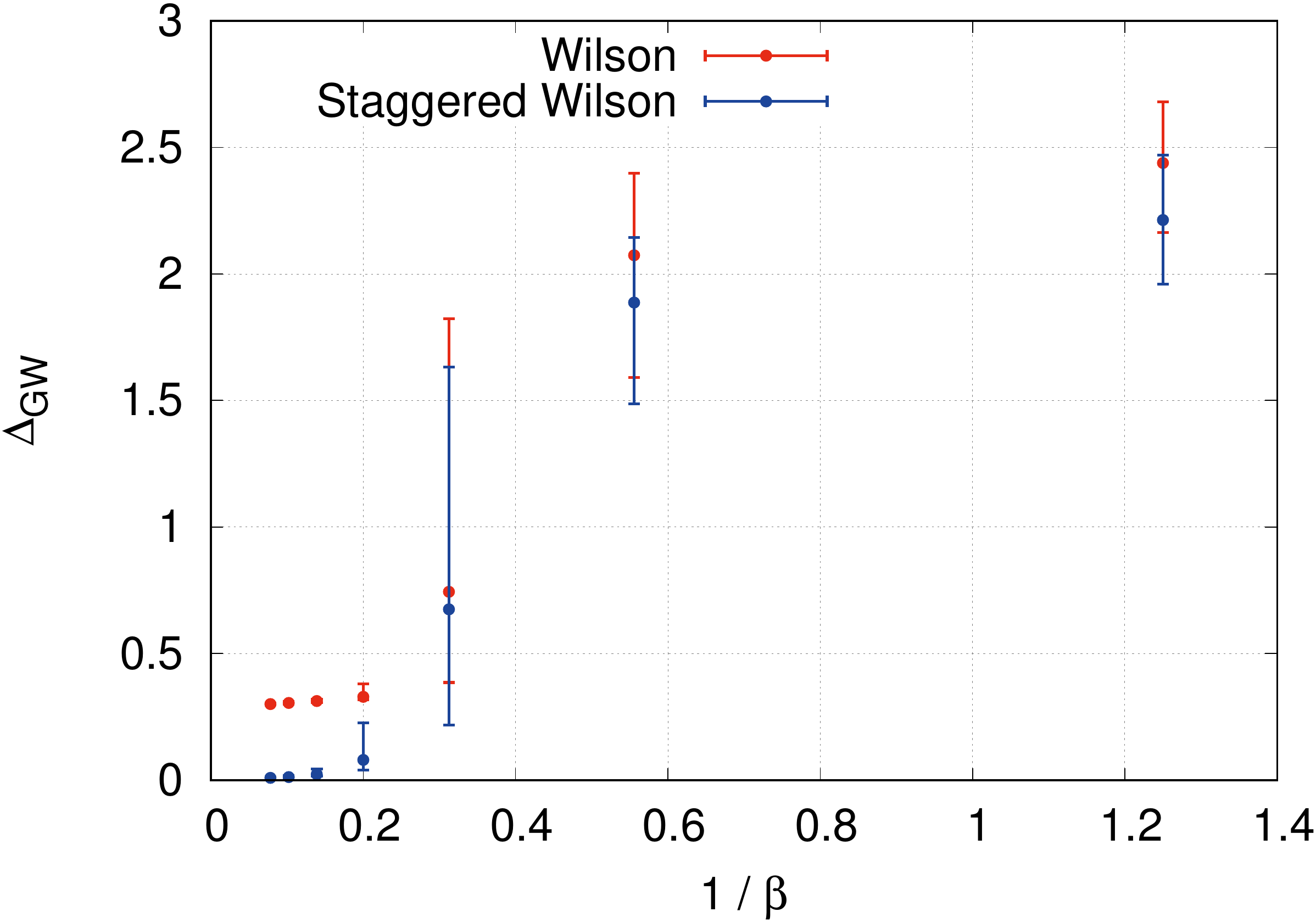}

}\hfill{}\subfloat[With smearing]{\includegraphics[width=0.49\textwidth]{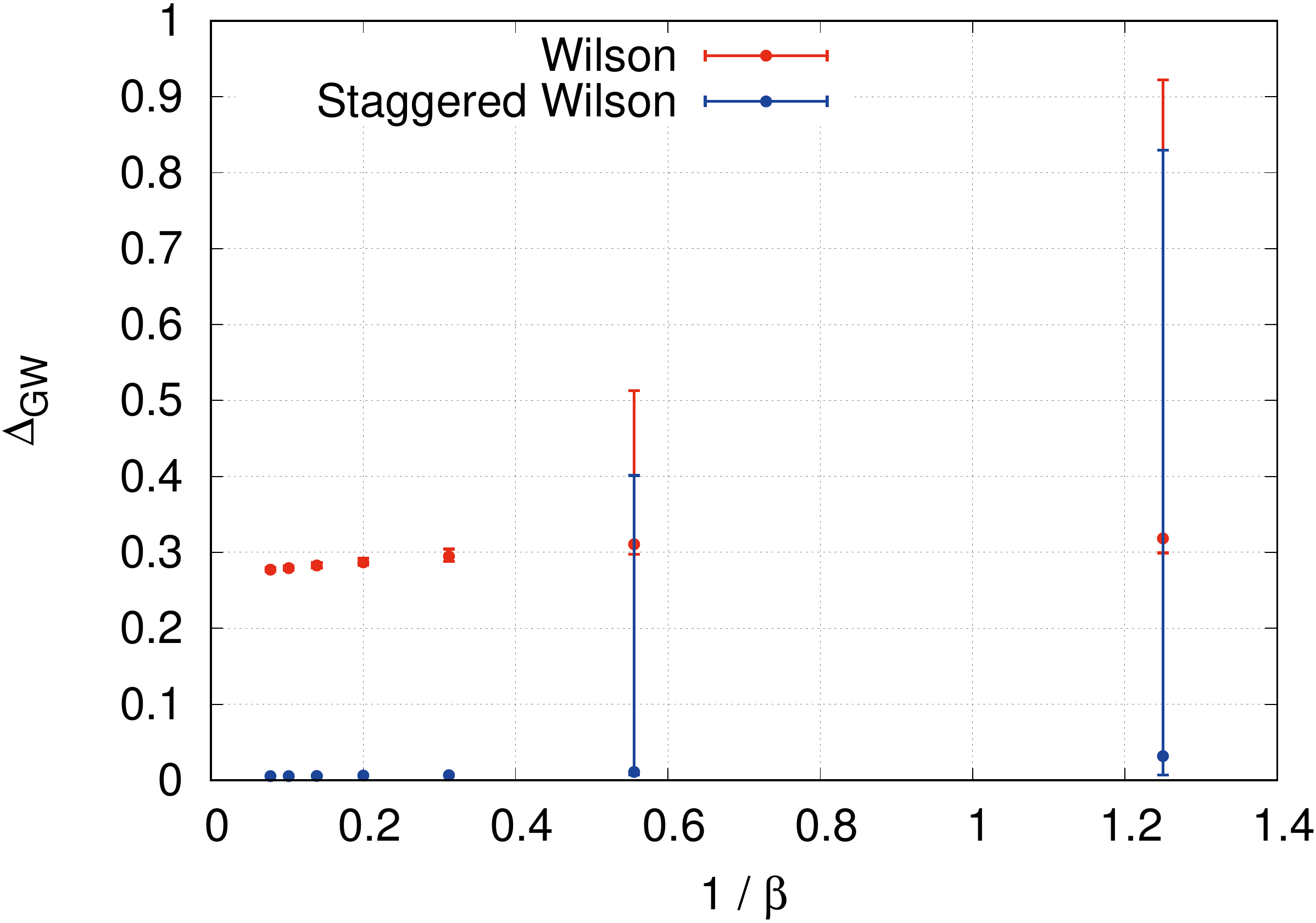}

}
\par\end{centering}
\caption{Violation of the Ginsparg-Wilson relation $\Delta_{\mathsf{GW}}$
as a function of $1/\beta$ for $\varrho D_{\mathsf{eff}}$ in Boriçi's
construction at $N_{s}=4$. \label{fig:cont-limit-gwr}}
\end{figure}

In general, we find that the standard construction is outperformed
by Boriçi's construction with respect to our chirality measures, while
optimal domain wall fermions show the best overall performance. A
notable exception is the effective mass $m_{\mathsf{eff}}$, where
the optimal construction with a Wilson kernel is not so clearly outperforming
the competing variants. This is not surprising, considering that the
optimal rational function approximation of the $\sign$-function deviates
the most at $\lambda_{\mathsf{min}}$ (and $\lambda_{\mathsf{max}}$)
and, thus, low-lying eigenvalues are typically not mapped very accurately.
However, as the size of the error decreases rapidly with increasing
values of $N_{s}$, this problem is only present for small values
of $N_{s}$.

In this context the distribution of the lowest eigenvalue
\begin{equation}
\lambda_{\mathsf{min}}=\min_{\lambda\in\spec H}\left|\lambda\right|
\end{equation}
of the respective kernel operators $H=H_{\mathsf{w}}$ and $H=H_{\mathsf{sw}}$
is of interest. In Fig.~\ref{fig:cdf-lmin}, we show the numerically
determined cumulative distribution functions (CDFs) of $\lambda_{\mathsf{min}}$
for both the Wilson and staggered Wilson kernel at different values
of $\beta$. When increasing $\beta$, we observe how the distribution
moves toward larger values of $\lambda_{\mathsf{min}}$ and pathologically
small values become extremely unlikely. This is important, as rational
approximations to the $\sign$-function become inaccurate for small
arguments. We also note that the corresponding CDFs for the Wilson
and staggered Wilson kernel are very similar.

Finally, in Figs.~\ref{fig:cont-limit-norm} and \ref{fig:cont-limit-gwr}
we show two examples of the $\beta$-dependence of our chiral symmetry
measures $\Delta_{\mathsf{N}}$ and $\Delta_{\mathsf{GW}}$. In the
plot we show the median value and the width of the distribution. Fixing
the value
\begin{equation}
q=\frac{1}{2}\left[1-\erf\left(\frac{1}{\sqrt{2}}\right)\right],
\end{equation}
where $\erf\left(x\right)$ refers to the error function, then we
define the width using the $q$-quantile and the $\left(1-q\right)$-quantile.
With our choice of $q$, we find $\unit[68.3]{\%}$ of the values
within the range. Our figures illustrate the superior chiral properties
of the effective operator with a staggered Wilson kernel at sufficiently
large values of $\beta$. With the use of smearing, the chiral properties
are even further improved.

\section{Quenched quantum chromodynamics \label{sec:DW-Quenched-QCD}}

\begin{figure}[t]
\begin{centering}
\subfloat[Wilson kernel]{\includegraphics[width=0.45\textwidth]{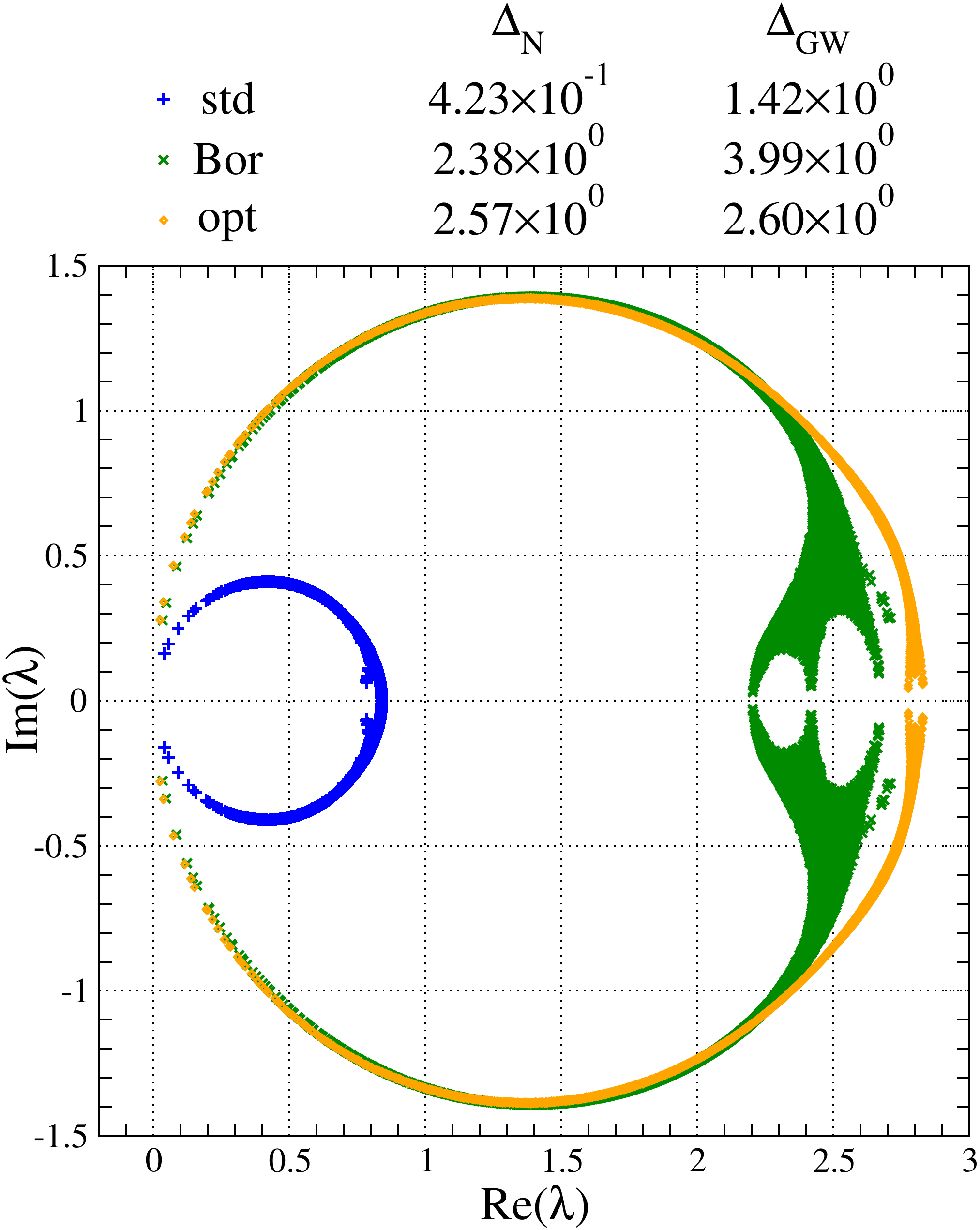}

}\hfill{}\subfloat[Staggered Wilson kernel]{\includegraphics[width=0.45\textwidth]{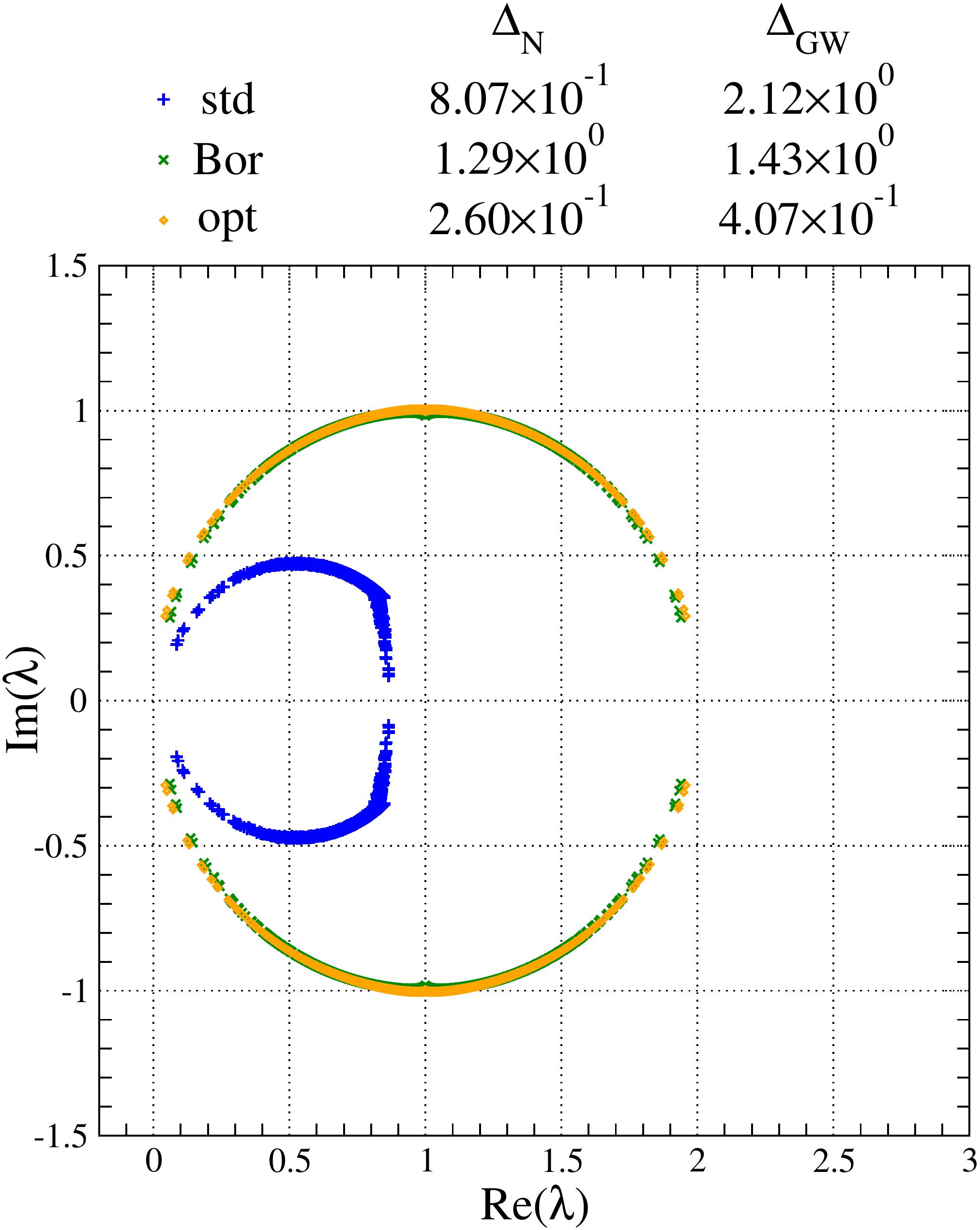}

}
\par\end{centering}
\caption{Spectrum of $\varrho D_{\mathsf{eff}}$ at $N_{s}=4$ on a smeared
$6^{4}$ configuration in QCD\protect\textsubscript{4} at $\beta=6$.
\label{fig:deff-qcd-Ns4}}
\end{figure}
\begin{figure}[t]
\begin{centering}
\subfloat[Wilson kernel]{\includegraphics[width=0.45\textwidth]{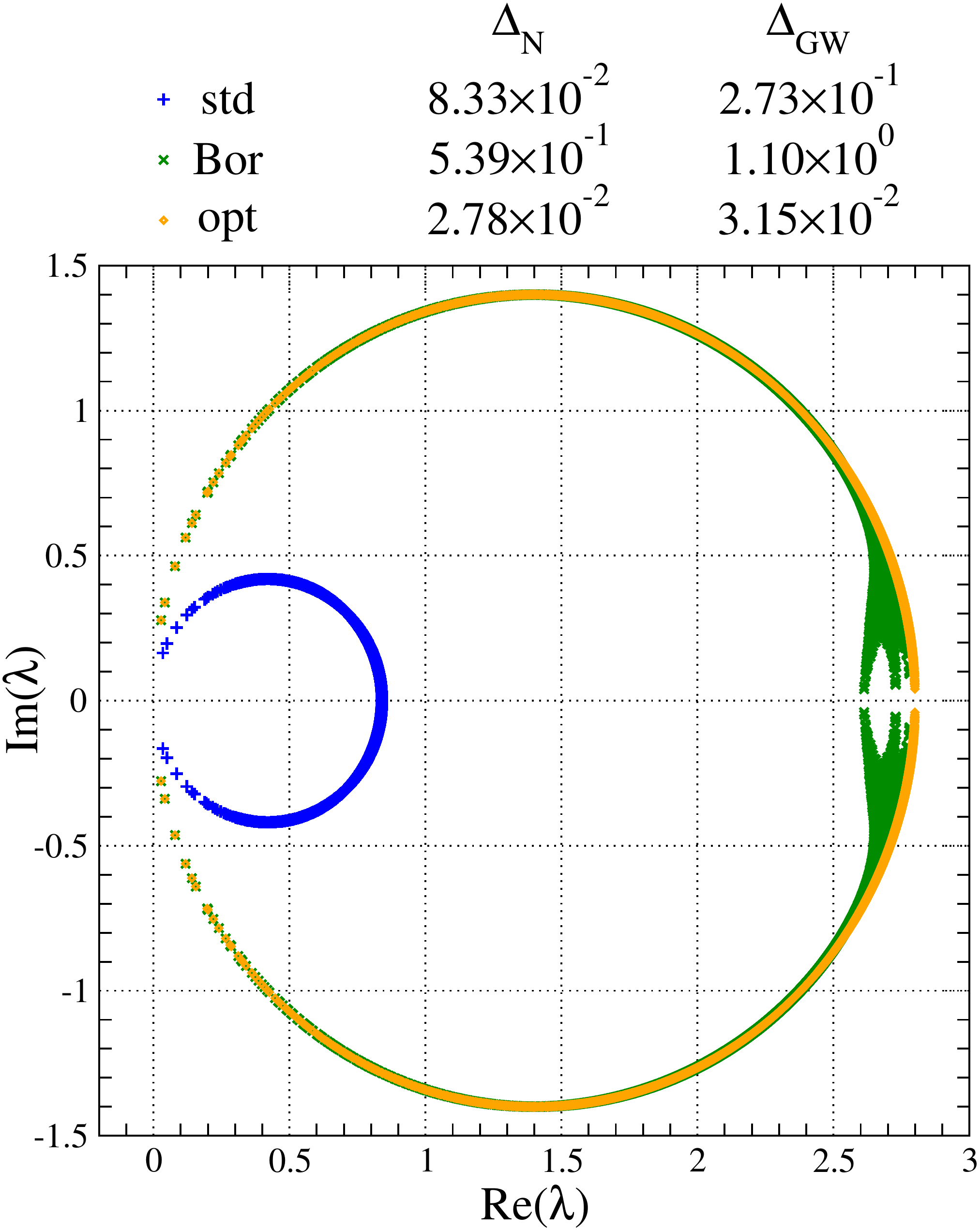}

}\hfill{}\subfloat[Staggered Wilson kernel]{\includegraphics[width=0.45\textwidth]{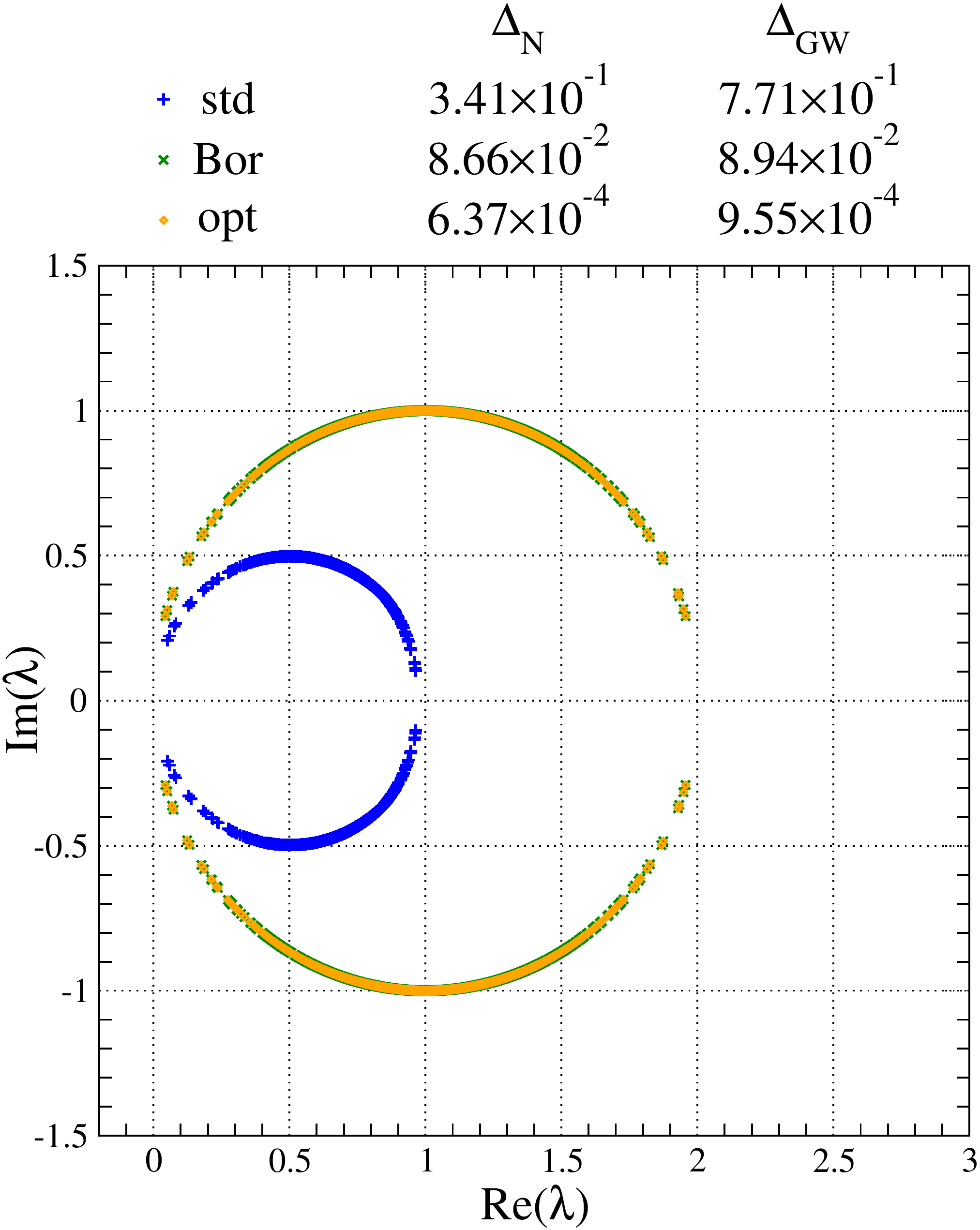}

}
\par\end{centering}
\caption{Spectrum of $\varrho D_{\mathsf{eff}}$ at $N_{s}=8$ on a smeared
$6^{4}$ configuration in QCD\protect\textsubscript{4} at $\beta=6$.
\label{fig:deff-qcd-Ns8}}
\end{figure}

Although our study focuses on the setting of the Schwinger model,
we also investigated the case of four-dimensional quantum chromodynamics.
Our results presented here are exploratory in nature, go beyond our
discussion in Ref.~\citep{Hoelbling:2016qfv} and were first presented
in Ref.~\citep{Hoelbling:2016dug}.

In Figs.~\ref{fig:deff-qcd-Ns4} and \ref{fig:deff-qcd-Ns8} we can
find the eigenvalue spectrum of the effective Dirac operator in the
standard, Boriçi's and the optimal construction at $N_{s}\in\left\{ 4,8\right\} $
on a smeared four-dimensional $6^{4}$ lattice. The $\SUthree$ gauge
configuration under consideration was generated at $\beta=6$ and
then smeared with one step of the \textsc{ape} smearing prescription
using a smearing parameter of $\alpha=0.65$. As it is a topologically
trivial configuration, the effective mass $m_{\mathsf{eff}}$ is omitted
in the figure labels. Taking smearing into account, for the domain
wall height we use $M_{0}=1.4$ in the case of the Wilson kernel and
$M_{0}=1$ for the staggered Wilson kernel.

While in the case of the standard construction with a Wilson kernel
we actually observe smaller chiral symmetry violations compared to
the staggered Wilson kernel, for Boriçi's and the optimal construction
the use of the staggered Wilson kernel results in clearly superior
chiral properties. We often observe a reduction in $\Delta_{\mathsf{N}}$
and $\Delta_{\mathsf{GW}}$ by more than an order of magnitude.

Although without considering more samples it is too early to judge
the potential benefits of using staggered domain wall fermions in
the setting of four-dimensional quantum chromodynamics, our first
results are very encouraging.

\section{Optimal weights \label{sec:DW-Optimal-weights}}

\begin{table}[t]
\begin{centering}
\begin{tabular}{>{\centering}m{0.05\columnwidth}>{\centering}m{0.4\columnwidth}>{\centering}m{0.4\columnwidth}}
\toprule 
$s$  & $\omega_{s}\left(\lambda_{\mathsf{min}}=1,\lambda_{\mathsf{max}}=3\right)$  & $\omega_{s}\left(\lambda_{\mathsf{min}}=1,\lambda_{\mathsf{max}}=\sqrt{2}\right)$\tabularnewline
\midrule
\midrule 
1 & \texttt{0.989011284192743} & \texttt{0.996659816028010}\tabularnewline
\midrule 
2 & \texttt{0.908522120246430} & \texttt{0.971110743060917}\tabularnewline
\midrule 
3 & \texttt{0.779520722603956} & \texttt{0.925723982088869}\tabularnewline
\midrule 
4 & \texttt{0.641124364053574} & \texttt{0.869740043520870}\tabularnewline
\midrule 
5 & \texttt{0.519919928211430} & \texttt{0.813009342796322}\tabularnewline
\midrule 
6 & \texttt{0.427613177773962} & \texttt{0.763841917102527}\tabularnewline
\midrule 
7 & \texttt{0.366896221792507} & \texttt{0.728142270322088}\tabularnewline
\midrule 
8 & \texttt{0.337036936444470} & \texttt{0.709476563432225}\tabularnewline
\bottomrule
\end{tabular}
\par\end{centering}
\caption{Weight factors $\omega_{s}$ for the optimal domain wall fermion construction.
\label{tab:Example-weights}}
\end{table}

Before ending this chapter with our conclusions, we quote some example
weight factors $\omega_{s}$ for optimal domain wall fermions as defined
in Sec.\ \ref{subsec:DW-Optimal-construction}. We consider the free-field
case in two dimensions and set $M_{0}=1$ and $N_{s}=8$. In the case
of the Wilson kernel we then fix $\lambda_{\mathsf{min}}=1$ and $\lambda_{\mathsf{max}}=3$,
while for the staggered Wilson kernel we let $\lambda_{\mathsf{min}}=1$
and $\lambda_{\mathsf{max}}=\sqrt{2}$. The resulting weights $\omega_{s}$
can be found in Table\ \ref{tab:Example-weights}.

\section{Conclusions \label{sec:DW-Conclusions}}

In this chapter, we explicitly constructed various formulations of
staggered domain wall fermions. We investigated their chiral properties
in the free-field case, on quenched thermalized background fields
in the Schwinger model and on smooth topological configurations. With
respect to our study, staggered domain wall fermions work as advertised
and generalize the domain wall fermion construction to staggered kernels.

We also went beyond Adams' original proposal of staggered domain wall
fermions and generalized existing variants to the staggered case,
such as Boriçi's and Chiu's construction. This resulted in the formulation
of truncated staggered domain wall fermions and optimal staggered
domain wall fermions, which were previously not considered in the
literature. Especially these new variants of staggered domain wall
fermions show significantly improved chiral properties compared to
the traditional Wilson-based constructions, at least in the context
of the Schwinger model. In a setting where chiral symmetry plays an
important role, the use of these novel lattice fermion formulations
can be potentially very advantageous.

While our results in the setting of a $\Uone$ gauge theory are very
encouraging, the case of a $\SUthree$ gauge group is not fully understood
yet. As our first results in four-dimensional quantum chromodynamics
are promising, they warrant further investigations.

\chapter{Conclusions \label{chap:Conclusions}}

In this thesis we aimed for a better understanding of the theoretical
foundations, computational properties and the practical applications
of staggered Wilson, staggered domain wall and staggered overlap fermions.

\section{Summary}

To summarize some of the core parts of this thesis we recall that
in Chapter\ \ref{chap:Staggered-Wilson-fermions}, we gave an extensive
review of the symmetries of staggered fermions and the construction
of flavored mass terms. We discussed both Adams' proposal for a two-flavor
and Hoelbling's proposal for a one-flavor staggered Wilson term. In
Sec.\ \ref{sec:SW-Generalized-mass-terms}, we built upon these ideas
and generalized the staggered Wilson term to allow for arbitrary mass
splittings in arbitrary even dimensions.

In Chapter\ \ref{chap:Computational-efficiency}, we estimated the
gains in computational efficiency when using staggered Wilson fermions
instead of usual Wilson fermions for inverting the Dirac matrix on
a source, where we found a speedup factor of $4\text{--}6$ in our
particular benchmark setting. We also analyzed the memory bandwidth
requirements of staggered Wilson fermions, quantified by the arithmetic
intensity, and found that they are comparable to other common staggered
formulations.

In Chapter\ \ref{chap:Pseudoscalar-mesons}, we demonstrated the
feasibility of spectroscopy calculations with staggered Wilson fermions
by studying the case of pseudoscalar mesons. We adapted the operators
used for spectroscopy with usual staggered fermions and did a numerical
study of the pion spectrum when turning the two-flavor staggered Wilson
term on.

In Chapter\ \ref{chap:Overlap-fermions}, we reviewed the overlap
construction and discussed Adams' generalization to the case of a
staggered kernel. In this context we did an analytical verification
of the correctness of the continuum limit of the index and axial anomaly
for the staggered overlap Dirac operator in Sec.\ \ref{sec:OV-Continuum-limit}.

In Chapter\ \ref{chap:Eigenvalue-spectra}, we analyzed the eigenvalue
spectrum of the staggered Wilson Dirac operator and related it to
the computational efficiency of staggered overlap fermions. We proposed
an explanation to resolve the apparent discrepancies in the speedup
factors for the staggered Wilson kernel and for the staggered overlap
formulation.

Finally, in Chapter\ \ref{chap:Staggered-domain-wall} we studied
Adams' proposal of staggered domain wall fermions. In the process
we generalized some variants of domain wall fermions to a staggered
kernel, investigated spectral properties and quantified chiral symmetry
violations of various domain wall formulations. We found that in the
setting of the Schwinger model the use of a staggered Wilson kernel
results in significantly improved chiral properties.

\section{Outlook}

While staggered Wilson fermions and their derived formulations have
both advantages and disadvantages compared to more established lattice
fermion formulations, we believe that their use can be beneficial
in some settings. In four dimensions, the two flavors of Adams' staggered
Wilson fermions can be naturally used as a discretization of the light
up and down quarks. Moreover, compared to Wilson fermions, staggered
Wilson fermions are at no disadvantage with respect to flavor-singlet
physics \citep{Adams:2013tya}. Possible applications include the
high precision determination of the $\eta^{\prime}$ mass and the
calculation of bulk quantities in thermodynamics \citep{Adams:2013tya,AdamsLat14}.
Furthermore, one can test universality, i.e.\ the independence of
physical results of the choice of the lattice action, with the help
of new lattice fermion formulations. 

While the staggered Wilson kernel defines an interesting lattice fermion
discretization in its own right, its derived chiral fermion formulations
bear the potential of reducing the computational costs of simulations
with chiral lattice fermions significantly. While staggered overlap
fermions obey an exact chiral symmetry, staggered domain wall fermions
implement an approximate chiral symmetry in a controlled manner. While
the achievable speedup factor of staggered overlap fermions compared
to Neuberger's overlap fermions has yet to be determined in a realistic
setting (cf.\ Chapter\ \ref{chap:Eigenvalue-spectra}), for staggered
domain wall fermions an improvement of chirality by often more than
an order of magnitude could be already shown in the Schwinger model.

Besides Adams' formulation, one may also ask about the potential of
Hoelbling's single flavor staggered action (see Sec.\ \ref{sec:SW-Hoelblings-mass-term}).
As already pointed out in Ref.~\citep{Hoelbling:2010jw}, it is important
to note that in this case the hypercubic rotational symmetry is broken,
although the double rotation symmetry is preserved. This results in
the need for two gluonic counterterms to restore rotational invariance
in unquenched simulations \citep{SharpeWorkshop} as the terms $G_{12}^{2}+G_{34}^{2}$,
$G_{13}^{2}+G_{24}^{2}$ and $G_{14}^{2}+G_{23}^{2}$ can all appear
with different coefficients, where $G_{\mu\nu}$ denotes the clover
version of the gauge field-strength tensor (see e.g.~Ref.~\citep{Gattringer:2010zz}).

While this fine-tuning is in principle possible, the simultaneous
fine-tuning of coefficients requires significant effort and makes
the use of this action unlikely in the context of practical applications
in the near future. Perturbative methods allow studying this fine-tuning
on an analytical level, although we expect the necessary calculations
to quickly become tedious due to the nontrivial structure of terms
of the form $G_{\alpha\beta}^{2}+G_{\gamma\delta}^{2}$ with $\alpha$,
$\beta$, $\gamma$, $\delta$ pairwise different. While these issues
do not render Hoelbling's formulation invalid, one can get an idea
of the challenges ahead from the many-year efforts needed in studying
similar issues in the context of minimally doubled fermions of the
Karsten-Wilczek kind \citep{Karsten:1981gd,Wilczek:1987kw}, both
on an analytical \citep{Capitani:2010nn} and a numerical level \citep{Weber:2013tfa}.

For the future, we hope that the practical potential of these new
novel fermion formulations is explored further, also in the case of
full dynamical simulations. As the technical properties of staggered
Wilson fermions mimic the ones of Wilson fermions (with the substitution
$D_{\mathsf{w}}\to D_{\mathsf{sw}}$ and $\gamma_{5}\to\epsilon$),
it is expected that the commonly employed algorithms for Wilson fermions
can be adapted to staggered Wilson fermions.

In particular rational and polynomial hybrid Monte Carlo methods \citep{Borici:1995np,Clark:2006fx,Clark:2006wq}
would then allow the generation of two-flavor full QCD configurations
and render any rooting procedure—together with any potential related
problems—unnecessary. However, practical implementations of these
algorithms together with in-depth studies of their theoretical and
computational properties are still outstanding and left for future
work.

\begin{singlespace}
\bibliographystyle{thesis_zielinski}
\phantomsection\addcontentsline{toc}{chapter}{\bibname}\bibliography{thesis_zielinski}
\end{singlespace}

\end{document}